%% file: main.tex
\documentclass[12pt,english,ms]{report}
\usepackage[T1]{fontenc}
\usepackage[latin9]{inputenc}
\usepackage{babel}
\usepackage{longtable}
\usepackage{float}
\usepackage{calc}
\usepackage{amsmath}
\usepackage{amsthm}
\usepackage{setspace}
\usepackage[unicode=true,pdfusetitle,
 bookmarks=true,bookmarksnumbered=false,bookmarksopen=false,
 breaklinks=false,pdfborder={0 0 1},backref=false,colorlinks=false]
 {hyperref}

\makeatletter
\usepackage{graphicx}
\usepackage{UTSAthesis}
\usepackage{times}
\usepackage{latexsym}
\usepackage{tikz}
\usetikzlibrary{quantikz2}
\makeatother

\begin{document}

\supervisor{Artyom M. Grigoryan, Ph.D.}
\committeeB{Ryan B. Casey, Ph.D.}
\committeeC{Claire Walton, Ph.D.}

\informationitems{Master of Science in Electrical Engineering}{M.S.}{B.S.}
{Department of Electrical and Computer Engineering}{Klesse College of Engineering and Integrated Design}{August}{2026}

\thesiscopyright{Copyright 2026 Alexis A. Gomez \\
All rights reserved.}
\newpage
\dedication{\emph{Dedicated to my grandfather, Gustavo Perez, whose perseverance and wisdom shaped who I am today, and to my cat, Fluffy, whose warm company carried me through the long nights.}}

\title{\textbf{NEW METHODS AND FRAMEWORKS FOR QUANTUM STATE PREPARATION IN MODERN QUANTUM SYSTEMS}}
\author{Alexis A. Gomez}

\maketitle

\begin{acknowledgements}
I owe my deepest gratitude to my advisor, Dr. Artyom M. Grigoryan, whose mentorship has shaped my development as a researcher since my junior year of undergraduate studies. He has always been generous with his time, answered every question with patience, and pushed me further than I thought I could go. Beyond research, the wisdom he shared along the way has stayed with me just as much as anything in this thesis. This work would not exist without his guidance.

I am sincerely grateful to Dr. Ryan B. Casey, not only for his role on my committee but for the two years of mentorship in my professional life. He has always made time to talk me through challenges at work, offering perspective that was practical and genuinely helpful. His advice on navigating the workplace and handling the situations that no course prepares you for has been as important as anything I learned in a classroom.

I also thank Dr. Claire Walton, whose graduate course opened an entirely new direction for me. She was the first instructor in my graduate studies to draw my attention beyond signal processing and into optimization, and I am grateful for her patience during countless office hours.

Finally, I would like to thank my family and friends for their unwavering support and understanding throughout this journey. I owe a great deal to Miles Salas, my closest friend across every chapter of my adult life: from being college classmates to coworkers. He introduced me to the gym and has spent more hours than I can count lecturing me on weightlifting and tennis. He has had a constant presence through it all, and at this point he is simply family. I am equally grateful to his family for welcoming me as one of their own.
\end{acknowledgements}

\begin{abstract}
This thesis studies exact, deterministic preparation of arbitrary dense $n$-qubit states, the fundamental data-loading step in quantum signal and image processing. It introduces and derives the Quantum Signal-induced Heap Transform (QsiHT) Fast Path Real Synthesis and the QsiHT Fast Path Complex Synthesis, two syntheses built on the Digital Signal-induced Heap Transform (DsiHT). They are placed within ten benchmark configurations spanning the UCR, isometry, multiplexor, Schmidt/SVD, QSD, and heap-transform construction families. Several configurations are practical realizations through Qiskit builders or compiler optimization, including the deployed \texttt{StatePreparation} by Qiskit, rather than independent from-scratch reimplementations. The two QsiHT syntheses remain the methods evaluated as contributions.

The $n=3$ noisy comparison spans \texttt{ibm\_fez}, \texttt{ibm\_kingston}, and \texttt{ibm\_marrakesh}, three $156$-qubit IBM Heron r2 processors, where single-submission nine-method jobs make within-session family-wide Benjamini--Hochberg-corrected comparisons legitimate: the frontier-versus-QSD separation reproduces within one calibration on every device, while the within-frontier order does not reproduce across devices or calibration days. A deep-circuit $n=8$ run on \texttt{ibm\_fez} shows the executed-count ordering re-emerge outside the run-to-run spread on the complex target, with Plesch--Brukner's lowest infidelity associated with its shallower circuit. Depth was not isolated from mapping, gate composition, calibration, or session effects.

All methods prove exact to machine precision and differ only in cost: under noise the coarse error tier tracks the executed (routed) two-qubit count, separating the $\Theta(2^n)$ frontier from QSD's $\Theta(4^n)$ and nothing finer. Plesch--Brukner attains the best from-scratch complex-data constant through its bipartite split. Both syntheses realize the deployed Qiskit \texttt{StatePreparation} floor of $2^n-n-1$ CNOTs, undercutting every other from-scratch method on the as-built CNOT axis: the Real Synthesis serves real (sign-bearing) targets and holds the lowest classical build cost among the exact loaders at large register sizes (one fast Walsh--Hadamard pass), and the Complex Synthesis serves arbitrary complex targets, tying the deployed \texttt{StatePreparation} from Qiskit for the lowest simulated sampled error. Both are competitive loaders for dense real-valued signal and image data on NISQ devices.
\end{abstract}

\chapter*{List of Abbreviations}
\addcontentsline{toc}{chapter}{List of Abbreviations}
\begin{center}
\begin{tabular}{ll}
CNOT & Controlled-NOT Gate \\
CX & Controlled-$X$ (CNOT) Gate \\
CZ & Controlled-$Z$ Gate \\
D2siHT & Discrete Two Signal-induced Heap Transform \\
DsiHT & Digital Signal-induced Heap Transform \\
ECR & Echoed Cross-Resonance Gate \\
iSWAP & Imaginary SWAP Gate \\
LAPACK & Linear Algebra Package \\
MSE & Mean-Square Error \\
NISQ & Noisy Intermediate-Scale Quantum \\
Q2siHT & Quantum Two Signal-induced Heap Transform \\
QSD & Quantum Shannon Decomposition \\
QsiHT & Quantum Signal-induced Heap Transform \\
SVD & Singular Value Decomposition \\
SWAP & Swap (Qubit-Exchange) Gate \\
UCR & Uniformly Controlled Rotation \\
ZYZ & $Z$--$Y$--$Z$ Euler Decomposition \\
\end{tabular}
\end{center}

\pageone{}

\input{chapters/chapter2.tex}

\input{chapters/chapter3.tex}

\input{chapters/chapter4.tex}

\input{chapters/chapter5.tex}

\input{chapters/chapter6.tex}

\appendix
\input{chapters/chapter1.tex}

\input{chapters/appendix_gates.tex}

\bibliographystyle{unsrt}
\bibliography{references}

\begin{vita}
Alexis A. Gomez earned his Bachelor of Science in Electrical Engineering from The University of Texas at San Antonio in 2024. As an undergraduate, he began research in signal and image processing under the supervision of Dr. Artyom M. Grigoryan. This work grew into quantum computation and imaging, and more specifically, the state-preparation methods developed in this thesis. He currently has 5 peer-reviewed journal articles and 7 SPIE conference papers in quantum image processing, quantum state preparation, and color image enhancement. He will begin doctoral studies in Electrical Engineering at The University of Texas at San Antonio in Fall 2026.
\end{vita}

\end{document}

%% file: chapters/chapter2.tex
\chapter{Background}
\label{ch:background}

This chapter establishes the formal background on which the remainder of the thesis builds. It fixes the notation for quantum states and the amplitude vector that any preparation procedure must reproduce, distinguishes the dense targets of interest from the sparse case, and defines deterministic state preparation precisely (Sections~\ref{sec:states}--\ref{sec:deterministic}). It then derives the parameter-counting lower bounds that make dense preparation exponentially expensive (Section~\ref{sec:lower_bounds}), specifies the resource metrics (gate counts, depth, ancillae, and fidelity) used to compare methods throughout (Sections~\ref{sec:gate_metrics}--\ref{sec:fidelity}), and closes by situating these costs in the Noisy Intermediate-Scale Quantum (NISQ) regime (Section~\ref{sec:nisq_tradeoffs}).

\section{Quantum States and State Vectors}
\label{sec:states}

The state of an isolated quantum system is described by a unit vector in a complex Hilbert space $\mathcal{H}$~\cite{nielsen_chuang}. For a single qubit, $\mathcal{H} = \mathbb{C}^2$, and the orthonormal computational basis $\{\ket{0}, \ket{1}\}$ satisfies $\braket{i}{j} = \delta_{ij}$. A pure single-qubit state is the linear combination
\begin{equation}
\ket{\psi} = \alpha\ket{0} + \beta\ket{1}, \qquad \alpha, \beta \in \mathbb{C},
\label{eq:single_qubit}
\end{equation}
where the complex amplitudes $\alpha$ and $\beta$ obey the normalization condition
\begin{equation}
\braket{\psi}{\psi} = |\alpha|^2 + |\beta|^2 = 1.
\label{eq:single_norm}
\end{equation}
By the Born rule, $|\alpha|^2$ and $|\beta|^2$ are the probabilities of measuring $\ket{0}$ and $\ket{1}$, respectively, in the computational basis~\cite{nielsen_chuang}.

When the system comprises $n$ qubits, the joint state space is the tensor product $\mathcal{H} = (\mathbb{C}^2)^{\otimes n} \cong \mathbb{C}^{2^n}$, spanned by the $2^n$ computational basis states $\ket{x}$ indexed by the binary strings $x = x_1 x_2 \cdots x_n \in \{0,1\}^n$, each identified with the integer it represents. A general pure $n$-qubit state vector is therefore
\begin{equation}
\ket{\Psi} = \sum_{x=0}^{2^n-1} \alpha_x \ket{x}, \qquad \alpha_x \in \mathbb{C}, \qquad \sum_{x=0}^{2^n-1} |\alpha_x|^2 = 1,
\label{eq:multi_qubit}
\end{equation}
which is fully specified by the complex amplitude vector $\boldsymbol{\alpha} = (\alpha_0, \alpha_1, \dots, \alpha_{2^n-1})^T \in \mathbb{C}^{2^n}$. The exponential growth of this amplitude vector with the qubit count $n$ is the root of both the computational power and the preparation difficulty of quantum systems~\cite{nielsen_chuang, biamonte2017quantum}.

Two pure states that differ only by a global phase $e^{i\gamma}$ are physically indistinguishable, since $|\braket{x}{\Psi}|^2$ is invariant under $\ket{\Psi} \mapsto e^{i\gamma}\ket{\Psi}$. Consequently, the $2^{n+1}$ real numbers defining $\boldsymbol{\alpha}$ are constrained by the single normalization equation~\eqref{eq:multi_qubit} and one global-phase freedom, leaving $2^{n+1} - 2$ independent real parameters that any state-preparation procedure must reproduce~\cite{plesch2011}.

\section{Dense Versus Sparse Quantum States}
\label{sec:dense_sparse}

A useful classification of target states for preparation is based on the \emph{sparsity} of the amplitude vector. Let $s$ denote the number of non-zero amplitudes of an $n$-qubit state $\ket{\Psi}$ from~\eqref{eq:multi_qubit}, i.e.\ the cardinality of its support,
\begin{equation}
s = \big|\{\, x \in \{0,1\}^n : \alpha_x \neq 0 \,\}\big|, \qquad 1 \le s \le 2^n.
\label{eq:sparsity}
\end{equation}
A state is said to be \emph{sparse} when only a small fraction of its amplitudes are non-zero, typically $s = O(\operatorname{poly}(n))$, and \emph{dense} when a constant fraction of all $2^n$ amplitudes are populated, $s = \Theta(2^n)$. The canonical maximally dense example is the uniform superposition
\begin{equation}
\ket{u} = \frac{1}{\sqrt{2^n}} \sum_{x=0}^{2^n-1} \ket{x},
\label{eq:uniform}
\end{equation}
for which $s = 2^n$, whereas a single computational basis state $\ket{x_0}$ has $s = 1$.

Sparse states admit efficient preparation: dedicated sparse-state algorithms synthesize a state with $s$ non-zero amplitudes using $O(sn)$ elementary gates, since their circuit cost scales with the support rather than the full dimension~\cite{gleinig2021sparse, malvetti2021sparse}. These sparse constructions are surveyed as related work in Chapter~\ref{ch:litreview} but are not benchmarked here, since this thesis targets the dense case. Dense states are fundamentally harder because the preparation circuit must encode, and therefore individually control, information proportional to all $2^n$ amplitudes. As established above via the $2^{n+1}-2$ independent real parameters of an arbitrary state~\eqref{eq:multi_qubit}, no fixed sub-exponential set of gate angles can reproduce a generic dense amplitude vector, so the gate count must grow as $\Omega(2^n)$~\cite{shende2006, mottonen2004}. This information-theoretic barrier, loading $O(2^n)$ classical numbers into the quantum register, is the central difficulty motivating the comparative study of dense-state-preparation methods in this thesis~\cite{aaronson2015read}. Aaronson's caveat about quantum machine learning proposals cuts both ways here: an efficient loader is a \emph{necessary} condition for any end-to-end quantum advantage on classical data, not a sufficient one, since the surrounding algorithm and its readout must also avoid exponential costs~\cite{aaronson2015read}. This thesis accordingly compares loaders on their own cost and makes no claim that a cheaper loader by itself delivers an application-level speedup.

\section{Deterministic Quantum State Preparation}
\label{sec:deterministic}

Quantum state preparation (QSP) is the task of constructing a quantum circuit that maps the reference ground state $\ket{0}^{\otimes n}$ to a desired target state $\ket{\Psi}$. In the \emph{deterministic} setting considered throughout this thesis, this map is realized by a single unitary operator $U_\Psi$ such that
\begin{equation}
U_\Psi \ket{0}^{\otimes n} = \ket{\Psi} = \sum_{x=0}^{2^n-1} \alpha_x \ket{x}.
\label{eq:qsp_unitary}
\end{equation}
Such a $U_\Psi$ always exists because any unit vector $\ket{\Psi}$ can be taken as the first column of a unitary matrix. The synthesis problem is to decompose $U_\Psi$ into a sequence of hardware-native gates~\cite{shende2006, mottonen2004}. Deterministic preparation is characterized by four properties:

\begin{enumerate}
    \item \textbf{No post-selection.} The output state is obtained on every run of the circuit. This contrasts with probabilistic schemes, where an auxiliary register is measured and the preparation is declared successful only on a particular outcome (e.g.\ measuring an ancilla in $\ket{0}$), succeeding with probability $p_{\text{succ}} < 1$ and otherwise requiring repetition or restart~\cite{nielsen_chuang}.
    \item \textbf{Fixed circuit construction.} The gate sequence and its connectivity are determined entirely by the target amplitudes $\boldsymbol{\alpha}$ ahead of execution. No run-time branching or feedback alters the structure.
    \item \textbf{Exact preparation.} The synthesized state equals the target up to a global phase, $U_\Psi \ket{0}^{\otimes n} = e^{i\gamma}\ket{\Psi}$, in the noiseless limit, as opposed to approximate or variational schemes that minimize a distance $\| | \widetilde{\Psi} \rangle - | \Psi \rangle \|$ to within a tolerance $\varepsilon$~\cite{cerezo2021variational}.
    \item \textbf{Measurement-free or measurement-corrected.} A purely unitary preparation uses no intermediate measurements. Some constructions employ ancilla qubits that are measured and \emph{deterministically corrected} via classically controlled Pauli operations, so that the success probability is restored to unity regardless of the measurement outcome.
\end{enumerate}

These criteria provide the scope of the comparison: the methods studied here all realize the exact unitary~\eqref{eq:qsp_unitary} with a fixed, post-selection-free circuit, which makes their resource costs directly and fairly comparable~\cite{plesch2011, sun2023asymptotically}.

\section{Lower Bounds and Exponential Cost}
\label{sec:lower_bounds}

The exponential difficulty of dense-state preparation can be made precise through parameter-counting and gate-counting arguments. As derived in Section~\ref{sec:states}, an arbitrary $n$-qubit pure state is specified by
\begin{equation}
D = 2^{n+1} - 2
\label{eq:dof}
\end{equation}
independent real parameters after removing the normalization constraint and the global phase~\cite{plesch2011}. Any circuit that prepares an \emph{arbitrary} state must contain at least $D$ tunable real parameters (e.g.\ rotation angles), since fewer parameters describe a manifold of states of dimension less than $D$ and therefore cannot reach every target. This forces the gate count to grow as $\Omega(2^n)$.

A sharper statement bounds the number of two-qubit entangling gates. A counting argument over the parameters supplied per controlled-NOT (CNOT) gate, combined with the dimension~\eqref{eq:dof}, yields a lower bound on the number of CNOT gates required to prepare a generic $n$-qubit state without ancillae~\cite{plesch2011, shende2006}:
\begin{equation}
\#\mathrm{CNOT} \;\ge\; \left\lceil \frac{1}{2}\left(2^n - n - 1\right) \right\rceil \;=\; \Omega(2^n).
\label{eq:cnot_lb}
\end{equation}
Constructive synthesis methods nearly saturate this bound: the uniformly controlled rotation (UCR) scheme of M\"ott\"onen \emph{et al.}~\cite{mottonen2004} and the isometry decomposition of Iten \emph{et al.}~\cite{iten2016quantum} both prepare arbitrary states with $O(2^n)$ CNOTs and circuit depth. For depth the situation is sharper: Sun \emph{et al.}\ prove that depth $\Theta(2^n/n)$ is both achievable and optimal \emph{without any ancilla qubits}, and that only the further reduction to depth $\Theta(n)$ requires spending $O(2^n)$ ancillae~\cite{sun2023asymptotically}. The ancilla-free methods studied in this thesis all realize depth $\Theta(2^n)$, a factor of $n$ above that ancilla-free optimum.

The bound~\eqref{eq:cnot_lb} counts parameters for a generic \emph{complex} target. Because real (sign-bearing) targets are central to this thesis, the same counting deserves to be repeated for the real case. A real $n$-qubit state carries $2^n - 1$ independent real parameters (the unit sphere in $\mathbb{R}^{2^n}$ modulo the global sign). A circuit built from real rotations and CNOTs, the class to which every sign-aware construction in this thesis belongs, supplies one parameter per $R_y$ rotation: at most $n$ rotations act before the first CNOT, and each CNOT admits at most two fresh rotations on its two output wires. Reaching every real target therefore requires $n + 2k \ge 2^n - 1$ for a circuit of $k$ CNOTs, that is,
\begin{equation}
\#\mathrm{CNOT} \;\ge\; \left\lceil \frac{1}{2}\left(2^n - n - 1\right) \right\rceil \qquad \text{(real targets, real-rotation circuits)},
\label{eq:cnot_lb_real}
\end{equation}
numerically the \emph{same} floor as the complex bound~\eqref{eq:cnot_lb}: halving the dimension of the target family and halving the parameters supplied per CNOT cancel exactly, so restricting to real targets does not lower the certified floor ($29$ at $n=6$). Two caveats qualify this statement. First,~\eqref{eq:cnot_lb_real} is proved for real-rotation circuits. If arbitrary complex single-qubit gates are allowed, the same counting certifies only the weaker floor $\lceil(2^n - 2n - 1)/4\rceil$ ($13$ at $n=6$), and no tight lower bound for real targets is known in either gate model. Second, both bounds are parameter counts, not constructions: the gap between these floors and the best known circuits cannot currently be closed from either side.

The practical consequence follows directly. Writing the per-CNOT error as $\epsilon_{\text{CNOT}}$ and the CNOT count as $G_{\text{CNOT}} \sim 2^n$, the accumulated error grows as $G_{\text{CNOT}}\,\epsilon_{\text{CNOT}} \sim 2^n \epsilon_{\text{CNOT}}$. Requiring this to remain below an order-unity error budget, $2^n \epsilon_{\text{CNOT}} \lesssim 1$, bounds the preparable system size by $n \lesssim \log_2(1/\epsilon_{\text{CNOT}})$ for fixed gate fidelity. This exponential scaling in both resource count and error is precisely the wall that approximate, structured, and hardware-specialized preparation strategies attempt to circumvent~\cite{aaronson2015read, preskill2018quantum}.

\section{Gate Metrics}
\label{sec:gate_metrics}

To compare state-preparation methods on a common basis, this thesis reports a fixed set of resource metrics for every benchmarked circuit. Let the synthesized circuit $\mathcal{C}$ consist of a sequence of gates $g_1, g_2, \dots, g_G$ acting on $n$ system qubits plus $a$ ancillae.

\begin{itemize}
    \item \textbf{Total gate count} ($G$): the number of elementary gates in $\mathcal{C}$ after decomposition into the target basis. It is a coarse measure of circuit size that combines both single- and two-qubit operations.
    \item \textbf{CNOT count} ($G_{\text{CNOT}}$): the number of two-qubit entangling CNOT gates. Because entangling gates are one to two orders of magnitude noisier than single-qubit rotations on NISQ hardware~\cite{ibm2022}, $G_{\text{CNOT}}$ is the dominant determinant of fidelity and is the primary cost metric in this work, lower-bounded by~\eqref{eq:cnot_lb}.
    \item \textbf{Circuit depth} ($d$): the length of the longest path through the circuit's dependency graph, equivalently the number of sequential time steps when independent gates are executed in parallel,
    \begin{equation}
    d = \max_{q} \big(\text{number of gates on the critical path ending at qubit } q\big).
    \label{eq:depth}
    \end{equation}
    Depth governs the elapsed execution time and must be kept below the coherence horizon $d \cdot t_g \lesssim T_2$, where $t_g$ is the mean gate duration and $T_2$ is the qubit coherence (dephasing) time beyond which the stored quantum information is lost.
    \item \textbf{Ancilla count} ($a$): the number of auxiliary qubits beyond the $n$ required to hold the target state. Ancillae can reduce depth (with $O(2^n)$ ancillae, to $\Theta(n)$, whereas the ancilla-free optimum is already $\Theta(2^n/n)$~\cite{sun2023asymptotically}) but increase the physical qubit footprint and the susceptibility to crosstalk.
    \item \textbf{Compilation time}: the classical elapsed time required by the synthesis and transpilation pipeline to produce the executable circuit from the target vector $\boldsymbol{\alpha}$. This captures the classical overhead of a method, which can itself scale as $O(2^n)$ or worse.
    \item \textbf{Transpiled depth} ($d_{\text{T}}$): the circuit depth after transpilation onto a specific hardware coupling map and native gate set. Limited qubit connectivity inserts SWAP (qubit-exchange) networks (each SWAP costing three CNOTs, see Appendix~\ref{ch:intro}), so generally $d_{\text{T}} \ge d$, and the gap quantifies a method's sensitivity to hardware topology.
    \item \textbf{Fidelity} ($F$): the quality of the prepared state relative to the ideal target, defined and discussed in Section~\ref{sec:fidelity}.
\end{itemize}

Reporting all of these metrics jointly is necessary because methods trade them against one another: for instance, reducing depth by adding ancillae, or reducing CNOT count at the expense of compilation time~\cite{shende2006, plesch2011}.

\section{Fidelity and Error Metrics}
\label{sec:fidelity}

The fidelity quantifies how closely a prepared state matches its target and is the principal quality metric used throughout this thesis~\cite{nielsen_chuang, jozsa1994fidelity}. For two pure states $\ket{\Psi}$ (ideal target) and $\lvert\widetilde{\Psi}\rangle$ (prepared), the state fidelity is the squared overlap
\begin{equation}
F(\Psi, \widetilde{\Psi}) = \lvert \langle \Psi | \widetilde{\Psi} \rangle \rvert^2,
\label{eq:pure_fidelity}
\end{equation}
which satisfies $0 \le F \le 1$, with $F = 1$ if and only if the states coincide up to a global phase, and $F = 0$ for orthogonal states. The associated \emph{infidelity}
\begin{equation}
\mathcal{I} = 1 - F
\label{eq:infidelity}
\end{equation}
measures the preparation error and is often more informative when $F$ is close to unity.

Because physical NISQ devices produce mixed states, the prepared register is in general described by a density operator $\rho$, and the comparison against the pure target $\ket{\Psi}$ uses the mixed-state (Uhlmann--Jozsa) fidelity~\cite{jozsa1994fidelity}, which for a pure reference reduces to
\begin{equation}
F(\ket{\Psi}, \rho) = \bra{\Psi} \rho \ket{\Psi}.
\label{eq:mixed_fidelity}
\end{equation}
A complementary distance measure is the trace distance $T(\rho, \sigma) = \tfrac{1}{2}\|\rho - \sigma\|_1$ between two density operators $\rho$ and $\sigma$, where $\|\cdot\|_1$ denotes the trace norm (the sum of the singular values of its argument), which bounds the maximum distinguishability of two states and is related to the fidelity for pure states by $T = \sqrt{1 - F}$~\cite{nielsen_chuang}.

For circuit-level error analysis, a convenient first-order model treats each gate as introducing an independent error of probability $\epsilon_i$, so that the overall preparation fidelity decays multiplicatively with the gate count $G$,
\begin{equation}
F \approx \prod_{i=1}^{G} (1 - \epsilon_i) \approx 1 - \sum_{i=1}^{G} \epsilon_i,
\label{eq:fidelity_decay}
\end{equation}
the last approximation holding when $\sum_i \epsilon_i \ll 1$. Equation~\eqref{eq:fidelity_decay} makes explicit why minimizing the CNOT count $G_{\text{CNOT}}$, whose error $\epsilon_{\text{CNOT}}$ dominates the sum, is the key to achieving high-fidelity dense-state preparation on near-term hardware~\cite{ibm2022, preskill2018quantum}.

\section{State Preparation Trade-offs in NISQ}
\label{sec:nisq_tradeoffs}

The metrics and bounds of the preceding sections interact most sharply in the NISQ regime~\cite{preskill2018quantum}, where circuits must run within limited coherence times ($T_2 \sim 100\text{--}300\,\mu s$), under gate error rates of roughly $0.1\%$ for single-qubit and $1\%$ for two-qubit operations (representative literature figures~\cite{ibm2022}), and on hardware with restricted qubit connectivity. Combining the first-order fidelity model~\eqref{eq:fidelity_decay} with the exponential gate scaling~\eqref{eq:cnot_lb} explains quantitatively why loading classical data dominates the cost budget, and demonstrates the trade-off between the three standard encoding strategies.

\paragraph{Amplitude encoding.} Mapping a normalized $N$-dimensional data vector onto the $2^n$ amplitudes of~\eqref{eq:multi_qubit} is maximally qubit-efficient, using only $n=\lceil\log_2 N\rceil$ qubits~\cite{schuld2021machine}. For $N=16$ features this is $n=4$ qubits, but the preparation circuit carries the $\Omega(2^n)$ CNOT cost of~\eqref{eq:cnot_lb}, giving a depth of order $N$ that quickly exceeds the coherence horizon $d\cdot t_g \lesssim T_2$. At the representative literature rate $\epsilon_{\text{CNOT}}=10^{-2}$, the fidelity model~\eqref{eq:fidelity_decay} predicts $F \approx 1 - \epsilon_{\text{CNOT}}\,G_{\text{CNOT}} \approx 1 - 0.01\times 16 = 0.84$ for this load. That rate is an assumption, not a measurement. The hardware benchmark of Chapter~\ref{ch:empirical} measures a smaller slope on an IBM Heron r2 processor, a two-point consistency estimate of $6.8$ to $8.0\times10^{-4}$ per two-qubit gate that agrees with a least-squares fit of $7.0\times10^{-4}$ over the simulated executed-count points, which moves the same $16$-gate load to $F\approx0.99$ and, as a rough extrapolation, pushes the register size at which loading consumes an $O(1)$ fraction of the fidelity budget from about $n\approx6.6$ to $n\approx10.5$. The linear form of~\eqref{eq:fidelity_decay} survives contact with hardware, but its measured coefficient makes the near-term loading penalty percent-level rather than the tens of percent the representative rate suggests. Loading remains the exponentially growing term in the budget either way.

\paragraph{Angle encoding.} Encoding each feature $x_i$ into an independent single-qubit rotation $R_y(x_i)\ket{0}$ collapses the circuit depth to a single layer, since the rotations commute and act in parallel~\cite{schuld2021machine, la2020quantum}. The cost reappears as register width: $N$ features require $N$ qubits, so the $N=16$ case needs $16$ qubits. Angle encoding thus trades the depth wall of amplitude encoding for a width wall set by the size of available hardware.

\paragraph{Variational preparation.} Parameterized circuits $U(\boldsymbol{\theta})\ket{0}^{\otimes n}$ optimized to approximate the target~\cite{cerezo2021variational, kandala2017hardware} can in principle reach a state with a shallow, hardware-native ansatz, but face two obstructions. First, for sufficiently expressive ans\"atze the variance of the gradient of the cost observable's expectation $\langle\hat{O}\rangle$ with respect to a circuit parameter $\theta_k$ vanishes exponentially in the qubit count, $\operatorname{Var}\!\left[\partial_{\theta_k}\langle\hat{O}\rangle\right]\sim 2^{-n}$, the barren-plateau phenomenon that flattens the optimization landscape~\cite{mcclean2018barren}. Second, convergence typically demands $O(10^4)$ classical optimization iterations per state, each requiring many circuit evaluations, and the approximate, non-deterministic result lies outside the exact-preparation scope of this thesis.

These strategies span the trade-off space mapped by the thesis: amplitude encoding minimizes width at the price of exponential depth and CNOT cost~\eqref{eq:cnot_lb}, angle encoding minimizes depth at the price of linear width, and variational methods relax exactness in exchange for shallow circuits, at the cost of trainability~\cite{preskill2018quantum}. Error-mitigation techniques can partially recover fidelity on the resulting noisy circuits~\cite{temme2017error}, but they do not remove the underlying scaling. Therefore, the remainder of this thesis focuses on exact, deterministic dense-state preparation, evaluated against the gate, depth, ancilla, and fidelity metrics established in this chapter.

%% file: chapters/chapter3.tex
\chapter{Literature Review}
\label{ch:litreview}

This chapter surveys the principal families of exact, deterministic methods for preparing an arbitrary dense $n$-qubit state $\ket{\Psi} = \sum_{x=0}^{2^n-1} \alpha_x \ket{x}$ from the reference register $\ket{0}^{\otimes n}$, in the sense formalized in Chapter~\ref{ch:background}. Each method realizes a unitary $U_\Psi$ satisfying $U_\Psi\ket{0}^{\otimes n} = \ket{\Psi}$~\eqref{eq:qsp_unitary}, and is assessed primarily by its two-qubit (CNOT) cost relative to the lower bound of~\eqref{eq:cnot_lb}. The methods divide naturally into four groups: \emph{direct} disentangling constructions (the QsiHT fast path, M\"ott\"onen, and multiplexor-based), \emph{isometry} reductions (Iten, general isometry decomposition), \emph{full-unitary} syntheses specialized to a single column (Quantum Shannon Decomposition), and the structure-exploiting Plesch--Brukner scheme that approaches the optimal constant factor. This fast path of the Quantum Signal-induced Heap Transform (QsiHT), drawn from recent work~\cite{gomez2025qsiht, grigoryan2014heap, grigoryan2025stateprep, grigoryan2025qrheap}, in the two synthesis forms derived in Section~\ref{sec:qsiht}, is the main method implemented and evaluated by this thesis, so it is presented first and in the greatest detail.

\paragraph{Running example.} To make the constructions concrete and directly comparable, every method below is derived and then applied to the \emph{same} dense two-qubit target
\begin{equation}
\ket{\Psi} = \frac{1}{\sqrt{30}}\big(\ket{00} + 2\ket{01} + 3\ket{10} + 4\ket{11}\big),
\label{eq:running}
\end{equation}
whose amplitudes, arranged as a $2\times2$ matrix with row index $q_0$ (top, most-significant qubit) and column index $q_1$, form
\begin{equation}
M = \begin{pmatrix} \alpha_{00} & \alpha_{01} \\ \alpha_{10} & \alpha_{11}\end{pmatrix} = \frac{1}{\sqrt{30}}\begin{pmatrix}1 & 2 \\ 3 & 4\end{pmatrix}.
\label{eq:running_matrix}
\end{equation}
All amplitudes of~\eqref{eq:running} are real and positive, so its phase stage is the identity and only the amplitude (magnitude) cost is incurred. This target isolates each method's \emph{magnitude} logic. To exercise the complementary \emph{phase} logic, a second running example carries the \emph{same magnitudes} but non-trivial phases,
\begin{equation}
\ket{\Phi} = \frac{1}{\sqrt{30}}\big(\ket{00} + 2i\,\ket{01} - 3\,\ket{10} + 4\,\ket{11}\big),
\label{eq:running_complex}
\end{equation}
that is, $\alpha_x = r_x\,e^{i\omega_x}$ with the magnitudes $r_x$ of~\eqref{eq:running_matrix} and the phase vector $\boldsymbol{\omega} = (\omega_{00},\omega_{01},\omega_{10},\omega_{11}) = \big(0,\tfrac{\pi}{2},\pi,0\big)$. Because the magnitudes coincide with~\eqref{eq:running}, the amplitude derivation is reused verbatim, so~\eqref{eq:running_complex} adds only the phase stage and isolates the cost of complex amplitudes. It is worked through for M\"ott\"onen in Section~\ref{sec:mottonen_complex} and for the QsiHT Complex Synthesis in Section~\ref{sec:qsiht_complex}. As a reference point, the lower bound~\eqref{eq:cnot_lb} for $n=2$ gives $\lceil\tfrac12(2^2-2-1)\rceil = 1$ CNOT for either target, so an optimal preparation uses a single entangling gate.

\section{Uniformly Controlled Rotations}
\label{sec:ucr}

Several methods in this chapter are built from a single entangling primitive, the \emph{uniformly controlled rotation} (UCR), also called a multiplexed rotation. A UCR with $m$ control qubits and a single target applies a \emph{different} axis rotation $R_a(\theta_j)$, $a\in\{y,z\}$, to the target for each of the $2^m$ computational-basis configurations $\ket{j}$ of the controls:
\begin{equation}
F_m^{a}(\theta_0,\dots,\theta_{2^m-1}) = \sum_{j=0}^{2^m-1} \ket{j}\!\bra{j} \otimes R_a(\theta_j).
\label{eq:ucr}
\end{equation}
A central result of~\cite{bergholm2005, mottonen2004} is that this operator decomposes into exactly $2^m$ single-qubit rotations interleaved with $2^m$ CNOTs, with no ancillae, by placing the CNOTs according to a Gray code. The rotation angles $\tilde{\theta}_k$ physically applied are obtained from the desired angles $\theta_j$ through a linear transform built from the Walsh--Hadamard matrix and the Gray-code permutation $G$,
\begin{equation}
\tilde{\boldsymbol{\theta}} = M\,\boldsymbol{\theta}, \qquad M_{kj} = 2^{-m}\,(-1)^{\,b_k \cdot g_j},
\label{eq:walsh}
\end{equation}
where $M$ is the $2^m\times 2^m$ angle-transform matrix (the Walsh--Hadamard matrix with its columns permuted by $G$, not the amplitude matrix of~\eqref{eq:running_matrix}), the row and column indices run over $k,j\in\{0,1,\ldots,2^m-1\}$, $b_k$ is the binary representation of $k$ and $g_j$ that of the $j$-th Gray-code integer, and $b_k\cdot g_j$ is their bitwise dot product taken modulo $2$. This precomputation is purely classical and runs in $O(m\,2^m)$ time. Every method below that disentangles one qubit at a time (the QsiHT fast path, the M\"ott\"onen construction, the isometry methods, and the multiplexor route) ultimately compiles to a cascade of these UCRs, differing only in how the angles $\theta_j$ are computed and in whether a separate phase cascade is required.

\section{Quantum Signal-Induced Heap Transform using the Fast Path}
\label{sec:qsiht}

The Quantum Signal-induced Heap Transform using the fast path, the main method of this thesis, prepares a state by inverting its classical analogue, the Digital Signal-induced Heap Transform (DsiHT)~\cite{grigoryan2014heap, gomez2025qsiht, grigoryan2025qrheap}, a structured sequence of Givens rotations that ``heaps'' the energy of a real amplitude vector into a single component. Where the other direct methods organize their rotations around a binary tree of subtree norms, the QsiHT organizes them around a \emph{generator} and a chosen \emph{path} through the data. The resulting recursive circuit is the \emph{fast path}, and, as shown in Sections~\ref{sec:qsiht_s2s}--\ref{sec:q2siht}, the same construction extends to transforming one arbitrary state into another and to preparing two states at once. This thesis does not introduce the transform, which is Grigoryan's~\cite{grigoryan2014heap}, and it does not claim the state-preparation circuit either: realizing the fast path as a quantum gate circuit is prior published work by Grigoryan, Gomez, and Agaian~\cite{gomez2025qsiht, grigoryan2025stateprep}. The contribution of this thesis is the derivation route and the evaluation. It re-derives the fast path as a cascade of uniformly controlled rotations compiled through the Gray-code decomposition of Section~\ref{sec:ucr}, which places the heap-transform construction on the same primitive as the standard baselines, and it benchmarks all of them uniformly. This re-derivation is a compilation and normalization step: expressing the published fast path on the same Gray-code UCR primitive as the baselines is what lets the benchmark compare like with like. On top of that shared primitive the chapter derives two state-preparation reductions of the cascade, the QsiHT Fast Path Real Synthesis (Section~\ref{sec:qsiht_real}) and the QsiHT Fast Path Complex Synthesis (Section~\ref{sec:qsiht_complex}), which defer one gate per stage so the entangling cost drops from the plain Gray-code count $2^n-2$ to the isometry floor $2^n-n-1$. These two syntheses are the forms this thesis prepares and benchmarks, and after their first use they are called the Real Synthesis and the Complex Synthesis.

\subsection{The Digital Signal-Induced Heap Transform}

The DsiHT is a unitary transform generated by a fixed real vector, the \emph{generator} $\boldsymbol{x} = (x_0, x_1, \ldots, x_{N-1})$ of length $N = 2^n$, together with a set of angular equations~\cite{grigoryan2014heap}. It is the product of $N-1$ elementary Givens rotations $T_{\vartheta_k}$ that successively combine pairs of components and, applied to the generator itself, collapse all of its energy into the first component,
\begin{equation}
H\boldsymbol{x} = \big(\lVert\boldsymbol{x}\rVert,\, 0,\, \ldots,\, 0\big)^{\!\top},
\label{eq:dsiht_heap}
\end{equation}
the single ``heap'' into which the generator's energy $E[\boldsymbol{x}] = \lVert\boldsymbol{x}\rVert$ is gathered. Each rotation acts on a component pair $(x_i, x_j)$ and zeroes the second through the angular equation $x_i\sin\vartheta + x_j\cos\vartheta = 0$, that is,
\begin{equation}
T_{\vartheta} = \begin{pmatrix}\cos\vartheta & -\sin\vartheta\\ \sin\vartheta & \cos\vartheta\end{pmatrix},
\qquad \vartheta = -\arctan\!\big(x_j / x_i\big).
\label{eq:givens_angle}
\end{equation}
Because the transform is fixed by its generator, a unit-norm generator is equivalent to its \emph{angular representation}, the list of rotation angles (the norm being $1$),
\begin{equation}
\boldsymbol{x} \;\longrightarrow\; A_{\boldsymbol{x}} = \{\vartheta_1, \vartheta_2, \ldots, \vartheta_{N-1}\}.
\label{eq:angular_rep}
\end{equation}
The construction is two-level: the angles $A_{\boldsymbol{x}}$ are first computed from the generator, and the \emph{same} angles, applied in the \emph{same} order, then define the transform $H$ acting on any input signal $\boldsymbol{z}$. Writing $R_{\vartheta;(i,j)}$ for the rotation $T_{\vartheta}$ embedded in components $i$ and $j$ of the $N$-dimensional space (the identity elsewhere), the DsiHT is the ordered product
\begin{equation}
H = R_{\vartheta_{N-1};(\,\cdot\,)}\,\cdots\,R_{\vartheta_2;(\,\cdot\,)}\,R_{\vartheta_1;(\,\cdot\,)},
\label{eq:dsiht_matrix}
\end{equation}
whose factor order and the pair $(i,j)$ each factor acts on are set by the \emph{path} of Section~\ref{sec:qsiht_path}. Taking the generator to be the target amplitude vector $\boldsymbol{\alpha}$ and inverting~\eqref{eq:dsiht_matrix} prepares $\boldsymbol{\alpha}$ from the ground state, since $H\boldsymbol{\alpha} = \ket{0}^{\otimes n}$ gives $H^{\dagger}\ket{0}^{\otimes n} = \boldsymbol{\alpha}$.

\subsection{Bit Planes and Angle Computation}
\label{sec:qsiht_planes}

In the quantum setting the $N = 2^n$ components are the computational basis states, indexed by their $n$-bit labels, and a rotation $R_{\vartheta;(i,j)}$ couples the two basis states, or \emph{bit planes}, $i$ and $j$. Two bit planes are \emph{adjacent} (nearest-neighbor) when their labels differ in exactly one bit. For instance, $0 = (00)$ and $1 = (01)$ are adjacent, whereas $1 = (01)$ and $2 = (10)$ are not. A rotation on adjacent planes is realizable as a single controlled rotation, one configuration of the uniformly controlled rotation~\eqref{eq:ucr}. A rotation on non-adjacent planes is not, and must first be routed onto adjacent planes by a permutation, as Section~\ref{sec:qsiht_perm} details.

The \emph{fast path} processes the amplitudes as a butterfly: at stage $s\in\{0,1,\ldots,n{-}1\}$, only pairs differing in bit $s$ are combined, and there are $2^{n-1-s}$ such independent pairs, one per configuration of bits $s{+}1,\ldots,n{-}1$. Every such pair lies on adjacent bit planes by construction. Each pair $(r_m, r_{m+2^s})$ is zeroed in its second element by a rotation with angle
\begin{equation}
u_{s,m} = -\arctan2\!\left(r_{m+2^s},\, r_m\right),
\label{eq:dsiht_angle}
\end{equation}
where $\arctan2(y,x)$ returns the signed angle in $(-\pi,\pi]$ and handles zero and negative inputs robustly. It is the numerically stable form of the angular equation~\eqref{eq:givens_angle}. The total number of angles across all stages is $\sum_{s=0}^{n-1}2^{n-1-s} = 2^n - 1$, matching the size of the angular representation~\eqref{eq:angular_rep}.

\subsection{Paths, Roadmaps, and the Fast Path}
\label{sec:qsiht_path}

\begin{figure}[H]
\centering
\begin{tikzpicture}[
  node distance=8mm,
  rot/.style={circle, draw, minimum size=7mm, inner sep=0pt, font=\small},
  inp/.style={font=\small},
  heap/.style={circle, fill=red, inner sep=1.6pt}
]
\node[inp] (a0) at (0,3) {$x_0$};
\node[inp] (a1) at (0,2) {$x_1$};
\node[inp] (a2) at (0,1) {$x_2$};
\node[inp] (a3) at (0,0) {$x_3$};
\node[rot] (A) at (1.5,3) {$1$};
\node[rot] (B) at (3.0,3) {$2$};
\node[rot] (C) at (4.5,3) {$3$};
\node[heap] (ah) at (5.5,3) {};
\draw[->] (a0) -- (A);
\draw[->] (a1) -- (A);
\draw[->] (A) -- (B);
\draw[->] (a2) -- (B);
\draw[->] (B) -- (C);
\draw[->] (a3) -- (C);
\draw[->] (C) -- (ah);
\node[font=\scriptsize] at (1.5,3.55) {$(0,1)$};
\node[font=\scriptsize] at (3.0,3.55) {$(0,2)$};
\node[font=\scriptsize, red] at (4.5,3.55) {$(0,3)$};
\node[font=\small] at (2.75,-0.9) {(a) weak (regular) path};
\begin{scope}[xshift=8cm]
\node[inp] (b0) at (0,3) {$x_0$};
\node[inp] (b1) at (0,2) {$x_1$};
\node[inp] (b2) at (0,1) {$x_2$};
\node[inp] (b3) at (0,0) {$x_3$};
\node[rot] (P) at (1.7,2.5) {$1$};
\node[rot] (Q) at (1.7,0.5) {$2$};
\node[rot] (Rn) at (3.4,1.8) {$3$};
\node[heap] (bh) at (4.4,1.8) {};
\draw[->] (b0) -- (P);
\draw[->] (b1) -- (P);
\draw[->] (b2) -- (Q);
\draw[->] (b3) -- (Q);
\draw[->] (P) -- (Rn);
\draw[->] (Q) -- (Rn);
\draw[->] (Rn) -- (bh);
\node[font=\scriptsize] at (1.7,3.1) {$(0,1)$};
\node[font=\scriptsize] at (1.7,1.1) {$(2,3)$};
\node[font=\scriptsize] at (3.4,2.4) {$(0,2)$};
\node[font=\small] at (2.2,-0.9) {(b) fast path};
\end{scope}
\end{tikzpicture}
\caption{Roadmaps for the $N=4$ ($n=2$) DsiHT. Each circle is a Givens rotation. Arrows show which components merge, and the label $(i,j)$ gives the bit planes the rotation couples. (a) The weak (regular) path is a sequential chain whose final merge couples the \emph{non-adjacent} planes $0=(00)$ and $3=(11)$ (red), which is not a single controlled rotation. (b) The fast path partitions the components into pairs: stage~1 merges $(x_0,x_1)$ and $(x_2,x_3)$, stage~2 merges the survivors, and every merge is on adjacent planes, so no permutation is needed. The red dot marks the heap $\lVert\boldsymbol{x}\rVert$.}
\label{fig:qsiht_roadmap}
\end{figure}
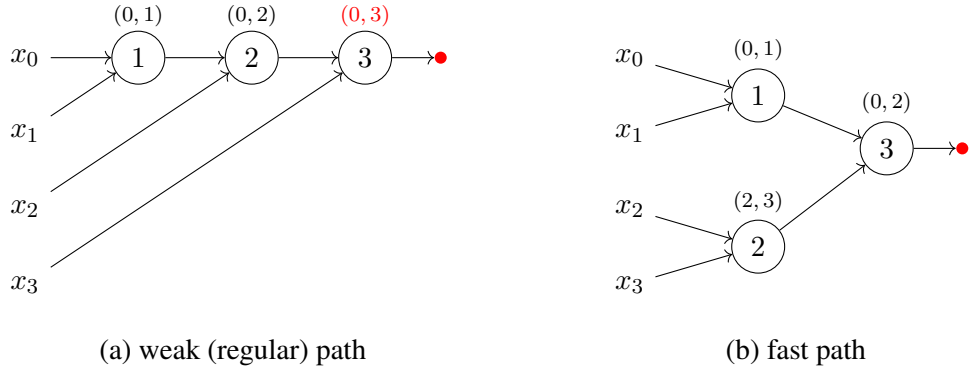

The defining freedom of the DsiHT is its \emph{path}: the order in which component pairs are merged as energy is heaped into the first component. Different paths give different (still unitary) transforms with the same generator, and a path is read most easily off a \emph{roadmap} that traces which pairs merge at each stage~\cite{gomez2025qsiht}. Two paths bracket the range. The \emph{weak two-wheel carriage} (the regular path) is a single sequential chain: combine $x_0$ with $x_1$, then the updated $x_0$ with $x_2$, then with $x_3$, and so on, gathering the heap one component at a time. The \emph{strong two-wheel carriage} instead works inward from the extreme components. Neither is, in general, permutation-free: the weak path's later merges land on non-adjacent bit planes, and for $N=4$ its final merge couples planes $0=(00)$ and $3=(11)$, which differ in two bits (Figure~\ref{fig:qsiht_roadmap}(a)).

The \emph{fast path} is the balanced, pair-partitioning roadmap that keeps every merge on adjacent bit planes. For $N=4$ it is the two-stage butterfly of Figure~\ref{fig:qsiht_roadmap}(b): stage~1 merges $(x_0,x_1)$ on planes $(0,1)$ and $(x_2,x_3)$ on planes $(2,3)$, and stage~2 merges the two survivors on planes $(0,2)$. All three pairs are adjacent, so the transform factors as
\begin{equation}
H_{4}^{\text{fast}} = R_{\vartheta_3;(0,2)}\,R_{\vartheta_2;(2,3)}\,R_{\vartheta_1;(0,1)},
\label{eq:fast_h4}
\end{equation}
three rotations on adjacent planes and no permutations. The fast path for larger $N$ is the recursion of this butterfly and coincides with the stage structure of Section~\ref{sec:qsiht_planes}. For each register size there are several fast paths, leaving freedom to match the hardware topology~\cite{gomez2025qsiht}.

\subsection{Permutation Maps and Permutation-Free Circuits}
\label{sec:qsiht_perm}

When a path places a rotation on non-adjacent bit planes $i$ and $j$ (labels differing in more than one bit), that rotation is not a single controlled gate. It is realized by first \emph{permuting} the planes so the rotation acts on adjacent labels, applying the adjacent-plane rotation, and permuting back. Writing $P$ for the permutation that carries the target pair onto an adjacent pair $(i',j')$, a non-adjacent rotation factors as
\begin{equation}
R_{\vartheta;(i,j)} = P^{\dagger}\,R_{\vartheta;(i',j')}\,P,
\label{eq:perm_decomp}
\end{equation}
and each permutation $P$ is itself built from controlled-NOT ($X$) gates. The weak and strong carriages pay for a number of these permutation gates. The strong three-qubit DsiHT, for example, needs eight controlled-NOTs on top of its seven rotations.

\begin{figure}[H]
\centering
\begin{quantikz}
\lstick{$q_0:\ket{0}$} & \gate{R_y(-\vartheta_3)} & \ctrl{1}                 & \octrl{1}                & \qw \\
\lstick{$q_1:\ket{0}$} & \octrl{-1}               & \gate{R_y(-\vartheta_2)} & \gate{R_y(-\vartheta_1)} & \qw
\end{quantikz}
\caption{Permutation-free inverse fast-path QsiHT for $n=2$, preparing $\ket{x}$ from $\ket{00}$. It is the three adjacent-plane rotations of~\eqref{eq:fast_h4} in reverse with negated angles. Filled ($\bullet$) and open ($\circ$) dots are controls on $\ket{1}$ and $\ket{0}$. The coupled bit planes are $(0,2)$, $(2,3)$, $(0,1)$. No CNOT appears: the circuit is controlled rotations only, until each gate is compiled to the hardware basis (Section~\ref{sec:qsiht_cost}).}
\label{fig:qsiht_permfree}
\end{figure}
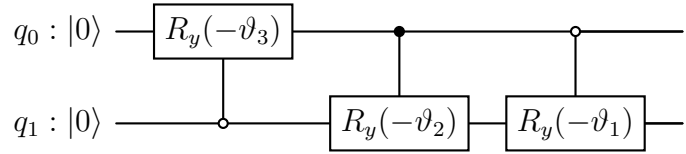

The fast path avoids permutations entirely. Because every merge is on adjacent bit planes, the abstract fast-path circuit consists \emph{only} of controlled rotation gates, $2^n-1$ of them, and contains no CNOTs at all before transpilation. This is the precise sense in which the QsiHT fast path is \emph{permutation-free}~\cite{gomez2025qsiht}. Figure~\ref{fig:qsiht_permfree} shows the two-qubit case: the inverse fast-path transform that prepares $\ket{x}$ from $\ket{00}$ is the three adjacent-plane rotations of~\eqref{eq:fast_h4} applied in reverse with negated angles, and not a single entangling gate appears. The CNOTs counted in the benchmarks arise only afterward, when each controlled rotation is compiled to a hardware-native $\{R_y,\,\text{CX}\}$ basis, where CX denotes the controlled-NOT (Section~\ref{sec:qsiht_cost}).

\subsection{Circuit Structure and Cost}
\label{sec:qsiht_cost}

The fast-path preparation circuit is best read at two levels. At the abstract level it is the inverse of~\eqref{eq:dsiht_matrix}: a cascade of
\begin{equation}
\mu(n) = 2^n - 1
\label{eq:qsiht_rotcount}
\end{equation}
controlled rotation gates on adjacent bit planes, permutation-free and CNOT-free. Grouping the rotations by their target qubit recovers the uniformly controlled form: stage $s$ rotates qubit $s$ for every configuration of the higher qubits $s{+}1,\ldots,n{-}1$, that is, a uniformly controlled $R_y$~\eqref{eq:ucr} with leaf angles $\{-2u_{s,m}\}$.

The factor of two in these leaf angles is a change of convention between the classical Givens rotation and the quantum gate that realizes it. Writing the hardware rotation as
\begin{equation}
R_y(\theta) = e^{-i\theta Y/2} = \begin{pmatrix}\cos\tfrac{\theta}{2} & -\sin\tfrac{\theta}{2}\\[2pt] \sin\tfrac{\theta}{2} & \cos\tfrac{\theta}{2}\end{pmatrix},
\label{eq:ry_gate}
\end{equation}
the gate is generated by $Y/2$, so it turns the plane it acts on through only half of its parameter $\theta$. The DsiHT angle $u_{s,m}$~\eqref{eq:dsiht_angle} is by contrast a full planar rotation angle, identical to the Givens angle $\vartheta$ of~\eqref{eq:givens_angle}, so that $T_{u_{s,m}} = R_y(2u_{s,m})$. Realizing the Givens rotation as a gate therefore doubles the angle. The negative sign is the inversion of~\eqref{eq:dsiht_matrix} that turns the disentangling transform into a preparation circuit, since $T_{u}^{-1} = T_{-u} = R_y(-2u)$. The two together give the leaf angles $-2u_{s,m}$, and the same doubling produces the preparation angles $\gamma_0,\gamma_1^{(m)}$ of the worked example in Section~\ref{sec:qsiht_example}.

At the transpiled level, each uniformly controlled $R_y$ is compiled to the native $\{R_y,\,\text{CX}\}$ basis by the Gray-code rule, so a UCR with $k$ control qubits contributes $2^k$ CNOTs. The cascade carries one UCR for each control count $k = 0, 1, \ldots, n{-}1$ (stage $s$ supplies $k = n{-}1{-}s$ controls), so it totals
\begin{equation}
G_{\text{CNOT}}^{\text{QsiHT}} = \sum_{k=1}^{n-1} 2^k = 2^n - 2,
\label{eq:qsiht_complexity_lit}
\end{equation}
the $k=0$ (bare) top-qubit rotation needing no CNOT. This $2^n-2$ is the plain, un-deferred Gray-code cost, and it is the starting point that the two syntheses of Sections~\ref{sec:qsiht_real} and~\ref{sec:qsiht_complex} improve on. For a real target, including sign-bearing states, the $\arctan2$ angles encode amplitude signs directly into the $R_y$ rotations, so no separate $R_z$ phase cascade is needed and $2^n-2$ is the complete un-deferred cost. A complex target instead appends a diagonal phase gate,
\begin{equation}
\Delta = \operatorname{diag}\!\big(e^{i\omega_0}, e^{i\omega_1}, \ldots, e^{i\omega_{2^n-1}}\big),
\label{eq:qsiht_diag}
\end{equation}
which adds $2^n-2$ further CNOTs via a uniformly controlled $R_z$ cascade, for the doubled un-deferred complex cost $2^{n+1}-4$, in the same $\Theta(2^n)$ band as M\"ott\"onen. The Complex Synthesis of Section~\ref{sec:qsiht_complex} removes this separate diagonal.

\subsection{Two-Qubit Derivation and Example}
\label{sec:qsiht_example}

Apply the DsiHT to $\boldsymbol{r} = \tfrac{1}{\sqrt{30}}(1,2,3,4)^\top$.

\paragraph{Stage $s{=}0$.} Two bit-0 pairs are processed:
\begin{equation}
u_{0,0} = -\arctan2(2,1) = -\arctan 2, \qquad
u_{0,1} = -\arctan2(4,3) = -\arctan\tfrac{4}{3}.
\label{eq:qsiht_s0}
\end{equation}
The working vector becomes $\tfrac{1}{\sqrt{30}}(\sqrt{5},0,5,0)^\top$.

\paragraph{Stage $s{=}1$.} One bit-1 pair remains:
\begin{equation}
u_{1,0} = -\arctan2\!\left(\tfrac{5}{\sqrt{30}},\tfrac{\sqrt{5}}{\sqrt{30}}\right) = -\arctan\sqrt{5}.
\label{eq:qsiht_s1}
\end{equation}
The vector collapses to $(1,0,0,0)^\top$, confirming the DsiHT maps $\boldsymbol{r}$ to its norm.

The preparation angles are the negatives applied in reverse stage order: the top qubit receives a bare $R_y$ with angle $\gamma_0 = 2\arctan\sqrt{5}\approx 2.3005$, and the one-control UCR on $q_1$ has leaf angles $\gamma_1^{(0)} = 2\arctan 2\approx 2.2143$ and $\gamma_1^{(1)} = 2\arctan\tfrac{4}{3}\approx 1.8546$, the rotations applied to $q_1$ when $q_0=0$ and $q_0=1$. Because all amplitudes of~\eqref{eq:running} are positive, $\arctan2$ coincides with the ordinary $\arctan$ here, and the same block-splitting angles reappear, by a different (binary-tree) route, in the M\"ott\"onen derivation of Section~\ref{sec:mottonen_example}.

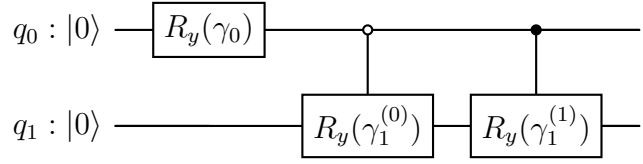
\begin{figure}[H]
\centering
\begin{quantikz}
\lstick{$q_0:\ket{0}$} & \gate{R_y(\gamma_0)} & \octrl{1}                  & \ctrl{1}                   & \qw \\
\lstick{$q_1:\ket{0}$} & \qw                  & \gate{R_y(\gamma_1^{(0)})} & \gate{R_y(\gamma_1^{(1)})} & \qw
\end{quantikz}
\caption{The one-control UCR of the running example drawn straight from its leaf angles, before Gray-code compilation. The bare $R_y(\gamma_0)$ loads the block norms on $q_0$, and the multiplexed rotation on $q_1$ is the open-controlled $R_y(\gamma_1^{(0)})$ (active when $q_0=0$) followed by the closed-controlled $R_y(\gamma_1^{(1)})$ (active when $q_0=1$). Each branch applies a different rotation, so the circuit is not yet in the native $\{R_y,\,\mathrm{CX}\}$ basis. The Walsh--Hadamard transform~\eqref{eq:wht1} turns it into the fixed-angle form of Figure~\ref{fig:qsiht_circ}.}
\label{fig:qsiht_uncompiled}
\end{figure}

Used straight from these leaf angles, the UCR is the pair of controlled rotations in Figure~\ref{fig:qsiht_uncompiled}: an open-controlled $R_y(\gamma_1^{(0)})$ that fires when $q_0=0$ and a closed-controlled $R_y(\gamma_1^{(1)})$ that fires when $q_0=1$. This form is not yet hardware-native, since it asks for a \emph{different} rotation on $q_1$ for each value of the control, whereas a physical circuit can only interleave \emph{fixed} single-qubit rotations with entangling gates.

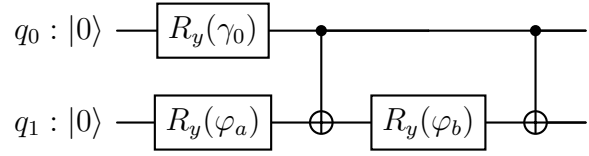
\begin{figure}[H]
\centering
\begin{quantikz}
\lstick{$q_0:\ket{0}$} & \gate{R_y(\gamma_0)}   & \ctrl{1} & \qw                   & \ctrl{1} & \qw \\
\lstick{$q_1:\ket{0}$} & \gate{R_y(\varphi_a)} & \targ{}  & \gate{R_y(\varphi_b)} & \targ{}  & \qw
\end{quantikz}
\caption{QsiHT fast-path preparation of the real target~\eqref{eq:running}, the Gray-code compilation of Figure~\ref{fig:qsiht_uncompiled}. The bare $R_y(\gamma_0)$ on $q_0$ loads the block norms. The one-control UCR on $q_1$ becomes the fixed rotations $R_y(\varphi_a)$, CNOT, $R_y(\varphi_b)$, CNOT with the applied angles of~\eqref{eq:qsiht_applied}. Two CNOTs, as predicted by~\eqref{eq:qsiht_complexity_lit} at $n=2$. This bare-rotation-then-UCR template recurs throughout the direct methods.}
\label{fig:qsiht_circ}
\end{figure}

The Gray-code decomposition removes the obstruction by writing the $m$-control UCR as $2^m$ fixed target rotations separated by $2^m$ CNOTs, and the order-$m$ Walsh--Hadamard transform~\eqref{eq:walsh} supplies those fixed angles from the leaf angles. For the single control here it is the $2\times2$ Hadamard,
\begin{equation}
M = \tfrac12\begin{pmatrix}1 & 1\\ 1 & -1\end{pmatrix},\qquad
\begin{pmatrix}\varphi_a\\[2pt] \varphi_b\end{pmatrix} = M\begin{pmatrix}\gamma_1^{(0)}\\[2pt] \gamma_1^{(1)}\end{pmatrix},
\label{eq:wht1}
\end{equation}
which gives the physically applied angles
\begin{equation}
\varphi_a = \tfrac{1}{2}\big(\gamma_1^{(0)} + \gamma_1^{(1)}\big) \approx 2.0344, \qquad
\varphi_b = \tfrac{1}{2}\big(\gamma_1^{(0)} - \gamma_1^{(1)}\big) \approx 0.1799,
\label{eq:qsiht_applied}
\end{equation}
so the preparation circuit (Figure~\ref{fig:qsiht_circ}) uses two CNOTs, matching~\eqref{eq:qsiht_complexity_lit} at $n=2$.

Why the Walsh--Hadamard transform is the right inverse can be read off the compiled circuit directly. Conjugating a target rotation by a CNOT flips the sign of its angle whenever the control is set, because $X$ anticommutes with $Y$,
\begin{equation}
X\,R_y(\varphi)\,X = e^{-i\varphi\,(XYX)/2} = e^{+i\varphi\,Y/2} = R_y(-\varphi),
\label{eq:xryx}
\end{equation}
with $R_y$ as defined in~\eqref{eq:ry_gate}. Tracing $R_y(\varphi_a),\ \mathrm{CNOT},\ R_y(\varphi_b),\ \mathrm{CNOT}$ for each value of the control recovers the leaf rotations of Figure~\ref{fig:qsiht_uncompiled}: with $q_0=0$ both CNOTs are inactive and $q_1$ sees $R_y(\varphi_a+\varphi_b)=R_y(\gamma_1^{(0)})$, while with $q_0=1$ the enclosed rotation flips sign through~\eqref{eq:xryx} and $q_1$ sees $R_y(\varphi_a-\varphi_b)=R_y(\gamma_1^{(1)})$. These two conditions are the rows of $M$, and inverting them returns the applied angles~\eqref{eq:qsiht_applied}.

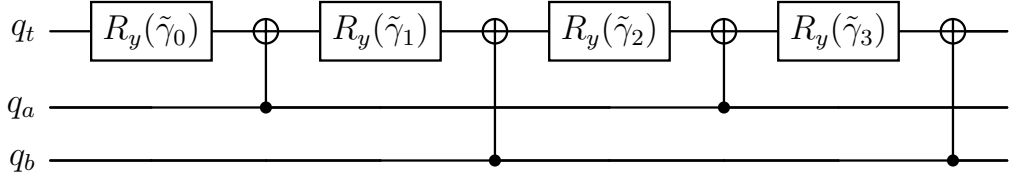
\begin{figure}[H]
\centering
\resizebox{0.82\textwidth}{!}{%
\begin{quantikz}
\lstick{$q_t$} & \gate{R_y(\tilde\gamma_0)} & \targ{} & \gate{R_y(\tilde\gamma_1)} & \targ{} & \gate{R_y(\tilde\gamma_2)} & \targ{} & \gate{R_y(\tilde\gamma_3)} & \targ{} & \qw \\
\lstick{$q_a$} & \qw & \ctrl{-1} & \qw & \qw & \qw & \ctrl{-1} & \qw & \qw & \qw \\
\lstick{$q_b$} & \qw & \qw & \qw & \ctrl{-2} & \qw & \qw & \qw & \ctrl{-2} & \qw
\end{quantikz}%
}
\caption[The two-control uniformly controlled rotation under the Gray-code rule]{The two-control uniformly controlled $R_y$ compiled by the Gray-code rule~\eqref{eq:wht2}, the smallest stage ($n=3$) at which the Gray code does visible work. The four fixed rotations $\tilde\gamma_0,\ldots,\tilde\gamma_3$, the applied angles of~\eqref{eq:wht2}, are interleaved with four CNOTs controlled in turn by the low control $q_a$ and the high control $q_b$ in the order low, high, low, high, the single-bit Gray toggles $00\to01\to11\to10\to00$. Every adjacent pair is one control apart, so no permutation network is needed.}
\label{fig:qsiht_twocontrol}
\end{figure}

The same rule scales to every deeper UCR of the cascade. Each fixed rotation $\tilde\gamma_k$ is seen by control configuration $\ket{j}$ with a sign set by the parity of the CNOTs that have fired, so the leaf angle it builds up to is
\begin{equation}
\gamma_j = \sum_{k=0}^{2^m-1} (-1)^{\,b_k\cdot g_j}\,\tilde\gamma_k ,
\label{eq:ucr_forward}
\end{equation}
the inverse of the transform~\eqref{eq:walsh} and computable in $O(m\,2^m)$ classical operations. The Gray-code ordering makes this sign matrix an orthogonal Walsh--Hadamard matrix: successive rotations are separated by a single control-bit toggle, so one CNOT, controlled by the qubit whose Gray-code bit changes, sits between them, and the closed cycle of $2^m$ toggles multiplies the CNOTs to the identity. The Gray code first does visible work at two controls, in the $n=3$ QsiHT, where a stage combines four leaf angles $\gamma_0,\ldots,\gamma_3$ through the order-2 transform
\begin{equation}
\begin{pmatrix}\tilde\gamma_0\\ \tilde\gamma_1\\ \tilde\gamma_2\\ \tilde\gamma_3\end{pmatrix}
= \tfrac14
\begin{pmatrix}1 & 1 & 1 & 1\\ 1 & -1 & -1 & 1\\ 1 & 1 & -1 & -1\\ 1 & -1 & 1 & -1\end{pmatrix}
\begin{pmatrix}\gamma_0\\ \gamma_1\\ \gamma_2\\ \gamma_3\end{pmatrix},
\label{eq:wht2}
\end{equation}
the $4\times4$ Walsh--Hadamard matrix with its columns reordered to $0,1,3,2$, the Gray sequence $00\to01\to11\to10$. The four CNOTs are controlled in turn by the low, high, low, and high control qubits (Figure~\ref{fig:qsiht_twocontrol}), and that pattern of single-bit toggles keeps every adjacent pair one control apart, so no permutation network is needed. A uniformly controlled $R_z$ compiles by the identical Gray-code rule with $R_z$ in place of $R_y$, since $X\,R_z(\varphi)\,X = R_z(-\varphi)$, which is what the plain complex baseline of Section~\ref{sec:qsiht_cost} appends as its phase cascade.

For a real sign-bearing state the advantage of $\arctan2$ over $\arctan$ appears: a negative amplitude, say $-3/\sqrt{30}$ in place of $+3/\sqrt{30}$, shifts $u_{0,1}$ from $-\arctan(4/3)$ to $\pi - \arctan(4/3)$ via $\arctan2$, absorbing the sign into the $R_y$ angle with no additional gate, whereas M\"ott\"onen would require a companion $R_z$ cascade to correct the phase.

\subsection{QsiHT Fast Path Real Synthesis}
\label{sec:qsiht_real}

The two CNOTs of Figure~\ref{fig:qsiht_circ} are the plain, un-deferred Gray-code cost~\eqref{eq:qsiht_complexity_lit}, and for state preparation they are one more than is needed. A preparation circuit acts on the single input $\ket{0}^{\otimes n}$, not on the whole $2^n$-dimensional space, and that slack lets one CNOT be dropped from every stage. Carrying the bookkeeping through gives the first synthesis this thesis prepares and benchmarks, an $R_y$-only circuit at the isometry CNOT count. Write $F_s$ for the stage-$s$ uniformly controlled $R_y$~\eqref{eq:ucr} on qubit $s$, so the Gray-code preparation cascade is $F_0 F_1 \cdots F_{n-1}$ with the bare top rotation $F_{n-1}$ applied first.

\paragraph{Truncation.} The reflected Gray walk of Section~\ref{sec:qsiht_example} closes by flipping its most significant control bit ($g_{2^k-1}=10\cdots0$ back to $g_0=0\cdots0$), so the trailing CNOT of every stage is controlled by the top qubit $q_{n-1}$. Removing the last gate of a circuit multiplies its unitary on the left by that gate's inverse, and $\mathrm{CX}^2=\mathbb{I}$, so dropping that trailing CNOT from stage $s\le n-2$ leaves
\begin{equation}
T_s = C_s\,F_s, \qquad C_s = \mathrm{CX}(q_{n-1}\!\to\! q_s),
\label{eq:qsiht_trunc}
\end{equation}
the truncated stage together with one stray fan-out CNOT $C_s$.

\paragraph{Conjugation.} Let $C$ permute the computational basis of a UCR's \emph{control} qubits while acting as the identity on its target, and let $\sigma$ be the induced permutation of control patterns. Then
\begin{equation}
C\,F(\boldsymbol{\theta})\,C = F(\boldsymbol{\theta}\circ\sigma),
\label{eq:qsiht_conj}
\end{equation}
a uniformly controlled rotation with its per-pattern angles relabeled and its cost unchanged: applied to $\ket{m}\ket{t}$ the left side gives $C F \ket{\sigma(m)}\ket{t} = C\ket{\sigma(m)}R_y(\theta_{\sigma(m)})\ket{t} = \ket{m}R_y(\theta_{\sigma(m)})\ket{t}$, using $C^2=\mathbb{I}$. Each stray $C_j=\mathrm{CX}(q_{n-1}\!\to\!q_j)$ is such a permutation for every stage $F_s$ with $s<j$, since both $q_j$ and $q_{n-1}$ are then controls of $F_s$, and its $\sigma$ flips pattern bit $j$ whenever pattern bit $n-1$ is set.

\paragraph{Collection.} The truncated preparation circuit is, in operator order,
\begin{equation}
U' = C_0 F_0\; C_1 F_1 \cdots C_{n-2}F_{n-2}\; F_{n-1}.
\label{eq:qsiht_collect}
\end{equation}
Moving every $C_j$ to the far left, each crosses the stages $F_s$ with $s<j$ and relabels them through~\eqref{eq:qsiht_conj}, while the $C_j$ commute among themselves as they share the control $q_{n-1}$. What is left standing on the left is their product,
\begin{equation}
U' = P\,\widetilde F_0 \widetilde F_1 \cdots \widetilde F_{n-1}, \qquad
P = \prod_{j=0}^{n-2}\mathrm{CX}(q_{n-1}\!\to\! q_j),
\label{eq:qsiht_fanout}
\end{equation}
where $P$ is the CNOT fan-out, the classical involution that complements bits $0,\ldots,n-2$ of every index whose top bit is $1$, and $\widetilde F_s$ is $F_s$ with its angles relabeled by the accumulated permutation $\pi_s$, which flips pattern bits $s+1,\ldots,n-2$ when pattern bit $n-1$ is set.

\paragraph{Solve.} Because $P^2=\mathbb{I}$, the requirement $U'\ket{0}^{\otimes n}=\ket{\Psi}$ is equivalent to asking the relabeled cascade to prepare $\boldsymbol{\chi}=P\ket{\Psi}$. That cascade is an ordinary adjacent-bit-plane UCR cascade with freely assignable per-pattern angles, so it prepares the real vector $\boldsymbol{\chi}$ by the DsiHT construction of Section~\ref{sec:qsiht_planes}: compute the DsiHT stage angles of $\boldsymbol{\chi}$, pre-compose stage $s$'s angle vector with the involution $\pi_s$, and map to physically applied angles by the Walsh--Hadamard transform~\eqref{eq:walsh}. Every step is an operator identity, so the preparation is exact.

\paragraph{Cost.} A stage with $k\ge1$ controls now emits $2^k$ rotations and $2^k-1$ CNOTs, one fewer than the plain cascade, and the $k=0$ stage a single bare rotation. Summing over the cascade,
\begin{equation}
G_{\text{CNOT}}^{\text{real}} = \sum_{k=1}^{n-1}\big(2^k-1\big) = 2^n - n - 1, \qquad
\mu(n) = 2^n - 1\ \text{$R_y$ rotations},
\label{eq:qsiht_real_complexity}
\end{equation}
with no other gate type. This is the isometry floor~\eqref{eq:iten_cost}, saving $n-1$ CNOTs against the plain Gray-code count~\eqref{eq:qsiht_complexity_lit}, and at $n=2$ it is a single CNOT, the lower bound~\eqref{eq:cnot_lb}. The permuted signal $\boldsymbol{\chi}=P\ket{\Psi}$ is a fan-out-reordered generator, a genuinely different heap-transform path, and because $P$ is a conditional complement rather than a relabeling of qubits, this reduction is not the layout freedom an optimizing compiler already owns.

\begin{figure}[H]
\centering
\begin{quantikz}
\lstick{$q_0:\ket{0}$} & \gate{R_y(\gamma_0)} & \ctrl{1} & \qw                & \qw \\
\lstick{$q_1:\ket{0}$} & \gate{R_y(\rho_a)}   & \targ{}  & \gate{R_y(\rho_b)} & \qw
\end{quantikz}
\caption{QsiHT Fast Path Real Synthesis of the running target~\eqref{eq:running}. Dropping the trailing CNOT of the plain Gray-code circuit (Figure~\ref{fig:qsiht_circ}) and re-solving against the fan-out-permuted signal $\boldsymbol{\chi}$~\eqref{eq:qsiht_chi} leaves a single CNOT. The bare $R_y(\gamma_0)$ on $q_0$ loads the block norms, and the one-control stage on $q_1$ carries the applied angles~\eqref{eq:qsiht_real_applied}, $\rho_a\approx1.7506$ and $\rho_b\approx0.4636$. One CNOT, pure $R_y$, attaining the lower bound~\eqref{eq:cnot_lb} at $n=2$.}
\label{fig:qsiht_real_circ}
\end{figure}
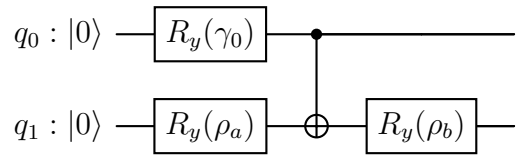

\paragraph{Two-qubit example.} In the running example's convention (the amplitude matrix~\eqref{eq:running_matrix} and the figures) $q_0$ is the top, most significant qubit, so the general formula's control $q_{n-1}$ is the example's $q_0$ and its target $q_0$ is the example's $q_1$, and~\eqref{eq:qsiht_fanout} specializes at $n=2$ to $P=\mathrm{CX}(q_0\!\to\!q_1)$ in the example's labels. For the running target~\eqref{eq:running} at $n=2$ the fan-out is therefore $P=\mathrm{CX}(q_0\!\to\!q_1)$, so
\begin{equation}
\boldsymbol{\chi} = P\ket{\Psi} = \tfrac{1}{\sqrt{30}}(1,\,2,\,4,\,3)^\top,
\label{eq:qsiht_chi}
\end{equation}
the running amplitudes with the last two swapped. Its blocks $(1,2)$ and $(4,3)$ carry the same norms $\sqrt5$ and $5$ as~\eqref{eq:running}, so the top rotation is unchanged, $\gamma_0 = 2\theta_3 = 2\arctan\sqrt5 \approx 2.3005$ with $\theta_3$ the stage-$1$ angle of~\eqref{eq:qsiht_s1}. The one-control stage on $q_1$ solves the swapped block $(4,3)$ in place of $(3,4)$, and its Walsh--Hadamard applied angles~\eqref{eq:walsh} are
\begin{equation}
\rho_a = \theta_1 + \tfrac{\pi}{2} - \theta_2 \approx 1.7506, \qquad
\rho_b = \theta_1 - \tfrac{\pi}{2} + \theta_2 \approx 0.4636,
\label{eq:qsiht_real_applied}
\end{equation}
with $\theta_1=\arctan2$ and $\theta_2=\arctan\tfrac43$ the stage-$0$ magnitudes of~\eqref{eq:qsiht_s0}. The preparation is the three-rotation, one-CNOT circuit of Figure~\ref{fig:qsiht_real_circ}. Built with Qiskit~2.4.2, the circuit contains only $R_y$ and CNOT, uses one CNOT as~\eqref{eq:qsiht_real_complexity} predicts at $n=2$, and reproduces~\eqref{eq:running} at fidelity $1$.

\subsection{QsiHT Fast Path Complex Synthesis}
\label{sec:qsiht_complex}

The plain fast path handles a complex target by preparing its magnitudes and then appending the diagonal~\eqref{eq:qsiht_diag}, at the doubled cost $2^{n+1}-4$ of Section~\ref{sec:qsiht_cost}. The second synthesis removes the separate diagonal by letting each butterfly rotation be a full $\mathrm{SU}(2)$ block instead of a real Givens rotation, so magnitude and phase are cleared together.

\paragraph{The complex butterfly.} Let a stage pair be $(a,b)=(A e^{i\alpha}, B e^{i\beta})$ with $A,B\ge0$ and $r=\sqrt{A^2+B^2}$. The block
\begin{equation}
U = \begin{pmatrix} \tfrac{A}{r}e^{-i\alpha} & \tfrac{B}{r}e^{-i\beta}\\[3pt] -\tfrac{B}{r}e^{i\beta} & \tfrac{A}{r}e^{i\alpha}\end{pmatrix},
\qquad U\begin{pmatrix}a\\ b\end{pmatrix} = \begin{pmatrix}r\\ 0\end{pmatrix}, \qquad \det U = 1,
\label{eq:qsiht_cbutterfly}
\end{equation}
zeroes the second component and sends the pair to its real, non-negative norm $r$. In ZYZ form, with $\theta=\arctan2(B,A)$ the same magnitude angle the real DsiHT~\eqref{eq:dsiht_angle} produces,
\begin{equation}
U = R_z(\alpha+\beta+\pi)\,R_y(2\theta)\,R_z(\alpha-\beta-\pi).
\label{eq:qsiht_cbutterfly_zyz}
\end{equation}
For a real signed target ($\alpha,\beta\in\{0,\pi\}$) the two $R_z$ factors collapse to signs and $U$ reduces to $\pm R_y(2\theta)$, so this is a strict generalization of the real butterfly, not a different transform. The survivor $r$ is real and non-negative, so by induction the working vector stays real after every stage and the recursion is the same heap transform with phase-decorated blocks. No separate diagonal ever appears.

\paragraph{Stage emission.} The disentangler is the cascade $F_{n-1}\cdots F_1 F_0$, where $F_s$ is now a uniformly controlled $\mathrm{U}(2)$ gate applying the block $U_m$ on qubit $s$ for each pattern $m$ of the higher qubits. A $k$-control uniformly controlled $\mathrm{U}(2)$ is synthesized \emph{up to a diagonal} at only $2^k-1$ CNOTs, one fewer than the $2^k$ a plain multiplexed rotation costs, by the construction of Bergholm, Vartiainen, M\"ott\"onen, and Salomaa~\cite{bergholm2005}, and the synthesis leaves a known residual diagonal $D_s$ on the qubits it touches. This same $2^k-1$ per-stage floor is what the deployed Qiskit \texttt{StatePreparation} and \texttt{Isometry} routines realize~\cite{iten2016quantum}.

\paragraph{Diagonal absorption.} The residual diagonals are cleared for free by working in the disentangling direction. After stage $s$ acts, qubit $s$ holds $\ket{0}$ across the whole support of the state, so only the bit-$s{=}0$ entries of $D_s$ act on anything, and those entries act exactly on the qubits $s+1,\ldots,n-1$ that stage $s+1$ touches. Folding them into the next stage's blocks,
\begin{equation}
U'_m \leftarrow U_m\,\operatorname{diag}\!\big(e^{i\delta_{2m}},\,e^{i\delta_{2m+1}}\big),
\label{eq:qsiht_absorb}
\end{equation}
keeps every block in $\mathrm{U}(2)$, which the uniformly controlled synthesis accepts, and costs nothing. The final stage is a single $2\times2$ unitary whose residual diagonal acts on $\ket{0}^{\otimes n}$ alone, an unobservable global phase, so no diagonal is ever paid as gates. The preparation circuit is the inverse of this disentangler. Direction matters: absorbing in the preparation direction instead strands the largest residual diagonal on a fully populated state, where clearing it costs $2^n-2$ CNOTs and erases the saving.

\paragraph{Cost.} Summing the per-stage counts,
\begin{equation}
G_{\text{CNOT}}^{\text{complex}} = \sum_{k=1}^{n-1}\big(2^k-1\big) = 2^n - n - 1, \qquad \text{depth } 2^{n+1}-2n-1,
\label{eq:qsiht_complex_complexity}
\end{equation}
for an \emph{arbitrary} complex target, the same isometry floor~\eqref{eq:iten_cost} as the Real Synthesis and matching the deployed Qiskit \texttt{StatePreparation}~\cite{iten2016quantum}. Against the plain complex fast path this removes the entire phase cascade, halving the entangling cost from $2^{n+1}-4$. At $n=2$ it is a single CNOT, the lower bound~\eqref{eq:cnot_lb}, which among the surveyed baselines only Plesch--Brukner reached.

\begin{figure}[H]
\centering
\begin{quantikz}
\lstick{$q_0:\ket{0}$} & \gate{A_0} & \ctrl{1} & \qw        & \qw \\
\lstick{$q_1:\ket{0}$} & \gate{B_0} & \targ{}  & \gate{B_1} & \qw
\end{quantikz}
\caption{QsiHT Fast Path Complex Synthesis of the complex target~\eqref{eq:running_complex}. Stage~$0$'s uniformly controlled $\mathrm{U}(2)$, carrying the pair phases of the blocks $U_0,U_1$, is demultiplexed up to a diagonal into single-qubit $\mathrm{U}(2)$ gates $A_0$ on $q_0$ and $B_0,B_1$ on $q_1$ around one CNOT~\cite{bergholm2005}, and stage~$1$ with every residual diagonal folds into these gates and the global phase. One CNOT prepares a complex target that the plain fast path~\eqref{eq:qsiht_diag} needs four to reach, so a complex target costs what a real one costs.}
\label{fig:qsiht_complex}
\end{figure}
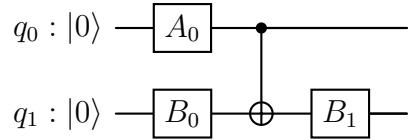

\paragraph{Two-qubit example.} Prepare the complex target~\eqref{eq:running_complex}, whose amplitudes $(1,2i,-3,4)/\sqrt{30}$ share the magnitudes of the running example but carry the phases $\boldsymbol{\omega}=(0,\tfrac{\pi}{2},\pi,0)$. Stage~$0$ clears the two bit-$0$ pairs with the phase-carrying blocks~\eqref{eq:qsiht_cbutterfly}. The pair $(\alpha_{00},\alpha_{01})=(1,\,2i)/\sqrt{30}$ has $A=1$, $\alpha=0$, $B=2$, $\beta=\tfrac{\pi}{2}$, hence $\theta=\arctan2\approx1.1071$ and real survivor $r=\sqrt5/\sqrt{30}$, giving
\begin{equation}
U_0 = \begin{pmatrix} 0.4472 & -0.8944\,i\\ -0.8944\,i & 0.4472\end{pmatrix},
\end{equation}
and the pair $(\alpha_{10},\alpha_{11})=(-3,\,4)/\sqrt{30}$ has $A=3$, $\alpha=\pi$, $B=4$, $\beta=0$, hence $\theta=\arctan\tfrac43\approx0.9273$ and survivor $r=5/\sqrt{30}$, giving the real block
\begin{equation}
U_1 = \begin{pmatrix} -0.6 & 0.8\\ -0.8 & -0.6\end{pmatrix}.
\end{equation}
Both input phases are consumed inside stage~$0$. The survivors $(\sqrt5,\,5)/\sqrt{30}$ are real, so stage~$1$ is the plain Givens rotation with $\theta_3=\arctan\sqrt5\approx1.1503$ of the real example, and its residual is a global phase. Synthesizing stage~$0$'s uniformly controlled $\mathrm{U}(2)$ up to its diagonal gives the single-CNOT circuit of Figure~\ref{fig:qsiht_complex}. Built with Qiskit~2.4.2, the circuit uses one CNOT at depth $3$ in the $\{u,\mathrm{cx}\}$ basis, or three $R_y$ and five $R_z$ around one CNOT once lowered to $\{R_z,R_y,R_x,\mathrm{cx}\}$, and reproduces~\eqref{eq:running_complex} at fidelity $1$. The plain complex fast path spends four CNOTs on the same target as built, four times as many, and optimization level~3 lowers it only to two, still twice the Complex Synthesis's single CNOT, with the extra gates all belonging to the phase cascade~\eqref{eq:qsiht_diag} that this synthesis folds away.

Both syntheses land at the isometry floor $2^n-n-1$ by the same idea, deferral. The Real Synthesis defers a classical fan-out permutation to the end of the circuit and absorbs it into the target, staying inside the real orthogonal group with a pure-$R_y$ gate set. The Complex Synthesis defers a phase diagonal from stage to stage and absorbs it into the next stage's blocks, ending in a global phase. The deferred object is Clifford in the first case and a diagonal in the second, and in both the state-preparation setting is what makes the deferral legal.

\subsection{State-to-State Transformation}
\label{sec:qsiht_s2s}

State \emph{preparation} is the special case $\ket{0}^{\otimes n}\!\to\!\ket{\Psi}$ of the more general \emph{state-to-state transformation} $\ket{x}\!\to\!\ket{y}$ between two arbitrary $n$-qubit superpositions, and the QsiHT realizes the general case at no extra asymptotic cost~\cite{gomez2025qsiht}. The general route maps both endpoints to the reference state, $U_x\ket{x}=\ket{0}^{\otimes n}$ and $U_y\ket{y}=\ket{0}^{\otimes n}$, and composes the inverse,
\begin{equation}
T = U_y^{\dagger}U_x, \qquad T\ket{x}=\ket{y},
\label{eq:s2s_twostage}
\end{equation}
running two full disentangling transforms and so paying roughly twice the rotation count. The QsiHT instead performs the transformation in a single pass: because the DsiHT is \emph{generated} by the data it processes, the same heap of Givens rotations that clears $\ket{x}$ onto the first axis is re-aimed at $\ket{y}$ by recomputing its angles from the target amplitudes, $\theta_k = \theta_k(\boldsymbol{y})$, so one transform of the form $U_x$ replaces the two in~\eqref{eq:s2s_twostage}. The result is an $n$-qubit state-to-state map in only $\mu(n) = 2^n - 1$ single-angle $R_y$ rotations~\eqref{eq:qsiht_rotcount}, half of the $2(2^n-1)$ rotations of the two-stage construction, with no Gray-code permutation network and at most $2^n-1$ CNOTs in the controlled realization~\cite{gomez2025qsiht}. A single controlled phase-shift gate fixes the sign of the leading amplitude, and even this can be absorbed into the final DsiHT angle by negating its parameter ($\theta\mapsto\theta+\pi$), so the sign correction costs no extra gate. Preparation from $\ket{0}^{\otimes n}$, used for every other method in this chapter, is the special case $\ket{x}=\ket{0}^{\otimes n}$, for which the leading-component clearing is trivial and the CNOT count drops to the un-deferred Gray-code baseline $2^n-2$ of~\eqref{eq:qsiht_complexity_lit}, which the Real and Complex Syntheses of Sections~\ref{sec:qsiht_real} and~\ref{sec:qsiht_complex} lower further to $2^n-n-1$.

\subsection{Simultaneous Two-State Preparation: The Q2siHT}
\label{sec:q2siht}

The single-generator DsiHT clears one superposition onto one basis state. A further extension, the \emph{Quantum Two Signal-induced Heap Transform} (Q2siHT), uses \emph{two} generators to prepare two different superpositions within a single circuit~\cite{grigoryan2025q2siht}. Its classical counterpart, the Discrete Two Signal-induced Heap Transform (D2siHT), is generated by a pair of vectors $\boldsymbol{x}$ and $\boldsymbol{y}$ and is built from elementary $3\times3$ Givens rotations acting on component triplets along a chosen path, closed by a single $2\times2$ rotation. An $N$-point D2siHT uses $N-2$ three-dimensional rotations and one two-dimensional rotation, for $2N-3$ angles in total. A single such transform simultaneously heaps the first generator onto $\ket{0}^{\otimes n}$ and the second onto a computational basis state $\ket{b}$ that differs from $\ket{0}^{\otimes n}$ in one selector qubit, so its inverse, the Q2siHT fast path, initializes both states at once,
\begin{equation}
\ket{0}^{\otimes n}\mapsto\ket{x}, \qquad \ket{b}\mapsto\ket{y},
\label{eq:q2siht_map}
\end{equation}
the two outputs distinguished by the value of that selector qubit. The $n$-qubit Q2siHT uses at most
\begin{equation}
n_R = 2\big(2^n-1\big)\ \text{controlled rotations}, \qquad n_X = 2\big(2^n-2\big)\ \text{controlled-NOT gates},
\label{eq:q2siht_cost}
\end{equation}
realized, for $n=3$, by $13$ rotation gates and $12$ CNOTs. The rotation count is the D2siHT's $2N-3$ angles, one below the $2(2^n-1)$ upper bound, while the CNOT count sits at the full $2(2^n-2)$. This is roughly twice the single-state QsiHT, the expected price for encoding two independent superpositions in one circuit rather than running two separate preparations. A useful property is \emph{multiplexing}: once the circuit is fixed, the second prepared state can be switched by retuning the angle of a single rotation gate, so a family of related states is reachable by one-parameter edits rather than full recompilation. Like the single-generator fast path, the construction needs no qubit-permutation network. The counts~\eqref{eq:q2siht_cost} are the un-deferred Gray-code costs of the two coupled cascades, since the trailing-CNOT deferral of the Real Synthesis and the diagonal deferral of the Complex Synthesis (Sections~\ref{sec:qsiht_real} and~\ref{sec:qsiht_complex}) each apply per single-generator cascade. Whether those deferrals carry over to the coupled two-generator construction is left to future work. The present formulation is restricted to real amplitudes, with a complex Q2siHT left to future work~\cite{grigoryan2025q2siht}.

\section{M\"ott\"onen State Preparation}
\label{sec:mottonen}

The method of M\"ott\"onen, Vartiainen, Bergholm, and Salomaa~\cite{mottonen2004, bergholm2005} is the canonical general-purpose direct construction and the conceptual reference for the disentangling methods that follow. It shares the uniformly controlled rotation~\eqref{eq:ucr} and Walsh--Hadamard angle transform~\eqref{eq:walsh} of Section~\ref{sec:ucr}. Rather than building $U_\Psi$ directly, it constructs the \emph{inverse} disentangling unitary $U_\Psi^{\dagger}$ that maps the target onto the ground state,
\begin{equation}
U_\Psi^{\dagger}\ket{\Psi} = \ket{0}^{\otimes n},
\label{eq:disentangle}
\end{equation}
and then takes $U_\Psi = (U_\Psi^{\dagger})^{\dagger}$ by reversing and inverting the gate sequence. The preparation factorizes into an amplitude (magnitude) stage and a phase stage by writing each amplitude in polar form,
\begin{equation}
\alpha_x = r_x\, e^{i\omega_x}, \qquad r_x = |\alpha_x| \ge 0, \qquad \omega_x = \arg \alpha_x.
\label{eq:polar}
\end{equation}

\subsection{Recursive Amplitude and Phase Preparation}

The amplitude stage equalizes the magnitudes $r_x$ qubit by qubit. Organize the $2^n$ magnitudes as the leaves of a binary tree. Each internal node stores the Euclidean norm of the amplitudes in its subtree. Disentangling the least-significant qubit requires, for every assignment of the remaining $n-1$ qubits, a $y$-rotation that rotates the two sibling amplitudes into their parent norm. For a node with child norms $a_L$ (suffix $0$) and $a_R$ (suffix $1$), the rotation angle is
\begin{equation}
\theta = 2\arctan\!\left(\frac{a_R}{a_L}\right),
\qquad
R_y(-\theta)\begin{pmatrix} a_L \\ a_R \end{pmatrix}
= \sqrt{a_L^2 + a_R^2}\,\begin{pmatrix} 1 \\ 0 \end{pmatrix}.
\label{eq:amp_angle}
\end{equation}
Collecting the $2^{k-1}$ angles needed to disentangle qubit $k$ into a single UCR controlled by the $k-1$ higher qubits, the amplitude stage is a cascade of $n$ uniformly controlled $R_y$ rotations. The phase stage then removes the relative phases $\omega_x$ with an analogous cascade of uniformly controlled $R_z$ rotations, whose angles are the finite-difference combinations of the $\omega_x$ obtained from the same Walsh--Hadamard transform~\eqref{eq:walsh}.

\subsection{Deterministic Exact Cost}

Summing the CNOT cost~\eqref{eq:ucr} of each UCR over both stages gives, for a general complex state,
\begin{equation}
G_{\text{CNOT}}^{\text{M\"ott}} = \underbrace{\sum_{k=1}^{n-1} 2^{k}}_{\text{amplitudes}} + \underbrace{\sum_{k=1}^{n-1} 2^{k}}_{\text{phases}}
\;=\; 2\big(2^{n} - 2\big) \;=\; 2^{n+1} - 4,
\label{eq:mottonen_cost}
\end{equation}
where each cascade contributes $2^n-2$ CNOTs, the top-qubit ($k=0$) rotation of each needing no CNOT. This as-built count $2^{n+1}-4$ ($124$ at $n=6$) is the one the from-scratch cascade of this thesis realizes. Merging the boundary CNOTs that abut between consecutive uniformly controlled rotations tightens the count to the $2^{n+1}-2n$ bound reported in the literature~\cite{bergholm2005}, but the as-built cascade performs no such merge. The construction is deterministic and exact: it uses no post-selection, no measurements, and no ancillae, reproducing $\ket{\Psi}$ up to global phase in the noiseless limit. Its cost is within roughly a factor of four of the lower bound~\eqref{eq:cnot_lb}, and it is the canonical baseline, reimplemented from its construction for the benchmarks of this thesis rather than invoked through a library routine.

\subsection{Two-Qubit Derivation and Example}
\label{sec:mottonen_example}

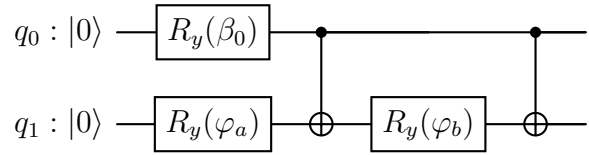
\begin{figure}[H]
\centering
\begin{quantikz}
\lstick{$q_0:\ket{0}$} & \gate{R_y(\beta_0)}   & \ctrl{1} & \qw                  & \ctrl{1} & \qw \\
\lstick{$q_1:\ket{0}$} & \gate{R_y(\varphi_a)} & \targ{}  & \gate{R_y(\varphi_b)}& \targ{}  & \qw
\end{quantikz}
\caption{M\"ott\"onen preparation of the target~\eqref{eq:running}. The boxed UCR on $q_1$ (two CNOTs) realizes the conditional angles $\beta_1^{(0)},\beta_1^{(1)}$.}
\label{fig:mottonen_circ}
\end{figure}

For $n=2$ the amplitude cascade is a bare rotation on the top qubit $q_0$ followed by a single one-control UCR on $q_1$. From the amplitude matrix~\eqref{eq:running_matrix}, the block (subtree) norms are
\begin{equation}
\nu_0 = \sqrt{r_{00}^2 + r_{01}^2} = \sqrt{\tfrac{1+4}{30}} = \tfrac{1}{\sqrt{6}}, \qquad
\nu_1 = \sqrt{r_{10}^2 + r_{11}^2} = \sqrt{\tfrac{9+16}{30}} = \sqrt{\tfrac{5}{6}},
\end{equation}
with $\nu_0^2+\nu_1^2=1$. The top rotation splits probability between the $q_0{=}0$ and $q_0{=}1$ blocks using~\eqref{eq:amp_angle}:
\begin{equation}
\beta_0 = 2\arctan(\nu_1/\nu_0) = 2\arctan\sqrt{5} \approx 2.3005~\text{rad}.
\end{equation}
The two conditional rotations that split each block into its leaves are
\begin{equation}
\beta_1^{(0)} = 2\arctan\!\frac{r_{01}}{r_{00}} = 2\arctan 2 \approx 2.2143, \qquad
\beta_1^{(1)} = 2\arctan\!\frac{r_{11}}{r_{10}} = 2\arctan\tfrac{4}{3} \approx 1.8546.
\end{equation}
These form a one-control UCR~\eqref{eq:ucr}. Its order-$1$ Walsh--Hadamard transform~\eqref{eq:walsh} yields the physically applied angles
\begin{equation}
\varphi_a = \tfrac12\big(\beta_1^{(0)}+\beta_1^{(1)}\big) \approx 2.0344, \qquad
\varphi_b = \tfrac12\big(\beta_1^{(0)}-\beta_1^{(1)}\big) \approx 0.1799.
\end{equation}
The preparation circuit (Figure~\ref{fig:mottonen_circ}) uses two CNOTs. Tracing the evolution verifies the construction: the top rotation produces $\nu_0\ket{00}+\nu_1\ket{10}$, after which the UCR applies $R_y(\beta_1^{(0)})$ in the $q_0{=}0$ branch and $R_y(\beta_1^{(1)})$ in the $q_0{=}1$ branch, giving
\begin{equation}
\tfrac{1}{\sqrt{6}}\big(\tfrac{1}{\sqrt5}\ket{00}+\tfrac{2}{\sqrt5}\ket{01}\big) + \sqrt{\tfrac56}\big(\tfrac{3}{5}\ket{10}+\tfrac{4}{5}\ket{11}\big) = \ket{\Psi}.
\end{equation}

The two-CNOT amplitude cost matches~\eqref{eq:mottonen_cost} restricted to the phase-free case. It exceeds the optimal single CNOT, illustrating that the UCR cascade is not entanglement-optimal at small $n$.

\subsection{Complex Two-Qubit Example: The Phase Stage}
\label{sec:mottonen_complex}

The complex target~\eqref{eq:running_complex} shares the magnitudes of~\eqref{eq:running}, so its amplitude cascade is \emph{identical} to Section~\ref{sec:mottonen_example}: the same $R_y(\beta_0)$ on $q_0$ and one-control $R_y$ UCR on $q_1$ produce the real magnitude state $\tfrac{1}{\sqrt{30}}(\ket{00}+2\ket{01}+3\ket{10}+4\ket{11})$. What remains is to imprint the phases $\boldsymbol{\omega}=\big(0,\tfrac{\pi}{2},\pi,0\big)$ with the analogous $R_z$ cascade.

\begin{figure}[H]
\centering
\begin{quantikz}
\lstick{$q_0:\ket{0}$} & \gate{R_y(\beta_0)} & \ctrl{1} & \qw & \ctrl{1} & \gate{R_z(\phi_0)} & \ctrl{1} & \qw & \ctrl{1} & \qw \\
\lstick{$q_1:\ket{0}$} & \gate{R_y(\varphi_a)} & \targ{} & \gate{R_y(\varphi_b)} & \targ{} & \gate{R_z(\psi_a)} & \targ{} & \gate{R_z(\psi_b)} & \targ{} & \qw
\end{quantikz}
\caption{M\"ott\"onen preparation of the complex target~\eqref{eq:running_complex}. The amplitude ($R_y$) cascade is identical to Figure~\ref{fig:mottonen_circ}. The appended $R_z$ cascade, a bare $R_z(\phi_0)$ on $q_0$ and a one-control uniformly controlled $R_z$ on $q_1$ (a second pair of CNOTs), gives the phases $\boldsymbol{\omega}=(0,\tfrac{\pi}{2},\pi,0)$.}
\label{fig:mottonen_complex}
\end{figure}
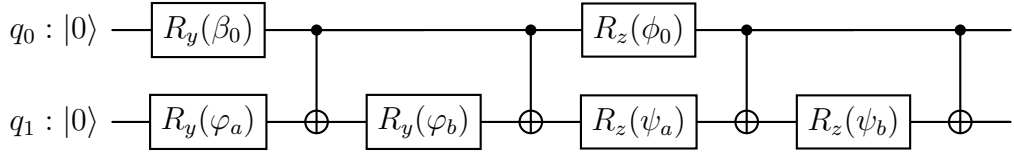

Writing each amplitude in polar form $\alpha_x = r_x e^{i\omega_x}$~\eqref{eq:polar}, the phase stage disentangles $q_1$ first. For each value of the control $q_0$ it removes the \emph{relative} phase of the sibling pair,
\begin{equation}
\phi_1^{(0)} = \omega_{01}-\omega_{00} = \tfrac{\pi}{2}, \qquad
\phi_1^{(1)} = \omega_{11}-\omega_{10} = -\pi,
\end{equation}
which together form a one-control uniformly controlled $R_z$ on $q_1$. Its order-1 Walsh--Hadamard transform~\eqref{eq:walsh} gives the physically applied angles
\begin{equation}
\psi_a = \tfrac12\big(\phi_1^{(0)}+\phi_1^{(1)}\big) = -\tfrac{\pi}{4}, \qquad
\psi_b = \tfrac12\big(\phi_1^{(0)}-\phi_1^{(1)}\big) = \tfrac{3\pi}{4}.
\end{equation}
The remaining top-qubit rotation removes the phase difference between the two block averages $\tfrac12(\omega_{00}+\omega_{01})=\tfrac{\pi}{4}$ and $\tfrac12(\omega_{10}+\omega_{11})=\tfrac{\pi}{2}$,
\begin{equation}
\phi_0 = \tfrac12(\omega_{10}+\omega_{11}) - \tfrac12(\omega_{00}+\omega_{01}) = \tfrac{\pi}{4},
\end{equation}
applied as a bare $R_z(\phi_0)$ on $q_0$, with the leftover uniform phase $\tfrac14\sum_x\omega_x = \tfrac{3\pi}{8}$ absorbed into the (unobservable) global phase. The full circuit (Figure~\ref{fig:mottonen_complex}) is the phase-free circuit of Figure~\ref{fig:mottonen_circ} followed by this $R_z$ cascade, for a total of
\begin{equation}
G_{\text{CNOT}} = \underbrace{2}_{\text{amplitude}} + \underbrace{2}_{\text{phase}} = 4 = \big(2^{n+1}-4\big)\big|_{n=2},
\end{equation}
the general complex cost~\eqref{eq:mottonen_cost}, exactly twice the phase-free count of~\eqref{eq:running}. The complex example thus exercises precisely the logic, the $R_z$ cascade and its second pair of CNOTs, that the real target~\eqref{eq:running} leaves dormant.

The remaining methods of this chapter prepare~\eqref{eq:running_complex} through their own phase mechanisms: a diagonal phase gate appended to the magnitude circuit (isometry, multiplexor), complex local unitaries in the Schmidt bases (Plesch--Brukner), or simply complex entries throughout the synthesized unitary (QSD). Their costs follow the asymptotic counts of Table~\ref{tab:method_cost} and Chapter~\ref{ch:theory}. Only M\"ott\"onen is worked out in full here, as the canonical illustration of the amplitude/phase split.

\section{Iten Isometry-Based State Preparation}
\label{sec:iten}

Iten, Colbeck, Kukuljan, Home, and Christandl~\cite{iten2016quantum} treat state preparation as the simplest instance of synthesizing an \emph{isometry}. An isometry is a norm-preserving linear map $V:\mathbb{C}^{2^m}\!\to\mathbb{C}^{2^n}$ with $m\le n$, characterized by
\begin{equation}
V^{\dagger}V = \mathbb{I}_{2^m},
\label{eq:isometry}
\end{equation}
i.e.\ $V$ embeds a $2^m$-dimensional input space into the larger $2^n$-dimensional Hilbert space while preserving inner products. State preparation is the boundary case $m=0$: the map $V:\mathbb{C}\to\mathbb{C}^{2^n}$ sends the scalar $1$ (equivalently $\ket{0}^{\otimes n}$) to the target $\ket{\Psi}$, so the isometry is a single column.

\subsection{Column-by-Column Decomposition}

Iten's algorithm reduces a general isometry to the identity by clearing one column at a time. To disentangle column $\ket{c}$, the method applies a sequence of two primitives: \emph{uniformly controlled rotations}~\eqref{eq:ucr} that merge amplitudes within a register, and multi-controlled single-qubit gates that map a partially reduced column onto a basis state. For state preparation the procedure collapses to exactly the disentangling cascade of Section~\ref{sec:mottonen}, but the angle bookkeeping is reorganized so that redundant rotations are eliminated, lowering the leading constant. The resulting CNOT count for a generic $n$-qubit state is
\begin{equation}
G_{\text{CNOT}}^{\text{Iten}} = 2^{n} - n - 1,
\label{eq:iten_cost}
\end{equation}
an improvement on the as-built cascade~\eqref{eq:mottonen_cost} by roughly a factor of two ($57$ against $124$ at $n=6$), and within a small factor of the lower bound~\eqref{eq:cnot_lb}~\cite{iten2016quantum}. Because it is exact, ancilla-free, and near-optimal, this merged-multiplexor decomposition underlies the deployed \texttt{StatePreparation} (and \texttt{Initialize}) routine in Qiskit~\cite{qiskit2024}, which realizes exactly $2^n-n-1$ CNOTs ($57$ at $n=6$) and is benchmarked as a deployed-library baseline in this work. The from-scratch Iten-style row of Chapter~\ref{ch:empirical} is a compiler reconstruction that applies optimization-level-3 peephole passes rather than the full merge, so it realizes the looser $\approx 2^{n+1}$ of the compressed cascade instead (\S\ref{sec:implementations}).

\subsection{Two-Qubit Derivation and Example}
\label{sec:iten_example}

Iten's algorithm is most transparent in the disentangling direction: it builds $U_\Psi^{\dagger}$ that clears the single occupied column of the isometry, mapping $\ket{\Psi}\to\ket{00}$, and then $U_\Psi = (U_\Psi^{\dagger})^{\dagger}$. For the target~\eqref{eq:running} the column is cleared in two reductions:
\begin{enumerate}
    \item \textbf{Clear the low qubit $q_1$.} Apply a one-control UCR $\big(F_1^{y}\big)^{\dagger}$ on $q_1$ with the inverse leaf angles $-\beta_1^{(0)}, -\beta_1^{(1)}$ from Section~\ref{sec:mottonen_example}. This merges each amplitude pair into its block norm, producing the disentangled intermediate
    \begin{equation}
    U^{\dagger}_{q_1}\ket{\Psi} = \big(\nu_0\ket{0} + \nu_1\ket{1}\big)_{q_0}\otimes\ket{0}_{q_1} = \tfrac{1}{\sqrt6}\ket{00} + \sqrt{\tfrac56}\ket{10}.
    \end{equation}
    \item \textbf{Clear the high qubit $q_0$.} Apply $R_y(-\beta_0)$ on $q_0$, rotating $\nu_0\ket{0}+\nu_1\ket{1}\to\ket{0}$, which yields $\ket{00}$.
    \end{enumerate}
Reversing and conjugating this sequence reproduces exactly the forward circuit of Figure~\ref{fig:mottonen_circ}. Thus for $n=2$ the plain column-clearing cascade coincides with M\"ott\"onen at \emph{two} CNOTs. The merged-multiplexor optimum of~\eqref{eq:iten_cost}, $2^n-n-1$, drops to a single CNOT here and is realized by the deployed Qiskit \texttt{StatePreparation} object. Its saving comes from eliminating redundant rotations across the deeper UCRs of larger registers ($n\ge 3$) and is not reproduced by the compiler-optimized cascade that stands in for it in Chapter~\ref{ch:empirical}. The example therefore illustrates the \emph{column-clearing} principle that generalizes to arbitrary isometries in the next section.

\begin{figure}[H]
\centering
\begin{quantikz}
\lstick[2]{$\ket{\Psi}$} & \ctrl{1} & \qw & \ctrl{1} & \qw & \gate{R_y(-\beta_0)} & \rstick[2]{$\ket{00}$} \\
 & \targ{} & \gate{R_y(-\varphi_b)} & \targ{} & \gate{R_y(-\varphi_a)} & \qw &
\end{quantikz}
\caption{Iten column-clearing (disentangling) circuit $U_\Psi^{\dagger}$ for the real target~\eqref{eq:running}. The one-control uniformly controlled $R_y$ on $q_1$ (inverse leaf angles, applied values $-\varphi_a,-\varphi_b$ from Section~\ref{sec:mottonen_example}) merges each amplitude pair into its block norm, after which $R_y(-\beta_0)$ collapses the top qubit, so that $\ket{\Psi}\mapsto\ket{00}$. The forward preparation $U_\Psi$ is this circuit reversed and negated, which for $n=2$ is identical to the M\"ott\"onen circuit of Figure~\ref{fig:mottonen_circ} at two CNOTs.}
\label{fig:iten_circ}
\end{figure}
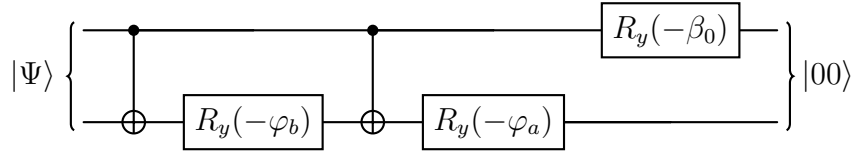

\subsection{Complex Two-Qubit Example}
\label{sec:iten_complex}

For the complex target~\eqref{eq:running_complex} the single occupied column carries the phases $\boldsymbol{\omega}=(0,\tfrac{\pi}{2},\pi,0)$, so a complex amplitude pair $(a_L,a_R)$ can no longer be merged into its norm by a real $R_y$ alone. Iten clears such a pair in two rotations: a uniformly controlled $R_z$ first cancels the relative phases $\phi_1^{(0)}=\tfrac{\pi}{2}$ and $\phi_1^{(1)}=-\pi$ (applied angles $\psi_a=-\tfrac{\pi}{4}$, $\psi_b=\tfrac{3\pi}{4}$, with the top phase $\phi_0=\tfrac{\pi}{4}$ removed on $q_0$, exactly the values of Section~\ref{sec:mottonen_complex}), after which the real magnitude clearing of Figure~\ref{fig:iten_circ} proceeds unchanged. The disentangling circuit is therefore the inverse of the M\"ott\"onen complex circuit of Figure~\ref{fig:mottonen_complex}, and the forward preparation again coincides with it at four CNOTs, the as-built value $2^{n+1}-4$ at $n=2$. The constant-factor saving of~\eqref{eq:iten_cost} over the as-built M\"ott\"onen cascade stays dormant at $n=2$ and emerges only for $n\ge3$, where the deeper uniformly controlled rotations contain the redundant angles that Iten's bookkeeping removes.

\section{General Isometry Decomposition}
\label{sec:general_isometry}

The isometry viewpoint of Section~\ref{sec:iten} is strictly more general than state preparation and is worth treating in its own right, because several compiler toolchains expose the full isometry primitive. A general isometry~\eqref{eq:isometry} maps an $m$-qubit input \emph{subspace} into an $n$-qubit output space. The special cases recover familiar tasks:
\begin{equation}
\begin{aligned}
m = 0 &: \quad \text{state preparation (one column)},\\
m = n &: \quad \text{full unitary synthesis } (V^{\dagger}V = VV^{\dagger} = \mathbb{I}),\\
0 < m < n &: \quad \text{partial embeddings, e.g.\ encoder circuits.}
\end{aligned}
\label{eq:isometry_cases}
\end{equation}
The decomposition proceeds analogously to a QR reduction: successive uniformly controlled gates triangularize the isometry, reducing the $2^m$ specified columns to standard basis vectors. The total cost interpolates between the state-preparation cost~\eqref{eq:iten_cost} and the full-unitary cost, scaling as $O(2^{n+m})$ entangling gates~\cite{iten2016quantum}. State preparation is recovered as the cheapest end of this spectrum, corresponding to clearing a single column. Implementations of the general primitive include the Qiskit \texttt{Isometry} class~\cite{qiskit2024} and the \texttt{UniversalQCompiler} package~\cite{iten2021introduction}, both of which can serve as exact state-preparation backends by setting $m=0$.

\subsection{Two-Qubit Derivation and Example}
\label{sec:isometry_example}

\begin{figure}[H]
\centering
\begin{quantikz}
\lstick[2]{$\ket{V_j}$} & \ctrl{1} & \qw & \ctrl{1} & \qw & \gate{R_y(-\beta_0)} & \octrl{1} & \rstick[2]{$\ket{0j}$} \\
 & \targ{} & \gate{R_y(-\varphi_b)} & \targ{} & \gate{R_y(-\varphi_a)} & \qw & \gate{R_y(\gamma)} &
\end{quantikz}
\caption{Gate-level column-clearing circuit $V^{\dagger}$ for the $m=1$ isometry~\eqref{eq:iso_example}. The first four gates clear the first column $\ket{V_0}=\ket{\Psi}$ exactly as in Figure~\ref{fig:iten_circ}, mapping it to $\ket{00}$. The open-controlled rotation $R_y(\gamma)$ on $q_1$ (active when $q_0=0$, shown by the open dot) then clears the transformed second column onto $\ket{01}$, completing the embedding $\ket{0}\mapsto\ket{00}$, $\ket{1}\mapsto\ket{01}$. The wire braces denote the generic input column $\ket{V_j}$ and its cleared image $\ket{0j}$ for $j\in\{0,1\}$, with $\ket{V_0}=\ket{\Psi}$. Removing this controlled rotation leaves the two-CNOT $m=0$ state-preparation circuit, which is why state preparation sits at the cheap end of the cost spectrum~\eqref{eq:isometry_cases}.}
\label{fig:isometry_circ}
\end{figure}
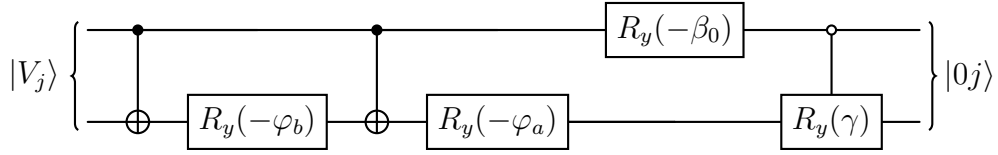

To exhibit the column-clearing mechanism in a non-degenerate setting, consider an $m=1$ isometry $V:\mathbb{C}^2\to\mathbb{C}^4$ whose \emph{first} column is the running target~\eqref{eq:running} and whose second column is any orthonormal completion, e.g.
\begin{equation}
V = \frac{1}{\sqrt{30}}\begin{pmatrix} 1 & 2 \\ 2 & -1 \\ 3 & 4 \\ 4 & -3 \end{pmatrix},
\qquad V^{\dagger}V = \mathbb{I}_2,
\label{eq:iso_example}
\end{equation}
since the two columns are orthogonal ($1\cdot2+2\cdot(-1)+3\cdot4+4\cdot(-3)=0$) and each has unit norm. The decomposition clears the columns in sequence, yielding the gate-level circuit of Figure~\ref{fig:isometry_circ}:
\begin{enumerate}
    \item Apply the disentangling sequence of Section~\ref{sec:iten_example} (a UCR on $q_1$ plus a rotation on $q_0$) to map the first column $\ket{V_0}=\ket{\Psi}$ to $\ket{00}$.
    \item The same gates transform the second column $\ket{V_1}$ into some residual vector orthogonal to $\ket{00}$. A final \emph{multi-controlled} single-qubit gate, controlled on $q_0=0$, rotates this residual onto $\ket{01}$, completing $V^{\dagger}V_1 = \ket{01}$.
\end{enumerate}
The full $V^{\dagger}$ then acts as $\ket{0}\mapsto\ket{00}$, $\ket{1}\mapsto\ket{01}$, i.e.\ an embedding of $q_1$. State preparation is recovered by deleting the second column entirely, leaving only step~1 and reducing the cost from the two-column isometry count to the single-column cost of Section~\ref{sec:iten_example}, the cheap $m=0$ end of the cost spectrum~\eqref{eq:isometry_cases}.

\paragraph{Complex target.} Within the isometry framework, complex amplitudes are absorbed exactly as in the state-preparation special case: the columns are first cleared in magnitude by the real $R_y$ cascade, and the residual relative phases are removed by appending a single diagonal phase gate, that is, a uniformly controlled $R_z$ cascade. For the $m=0$ reduction applied to the complex target~\eqref{eq:running_complex}, this diagonal is
\begin{equation}
\Delta = \operatorname{diag}\big(e^{i\omega_{00}},e^{i\omega_{01}},e^{i\omega_{10}},e^{i\omega_{11}}\big) = \operatorname{diag}(1,\,i,\,-1,\,1),
\label{eq:iso_diag}
\end{equation}
since $\Delta\,(1,2,3,4)^{\!\top}/\sqrt{30} = (1,2i,-3,4)^{\!\top}/\sqrt{30} = \ket{\Phi}$. Up to an unobservable global phase, $\Delta$ factors into the same $R_z$ cascade as Section~\ref{sec:mottonen_complex} (applied angles $\psi_a=-\tfrac{\pi}{4}$, $\psi_b=\tfrac{3\pi}{4}$, $\phi_0=\tfrac{\pi}{4}$) for two further CNOTs, a four-CNOT total at $n=2$. Because the diagonal is appended once, after every column has been cleared, its cost is independent of the isometry rank $m$.

\section{Quantum Shannon Decomposition}
\label{sec:qsd}

The Quantum Shannon Decomposition (QSD) of Shende, Bullock, and Markov~\cite{shende2006} is a method for synthesizing an \emph{arbitrary} $n$-qubit unitary $U$, and it can prepare a state by constructing any unitary whose first column equals $\ket{\Psi}$. It is built on the cosine--sine decomposition (CSD) of linear algebra: any $2^n\times 2^n$ unitary, partitioned into four $2^{n-1}\times 2^{n-1}$ blocks, factors as
\begin{equation}
U =
\begin{pmatrix} L_0 & 0 \\ 0 & L_1 \end{pmatrix}
\begin{pmatrix} C & -S \\ S & C \end{pmatrix}
\begin{pmatrix} R_0 & 0 \\ 0 & R_1 \end{pmatrix},
\qquad
C = \operatorname{diag}(\cos\theta_i),\; S = \operatorname{diag}(\sin\theta_i),
\label{eq:csd}
\end{equation}
with $L_0,L_1,R_0,R_1$ themselves $(n{-}1)$-qubit unitaries. The block-diagonal factors are uniformly controlled $(n{-}1)$-qubit unitaries (multiplexors controlled by the top qubit), while the central cosine--sine factor is exactly a uniformly controlled $R_y$ rotation~\eqref{eq:ucr}. Recursing on the smaller unitaries and applying multiplexor-diagonalization identities yields the optimized count
\begin{equation}
G_{\text{CNOT}}^{\text{QSD}} = \frac{23}{48}\,4^{n} - \frac{3}{2}\,2^{n} + \frac{4}{3},
\label{eq:qsd_cost}
\end{equation}
which is the best known leading constant for generic unitary synthesis~\cite{shende2006}. This cost is $\Theta(4^n)$, quadratically larger than the $\Theta(2^n)$ direct state-preparation cost of~\eqref{eq:mottonen_cost} and~\eqref{eq:iten_cost}: QSD wastefully synthesizes all $2^n$ columns when only one is constrained. It is therefore included as an \emph{upper-baseline} that quantifies the penalty of treating state preparation as full-unitary synthesis, and it remains relevant whenever the surrounding algorithm already requires a complete unitary.

\subsection{Two-Qubit Derivation and Example}
\label{sec:qsd_example}

\begin{figure}[H]
\centering
\begin{quantikz}
\lstick{$q_0:\ket{0}$} & \gate{A_1} & \ctrl{1} & \gate{A_2} & \ctrl{1} & \gate{A_3} & \ctrl{1} & \gate{A_4} & \qw \\
\lstick{$q_1:\ket{0}$} & \gate{B_1} & \targ{} & \gate{B_2} & \targ{} & \gate{B_3} & \targ{} & \gate{B_4} & \qw
\end{quantikz}
\caption{Gate-level realization of the QSD output for $n=2$. After the cosine--sine factorization~\eqref{eq:csd} and the QSD diagonalization identities fuse the three multiplexors, an arbitrary two-qubit unitary (hence any state-preparation completion) is realized by three CNOTs interleaved with single-qubit gates $A_i,B_i$~\cite{shende2006}, the optimum for a generic two-qubit operator. The single-qubit gates are fixed by the cosine--sine factors and the free completion, with the running target setting the first cosine--sine angle to $\theta_1=\beta_0$. Each single-qubit gate $A_i,B_i$ is a $\mathrm{U}(2)$ operation, compiled by the ZYZ Euler decomposition of Section~\ref{sec:plesch_zyz}. Only the three CNOTs are unavoidable, one more than the direct methods because all four columns are synthesized.}
\label{fig:qsd_circ}
\end{figure}
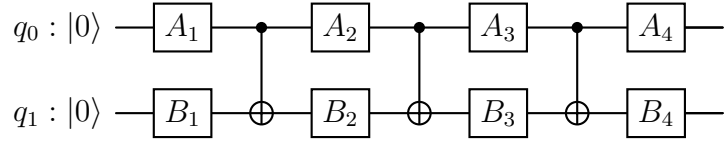

To prepare the target~\eqref{eq:running} via QSD one must first \emph{complete} it to a full $4\times4$ unitary $U$ whose first column is $\ket{\Psi}$. The remaining three columns are free and are chosen orthonormal, e.g.\ by Gram--Schmidt. Partitioning the first column by the top qubit gives the two half-vectors
\begin{equation}
U e_0 = \begin{pmatrix} \mathbf{a} \\ \mathbf{b}\end{pmatrix},\qquad
\mathbf{a} = \tfrac{1}{\sqrt{30}}\begin{pmatrix}1\\2\end{pmatrix},\quad
\mathbf{b} = \tfrac{1}{\sqrt{30}}\begin{pmatrix}3\\4\end{pmatrix},
\end{equation}
with $\|\mathbf{a}\| = \nu_0 = 1/\sqrt6$ and $\|\mathbf{b}\| = \nu_1 = \sqrt{5/6}$. Applying the cosine--sine decomposition~\eqref{eq:csd}, the central uniformly controlled $R_y$ factor carries angles $\theta_1,\theta_2$ whose first entry is fixed by the column norms,
\begin{equation}
\cos(\theta_1/2) = \nu_0 = \tfrac{1}{\sqrt6}, \qquad \sin(\theta_1/2) = \nu_1 = \sqrt{\tfrac56}\;\;\Rightarrow\;\; \theta_1 = \beta_0 \approx 2.3005,
\end{equation}
recovering the same top-level split angle $\beta_0$ as the direct methods, while $\theta_2$ and the four single-qubit blocks $L_0,L_1,R_0,R_1$ are determined by the (arbitrary) completion. The optimized QSD then realizes $U$ with
\begin{equation}
G_{\text{CNOT}}^{\text{QSD}}(n{=}2) = \tfrac{23}{48}\cdot 16 - \tfrac{3}{2}\cdot 4 + \tfrac{4}{3} = 3
\end{equation}
CNOTs, matching the formula~\eqref{eq:qsd_cost} and the known optimum for a \emph{generic} two-qubit unitary. This is one CNOT more than M\"ott\"onen (Figure~\ref{fig:mottonen_circ}) and \emph{three times} the single-CNOT optimum for the state alone, the price of synthesizing all four columns when only one is constrained. Each of the resulting single-qubit gates $A_i,B_i$ in Figure~\ref{fig:qsd_circ} is an element of $\mathrm{U}(2)$ and is compiled to the hardware basis by the same ZYZ Euler decomposition used for the Plesch--Brukner local unitaries, which is developed in full with a worked example in Section~\ref{sec:plesch_zyz}.

\paragraph{Complex target.} Unlike the direct methods, QSD never separates magnitude from phase: it synthesizes a complete unitary whose first column is the target, so complex amplitudes are simply absorbed into the (generically complex) single-qubit gates $A_i,B_i$ and the free completion. The complex target~\eqref{eq:running_complex} is therefore prepared by the \emph{same} three-CNOT circuit of Figure~\ref{fig:qsd_circ}, only with complex single-qubit gates. QSD is thus the one surveyed method whose cost does not grow for a complex target, since a generic two-qubit unitary already requires three CNOTs whether its entries are real or complex. This phase-agnostic count is the flip side of its wastefulness: it pays for all four columns regardless of how many phases the target carries.

\section{Multiplexor-Based State Preparation}
\label{sec:multiplexor}

Multiplexor-based preparation is, mathematically, the uniformly controlled rotation construction of Section~\ref{sec:mottonen} repackaged as a single composite gate. A \emph{multiplexed rotation} is precisely the operator $F_m^{a}$ of~\eqref{eq:ucr}. A preparation circuit is the cascade of such multiplexors, and a generic multiplexor over $m$ controls compiles to $2^m$ CNOTs by the Gray-code rule. The distinction from M\"ott\"onen's presentation is one of software abstraction rather than asymptotic cost: the compiler is handed the angle data $\{\theta_j\}$ for each multiplexor and is free to fuse adjacent rotations, cancel CNOTs across multiplexor boundaries, and re-route under hardware connectivity constraints during transpilation. The benchmark of this thesis realizes this route with Qiskit's uniformly controlled rotation gate objects, handing the angle data to the transpiler and letting its optimization-level-3 peephole passes fuse adjacent rotations and cancel CNOTs across multiplexor boundaries, so that the realized CNOT count can fall below the nominal $2^{n+1}-4$ of~\eqref{eq:mottonen_cost} after compilation. This makes the multiplexor route a useful probe of how much practical benefit a NISQ-oriented optimizing compiler extracts on top of the textbook construction.

\subsection{Two-Qubit Derivation and Example}
\label{sec:multiplexor_example}

\begin{figure}[H]
\centering
\begin{quantikz}
\lstick{$q_0:\ket{0}$} & \gate{R_y(\beta_0)} & \ctrl{1} & \qw & \ctrl{1} & \qw \\
\lstick{$q_1:\ket{0}$} & \gate{R_y(\varphi_a)} & \targ{} & \gate{R_y(\varphi_b)} & \targ{} & \qw
\end{quantikz}
\caption{Gate-level (Gray-code) expansion of the multiplexor preparation of the real target~\eqref{eq:running}: the one-control multiplexed rotation $F_1^{y}$ (leaf angles $\beta_1^{(0)},\beta_1^{(1)}$) unfolds into the two single-qubit rotations $R_y(\varphi_a),R_y(\varphi_b)$ and two CNOTs, coinciding gate-for-gate with the M\"ott\"onen circuit of Figure~\ref{fig:mottonen_circ}. The multiplexor view differs only in that the compiler receives $R_y(\varphi_a),\,\mathrm{CNOT},\,R_y(\varphi_b),\,\mathrm{CNOT}$ as one composite gate that it is then free to optimize across boundaries.}
\label{fig:multiplexor_circ}
\end{figure}
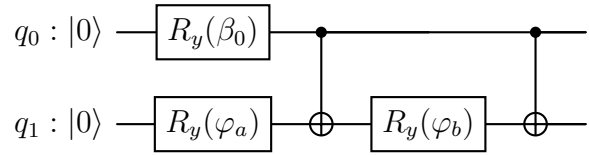

For the target~\eqref{eq:running} this route emits exactly the multiplexor cascade derived in Section~\ref{sec:mottonen_example}: a bare $R_y(\beta_0)$ on $q_0$ and the one-control multiplexed $R_y$ on $q_1$ with leaf angles $\beta_1^{(0)},\beta_1^{(1)}$, i.e.\ the circuit of Figure~\ref{fig:mottonen_circ} with a nominal two CNOTs, shown at gate level in Figure~\ref{fig:multiplexor_circ}. The distinction is what the compiler does next. During transpilation at optimization level~3, Qiskit applies optimization rules that
\begin{equation}
\begin{aligned}
&\text{(i) fuse adjacent } R_y \text{ rotations},\\
&\text{(ii) cancel } \mathrm{CNOT}\cdot\mathrm{CNOT}=\mathbb{I} \text{ across boundaries},\\
&\text{(iii) re-route under coupling maps},
\end{aligned}
\end{equation}
so that for this two-qubit instance the realized count meets, but does not beat, the nominal $2^{n+1}-4$ of~\eqref{eq:mottonen_cost}. Genuine reductions from fusion only emerge when several multiplexors abut in larger circuits. The example thus isolates the contribution of compiler optimization from that of the underlying mathematical construction, which is identical to M\"ott\"onen's.

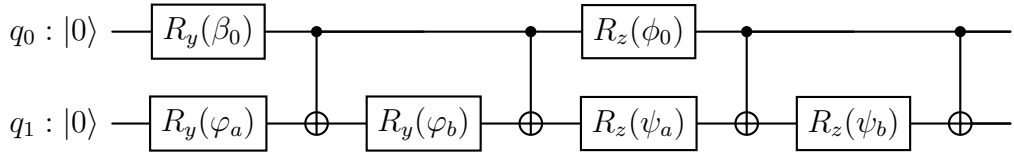
\begin{figure}[H]
\centering
\begin{quantikz}
\lstick{$q_0:\ket{0}$} & \gate{R_y(\beta_0)} & \ctrl{1} & \qw & \ctrl{1} & \gate{R_z(\phi_0)} & \ctrl{1} & \qw & \ctrl{1} & \qw \\
\lstick{$q_1:\ket{0}$} & \gate{R_y(\varphi_a)} & \targ{} & \gate{R_y(\varphi_b)} & \targ{} & \gate{R_z(\psi_a)} & \targ{} & \gate{R_z(\psi_b)} & \targ{} & \qw
\end{quantikz}
\caption{Gate-level expansion of the multiplexor preparation of the complex target~\eqref{eq:running_complex}: the magnitude multiplexor $F_1^{y}$ and the phase multiplexor $F_1^{z}$ each unfold into two single-qubit rotations and two CNOTs, reproducing the M\"ott\"onen complex circuit of Figure~\ref{fig:mottonen_complex}. The boundary between the abutting $R_y$ and $R_z$ cascades (the second and third entangling layers) is where adjacent multiplexors share and cancel CNOTs once several meet in a larger circuit.}
\label{fig:multiplexor_complex}
\end{figure}

\paragraph{Complex target.} The complex target~\eqref{eq:running_complex} appends a second multiplexor: the phase rotation $F_1^{z}$ on $q_1$, preceded by a bare $R_z(\phi_0)$ on $q_0$. Unfolded to gate level (Figure~\ref{fig:multiplexor_complex}), the magnitude cascade $R_y(\beta_0),R_y(\varphi_a),\mathrm{CNOT},R_y(\varphi_b),\mathrm{CNOT}$ is followed by the phase cascade $R_z(\phi_0),R_z(\psi_a),\mathrm{CNOT},R_z(\psi_b),\mathrm{CNOT}$ with the angles of Section~\ref{sec:mottonen_complex}. Each multiplexor contributes two CNOTs, for the nominal four-CNOT count $2^{n+1}-4$ at $n=2$ that the compiler then tries to reduce. The merit of the multiplexor abstraction is precisely that handing these rotations to the compiler as composite gates keeps the fusion opportunities visible, such as the boundary CNOTs where the $R_y$ and $R_z$ cascades meet, which adjacent multiplexors share and cancel once several abut in larger circuits.

\section{Plesch--Brukner Universal-Gate-Decomposition Method}
\label{sec:plesch}

Plesch and Brukner~\cite{plesch2011} give an exact construction that exploits the bipartite structure of the target state to reduce the leading constant toward the optimum of~\eqref{eq:cnot_lb}. Split the $n$ qubits into two registers $A$ and $B$ of $\lceil n/2\rceil$ and $\lfloor n/2\rfloor$ qubits and write the Schmidt decomposition of the target,
\begin{equation}
\ket{\Psi} = \sum_{i=0}^{\chi-1} \lambda_i \, \ket{u_i}_A \otimes \ket{v_i}_B,
\qquad \lambda_i \ge 0,\quad \sum_i \lambda_i^2 = 1,
\label{eq:schmidt}
\end{equation}
where the Schmidt rank obeys $\chi \le 2^{\lfloor n/2\rfloor}$ and $\{\ket{u_i}\}$, $\{\ket{v_i}\}$ are orthonormal. The preparation then proceeds in three structured steps:
\begin{enumerate}
    \item \textbf{Schmidt-coefficient loading.} Prepare the much smaller $\lceil n/2\rceil$-qubit state $\sum_i \lambda_i \ket{i}_A \ket{0}_B$, which depends only on the $\chi$ real coefficients $\lambda_i$.
    \item \textbf{Entangling copy.} Apply a transversal ladder of $\lfloor n/2 \rfloor$ CNOTs from $A$ to $B$ to produce the maximally correlated intermediate $\sum_i \lambda_i \ket{i}_A \ket{i}_B$.
    \item \textbf{Local basis rotations.} Apply the local unitaries $U_A$ and $V_B$, with $U_A\ket{i}=\ket{u_i}$ and $V_B\ket{i}=\ket{v_i}$, each an arbitrary $\lceil n/2\rceil$- or $\lfloor n/2\rfloor$-qubit operation, to rotate the computational basis into the Schmidt bases and recover~\eqref{eq:schmidt}.
\end{enumerate}
Because the two local unitaries act on only $n/2$ qubits each, their synthesis cost scales as $O(4^{n/2}) = O(2^{n})$ rather than $O(4^n)$, and adding the linear entangling ladder yields a total approaching
\begin{equation}
G_{\text{CNOT}}^{\text{P--B}} \approx 2^{n+1},
\label{eq:plesch_cost}
\end{equation}
roughly a factor of two above the lower bound~\eqref{eq:cnot_lb} and competitive with the isometry method \eqref{eq:iten_cost}~\cite{plesch2011}. This $2^{n+1}$ is a loose estimate. The non-truncating reference implementation benchmarked in Chapter~\ref{ch:empirical} realizes a leading constant near $1.5\cdot 2^{n}$, the value tabulated in Table~\ref{tab:method_cost}. The Plesch--Brukner scheme is significant as the construction that most transparently connects the entanglement structure (Schmidt rank) of a target state to its preparation cost, and it motivates structure-aware methods for states of low Schmidt rank.

\subsection{Two-Qubit Derivation and Example}
\label{sec:plesch_example}

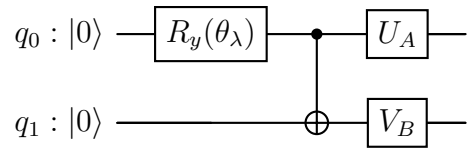
\begin{figure}[H]
\centering
\begin{quantikz}
\lstick{$q_0:\ket{0}$} & \gate{R_y(\theta_\lambda)} & \ctrl{1} & \gate{U_A} & \qw \\
\lstick{$q_1:\ket{0}$} & \qw                        & \targ{}  & \gate{V_B} & \qw
\end{quantikz}
\caption{Plesch--Brukner preparation of the target~\eqref{eq:running} using the Schmidt (SVD) decomposition. Only one CNOT is required, attaining the two-qubit optimum.}
\label{fig:plesch_circ}
\end{figure}

For $n=2$ the registers are $A=\{q_0\}$ and $B=\{q_1\}$, and the Schmidt decomposition~\eqref{eq:schmidt} is precisely the singular value decomposition (SVD) of the amplitude matrix~\eqref{eq:running_matrix}, $M = U_A\,\Sigma\,V_B^{\dagger}$. From $M^{\dagger}M = \tfrac{1}{30}\left(\begin{smallmatrix}10 & 14\\ 14 & 20\end{smallmatrix}\right)$, the singular (Schmidt) values are
\begin{equation}
\lambda_{1,2} = \sqrt{\frac{30 \pm \sqrt{884}}{60}}, \qquad \lambda_1 \approx 0.99776,\quad \lambda_2 \approx 0.06683, \quad \lambda_1^2+\lambda_2^2=1,
\end{equation}
so the Schmidt rank is $2$. The corresponding singular vectors give the local bases
\begin{equation}
U_A \approx \begin{pmatrix}0.405 & -0.915\\ 0.915 & 0.405\end{pmatrix},\qquad
V_B \approx \begin{pmatrix}0.576 & 0.817\\ 0.817 & -0.576\end{pmatrix}.
\end{equation}
The three-step construction then reads:
\begin{enumerate}
    \item \textbf{Load coefficients} on $q_0$: $R_y(\theta_\lambda)$ with $\theta_\lambda = 2\arctan(\lambda_2/\lambda_1) \approx 0.1338$, producing $\lambda_1\ket{00}+\lambda_2\ket{10}$.
    \item \textbf{Entangle} with a single $\text{CNOT}(q_0\!\to\!q_1)$, giving the Schmidt-diagonal state $\lambda_1\ket{00}+\lambda_2\ket{11}$.
    \item \textbf{Rotate bases} by applying $U_A$ on $q_0$ and $V_B$ on $q_1$, yielding $\lambda_1\ket{u_1}\ket{v_1}+\lambda_2\ket{u_2}\ket{v_2}=\ket{\Psi}$.
\end{enumerate}
The circuit is shown in Figure~\ref{fig:plesch_circ}. Because $\det M = -\tfrac{1}{15} < 0$, the SVD forces $\det U_A \cdot \det V_B = -1$, so exactly one local factor is a reflection, implemented as a single-qubit rotation composed with a $Z$ gate, no extra entangling cost. The total is therefore a \emph{single} CNOT, which exactly attains the lower bound $\lceil\tfrac12(2^2-2-1)\rceil=1$ for two qubits. Plesch--Brukner thus prepares the running target optimally, beating the two-CNOT direct methods, because it exploits the rank-$2$ Schmidt structure that all two-qubit states necessarily possess.

\subsection{Realizing the Local Unitaries: ZYZ Euler Decomposition}
\label{sec:plesch_zyz}

Step~3 leaves two single-qubit unitaries, $U_A$ on $q_0$ and $V_B$ on $q_1$, still to be turned into hardware gates. At $n=2$ each acts on a single qubit, so each is compiled directly by the \emph{ZYZ Euler decomposition}: every $U\in\mathrm{U}(2)$ factors as
\begin{equation}
U = e^{i\alpha}\,R_z(\beta)\,R_y(\gamma)\,R_z(\delta),
\qquad
R_z(t) = \begin{pmatrix}e^{-it/2} & 0\\ 0 & e^{it/2}\end{pmatrix},
\quad
R_y(t) = \begin{pmatrix}\cos\tfrac{t}{2} & -\sin\tfrac{t}{2}\\ \sin\tfrac{t}{2} & \cos\tfrac{t}{2}\end{pmatrix}.
\label{eq:zyz}
\end{equation}
The four real parameters, a global phase $\alpha$ plus three angles, match the four real dimensions of $\mathrm{U}(2)$ (three once the unobservable phase is dropped, i.e.\ $\mathrm{SU}(2)$). Multiplying the three factors out shows how the parameters split, the magnitudes carrying $\gamma$ and the phases carrying $\beta$ and $\delta$,
\begin{equation}
R_z(\beta)\,R_y(\gamma)\,R_z(\delta) =
\begin{pmatrix}
\cos\tfrac{\gamma}{2}\,e^{-i(\beta+\delta)/2} & -\sin\tfrac{\gamma}{2}\,e^{-i(\beta-\delta)/2}\\[4pt]
\sin\tfrac{\gamma}{2}\,e^{\,i(\beta-\delta)/2} & \cos\tfrac{\gamma}{2}\,e^{\,i(\beta+\delta)/2}
\end{pmatrix}.
\label{eq:zyz_expanded}
\end{equation}
Given a target $U$ with entries $u_{jk}$, only the left column is needed since unitarity fixes the rest. Stripping the phase with $V = e^{-i\alpha}U$ (which has $\det V = 1$) and reading~\eqref{eq:zyz_expanded} off its entries $v_{jk}$ gives
\begin{equation}
\begin{gathered}
\alpha = \tfrac12\arg(\det U), \qquad
\gamma = 2\arctan2\big(|v_{10}|,\,|v_{00}|\big),\\[3pt]
\beta = \arg v_{10} - \arg v_{00}, \qquad
\delta = -\arg v_{10} - \arg v_{00}.
\end{gathered}
\label{eq:zyz_extract}
\end{equation}
When $v_{00}\approx 0$ (so $\gamma = \pi$) only the combination $\beta-\delta$ is fixed, the gimbal-lock degeneracy of Euler angles, and one sets $\delta = 0$ and reads $\beta$ from $v_{10}$.

\paragraph{The proper rotation $U_A$ ($\det = +1$).} Here $\det U_A = +1$, so $\alpha = 0$ and $V = U_A$ is already real with a positive left column. Then $\arg v_{10} = \arg v_{00} = 0$, so $\beta = \delta = 0$ and~\eqref{eq:zyz_extract} leaves only
\begin{equation}
\gamma_U = 2\arctan2(0.915,\,0.405) \approx 2.309,\qquad U_A = R_y(\gamma_U),
\end{equation}
a bare $R_y$: a real proper rotation carries no phase content, so both $R_z$ layers vanish.

\paragraph{The reflection $V_B$ ($\det = -1$).} Now $\det V_B = -1$, so $\alpha = \tfrac12\arg(-1) = \tfrac{\pi}{2}$, and dividing by $e^{i\alpha}$ restores $\det V = 1$. The recipe~\eqref{eq:zyz_extract} then gives $\gamma_V \approx 1.914$, $\beta = 0$, and $\delta = \pi$, so
\begin{equation}
V_B = e^{i\pi/2}\,R_y(\gamma_V)\,R_z(\pi),\qquad
R_z(\pi) = \operatorname{diag}(-i,\,i) = -i\,Z.
\end{equation}
The two leftover phases combine as $e^{i\pi/2}R_z(\pi) = i\cdot(-iZ) = Z$, collapsing the reflection to a rotation followed by a $Z$,
\begin{equation}
V_B = R_y(1.914)\,Z \quad (\text{up to global phase}),
\end{equation}
which is exactly the ``single-qubit rotation composed with a $Z$ gate'' anticipated in Section~\ref{sec:plesch_example}. Both local unitaries thus reduce to one $R_y$ each (with a fixed $Z$ on $q_1$), adding no entangling gate, so the single CNOT of Figure~\ref{fig:plesch_circ} is the whole two-qubit cost. For $n > 2$ the same idea recurses one level up: $U_A$ and $V_B$ become arbitrary $\lceil n/2\rceil$- and $\lfloor n/2\rfloor$-qubit unitaries, compiled by a QSD or UCR cascade at $O(4^{n/2}) = O(2^n)$ rather than a single ZYZ, which is the source of the $\approx 2^{n+1}$ leading term in~\eqref{eq:plesch_cost}.

\section{Related Work Outside the Benchmark Scope}
\label{sec:related_out_of_scope}

The method set surveyed above and benchmarked in Chapter~\ref{ch:empirical} is deliberately narrow: exact, deterministic, ancilla-free preparation of \emph{dense} targets. Its youngest member is Iten's 2016 isometry method, so claims in this thesis about ``the field'' should be read as claims about this pre-2017 ancilla-free dense method set. Three newer lines of work fall outside that scope and are cited here for orientation, not benchmarked.

\paragraph{Ancilla-assisted and depth-optimal constructions.} Araujo \emph{et al.}'s divide-and-conquer scheme trades an exponential ancilla register for polylogarithmic depth~\cite{araujo2021divide}, Rosenthal obtains low-depth preparation through Grover-style constructions with large ancilla counts~\cite{rosenthal2021query}, and Sun \emph{et al.}\ settle the depth question: $\Theta(2^n/n)$ is optimal without ancillae, and $O(2^n)$ ancillae buy depth $\Theta(n)$~\cite{sun2023asymptotically}. These methods answer a different resource question (depth under a qubit surplus) than the qubit-frugal one posed here.

\paragraph{Sparse-state preparation.} For a target with $s$ non-zero amplitudes, the algorithms of Gleinig and Hoefler~\cite{gleinig2021sparse} and Malvetti, Iten, and Colbeck~\cite{malvetti2021sparse} synthesize circuits whose cost scales with $s$ rather than $2^n$, roughly $O(sn)$ gates. None of the dense methods compared in this thesis shares that scaling. In particular, the QsiHT fast path carries no sparse-cost scaling: its uniformly controlled cascade emits its worst-case dense CNOT budget on a full-support, full-Schmidt-rank real target, the plain Gray cascade's $2^n-2$ or the two syntheses' $2^n-n-1$, and reads no support information to prune gates, so unlike the dedicated sparse algorithms its cost does not scale with the support $s$. Zero-valued rotation angles do drop out on degenerate inputs, lowering the realized count below $2^n-2$ there (the stability study of Chapter~\ref{ch:empirical} records $36$ CNOTs at an exactly degenerate spectrum and $20$ after optimization on a rank-1 product state), but this is an incidental effect of vanishing angles and peephole compression rather than a support-aware construction. No sparse-cost claim is made for it anywhere in this thesis, and whether a heap-transform route can exploit sparsity is an open question rather than a result.

\paragraph{Approximate and tensor-network loaders.} Matrix-product-state encodings~\cite{ran2020encoding} and approximate amplitude encoding in shallow variational circuits~\cite{nakaji2022approximate} trade a controlled infidelity for polynomially shallow circuits. They break the exact-deterministic contract of Chapter~\ref{ch:background} that defines this comparison, which is why they are out of scope, but for near-term registers beyond $n\approx 10$ they are the practically relevant competition for any exact loader.

\section{Summary}

Table~\ref{tab:method_cost} consolidates the asymptotic two-qubit cost of all surveyed methods, including the two QsiHT syntheses. The direct, isometry, multiplexor, Plesch--Brukner, and QsiHT methods all achieve the optimal $\Theta(2^n)$ scaling, differing only in the leading constant, whereas QSD pays a quadratic $\Theta(4^n)$ overhead for synthesizing an entire unitary. These analytic costs frame the empirical comparison carried out in the following chapters, where each method is implemented, transpiled to realistic hardware, and benchmarked on identical dense target states against the gate, depth, ancilla, and fidelity metrics of Chapter~\ref{ch:background}.

\begin{table}[H]
\centering
\caption{Asymptotic CNOT cost of exact arbitrary dense-state preparation methods (no ancillae), with the count realized on the worked two-qubit target~\eqref{eq:running}. The M\"ott\"onen entry is the as-built two-cascade count $2^{n+1}-4$ ($124$ at $n=6$). The boundary-merged $2^{n+1}-2n$ of the literature~\cite{bergholm2005} is not realized by the from-scratch cascade. The Iten entry is the merged-multiplexor optimum $2^n-n-1$ ($57$ at $n=6$), realized by the deployed Qiskit \texttt{StatePreparation}, while the from-scratch Iten-style and multiplexor reconstructions instead realize $\approx 2^{n+1}$ (\S\ref{sec:scaling}), the value shown in the $n=2$ column. The two QsiHT synthesis rows are the trailing-CNOT (real, Section~\ref{sec:qsiht_real}) and diagonal (complex, Section~\ref{sec:qsiht_complex}) deferrals of the Gray-code cascade, each reaching the isometry floor $2^n-n-1$ and the single-CNOT optimum at $n=2$. The Quantum Shannon $n=2$ entry is the realized count of the constant-unoptimized completion. The formula's value $3$, the Vatan--Williams two-qubit optimum~\cite{vatan2004optimal}, is not reached without the diagonal-merging pass (Section~\ref{sec:qsd_example}).}
\label{tab:method_cost}
\begin{tabular}{llcl}
\hline
\textbf{Method} & \textbf{CNOT count} & \textbf{$n=2$ example} & \textbf{Reference} \\
\hline
Lower bound & $\left\lceil\tfrac{1}{2}(2^n - n - 1)\right\rceil$ & $1$ & \cite{plesch2011, shende2006} \\
QsiHT Real Synthesis & $2^n - n - 1$ & $\mathbf{1}$ & \cite{gomez2025qsiht} \\
QsiHT Complex Synthesis & $2^n - n - 1$ & $\mathbf{1}$ & \cite{gomez2025qsiht} \\
M\"ott\"onen (UCR) & $2^{n+1} - 4$ (as built) & $2$ & \cite{mottonen2004, bergholm2005} \\
Iten isometry & $2^{n} - n - 1$ (merged) & $2$ & \cite{iten2016quantum} \\
Multiplexor (Qiskit UCR) & $\approx 2^{n+1}$ (as built) & $2$ & \cite{qiskit2024} \\
Plesch--Brukner & $\approx 1.5\cdot 2^{n}$ & $\mathbf{1}$ & \cite{plesch2011} \\
Quantum Shannon Decomp. & $\tfrac{23}{48}\,4^{n} - \tfrac{3}{2}2^{n} + \tfrac{4}{3}$ & $6$ & \cite{shende2006} \\
\hline
\end{tabular}
\end{table}

On the running two-qubit example the asymptotic ranking is partially inverted. Plesch--Brukner and the two QsiHT syntheses attain the optimal single CNOT, Plesch--Brukner by exploiting the guaranteed rank-$2$ Schmidt structure and the QsiHT syntheses by deferring one CNOT per stage, while the direct M\"ott\"onen, isometry, and multiplexor methods use two and the unoptimized QSD completion uses six. This small-$n$ behavior foreshadows a recurring theme of the empirical study: constant factors, structure-awareness, and angle-pipeline robustness govern performance on the modest register sizes accessible to NISQ hardware.

%% file: chapters/chapter4.tex
\chapter{Theoretical Comparison of Methods}
\label{ch:theory}

Chapter~\ref{ch:litreview} derived, for each exact dense-state preparation method, both its general construction and an explicit two-qubit realization. This chapter lifts those isolated derivations into a unified theoretical comparison. The goal is to characterize \emph{a priori}, before any hardware execution, how the methods of Chapter~\ref{ch:litreview} trade resources against one another as the register size $n$ grows, and to identify the routines in which each is preferable. The analysis is organized around a fixed comparison framework (\S\ref{sec:framework}), applied method by method (\S\ref{sec:method_analysis}), and consolidated into asymptotic complexity tables (\S\ref{sec:complexity_table}).

\section{Comparison Framework}
\label{sec:framework}

This chapter and the two that follow share one comparison framework with three layers, whose relationship should be fixed explicitly at the outset. The first layer is the \emph{measurement set}: the seven per-circuit resource metrics defined in Chapter~\ref{ch:background} (total gate count, CNOT count, depth, ancilla count, compilation time, transpiled depth, and fidelity), which the benchmark of Chapter~\ref{ch:empirical} records identically for every method. The second layer is the \emph{comparison criteria} of this chapter: nine attributes that wrap the asymptotic form of the measurement set in the qualitative screens needed to delimit the method set. The third layer is the set of \emph{results axes} into which Chapter~\ref{ch:empirical} condenses the measurements for its final verdict (Table~\ref{tab:method_verdict}). The layers are nested by construction, not three competing metric sets: the criteria below restate the measured quantities asymptotically, and the verdict axes summarize them after measurement.

Every method is assessed against nine criteria. The first three are \emph{qualitative} (Boolean or categorical), the next four are \emph{asymptotic complexities} in the qubit count $n$ (where the state dimension is $N = 2^n$), and the last two are \emph{practical} attributes. Formally, to each method $\mathcal{M}$ we associate the cost tuple
\begin{equation}
\mathcal{C}(\mathcal{M}) = \big(\, G(\mathcal{M}),\; G_{\text{CNOT}}(\mathcal{M}),\; d(\mathcal{M}),\; a(\mathcal{M}) \,\big),
\label{eq:cost_tuple}
\end{equation}
whose entries are the total gate count, the CNOT count, the depth, and the ancilla count introduced in Chapter~\ref{ch:background}, each reported in $\Theta(\cdot)$ asymptotics. The criteria are:

\begin{enumerate}
    \item \textbf{Exactness.} Whether $\mathcal{M}$ realizes the unitary $U_\Psi\ket{0}^{\otimes n} = \ket{\Psi}$ of~\eqref{eq:qsp_unitary} exactly (up to global phase) in the noiseless limit, as opposed to producing an $\varepsilon$-approximation. All methods surveyed here are exact.
    \item \textbf{Determinism.} Whether the output state is obtained on every shot without post-selection, per the criteria of Chapter~\ref{ch:background}. All surveyed methods are deterministic and measurement-free.
    \item \textbf{Generality.} Whether $\mathcal{M}$ handles an \emph{arbitrary} complex dense amplitude vector $\boldsymbol{\alpha}\in\mathbb{C}^{2^n}$, rather than only real, sparse, or structured states.
    \item \textbf{Gate complexity} $G$: the asymptotic growth of the total elementary-gate count.
    \item \textbf{CNOT complexity} $G_{\text{CNOT}}$: the growth of the two-qubit gate count, the dominant NISQ cost driver and the quantity lower-bounded by~\eqref{eq:cnot_lb}.
    \item \textbf{Depth complexity} $d$: the growth of the circuit depth~\eqref{eq:depth}, which determines whether the circuit fits within the coherence horizon $d\cdot t_g \lesssim T_2$.
    \item \textbf{Ancilla requirement} $a$: the number of auxiliary qubits, which can be traded against depth.
    \item \textbf{Numerical stability}: the sensitivity of the classical angle-computation pipeline to small or vanishing amplitudes.
    \item \textbf{Implementation availability}: whether a maintained open-source implementation exists, which conditions reproducibility and the empirical study of later chapters.
\end{enumerate}

Most of these criteria do not discriminate among the methods. Every surveyed method is exact, deterministic, and ancilla-free, and every method except the QsiHT Real Synthesis is fully general over complex amplitudes. The Real Synthesis is restricted to real, sign-bearing targets, as Table~\ref{tab:qualitative} records. Exactness, determinism, and the ancilla requirement therefore act as admission screens, while generality separates the real-only builder from the arbitrary-complex configurations. Of the remaining criteria, the gate count, CNOT count, and depth of these ancilla-free constructions rise and fall together (every method is $\Theta(2^n)$ on all three except QSD, which is $\Theta(4^n)$ on all three), and the numerical-stability distinctions of \S\ref{sec:stability} are conditioning arguments that do not surface as accuracy differences on the Haar-random benchmark targets, where every applicable method is exact to machine precision, separating builders only on near-degenerate inputs (\S\ref{sec:exactness}). The framework is therefore \emph{not} a nine-axis aggregation and this thesis does not score methods across nine dimensions. It operates as a CNOT-led screen, ranking methods primarily by their two-qubit count, followed by a Pareto view of the other measured axes (depth, total gates, and compilation time) in which QSD is dominated on every measured axis at once on both target families, while the frontier methods trade constant factors and structure sensitivity against one another. How much of the frontier survives depends on the target family: on complex targets the compilation-time axis keeps the compiler-optimized cascades on the frontier, since the Complex Synthesis compiles more slowly than they do while dominating them on the structural axes (counts, depth, total gates), whereas on real targets the Real Synthesis is at the benchmarked sizes also the fastest of the ten builders, so it dominates the multiplexor-family cascades on every measured axis at once (\S\ref{sec:scaling}).

\subsection{On Numerical Stability}
\label{sec:stability}

Numerical stability deserves explicit treatment because it distinguishes methods that are asymptotically identical. The direct constructions compute rotation angles through inverse trigonometric functions of amplitude ratios, e.g.\ the M\"ott\"onen leaf angle
\begin{equation}
\theta = 2\arctan\!\left(\frac{a_R}{a_L}\right), \qquad \omega = \arg(\alpha_x),
\label{eq:angle_recall}
\end{equation}
where $a_L, a_R$ are sibling subtree norms. Two failure modes arise. First, when an entire subtree is empty ($a_L = a_R = 0$) the ratio is the indeterminate $0/0$. The angle is conventionally set to zero, which is exact but requires guarded code. Second, and more insidiously, when a magnitude $|\alpha_x|$ is small but nonzero, its phase $\omega = \arg(\alpha_x)$ is poorly conditioned: a perturbation $\delta\alpha$ induces a phase error $\delta\omega \sim |\delta\alpha|/|\alpha_x|$ that diverges as $|\alpha_x|\to 0$. Direct methods therefore lose phase precision on states with many small amplitudes. By contrast, the Plesch--Brukner method routes the amplitude data through a singular value decomposition (SVD), which is backward stable: the computed factors are the exact SVD of a matrix within $O(\epsilon_{\text{mach}}\|M\|)$ of the input amplitude matrix $M$~\cite{golub2013matrix}, where $\epsilon_{\text{mach}}$ is the machine epsilon (the unit roundoff of the floating-point format). This would confer a conditioning advantage in the worst case, one a pure gate-count analysis does not see. Whether it matters in practice is an empirical question, and on these benchmarks it does not: the advantage is never observed, and where a measurable separation does appear the measured direction runs the other way. On the real Haar family of Chapter~\ref{ch:empirical} every builder's measured statevector residual at $n=6$ stays below $10^{-13}$, and the fine grading among those residuals reflects each implementation's numerical path (the SVD, the multiplexor resynthesis, the Walsh--Hadamard angle pass), not the conditioning tiers above and not the circuit length, so the advantage does not surface there (Table~\ref{tab:qualitative}). It takes a near-degenerate or near-sparse (interior-zero) input to produce a measurable separation, and the separation that then appears is between the opt-3-peephole and as-built realizations, with residuals up to about $1.2\times10^{-6}$ for the former, rather than between angle pipelines (\S\ref{sec:exactness}). On that degenerate family the QsiHT Real Synthesis in fact carries max residuals more than two orders of magnitude below Plesch--Brukner's ($1.5\times10^{-15}$ against $5.0\times10^{-13}$, about $335\times$), the Complex Synthesis about $5\times$ below ($9.1\times10^{-14}$), and the M\"ott\"onen cascades sit at Plesch--Brukner's level ($4.9\times10^{-13}$), all at machine precision, while the opt-3 peephole builders (Iten, the isometry, the multiplexor, and \texttt{StatePreparation}) are the worst on this family, the Iten/isometry pair reaching $1.2\times10^{-6}$, so at the level the data resolve the predicted SVD conditioning edge is invisible against the cascades and reversed against the two syntheses' \texttt{arctan2} pipelines.

\section{Method-by-Method Theoretical Analysis}
\label{sec:method_analysis}

\subsection{M\"ott\"onen / Uniformly Controlled Rotations}
\label{sec:an_mottonen}

\paragraph{Construction.} As derived in \S\ref{sec:mottonen}, the method disentangles the target qubit by qubit using two cascades of uniformly controlled rotations~\eqref{eq:ucr}: an amplitude cascade of $R_y$ rotations and a phase cascade of $R_z$ rotations. The angles follow from the binary-tree recursion~\eqref{eq:amp_angle} and the Walsh--Hadamard transform~\eqref{eq:walsh}, both computable classically in $O(2^n)$ time.

\paragraph{Complexity.} The $k$-control UCR costs $2^k$ CNOTs and $2^k$ rotations, so each cascade (control counts $k=1,\dots,n-1$, the top rotation needing none) contributes $2^n-2$ CNOTs, and summing the amplitude and phase cascades gives
\begin{equation}
G_{\text{CNOT}} = 2^{n+1} - 4 = \Theta(2^n), \qquad G = \Theta(2^n), \qquad d = \Theta(2^n), \qquad a = 0.
\label{eq:mottonen_complexity}
\end{equation}
This is the as-built count ($124$ at $n=6$). The boundary-merged bound $2^{n+1}-2n$~\cite{bergholm2005} is lower but is not realized by the as-built cascade benchmarked here (\S\ref{sec:scaling}). The depth is $\Theta(2^n)$ because the cascades are inherently sequential: the deepest UCR alone has depth $\Theta(2^{n-1})$.

\paragraph{Strengths and weaknesses.} The method is exact, fully general (handles arbitrary complex states), ancilla-free, and conceptually transparent, which makes it the canonical baseline. Its weaknesses are the factor-of-four gap to the lower bound~\eqref{eq:cnot_lb}, its strictly sequential $\Theta(2^n)$ depth, and the phase-conditioning issue of \S\ref{sec:stability}. It is available as an amplitude-embedding routine in PennyLane~\cite{mottonen2004, bergholm2005} and is reimplemented from its construction for the benchmarks of this thesis.

\subsection{Qiskit \texttt{StatePreparation} / Iten Isometry}
\label{sec:an_iten}

\paragraph{Construction.} Iten's method treats state preparation as the $m=0$ isometry~\eqref{eq:isometry} and clears the single occupied column using uniformly controlled gates and multi-controlled rotations~\cite{iten2016quantum}. Structurally the circuit is the same disentangling cascade as M\"ott\"onen, but the angle bookkeeping is reorganized to remove rotations that are provably redundant across successive columns.

\paragraph{Complexity.} The reorganization merges the amplitude and phase reductions into uniformly controlled single-qubit gates, lowering the leading constant to
\begin{equation}
G_{\text{CNOT}} = 2^{n} - n - 1 = \Theta(2^n), \qquad d = \Theta(2^n), \qquad a = 0~\text{(optional)},
\label{eq:iten_complexity}
\end{equation}
roughly a factor of two below the as-built M\"ott\"onen cascade~\eqref{eq:mottonen_complexity} ($57$ against $124$ at $n=6$) and within a small factor of the optimum~\eqref{eq:cnot_lb}. This merged count is the one Qiskit's deployed \texttt{StatePreparation} realizes. The from-scratch Iten-style cascade benchmarked in Chapter~\ref{ch:empirical} applies optimization-level-3 peephole passes rather than the full merge and realizes the looser $\approx 2^{n+1}$ instead (\S\ref{sec:scaling}).

\paragraph{Practical compiler behavior.} Because it is exact, general, and near-optimal, this decomposition is the engine behind Qiskit's \texttt{StatePreparation}/\texttt{Initialize}~\cite{qiskit2024}. In practice the transpiler then re-expresses the abstract circuit in the device basis and inserts SWAP networks for the coupling map, so the realized transpiled depth $d_{\text{T}}$ can substantially exceed the logical depth~\eqref{eq:iten_complexity}, especially on sparsely connected hardware.

\subsection{General Isometry Decomposition}
\label{sec:an_isometry}

\paragraph{State preparation versus isometry synthesis.} An isometry $V:\mathbb{C}^{2^m}\!\to\mathbb{C}^{2^n}$ specifies $2^m$ orthonormal columns. State preparation is the single-column case $m=0$. Clearing $2^m$ columns instead of one multiplies the work, giving the interpolating cost
\begin{equation}
G_{\text{CNOT}}^{\text{iso}}(m,n) = \Theta\!\big(2^{n+m}\big),
\label{eq:iso_complexity}
\end{equation}
which reduces to $\Theta(2^n)$ at $m=0$ and to the full-unitary $\Theta(4^n)$ at $m=n$~\cite{iten2016quantum}.

\paragraph{Resource overhead and usefulness.} Invoking the general isometry primitive purely to prepare a state wastes the column-clearing logic and incurs no benefit over the specialized $m=0$ path. Its value appears when the state-preparation step is genuinely \emph{embedded} in a larger transformation, for instance, loading data into a subspace that is subsequently rotated by an algorithmic block, so that the isometry $V$ encodes both the embedding and part of the computation in one synthesized object. Implementations include Qiskit \texttt{Isometry} and \texttt{UniversalQCompiler}~\cite{qiskit2024, iten2021introduction}.

\subsection{Quantum Shannon Decomposition}
\label{sec:an_qsd}

\paragraph{Construction.} QSD synthesizes an \emph{arbitrary} $n$-qubit unitary by recursive cosine--sine decomposition~\eqref{eq:csd} into multiplexed single-qubit gates and a central multiplexed $R_y$~\cite{shende2006}. A state is prepared by completing $\ket{\Psi}$ to any unitary $U$ with $U e_0 = \ket{\Psi}$ and synthesizing $U$.

\paragraph{Why it is overkill, and when it helps.} The optimized cost
\begin{equation}
G_{\text{CNOT}}^{\text{QSD}} = \tfrac{23}{48}\,4^{n} - \tfrac{3}{2}\,2^{n} + \tfrac{4}{3} = \Theta(4^n)
\label{eq:qsd_complexity}
\end{equation}
is \emph{quadratically} larger than the $\Theta(2^n)$ of the direct methods, because QSD constrains all $2^n$ columns when state preparation constrains only one. The remaining $2^n-1$ columns are pure overhead. Equation~\eqref{eq:qsd_complexity} is the diagonal-merged optimum, while the from-scratch benchmark of Chapter~\ref{ch:empirical} uses the constant-unoptimized recursion, whose as-built leading constant is $\approx 0.727\cdot4^n$ ($2976$ at $n=6$, Table~\ref{tab:complexity}) rather than the $\tfrac{23}{48}\approx0.479$ of~\eqref{eq:qsd_complexity}, a penalty of about $24$ to $52\times$ over the $\Theta(2^n)$ frontier depending on which frontier method it is set against. At $n=2$ this is the difference between three CNOTs and the single-CNOT optimum (\S\ref{sec:qsd_example}). QSD is nonetheless the right tool when the surrounding algorithm already requires a full unitary (e.g.\ block-encodings or simulation operators), in which case the state is prepared ``for free'' as the first column and the $\Theta(4^n)$ cost is amortized against the unitary that was needed anyway. For two-qubit blocks it coincides with the optimal three-CNOT construction of Vatan and Williams~\cite{vatan2004optimal}.

\subsection{Multiplexor-Based Preparation}
\label{sec:an_multiplexor}

\paragraph{Relationship to M\"ott\"onen.} A multiplexor is precisely a uniformly controlled rotation~\eqref{eq:ucr}. Multiplexor-based preparation is the M\"ott\"onen cascade exposed as a built-in compiler object rather than a hand-built gate sequence. Asymptotically the two are identical,
\begin{equation}
G_{\text{CNOT}} = \Theta(2^n), \qquad d = \Theta(2^n), \qquad a = 0,
\label{eq:mux_complexity}
\end{equation}
matching~\eqref{eq:mottonen_complexity}.

\paragraph{Library differences.} The practical distinction is the optimization layer. Compilers vary in how aggressively they fuse adjacent rotations, cancel CNOT pairs across multiplexor boundaries, and re-route for connectivity. The benchmark of this thesis realizes the route with Qiskit's uniformly controlled rotation gate objects and compiles them at optimization level~3, whose peephole passes fuse adjacent rotations and cancel CNOTs across multiplexor boundaries, so the realized count can dip below the nominal $2^{n+1}-4$ when multiplexors abut. These differences are immaterial to the asymptotics but can shift the constant factor measured empirically.

\subsection{Plesch--Brukner / SVD-Style Preparation}
\label{sec:an_plesch}

\paragraph{Recursive decomposition.} Plesch and Brukner exploit the bipartite Schmidt decomposition~\eqref{eq:schmidt}: split the register into halves $A$ and $B$, load the $\le 2^{n/2}$ Schmidt coefficients on $A$, copy them onto $B$ with a CNOT ladder, and rotate both halves into the Schmidt bases with two $\tfrac{n}{2}$-qubit unitaries~\cite{plesch2011}. The half-size local unitaries can themselves be synthesized recursively (e.g.\ via QSD on $n/2$ qubits).

\paragraph{Expected savings.} Because the local unitaries act on only $n/2$ qubits, each costs $\tfrac{23}{48}4^{n/2}=\tfrac{23}{48}2^n$ via QSD rather than $O(4^n)$, so the two local unitaries plus the linear entangling ladder yield
\begin{equation}
G_{\text{CNOT}} \approx 2^{n} = \Theta(2^n), \qquad d = \Theta(2^n), \qquad a = 0,
\label{eq:plesch_complexity}
\end{equation}
a small constant factor above the lower bound~\eqref{eq:cnot_lb}. This $\tfrac{23}{48}2^{n+1}$ figure (the two local unitaries summed) is an optimistic estimate that assumes the local unitaries are synthesized at the QSD optimum. The looser estimate~\eqref{eq:plesch_cost} of Chapter~\ref{ch:litreview} is $\approx 2^{n+1}$, and the reference implementation of Chapter~\ref{ch:empirical}, whose local blocks use the unoptimized recursion, realizes a leading constant near $1.5\cdot 2^{n}$, between the two. All three are $\Theta(2^{n})$ and differ only by a constant factor. The construction is the most transparent link between a state's entanglement (Schmidt rank $\chi$) and its cost: in principle low-rank states ($\chi \ll 2^{n/2}$) are dramatically cheaper, since only $\chi$ coefficients need loading. As \S\ref{sec:plesch_example} showed, at $n=2$ the rank-$2$ structure makes the method \emph{optimal} at a single CNOT. Realizing that saving, however, requires a truncating implementation that reads the Schmidt rank and loads only $\chi$ coefficients. The non-truncating reference implementation benchmarked in Chapter~\ref{ch:empirical} does not: at $n=6$ on real targets its CNOT count stays nearly flat across the Schmidt rank (about $75$ at $\chi=1$ rising to $81$ at full rank), so it is rank-blind and captures none of the low-rank reduction. A rank-adaptive Plesch--Brukner variant, implemented but outside the main benchmark set, does capture it, dropping to about $24$ CNOTs at $\chi=1$, with its systematic low-rank evaluation left to future work (\S\ref{sec:future}).

\paragraph{Numerical structure.} Routing data through the SVD carries the backward stability of LAPACK (the Linear Algebra Package)~\cite{golub2013matrix}, giving Plesch--Brukner a well-conditioned angle pipeline (\S\ref{sec:stability}). This is a conditioning argument, not a measured accuracy gap: on the benchmark targets of Chapter~\ref{ch:empirical} every method, Plesch--Brukner included, reproduces the target to machine precision (\S\ref{sec:exactness}).

\subsection{QsiHT Fast Path Real Synthesis}
\label{sec:an_qsiht}

\paragraph{Construction.} The Real Synthesis~\cite{gomez2025qsiht, grigoryan2014heap} prepares a real (sign-bearing) unit target with a single cascade of uniformly controlled $R_y$ rotations~\eqref{eq:ucr} and no other gate type. As Chapter~\ref{ch:litreview} derives, it inverts the Digital Signal-induced Heap Transform (DsiHT), a butterfly of Givens rotations that maps the amplitude vector to $(\lVert\boldsymbol{\alpha}\rVert, 0, \dots, 0)$, but truncates each Gray-code stage's trailing CNOT and folds the resulting basis permutations, by an involution, into a fan-out-reordered signal $\chi = P\boldsymbol{\alpha}$ whose ordinary DsiHT angles then drive the cascade~\eqref{eq:qsiht_real_complexity}. The $2^n-1$ rotation angles come from the heap as $\arctan2$ of successive amplitude pairs, a $\Theta(N)$ heap, and are mapped to the cascade by a fast Walsh--Hadamard transform in $O(N\log N)$ time (the same angle map the cascade baselines compute densely at $O(N^2)$). Because the amplitude signs ride entirely in these $R_y$ angles, no separate $R_z$ phase cascade is incurred and the realized circuit contains only $R_y$ and CNOT gates.

\subsection{QsiHT Fast Path Complex Synthesis}
\label{sec:an_qsiht_complex}

\paragraph{Construction.} The Complex Synthesis~\cite{gomez2025qsiht, grigoryan2014heap} prepares an arbitrary complex target with a single cascade of uniformly controlled $U(2)$ gates and no separate diagonal. As Chapter~\ref{ch:litreview} derives, each complex DsiHT butterfly maps an amplitude pair to a real non-negative survivor through a $\mathrm{ZYZ}$-form $U(2)$ whose central $R_y$ angle is the same magnitude angle the real DsiHT produces, so the survivor stays real and the recursion is self-similar~\eqref{eq:qsiht_complex_complexity} (real inputs reduce to $\pm R_y$, so this strictly generalizes the Real Synthesis). Disentangling from the least significant qubit lets each stage's residual phases be absorbed blockwise into the next stage's $U(2)$ gates at no gate cost, so the only leftover phase acts on $\ket{0}^{\otimes n}$ as a global phase and no diagonal phase gate ever appears. The $2^n-1$ block angles are again obtained from the $\Theta(N)$ heap and mapped by the fast Walsh--Hadamard transform.

\paragraph{Complexity.} Both syntheses emit the same cascade cost, so a single equation covers them. A uniformly controlled $U(2)$ gate, or in the real case a uniformly controlled $R_y$, with $k$ controls compiles to $2^k-1$ CNOTs up to a diagonal that merges into the next stage~\cite{bergholm2005}, and summing the CNOT-bearing stages over the control counts $k=1,\dots,n-1$ (the top-qubit gate needing none) gives
\begin{equation}
G_{\text{CNOT}} = \sum_{k=1}^{n-1}\!\big(2^k-1\big) = 2^{n}-n-1 = \Theta(2^n), \qquad d = 2^{n+1}-2n-1 = \Theta(2^n), \qquad a = 0,
\label{eq:qsiht_complexity}
\end{equation}
with $2^n-1$ single-qubit rotations and the depth $d$ counted in the two-qubit-native $\{u,\mathrm{cx}\}$ basis (lowering the $U(2)$ blocks to elementary rotations expands the depth by a constant factor without changing the CNOT count). This count holds for the Real Synthesis (real targets, $R_y$ only) and the Complex Synthesis (arbitrary targets, $U(2)$) alike, $57$ CNOTs at $n=6$ for both, and places the fast path on the same $\Theta(2^n)$ frontier as the direct methods. The blockwise phase absorption of the Complex Synthesis is what keeps it at $2^n-n-1$ rather than paying a separate diagonal: absorbing in the preparation direction would strand the last diagonal on a full state and cost $2^n-2$ instead.

\paragraph{Placement against the floor and the lower bound.} The count $2^n-n-1$ is the shared optimum of the multiplexor family, not a QsiHT-only result. It is exactly the count Qiskit's deployed \texttt{StatePreparation} realizes and the theoretical $m{=}0$ isometry bound gives (\S\ref{sec:an_iten}), so the two syntheses tie the deployed floor rather than beat it. The benchmarked General-isometry configuration, however, aliases the compiler-optimized Iten builder and records $120$ CNOTs on the real target and $122$ on the complex target at $n=6$, not $57$. The benchmark of Chapter~\ref{ch:empirical} confirms the two QsiHT counts are as-built and optimization-level independent: re-transpiling at optimization level~3 leaves them unchanged at every $n$ in $3,\dots,10$ (\S\ref{sec:scaling}). Against the parameter-counting lower bound $\lceil\tfrac12(2^n-n-1)\rceil$ ($29$ at $n=6$), the count sits a factor of about $2$ above ($57/29 = 1.97$ at $n=6$), which is where every known exact ancilla-free construction sits, and it saturates the bound at $n=2$, where $2^n-n-1 = 1$ equals the single-CNOT optimum~\eqref{eq:cnot_lb}. Three caveats bound this placement. First, $2^n-n-1$ is a family optimum, the best a uniformly controlled cascade attains, and not the problem's lower bound, so meeting it certifies no two-sided optimality. Second, Schmidt-rank constructions of the Plesch--Brukner type~\cite{plesch2011} run asymptotically a few percent below it, so the fast path does not hold the smallest known constant. Third, the Real Synthesis reaches this count only on real targets, since its $R_y$-only cascade cannot carry arbitrary phases, whereas the Complex Synthesis reaches it on any target.

\paragraph{Numerical structure.} Both syntheses route amplitude pairs through the two-argument $\arctan2$, which is well-conditioned even when an amplitude vanishes (it returns $\pm\pi/2$ rather than dividing by a small number). This is the numerically correct form, and it is the one every direct method in the benchmark uses: the shipped M\"ott\"onen and isometry angle passes compute their amplitude angles through the same two-argument arctangent, not a naive ratio, so $\arctan2$ conditioning is a shared implementation choice rather than a QsiHT discriminator (\S\ref{sec:stability}). The classical pipelines do not share an asymptotic order: Plesch--Brukner's SVD acts on the $\sqrt{N}\times\sqrt{N}$ reshape of the amplitude vector, an $O(N^{1.5})$ computation, the cascade baselines compute their Walsh--Hadamard angle map as a dense $O(N^2)$ pass, and the QsiHT route runs a $\Theta(N)$ heap of $2^n-1$ Givens rotations followed by that same Walsh--Hadamard map done fast at $O(N\log N)$, all completing in milliseconds at the benchmark sizes, and no classical-cost ranking among the frontier methods is claimed (Chapter~\ref{ch:empirical}). The Complex Synthesis carries the per-amplitude phases in the $\mathrm{ZYZ}$ angles of its $U(2)$ blocks rather than in a separate diagonal, so those phase angles pass through the same $\arg$ evaluation and share the phase-conditioning sensitivity of the direct methods. On the benchmark targets the conditioning of any of these pipelines produces no measurable accuracy separation, since every method is exact to machine precision (\S\ref{sec:exactness}).

\section{Expected Complexity Comparison}
\label{sec:complexity_table}

Table~\ref{tab:qualitative} summarizes the qualitative and practical criteria, and Table~\ref{tab:complexity} the asymptotic complexities. The unifying observation is that \emph{every} ancilla-free exact method shares the $\Theta(2^n)$ CNOT and depth scaling mandated by the parameter count $2^{n+1}-2$ of an arbitrary state, except QSD, which pays an extra factor of $2^n$ for synthesizing an entire unitary. The methods are therefore distinguished not by asymptotic order but by (i) the leading constant, (ii) numerical conditioning of the classical angle pipeline, which does not surface as measured accuracy on the Haar families (Table~\ref{tab:qualitative}), (iii) sensitivity to entanglement structure, and (iv) the availability of an ancilla--depth trade-off. Regarding the last point, depth $\Theta(2^n/n)$ is achievable, and optimal, \emph{without any ancillae}, and only the further reduction to depth $\Theta(n)$ spends $O(2^n)$ ancilla qubits~\cite{sun2023asymptotically}. The methods compared here all realize depth $\Theta(2^n)$, a factor of $n$ above the ancilla-free optimum, including the QsiHT fast path. They sit at the qubit-frugal, depth-heavy end of that trade-off.

\begin{table}[H]
\centering
\caption{Qualitative and practical comparison of exact dense-state preparation methods. This table once carried conditioning tiers, and later graded residuals, in a stability column. Both are gone because neither discriminates on the benchmark families: on the real Haar targets of Chapter~\ref{ch:empirical} every builder's measured statevector $2$-norm error at $n=6$ lies below $10^{-13}$ (the largest, QSD's, is $2.1\times10^{-14}$), reflecting each implementation's numerical path rather than any property of the loader, so among these builders numerical stability is not an independent discriminator \emph{on that family}. The measured stability separations appear off the Haar family, on near-degenerate and near-sparse (interior-zero) inputs, where the four builders whose elementary $u$/\textsc{cx} cascade is handed to the opt-3 peephole (Iten, the $m{=}0$ isometry, the multiplexor, and Qiskit \texttt{StatePreparation}) leave residuals up to about $1.2\times10^{-6}$ near the degeneracy, that ceiling set by the Iten/isometry pair, while the six as-built rows (both QsiHT syntheses, both M\"ott\"onen variants, Plesch--Brukner, and QSD) hold an off-grid maximum of about $5\times10^{-13}$ (\S\ref{sec:exactness}). The rows are benchmark configurations spanning the construction families, not independent reimplementations: the Iten row is the M\"ott\"onen UCR cascade after Qiskit optimization level~3, the empirical General-isometry row shares that builder, the multiplexor row uses Qiskit uniformly controlled rotation primitives, and the QsiHT rows use the project's reference implementation. The empirical study separately includes Qiskit's deployed \texttt{StatePreparation} library baseline (\S\ref{sec:implementations}).}
\label{tab:qualitative}
\begin{tabular}{lccl}
\hline
\textbf{Method} & \textbf{Exact} & \textbf{Arb.\ complex} & \textbf{Reference construction} \\
\hline
M\"ott\"onen (UCR)      & Yes & Yes & UCR cascade \\
Iten / Qiskit           & Yes & Yes & opt-3 UCR cascade \\
General isometry        & Yes & Yes & $m{=}0$ isometry \\
Quantum Shannon         & Yes & Yes & recursive cosine--sine \\
Multiplexor-based       & Yes & Yes & Qiskit UC-rotations \\
Plesch--Brukner / SVD   & Yes & Yes & Schmidt / SVD \\
QsiHT Real Synthesis    & Yes & No  & DsiHT heap, $R_y{+}$CX \\
QsiHT Complex Synthesis & Yes & Yes & DsiHT heap, $U(2)$ \\
\hline
\end{tabular}
\end{table}

\begin{table}[H]
\centering
\caption{Asymptotic resource complexity of exact dense-state preparation methods, giving the as-built leading CNOT constant on real and complex targets. Here $N=2^n$. The lower-bound row is the parameter-counting floor $\lceil\tfrac12(2^n-n-1)\rceil$, the same value for complex targets~\eqref{eq:cnot_lb} and for real targets under real-rotation circuits~\eqref{eq:cnot_lb_real} ($29$ at $n=6$). A weaker generic-gate floor $\lceil\tfrac14(2^n-2n-1)\rceil$ ($13$ at $n=6$) applies if arbitrary single-qubit gates are allowed. The lower-bound row's depth entry $\Omega(2^n/n)$ is the ancilla-free optimum of~\cite{sun2023asymptotically}. The M\"ott\"onen count $2^{n+1}-4$ ($124$ at $n=6$) is the as-built two-cascade value benchmarked here. The boundary-merged $2^{n+1}-2n$ is a cited theoretical bound not realized by that cascade. The Iten-style cascade and multiplexor reconstructions realize $\approx 2^{n+1}$ ($120$ and $120$--$122$ at $n=6$), while the merged-multiplexor construction deployed as Qiskit \texttt{StatePreparation} realizes $2^n-n-1$ ($57$ at $n=6$). The two QsiHT syntheses each emit $2^n-n-1$ ($57$ at $n=6$), tying the Qiskit \texttt{StatePreparation} row and sitting a factor $\approx 1.97$ above the floor ($57/29$). The Real Synthesis reaches this count on real targets with an $R_y{+}$CX cascade, so its complex cell (--) is not applicable, while the Complex Synthesis reaches it on any target. The Plesch--Brukner cells are the as-built constant of the non-truncating reference implementation, $\approx 1.5\cdot2^n$ and parity-dependent in $n$ ($81$ at $n=6$, not falling with rank). The Quantum Shannon cells are the as-built constant of the constant-unoptimized recursion, $\approx 0.727\cdot4^n$ ($2976$ at $n=6$, opt-3 $2208$), a penalty of about $24$ to $52\times$ over the frontier, rather than the diagonal-merged optimum $\tfrac{23}{48}4^n$ of~\eqref{eq:qsd_complexity}. $^{\dagger}$The general-isometry row is analytic only: the benchmarked $m{=}0$ instance is literally the Iten build (\S\ref{sec:implementations}), so it contributes no distinct measured circuit.}
\label{tab:complexity}
\resizebox{\textwidth}{!}{%
\begin{tabular}{lcccc}
\hline
\textbf{Method} & \textbf{$G_{\text{CNOT}}$ real} & \textbf{$G_{\text{CNOT}}$ complex} & \textbf{Order} & \textbf{Depth} \\
\hline
Lower bound                & $\lceil\tfrac12(2^n{-}n{-}1)\rceil$ & $\lceil\tfrac12(2^n{-}n{-}1)\rceil$ & $\Theta(2^n)$ & $\Omega(2^n/n)$ \\
M\"ott\"onen (UCR)         & $2^{n+1}-4$          & $2^{n+1}-4$          & $\Theta(2^n)$ & $\Theta(2^n)$ \\
M\"ott\"onen (sign-aware)  & $2^{n}-2$            & $2^{n+1}-4$          & $\Theta(2^n)$ & $\Theta(2^n)$ \\
Iten-style cascade (opt-3) & $\approx 2^{n+1}$    & $\approx 2^{n+1}$    & $\Theta(2^n)$ & $\Theta(2^n)$ \\
Qiskit \texttt{StatePreparation} & $2^{n}-n-1$    & $2^{n}-n-1$          & $\Theta(2^n)$ & $\Theta(2^n)$ \\
Multiplexor (Qiskit UCR)   & $\approx 2^{n+1}$    & $\approx 2^{n+1}$    & $\Theta(2^n)$ & $\Theta(2^n)$ \\
QsiHT Real Synthesis       & $2^{n}-n-1$         & --                  & $\Theta(2^n)$ & $\Theta(2^n)$ \\
QsiHT Complex Synthesis    & $2^{n}-n-1$         & $2^{n}-n-1$         & $\Theta(2^n)$ & $\Theta(2^n)$ \\
Plesch--Brukner / SVD      & $\approx 1.5\cdot 2^{n}$ & $\approx 1.5\cdot 2^{n}$ & $\Theta(2^n)$ & $\Theta(2^n)$ \\
General isometry ($m$)$^{\dagger}$ & $\Theta(2^{n+m})$    & $\Theta(2^{n+m})$    & $\Theta(2^{n+m})$ & $\Theta(2^{n+m})$ \\
Quantum Shannon            & $\approx 0.727\cdot 4^{n}$ & $\approx 0.727\cdot 4^{n}$ & $\Theta(4^n)$ & $\Theta(4^n)$ \\
\hline
\end{tabular}}
\end{table}

Overall, the theoretical analysis predicts that the direct, isometry, multiplexor, Plesch--Brukner, and QsiHT methods will be competitive at the $\Theta(2^n)$ frontier. Plesch--Brukner should enjoy the best constant among the from-scratch constructions on complex data, running a few percent below the multiplexor-family floor. The two QsiHT syntheses should meet that floor exactly: both should realize $2^n-n-1$ CNOTs and depth $2^{n+1}-2n-1$ at every $n$~\eqref{eq:qsiht_complexity}, tying the deployed Qiskit \texttt{StatePreparation} and the $m{=}0$ isometry on both counts rather than lying above or below them, with a well-conditioned $\arctan2$ angle pipeline shared with the other direct methods. Two observables should distinguish the fast path within that tie: the Real Synthesis should build its real-target circuit from an $R_y{+}$CX gate set with no $R_z$ stage, and both syntheses should saturate the parameter-counting lower bound~\eqref{eq:cnot_lb} at $n=2$, where $2^n-n-1 = 1$. Neither count is optimal in a two-sided sense, since $2^n-n-1$ is a family optimum roughly a factor of two above the parameter-counting floors~\eqref{eq:cnot_lb} and~\eqref{eq:cnot_lb_real}, not the problem's lower bound. The general isometry primitive is justified only when state preparation is part of a larger transformation, and QSD should be reserved for cases where a full unitary is independently required. These predictions are tested empirically against the gate, depth, and fidelity metrics of Chapter~\ref{ch:background} in the chapters that follow.

%% file: chapters/chapter5.tex
\chapter{Empirical Evaluation}
\label{ch:empirical}

This chapter turns the \emph{a priori} predictions of Chapter~\ref{ch:theory} into measured quantities. The theoretical comparison forecast that the direct, isometry, multiplexor, Plesch--Brukner, and QsiHT methods would share the $\Theta(2^n)$ frontier and differ only in their leading constants, that the general isometry primitive would be justified only when embedded in a larger transformation, and that the Quantum Shannon Decomposition (QSD) would pay an extra $\Theta(2^n)$ penalty for synthesizing an entire unitary. This chapter tests those forecasts by implementing every method, applying each to identical dense target states, and measuring the gate, depth, fidelity, and sampled-error metrics defined in Chapter~\ref{ch:background}. The two state-preparation reductions built on the Quantum Signal-induced Heap Transform fast path, the Real Synthesis and the Complex Synthesis, are benchmarked here against the ancilla-free exact dense methods of Chapter~\ref{ch:litreview} on the same baseline~\cite{grigoryan2025stateprep}. The scope caveat of Section~\ref{sec:related_out_of_scope} applies throughout: the youngest of these methods dates to 2016, and the ancilla-assisted, sparse, and approximate lines of work are not represented.

\section{Experimental Methodology}
\label{sec:methodology}

\subsection{Implementations}
\label{sec:implementations}

To keep the comparison transparent, Table~\ref{tab:implementations} treats the rows as benchmark configurations spanning the UCR, cosine--sine, Schmidt, isometry, multiplexor, and heap-transform construction families, not as independent from-scratch reimplementations. The direct UCR, cosine--sine, Schmidt, and heap-transform paths are exposed in NumPy and elementary Qiskit gates, while several practical configurations deliberately use shared or Qiskit-provided realizations: the Iten row is the M\"ott\"onen cascade after Qiskit optimization level~3, the General-isometry row calls that same builder, and the multiplexor row uses Qiskit's uniformly controlled rotation primitives. The two QsiHT rows, the Fast Path Real Synthesis and the Fast Path Complex Synthesis, are the thesis' own implementation. Qiskit's deployed \texttt{State\-Preparation} object is benchmarked \emph{as shipped} as a separate library baseline (\S\ref{sec:scaling}). The eight evaluated method configurations are summarized in Table~\ref{tab:implementations}. A sign-aware M\"ott\"onen variant and the deployed \texttt{State\-Preparation} bring the tables to ten builder rows. The correctness of the vendored syntheses is carried instead by the test suite, which checks them directly for unit fidelity on every applicable target and for their predicted gate counts.

\begin{table}[H]
\centering
\caption{State-preparation configurations evaluated in this chapter. The rows span construction families and combine direct NumPy/Qiskit implementations~\cite{harris2020numpy, qiskit2024} with practical Qiskit realizations. The Iten-style row is the M\"ott\"onen cascade re-optimized at compiler optimization level~3, the General-isometry row shares that builder, and the multiplexor route delegates to Qiskit's uniformly controlled rotation primitives. Qiskit's \texttt{StatePreparation} object is benchmarked separately as a deployed-library baseline. The two QsiHT syntheses are the main methods of this thesis, taken from the \texttt{fastpath\_synthesis} reference implementation.}
\label{tab:implementations}
\begin{tabular}{lll}
\hline
\textbf{Label} & \textbf{Construction} & \textbf{Reference} \\
\hline
QsiHT Fast Path Real Synthesis & $R_y$-only cascade, trailing CNOTs deferred & \S\ref{sec:qsiht_real} \\
QsiHT Fast Path Complex Synthesis & $U(2)$ cascade, phases absorbed blockwise & \S\ref{sec:qsiht_complex} \\
M\"ott\"onen (UCR) & Gray-code $R_y$/$R_z$ cascade, as built & \S\ref{sec:mottonen} \\
Iten-style cascade (opt-3) & UCR cascade, compiler-optimized & \S\ref{sec:iten} \\
Multiplexor       & Qiskit UC primitives + compiler & \S\ref{sec:multiplexor} \\
General isometry   & Single-column ($m{=}0$) isometry & \S\ref{sec:general_isometry} \\
Plesch--Brukner   & Schmidt/SVD + CNOT ladder + local QSD & \S\ref{sec:plesch} \\
Quantum Shannon   & Recursive cosine--sine (SciPy~\cite{virtanen2020scipy}) & \S\ref{sec:qsd} \\
\hline
\end{tabular}
\end{table}

The shared building block is the Gray-code uniformly controlled rotation~\eqref{eq:ucr}: a UCR with $k$ controls is emitted as $2^k$ single-qubit rotations interleaved with $2^k$ CNOTs, the physically applied angles being the Walsh--Hadamard transform~\eqref{eq:walsh} of the per-pattern angles. The M\"ott\"onen backend builds the amplitude ($R_y$) and phase ($R_z$) cascades directly from this primitive. A UCR whose angles all vanish is omitted entirely, so a phase-free target never pays for the $R_z$ cascade. The Iten and multiplexor backends represent the same construction as realized in practice: the Iten-style cascade row is the M\"ott\"onen cascade after the Qiskit transpiler's post-optimization (optimization level~3), an author reconstruction of the compiler-optimized cascade rather than a call to any library loader (Qiskit's deployed \texttt{StatePreparation}, built on the merged-multiplexor construction with cost $2^n-n-1$, is measured separately as a baseline), while the multiplexor row builds the cascade from Qiskit's uniformly controlled rotation \emph{gate objects} and lets the compiler choose their decomposition. As established in Chapter~\ref{ch:litreview}, the general isometry primitive at $m{=}0$ recovers the Iten single-column method exactly, so it shares that realization. The ten builder rows are benchmark configurations rather than ten independent constructions, and the General-isometry row is a copy of the Iten build rather than an independent measurement. The QSD backend applies SciPy's cosine--sine decomposition~\eqref{eq:csd} recursively about the most-significant qubit, demultiplexing the block-diagonal factors through the eigen-identity $\mathrm{blockdiag}(A,B)=(\mathbb{I}\otimes V)\,\mathrm{blockdiag}(D,D^\dagger)\,(\mathbb{I}\otimes W)$ down to a single-qubit $Z$--$Y$--$Z$ (ZYZ) base case. Here $A$ and $B$ are the two lower-register blocks of the top-qubit-controlled multiplexor, $V$ and $W$ are the unitaries that diagonalize $AB^\dagger$ and $B^\dagger A$, $D$ is the diagonal matrix of eigenphases, and $\mathbb{I}$ is the single-qubit identity acting on the control qubit. It is the standard (constant-unoptimized) recursion, so its leading constant sits above the optimized $\tfrac{23}{48}4^n$ of~\eqref{eq:qsd_cost} while retaining the characteristic $\Theta(4^n)$ order. The Plesch--Brukner backend computes the Schmidt decomposition by SVD of the reshaped amplitude matrix, loads the Schmidt coefficients on the upper register with the M\"ott\"onen cascade, copies them with a CNOT ladder, and synthesizes the two local Schmidt-basis unitaries with the from-scratch QSD. The two QsiHT syntheses produce their $2^n-1$ rotation angles with the DsiHT, a $\Theta(N)$ heap of Givens rotations, and map them to the uniformly controlled cascade through a fast Walsh--Hadamard transform ($O(N\log N)$, the same angle map the cascade baselines compute densely at $O(N^2)$). The Real Synthesis applies each stage as a uniformly controlled $R_y$ (Givens) rotation on the corresponding bit plane with one entangler per stage deferred, giving an $R_y{+}$CX circuit with no phase stage on real targets. The Complex Synthesis carries arbitrary phases by promoting each stage to a uniformly controlled $U(2)$ and absorbing the residual diagonal into the next stage's blocks, so both syntheses land at $2^n-n-1$ CNOTs without a separate diagonal gate (\S\ref{sec:an_qsiht}).

The software stack is Qiskit~2.4.2 with Qiskit Aer~0.17.2 for simulation, NumPy~2.4 and SciPy~1.17 for the classical linear algebra, executed under Python~3.12. Every transpiler invocation in the benchmark, including the optimization-level-3 passes inside the compiler-optimized builders, carries a pinned seed, so all structural tables in this chapter are exactly reproducible on this software stack. Structural metrics and the sampled-error analysis are computed with the thesis' \texttt{measurement} python package so that every method is scored by an identical, independently specified procedure.

\subsection{Target States}
\label{sec:targets}

Two families of target states are used. The first is the Chapter~\ref{ch:litreview} worked example $\ket{\Psi}=\tfrac{1}{\sqrt{30}}(\ket{00}+2\ket{01}+3\ket{10}+4\ket{11})$ of~\eqref{eq:running}, which anchors the empirical counts to the hand-derived predictions. The second is a family of \emph{Haar-random dense} states: for each qubit count $n\in\{2,\dots,6\}$, an amplitude vector is drawn from the complex (or real) standard normal distribution and normalized, giving a uniformly random pure state with support $s=2^n$. Both a real and a complex random family are used, with fixed seeds for reproducibility, because the phase handling of every method is exercised only by complex (or sign-bearing) amplitudes, and the real/complex gap is exactly where the methods are predicted to separate.

\subsection{Metrics}
\label{sec:metrics_used}

For each (method, target) pair the following are recorded, all per the definitions of Chapter~\ref{ch:background}:
\begin{itemize}
    \item \textbf{Exactness.} The state fidelity $F=\lvert\langle\Psi|\widetilde{\Psi}\rangle\rvert^2$ of~\eqref{eq:pure_fidelity} between the ideal target and the statevector the circuit actually produces, confirming that the synthesized unitary realizes~\eqref{eq:qsp_unitary} up to global phase.
    \item \textbf{Structural cost.} The CNOT count $G_{\text{CNOT}}$, total gate count $G$, and depth $d$, obtained by transpiling each circuit to the abstract basis $\{u,\textsc{cx}\}$ and counting elementary operations. The $\{u,\textsc{cx}\}$ basis isolates the two-qubit cost~\eqref{eq:cnot_lb} from single-qubit bookkeeping.
    \item \textbf{Hardware portability.} The transpiled depth $d_{\text{T}}$ and native two-qubit (entangling) count after compilation to each of the representative NISQ native gate sets of Table~\ref{tab:nisq_platforms}, revealing each method's sensitivity to the entangler and connectivity an actual device exposes.
    \item \textbf{Classical cost.} The elapsed real-world time to synthesize the circuit from the amplitude vector, capturing the angle-computation and decomposition overhead.
    \item \textbf{Sampled error.} The gap between the sampled and ideal output distributions is measured by the fidelity error and the mean-square error (MSE). Writing $p_x$ for the ideal probability and $\hat p_x$ for the sampled frequency over the $N=2^n$ basis states, the sampled (classical) fidelity is the squared Bhattacharyya overlap~\cite{bhattacharyya1943measure} $F=\big(\sum_x\sqrt{p_x\hat p_x}\big)^2$, so the fidelity error is the infidelity $1-F$ (cf.~\eqref{eq:infidelity}), and
    \begin{equation}
    1-F = 1-\Big(\textstyle\sum_x\sqrt{p_x\,\hat p_x}\Big)^{2},\qquad
    \mathrm{MSE}=\tfrac{1}{N}\textstyle\sum_x (p_x-\hat p_x)^2.
    \label{eq:error_metrics}
    \end{equation}
    Both are collected over $R=20$ independent seeded sampling runs and summarized by their mean. For the infidelity the run-to-run spread is additionally reported as the sample variance $\tfrac{1}{R-1}\sum_r (e_r-\bar e)^2$ about the mean $\bar e$ of the per-run values $e_r$. Errors are measured both noise-free (the finite-shot sampling floor) and under a noisy backend (tying the realized error to the gate count via the fidelity-decay model~\eqref{eq:fidelity_decay}).
\end{itemize}
Sampling uses the Qiskit Aer \texttt{SamplerV2} primitive for the noise-free runs, and a \texttt{GenericBackendV2} noise model otherwise. To make the noisy comparison fair and reproducible, all methods are evaluated on a \emph{single seeded} noise model with the per-run simulator seeds paired across methods, so every method receives the same $20$ shot-noise draws and builders that emit the identical circuit print identical errors. Pairing removes the per-call randomness of a fresh noise model and shot-noise differences between methods. What remains is each circuit's interaction with that one calibration draw: gate count and depth foremost, but also the placement and readout assignment the draw happens to contain, which is why the calibration ensemble of \S\ref{sec:noise} is required before any fine ordering is read. A calibration-ensemble sweep (\S\ref{sec:noise}) additionally repeats the noisy experiment over $20$ independently seeded calibrations with shot seeds paired across methods. Where the fidelity-decay model~\eqref{eq:fidelity_decay} enters the analysis, its coefficient is fitted from the measured data rather than assumed: the $\epsilon_{\text{CNOT}}\sim10^{-2}$ of Chapter~\ref{ch:background}'s motivating estimate is a representative literature figure, while the hardware runs of \S\ref{sec:noise} give the realized slope directly. On the \texttt{ibm\_fez} real target the infidelity difference between the QSD circuit and the QsiHT Real Synthesis circuit, divided by their executed two-qubit gap ($34$ against $7$ routed gates, a $27$-gate span), is $(2.574\times10^{-2}-4.035\times10^{-3})/27\approx8.0\times10^{-4}$ per executed two-qubit gate, the shared state-preparation-and-measurement and idle contributions cancelling in the difference (on the abstract $32$-CNOT gap the same endpoints give $6.8\times10^{-4}$ per CNOT). This single hardware difference is a two-point consistency check, not a fitted slope, and because the two circuits were submitted on different calibration sessions (\S\ref{sec:hw_limitations}) it is only indicative. It sits about $12$ times below the representative literature rate $\epsilon_{\text{CNOT}}\sim10^{-2}$. A separate least-squares fit over the eight distinct simulated executed-count points, computed in \texttt{python/fidelity\_decay\_fit.py}, gives a consistent $7.0\times10^{-4}$ per gate at $R^2=0.98$, so the linear form of~\eqref{eq:fidelity_decay} is supported by an actual fit rather than by assertion, with a coefficient more than an order of magnitude below the representative literature rate, which softens, without removing, the loading-dominates-the-budget motivation of Chapter~\ref{ch:background}.

\section{Exactness Verification}
\label{sec:exactness}

Every method, on every applicable target, reproduces the intended state up to global phase: across all ten builders on the real family, the nine arbitrary-complex builders on the complex family, and $n=2,\dots,6$, the measured fidelity~\eqref{eq:pure_fidelity} equals $1.000000$ to six decimal places (and, on these two families, the prepared statevector matches the target to better than $10^{-9}$ in norm once the global phase is fixed, the two documented exceptions being the near-degenerate family of \S\ref{sec:stability} and the reproducibility suite's near-sparse (interior-zero) edge-case target, where the opt-3 peephole leaves $\varepsilon$-scale residuals: a fine-grid peak of about $1.2\times10^{-6}$ near the degeneracy, quantified below, and about $5.5\times10^{-9}$ on the near-sparse target, measured by the edge-case battery of the reproducibility suite, while on the degenerate family the six as-built rows hold a maximum of about $5\times10^{-13}$). This confirms the qualitative claims of Tables~\ref{tab:qualitative} and~\ref{tab:complexity}: all methods are exact and deterministic on their applicable target families, while the QsiHT Real Synthesis alone is restricted to real amplitudes.

Two refinements bound this statement. First, at full precision the statevector $2$-norm residuals are not literally zero: on the real Haar family at $n=6$ they range from about $6\times10^{-16}$ (the sign-aware M\"ott\"onen) to $2.1\times10^{-14}$ (QSD), all below $10^{-13}$. The fine grading reflects each implementation's numerical path, the SVD, the multiplexor resynthesis, the Walsh--Hadamard angle pass, and not the circuit length: two of the three shortest $57$-CNOT circuits, Qiskit \texttt{StatePreparation} and the QsiHT Complex Synthesis, carry the largest non-QSD residuals (about $1.0\times10^{-14}$ and $1.1\times10^{-14}$), while the equally short QsiHT Real Synthesis ($7.6\times10^{-16}$) and the $120$-CNOT cascades sit an order of magnitude lower, so the residual is not a property of the loader and does not discriminate among the methods. Second, the conditioning distinctions of \S\ref{sec:stability} do become measurable off the Haar families. Perturbing a target with an exactly degenerate Schmidt spectrum by $\varepsilon$ along a fixed random direction (a $401$-point grid in $\varepsilon$ from $10^{-13}$ to $10^{-2}$ at $n=6$), the four builders whose elementary $u$/\textsc{cx} cascade is handed to the opt-3 peephole (Iten, the $m{=}0$ isometry, the multiplexor, and Qiskit \texttt{StatePreparation}) leave residuals of order $\varepsilon$ near the degeneracy, the Iten/isometry pair peaking at about $1.2\times10^{-6}$ at an off-decade $\varepsilon\approx2.8\times10^{-7}$ that the fine grid resolves (Qiskit \texttt{StatePreparation} at about $4.2\times10^{-10}$, the multiplexor at about $4.7\times10^{-11}$), and recovering machine precision once the spectrum splits strongly, while the six as-built rows (both QsiHT syntheses, both M\"ott\"onen variants, Plesch--Brukner, and QSD) stay below $5\times10^{-13}$ at every $\varepsilon$. The same split reappears on the reproducibility suite's near-sparse (interior-zero) edge-case target, not on the perturbation grid of this section: there the peephole makes a gate-cancellation decision on the $10^{-8}$ near-support amplitude and leaves the four peephole builders residuals at that scale (about $5.5\times10^{-9}$ at worst), the Complex Synthesis shares the exception at the same scale because its butterfly absorbs phases through that amplitude, and the remaining as-built rows hold machine precision. The separation belongs to the compiler's peephole, which absorbs $\varepsilon$-scale structure, not to any method's angle pipeline, and the fidelity is second-order blind to it ($1-F\sim\varepsilon^2$, at most $\sim10^{-12}$ even at the $1.2\times10^{-6}$ peak, far below any finite-shot floor), so it can never surface in a sampled experiment. Numerical stability therefore discriminates only on near-degenerate and near-sparse inputs, and there it separates opt-3-peephole from as-built realizations rather than one loader from another. Exactness having been established uniformly, the remainder of the chapter concerns \emph{cost}, where the methods differ sharply.

\section{The Worked Two-Qubit Example}
\label{sec:running_results}

\begin{figure}[H]
\centering
\includegraphics[width=0.82\textwidth]{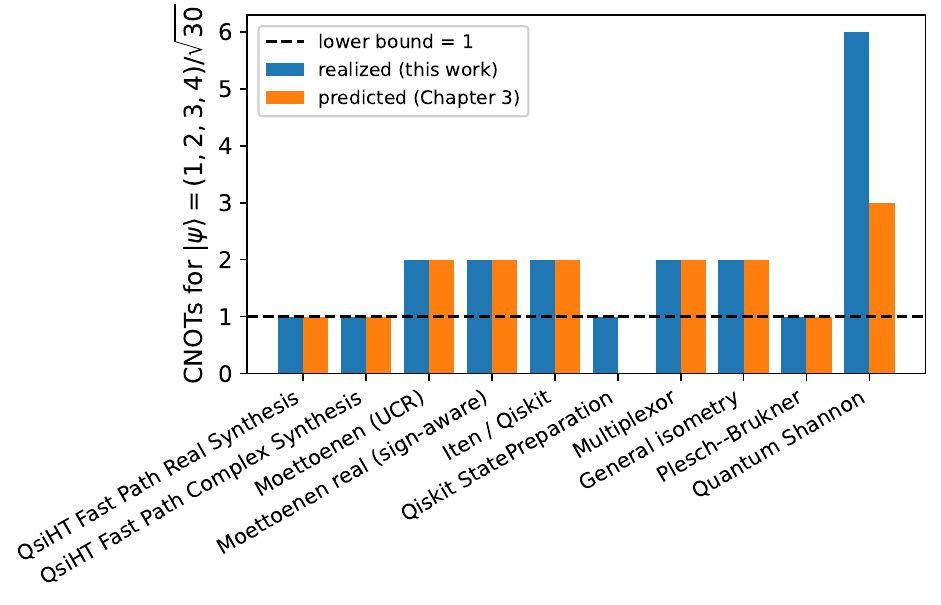}
\caption{Realized CNOT count on the running target~\eqref{eq:running} versus the theoretical prediction. The two QsiHT syntheses, Qiskit \texttt{StatePreparation}, and Plesch--Brukner attain the single-CNOT lower bound, the UCR family (M\"ott\"onen, Iten, multiplexor, isometry) meets the predicted two CNOTs, and the unoptimized QSD recursion uses six rather than three.}
\label{fig:running_example}
\end{figure}

Table~\ref{tab:running_results} reports the realized resource counts for the running target~\eqref{eq:running}, alongside the theoretical predictions of Tables~\ref{tab:method_cost} and~\ref{tab:complexity}. The agreement is exact across the frontier. The two QsiHT syntheses each realize a single CNOT at depth $3$, matching their predicted $2^n-n-1=1$~\eqref{eq:qsiht_complexity} and saturating the single-CNOT lower bound~\eqref{eq:cnot_lb} already at $n=2$, tied with the deployed Qiskit \texttt{StatePreparation} (also one CNOT) and with Plesch--Brukner, which reaches the bound by exploiting the guaranteed rank-2 Schmidt structure exactly as derived in~\S\ref{sec:plesch_example}. The UCR family, M\"ott\"onen, Iten, the multiplexor route, and the $m{=}0$ isometry, each realize two CNOTs. The QSD backend uses six CNOTs rather than the three of the optimized formula~\eqref{eq:qsd_cost}, the expected signature of the constant-unoptimized recursion, which does not perform the diagonal-merging that lowers the two-qubit case to the Vatan--Williams optimum~\cite{vatan2004optimal}. The two syntheses therefore sit at the lower-bound optimum already at $n=2$, and their behavior across register sizes is studied in~\S\ref{sec:scaling}. Figure~\ref{fig:running_example} visualizes the realized-versus-predicted counts.

\begin{table}[H]
\centering
\caption{Realized resources for the running target~\eqref{eq:running}, $\ket{\Psi}=\tfrac{1}{\sqrt{30}}(\ket{00}+2\ket{01}+3\ket{10}+4\ket{11})$, versus the theoretical prediction (Tables~\ref{tab:method_cost} and~\ref{tab:complexity}). The CNOT lower bound for $n=2$ is~$1$, met by the two QsiHT syntheses, Qiskit \texttt{StatePreparation}, and Plesch--Brukner (bold).}
\label{tab:running_results}
\resizebox{\textwidth}{!}{%
\begin{tabular}{lccccc}
\hline
\textbf{Method} & \textbf{Fidelity} & \textbf{CNOTs} & \textbf{Depth} & \textbf{Total gates} & \textbf{Predicted CNOTs} \\
\hline
QsiHT Fast Path Real Synthesis    & $1.000000$ & $\mathbf{1}$ & $3$ & $4$ & $\mathbf{1}$ \\
QsiHT Fast Path Complex Synthesis & $1.000000$ & $\mathbf{1}$ & $3$ & $4$ & $\mathbf{1}$ \\
M\"ott\"onen (UCR) & $1.000000$ & $2$ & $4$  & $5$  & $2$ \\
Iten-style cascade     & $1.000000$ & $2$ & $4$  & $5$  & $2$ \\
Multiplexor       & $1.000000$ & $2$ & $4$  & $5$  & $2$ \\
General isometry   & $1.000000$ & $2$ & $4$  & $5$  & $2$ \\
Qiskit \texttt{StatePreparation} & $1.000000$ & $\mathbf{1}$ & $3$ & $4$ & $\mathbf{1}$ \\
Plesch--Brukner   & $1.000000$ & $\mathbf{1}$ & $4$ & $6$ & $\mathbf{1}$ \\
Quantum Shannon   & $1.000000$ & $6$ & $21$ & $24$ & $3$ \\
\hline
\end{tabular}}
\end{table}

\section{Gate-Count and Depth Scaling}
\label{sec:scaling}

\begin{figure}[H]
\centering
\includegraphics[width=0.82\textwidth]{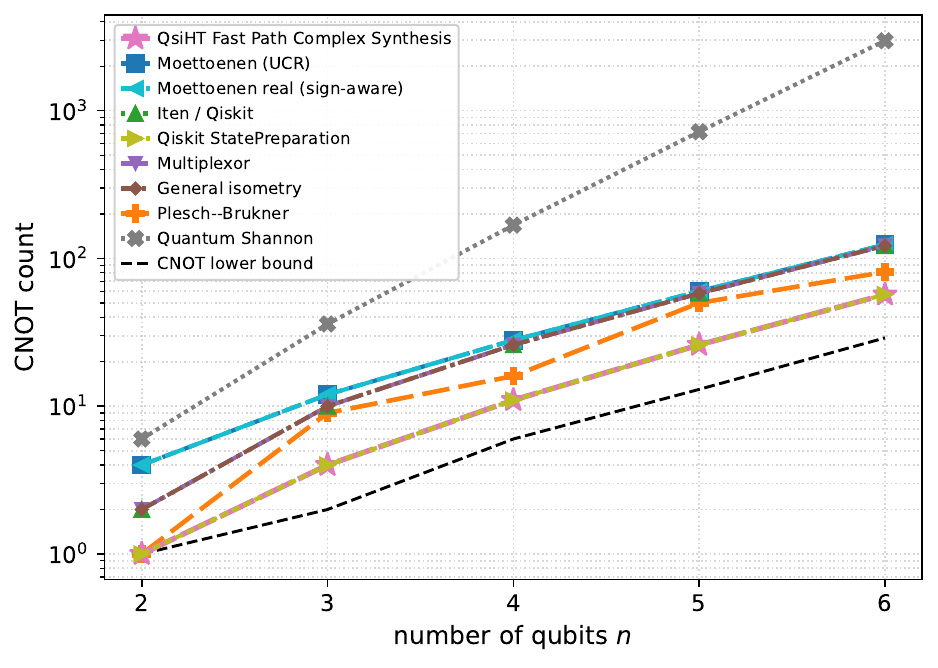}
\caption[CNOT count versus qubit count on complex Haar-random targets]{CNOT count versus qubit count $n$ (logarithmic axis) on complex Haar-random targets. The QsiHT Complex Synthesis (tied with Qiskit \texttt{StatePreparation} at $2^n-n-1$), the UCR family, and Plesch--Brukner occupy the $\Theta(2^n)$ frontier. Only QSD grows quadratically faster. The dashed line is the lower bound~\eqref{eq:cnot_lb}.}
\label{fig:cnot_scaling}
\end{figure}

Table~\ref{tab:cnot_scaling} reports the realized CNOT counts on the complex Haar-random family for $n=2,\dots,6$, with the lower bound~\eqref{eq:cnot_lb} for reference, and Figure~\ref{fig:cnot_scaling} plots them on a logarithmic axis. The QsiHT Complex Synthesis, the UCR family, and Plesch--Brukner all track the $\Theta(2^n)$ frontier, while QSD grows quadratically faster, reaching $2976$ CNOTs at $n=6$, a $24$-to-$52\times$ penalty over the frontier depending on which frontier method it is set against, an empirical confirmation of the $\Theta(4^n)$ penalty of synthesizing all $2^n$ columns when only one is constrained~\eqref{eq:qsd_complexity}. Within the frontier, the QsiHT Complex Synthesis realizes $2^n-n-1$ CNOTs~\eqref{eq:qsiht_complexity} ($57$ at $n=6$), tying the deployed \texttt{StatePreparation}'s merged-multiplexor route exactly and holding the lowest count above the bound. It sits a factor of about $1.97$ above the lower bound ($57/29$ at $n=6$), which is where the multiplexor family sits. The from-scratch builders are higher: the M\"ott\"onen cascade at $124$, the compiler-optimized Iten/multiplexor/isometry realizations at $122$, and Plesch--Brukner at $81$, its half-size local unitaries costing $O(4^{n/2})=O(2^n)$ rather than the cascade's full $O(2^n)$ with the larger prefactor (\S\ref{sec:an_plesch}). The $124$ read here for the M\"ott\"onen cascade is the as-built $2^{n+1}-4$ count~\eqref{eq:mottonen_cost}, not the lower $2^{n+1}-2n$ ($116$ at $n=6$) frequently quoted for M\"ott\"onen, which assumes boundary-CNOT cancellations between consecutive uniformly controlled cascades that the realized circuits do not perform and so is a cited theoretical bound rather than the count measured here. The synthesis count is optimization-level independent: re-transpiling the QsiHT circuits at optimization level~3 with a fixed seed leaves them unchanged at every $n$, the same as \texttt{StatePreparation}, whereas the pass compresses Plesch--Brukner from $81$ to $57$ and M\"ott\"onen from $124$ to $122$ at $n=6$. So the QsiHT Complex Synthesis meets the deployed floor at build time without an optimizer, and it undercuts every from-scratch construction on the as-built CNOT axis.

\begin{table}[H]
\centering
\caption{Realized CNOT count $G_{\text{CNOT}}$ on complex Haar-random targets, $n=2,\dots,6$, against the lower bound~\eqref{eq:cnot_lb}. Counts are after transpilation to the $\{u,\textsc{cx}\}$ basis. The QsiHT Complex Synthesis realizes $2^n-n-1$ ($57$ at $n=6$), tying the deployed Qiskit \texttt{StatePreparation} control on the merged-multiplexor construction and holding the lowest count above the bound (bold, tied minima marked). Both are optimization-level independent. The counts here are unchanged at optimization level~3.}
\label{tab:cnot_scaling}
\begin{tabular}{lrrrrr}
\hline
\textbf{Method} & $n{=}2$ & $n{=}3$ & $n{=}4$ & $n{=}5$ & $n{=}6$ \\
\hline
QsiHT Fast Path Complex Synthesis & $\mathbf{1}$ & $\mathbf{4}$ & $\mathbf{11}$ & $\mathbf{26}$ & $\mathbf{57}$ \\
M\"ott\"onen (UCR) & $4$  & $12$  & $28$  & $60$   & $124$  \\
Iten-style cascade     & $2$  & $10$  & $26$  & $58$   & $122$  \\
Qiskit \texttt{StatePreparation} & $\mathbf{1}$ & $\mathbf{4}$ & $\mathbf{11}$ & $\mathbf{26}$ & $\mathbf{57}$ \\
Multiplexor       & $2$  & $10$  & $26$  & $58$   & $122$  \\
General isometry   & $2$  & $10$  & $26$  & $58$   & $122$  \\
Plesch--Brukner   & $\mathbf{1}$ & $9$ & $16$ & $50$ & $81$ \\
Quantum Shannon   & $6$  & $36$  & $168$ & $720$  & $2976$ \\
\hline
Lower bound~\eqref{eq:cnot_lb} & $1$ & $2$ & $6$ & $13$ & $29$ \\
\hline
\end{tabular}
\end{table}

The same holds on \emph{real} (sign-bearing) targets, where both QsiHT syntheses are the cheapest of the main and from-scratch implementations in Table~\ref{tab:implementations}, tied with the deployed Qiskit \texttt{StatePreparation}. Table~\ref{tab:real_cnot} reports the real-target CNOT counts: the two syntheses each use $2^n-n-1$ CNOTs ($57$ at $n=6$), undercutting the sign-aware M\"ott\"onen cascade's $2^n-2$ ($62$), the M\"ott\"onen cascade ($124$), and Plesch--Brukner ($81$). The Real Synthesis reaches this count with an $R_y{+}$CX cascade and no phase stage, the DsiHT encoding the amplitude signs directly through the \texttt{arctan2} angle, and the Complex Synthesis reaches the same count on real targets by its blockwise phase absorption. The count coincides with Qiskit's deployed \texttt{StatePreparation} object, built on the merged-multiplexor construction with the same $2^n-n-1$ cost (\S\ref{sec:an_qsiht}), which realizes $1$, $4$, $11$, $26$, $57$ as well: at build time the two syntheses tie the deployed floor rather than sitting above it. For context, the sign-aware M\"ott\"onen variant, the cascade with signs folded into its $R_y$ angles and the phase stage disabled, realizes $2$, $6$, $14$, $30$, $62$, one full uniformly controlled stage above the syntheses, so the extra saving over that cascade is the deferred entangler per stage rather than sign-aware angle computation alone. The synthesis counts are optimization-level independent: re-transpiled at optimization level~3 they are unchanged at every $n$, as is \texttt{StatePreparation}. The same setting compresses Plesch--Brukner from $81$ to $56$ and M\"ott\"onen from $124$ to $120$ at $n=6$, which slips Plesch--Brukner's opt-3 count ($56$) just below the syntheses' opt-3-invariant $57$ at that size. On real dense data at build time, then, the two syntheses hold the fewest CNOTs of any benchmarked method alongside the deployed \texttt{StatePreparation}, and the Real Synthesis does so with a single $R_y$ cascade and $O(N\log N)$ angle synthesis.

\begin{table}[H]
\centering
\caption{Realized CNOT count on \emph{real} (sign-bearing) Haar-random targets. Both QsiHT syntheses realize $2^n-n-1$ ($57$ at $n=6$), tied with the deployed Qiskit \texttt{StatePreparation} for the fewest above the lower bound (bold, tied minima marked), one uniformly controlled stage below the sign-aware M\"ott\"onen cascade's $2^n-2$ ($62$). The synthesis counts are unchanged at optimization level~3. At that setting Plesch--Brukner compresses to $56$ and M\"ott\"onen to $120$ at $n=6$.}
\label{tab:real_cnot}
\begin{tabular}{lrrrrr}
\hline
\textbf{Method} & $n{=}2$ & $n{=}3$ & $n{=}4$ & $n{=}5$ & $n{=}6$ \\
\hline
QsiHT Fast Path Real Synthesis    & $\mathbf{1}$ & $\mathbf{4}$ & $\mathbf{11}$ & $\mathbf{26}$ & $\mathbf{57}$ \\
QsiHT Fast Path Complex Synthesis & $\mathbf{1}$ & $\mathbf{4}$ & $\mathbf{11}$ & $\mathbf{26}$ & $\mathbf{57}$ \\
M\"ott\"onen (UCR) & $4$ & $12$ & $28$ & $58$ & $124$ \\
M\"ott\"onen (sign-aware) & $2$ & $6$ & $14$ & $30$ & $62$ \\
Iten-style cascade     & $2$ & $7$  & $26$ & $58$ & $120$ \\
Qiskit \texttt{StatePreparation} & $\mathbf{1}$ & $\mathbf{4}$ & $\mathbf{11}$ & $\mathbf{26}$ & $\mathbf{57}$ \\
Multiplexor       & $2$ & $7$  & $26$ & $58$ & $120$ \\
General isometry   & $2$ & $7$  & $26$ & $58$ & $120$ \\
Plesch--Brukner   & $\mathbf{1}$ & $9$ & $16$ & $50$ & $81$ \\
Quantum Shannon   & $6$ & $36$ & $168$ & $720$ & $2976$ \\
\hline
Lower bound~\eqref{eq:cnot_lb} & $1$ & $2$ & $6$ & $13$ & $29$ \\
\hline
\end{tabular}
\end{table}

\begin{figure}[H]
\centering
\includegraphics[width=0.82\textwidth]{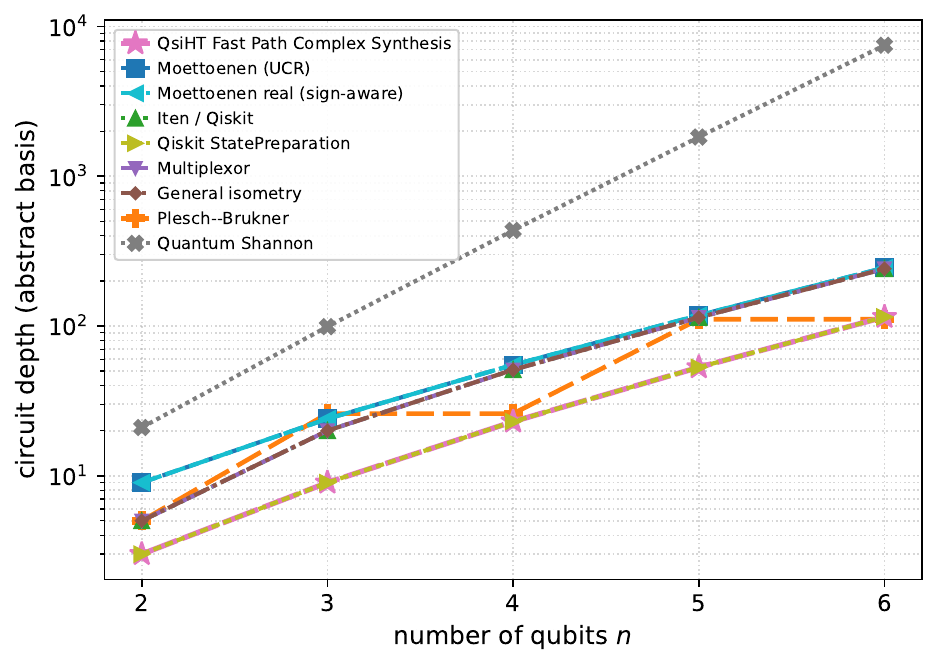}
\caption{Circuit depth versus $n$ (logarithmic axis), complex Haar-random targets, in the $\{u,\textsc{cx}\}$ basis. Plesch--Brukner's stepwise depth reflects the parity of the half-register split $\lceil n/2\rceil,\lfloor n/2\rfloor$.}
\label{fig:depth_scaling}
\end{figure}

The depth results, summarized in Table~\ref{tab:depth_scaling} for complex targets (plotted in Figure~\ref{fig:depth_scaling}) and Table~\ref{tab:depth_real} for real targets, mirror the CNOT trends because all of these constructions are predominantly sequential (\S\ref{sec:an_mottonen}). On complex data the QsiHT Complex Synthesis and the deployed \texttt{StatePreparation} share the shallowest exact depth, $2^{n+1}-2n-1$ ($53$ at $n=5$, $115$ at $n=6$, against the UCR cascades' $114$--$118$ at $n=5$ and $241$--$245$ at $n=6$), Plesch--Brukner edging ahead only at $n=6$ ($111$ versus $115$), while QSD is more than an order of magnitude deeper ($7491$ at $n=6$). On real data both syntheses match \texttt{StatePreparation} at the same depths, leading at $n=2$ through $5$ (depth $53$ at $n=5$ against the sign-aware M\"ott\"onen's $57$ and the M\"ott\"onen cascade's $108$), with Plesch--Brukner edging ahead at $n=6$ ($109$ versus $115$).

\begin{table}[H]
\centering
\caption[Realized circuit depth on complex Haar-random targets]{Realized circuit depth $d$ on complex Haar-random targets, after transpilation to $\{u,\textsc{cx}\}$. The QsiHT Complex Synthesis and Qiskit \texttt{StatePreparation} share the shallowest exact depth (bold), Plesch--Brukner edging ahead at $n=6$.}
\label{tab:depth_scaling}
\begin{tabular}{lrrrrr}
\hline
\textbf{Method} & $n{=}2$ & $n{=}3$ & $n{=}4$ & $n{=}5$ & $n{=}6$ \\
\hline
QsiHT Fast Path Complex Synthesis & $\mathbf{3}$ & $\mathbf{9}$ & $\mathbf{23}$ & $\mathbf{53}$ & $115$ \\
M\"ott\"onen (UCR) & $9$  & $24$  & $55$  & $118$  & $245$  \\
Iten-style cascade     & $5$  & $20$  & $51$  & $114$  & $241$  \\
Qiskit \texttt{StatePreparation} & $\mathbf{3}$ & $\mathbf{9}$ & $\mathbf{23}$ & $\mathbf{53}$ & $115$ \\
Multiplexor       & $5$  & $20$  & $51$  & $114$  & $241$  \\
General isometry   & $5$  & $20$  & $51$  & $114$  & $241$  \\
Plesch--Brukner   & $5$  & $26$  & $26$ & $111$  & $\mathbf{111}$ \\
Quantum Shannon   & $21$ & $99$  & $435$ & $1827$ & $7491$ \\
\hline
\end{tabular}
\end{table}

\begin{table}[H]
\centering
\caption[Realized circuit depth on real Haar-random targets]{Realized circuit depth $d$ on \emph{real} (sign-bearing) Haar-random targets, after transpilation to $\{u,\textsc{cx}\}$. Both QsiHT syntheses match Qiskit \texttt{StatePreparation} at the shallowest exact depth (bold), Plesch--Brukner edging ahead at $n=6$.}
\label{tab:depth_real}
\begin{tabular}{lrrrrr}
\hline
\textbf{Method} & $n{=}2$ & $n{=}3$ & $n{=}4$ & $n{=}5$ & $n{=}6$ \\
\hline
QsiHT Fast Path Real Synthesis    & $\mathbf{3}$ & $\mathbf{9}$ & $\mathbf{23}$ & $\mathbf{53}$ & $115$ \\
QsiHT Fast Path Complex Synthesis & $\mathbf{3}$ & $\mathbf{9}$ & $\mathbf{23}$ & $\mathbf{53}$ & $115$ \\
M\"ott\"onen (UCR) & $7$  & $20$  & $50$  & $108$  & $234$  \\
M\"ott\"onen (sign-aware) & $4$ & $11$ & $26$ & $57$ & $120$ \\
Iten-style cascade     & $5$  & $14$  & $46$  & $107$  & $228$  \\
Qiskit \texttt{StatePreparation} & $\mathbf{3}$ & $\mathbf{9}$ & $\mathbf{23}$ & $\mathbf{53}$ & $115$ \\
Multiplexor       & $5$  & $14$  & $46$  & $107$  & $228$  \\
General isometry   & $5$  & $14$  & $46$  & $107$  & $228$  \\
Plesch--Brukner   & $5$  & $24$  & $24$ & $105$  & $\mathbf{109}$ \\
Quantum Shannon   & $21$ & $99$  & $435$ & $1827$ & $7491$ \\
\hline
\end{tabular}
\end{table}

Another view for comparison is the \emph{classical} cost of synthesis, recorded as the end-to-end circuit-assembly time. In the $n\le 6$ structural sweep every method is timed as one build plus exactly one optimization-level-3 pass: the four practical library rows (Iten, \texttt{StatePreparation}, multiplexor, and isometry) run that pass inside their builder, and the others have the identical pass applied once after the build, so every row carries a single opt-3 pass and none is credited a raw-build-only time. On that common basis every method except QSD assembles its circuit in at most a few tens of milliseconds at $n=6$, the two syntheses sitting within the same band as the cascade methods and Plesch--Brukner, while QSD's $\Theta(4^n)$ recursion spends a few hundred milliseconds. The two syntheses separate only at larger registers, where the shipped angle pipelines' differing asymptotic orders begin to tell (the reshape SVD is $O(N^{1.5})$, the cascade's and merged-multiplexor's dense Walsh--Hadamard angle pass $O(N^2)$, and the QsiHT route's $\Theta(N)$ heap followed by its fast Walsh--Hadamard map at $O(N\log N)$). Table~\ref{tab:build_time} extends the build-time measurement to $n=10$--$12$. The Real Synthesis, whose fast Walsh--Hadamard transform (FWHT) pass carries no phase pipeline, assembles its $n=12$ circuit in about $37$ ms, roughly $80$ times faster than the deployed \texttt{StatePreparation} ($2.9$ s) and about $180$ times faster than the Iten cascade ($6.9$ s), an order-of-magnitude-plus separation that widens with $n$. The Complex Synthesis ($2.0$ s at $n=12$) carries the per-amplitude phases through its $U(2)$ blocks and so sits in the \texttt{StatePreparation} band rather than with the Real Synthesis. QSD is off this scale entirely, its $\Theta(4^n)$ recursion taking $83$ s at $n=10$ and about $6$ minutes at $n=11$. So the Real Synthesis is cheap both to \emph{specify} and to \emph{compile} at register sizes where the other exact loaders are already spending seconds, while the Complex Synthesis matches the deployed \texttt{StatePreparation}'s build cost.

\begin{table}[H]
\centering
\caption{Circuit-assembly time (milliseconds, real Haar-random target) at large registers, from the \texttt{scaling\_7to12} sweep. The Real Synthesis's $O(N\log N)$ angle pipeline pulls an order of magnitude or more below the merged-multiplexor and cascade routes by $n=11$--$12$, while the Complex Synthesis tracks the deployed \texttt{StatePreparation}. QSD (not shown) is off scale, taking $8.3\times10^{4}$ ms at $n=10$ and $3.8\times10^{5}$ ms at $n=11$.}
\label{tab:build_time}
\begin{tabular}{lrrr}
\hline
\textbf{Method} & $n{=}10$ & $n{=}11$ & $n{=}12$ \\
\hline
QsiHT Fast Path Real Synthesis    & $\mathbf{9.7}$ & $\mathbf{18.4}$ & $\mathbf{37.1}$ \\
QsiHT Fast Path Complex Synthesis & $302$ & $706$ & $1982$ \\
M\"ott\"onen (UCR) & $220$ & $925$ & $3685$ \\
Iten-style cascade     & $434$ & $1642$ & $6873$ \\
Qiskit \texttt{StatePreparation} & $1026$ & $1537$ & $2949$ \\
Plesch--Brukner   & $133$ & $340$ & $573$ \\
\hline
\end{tabular}
\end{table}

\begin{figure}[H]
\centering
\includegraphics[width=0.82\textwidth]{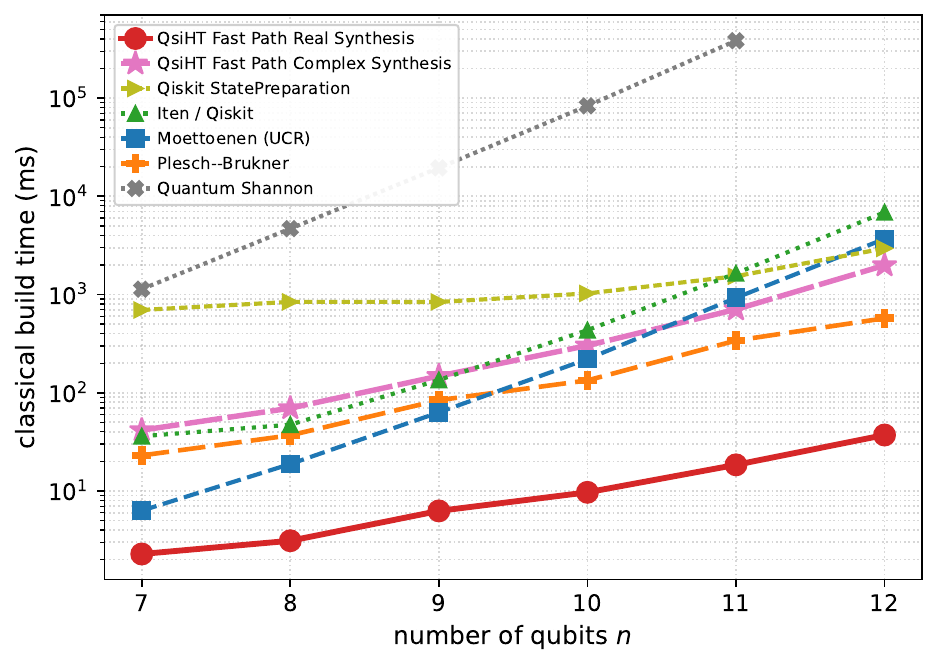}
\caption{Classical circuit-assembly time versus register size $n$ (log scale), real Haar-random target, over the $n=7$--$12$ scaling sweep. The Real Synthesis's $O(N \log N)$ FWHT-only pipeline pulls one to two orders of magnitude below the merged-multiplexor and cascade routes (including Qiskit StatePreparation) by $n=11$--$12$, while the Complex Synthesis, dominated by assembling its $U(2)$ block cascade, tracks the StatePreparation band. Both sit far below QSD's $\Theta(4^n)$ recursion.}
\label{fig:build_time_scaling}
\end{figure}

No finer classical-cost ranking among the frontier methods is claimed at the small sizes, since there the angle pipelines all complete in milliseconds and the differences are dominated by fixed overheads. What the sweep establishes is the large-register separation: the Real Synthesis is the cheapest exact loader to synthesize, the Complex Synthesis matches the deployed \texttt{StatePreparation}, and both sit far below QSD. Figure~\ref{fig:build_time_scaling} visualizes the scaling across the $n=7$--$12$ sweep.

\section{Cross-Platform NISQ Comparison}
\label{sec:nisq}

\begin{figure}[H]
\centering
\includegraphics[width=0.92\textwidth]{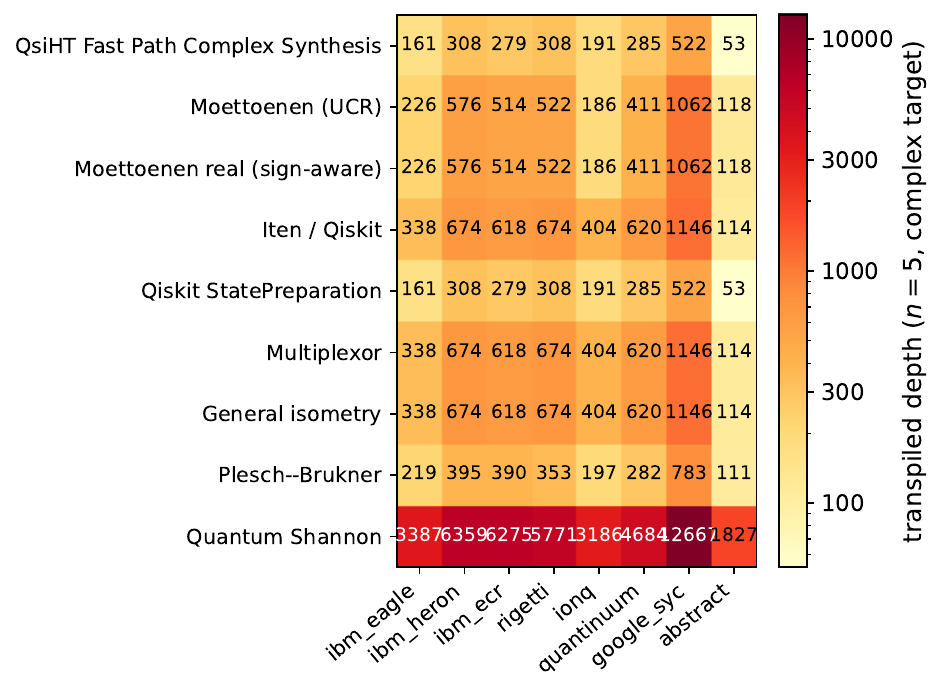}
\caption{Transpiled depth at $n=5$ (complex target) for every method (rows) across the NISQ native gate sets (columns). Cell values are the depth on a logarithmic color scale. The right-most column is the connectivity-free abstract depth.}
\label{fig:nisq_depth}
\end{figure}

\begin{figure}[H]
\centering
\includegraphics[width=0.92\textwidth]{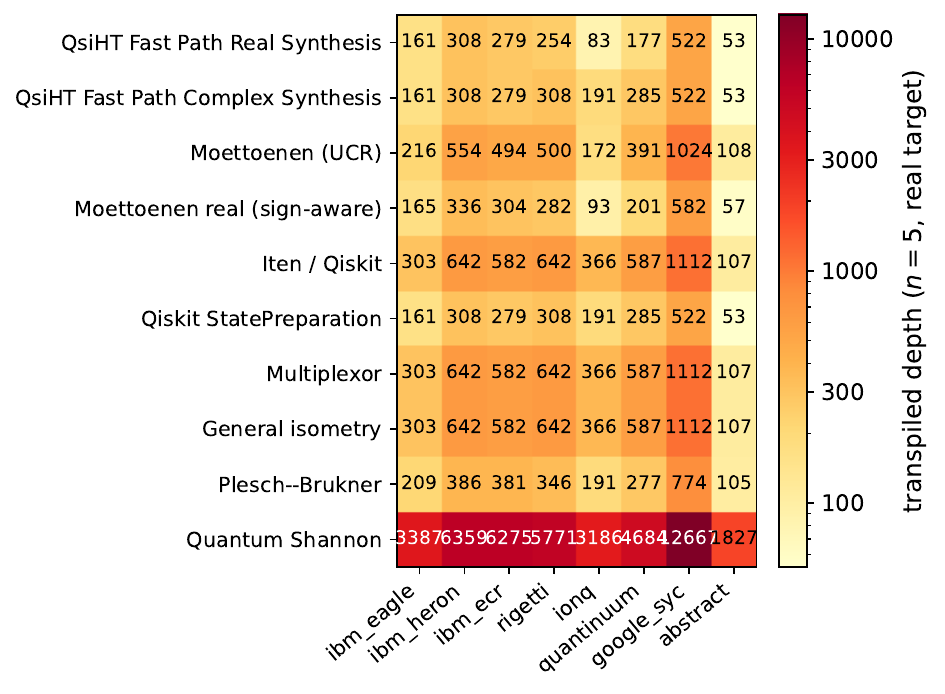}
\caption{Transpiled circuit depth for the real $n=5$ benchmark target across the eight native gate sets at optimization level~3, on a logarithmic color scale. The QsiHT Real Synthesis ($26$ two-qubit gates, its $R_y$-only single-qubit structure compiling tightest) is the shallowest on every gate set, uniquely leading on Rigetti, IonQ, and Quantinuum and tying the Complex Synthesis and Qiskit \texttt{StatePreparation} (also $26$ two-qubit gates) on the other five, including the abstract basis ($53$). The complex-target counterpart (Figure~\ref{fig:nisq_depth}) splits the lead differently, so neither target's fine ranking transfers to the other.}
\label{fig:nisq_depth_real}
\end{figure}

Logical $\{u,\textsc{cx}\}$ counts understate hardware cost because real devices expose different native entanglers and restricted connectivity (\S\ref{sec:gate_metrics}). Each method was therefore transpiled, at $n=5$, to the eight representative native gate sets of Table~\ref{tab:nisq_platforms}, and the resulting transpiled depth $d_{\text{T}}$ recorded for the complex target in Table~\ref{tab:nisq_depth} and Figure~\ref{fig:nisq_depth}, and for the real target in Table~\ref{tab:nisq_real} and Figure~\ref{fig:nisq_depth_real}. No coupling-map routing is applied in this comparison (every platform is treated as all-to-all, so zero SWAPs are inserted), which isolates the native-gate-set effect from connectivity. The connectivity effect is treated separately by the routed noise model of \S\ref{sec:noise} and the hardware runs of \S\ref{sec:hw_limitations}.

\begin{table}[H]
\centering
\caption{Representative NISQ native gate sets used for the cross-platform comparison, after~\cite{ibm2022, qiskit2024}. The two-qubit entangler is the dominant cost driver~\eqref{eq:fidelity_decay}. Here CZ is the controlled-$Z$ gate, ECR the echoed cross-resonance, and iSWAP the imaginary SWAP.}
\label{tab:nisq_platforms}
\begin{tabular}{lll}
\hline
\textbf{Platform} & \textbf{Single-qubit set} & \textbf{Entangler} \\
\hline
IBM Eagle     & $\textsc{sx}$, $R_z$, $X$        & CX \\
IBM Heron     & $\textsc{sx}$, $R_z$, $X$        & CZ \\
IBM ECR       & $\textsc{sx}$, $R_z$, $X$        & ECR \\
Rigetti       & $R_x$, $R_z$                     & CZ \\
IonQ          & $R_x$, $R_y$, $R_z$             & XX (M{\o}lmer--S{\o}rensen) \\
Quantinuum    & $R_x$, $R_y$, $R_z$             & $R_{zz}$ (ZZPhase) \\
Google Sycamore & $\textsc{sx}$, $R_z$, $X$      & iSWAP \\
Abstract       & $u$                             & CX \\
\hline
\end{tabular}
\end{table}

\begin{table}[H]
\centering
\caption{Transpiled depth $d_{\text{T}}$ at $n=5$ (complex Haar-random target) across the native gate sets of Table~\ref{tab:nisq_platforms}. The right-most column is the connectivity-free abstract depth for reference. The QsiHT Complex Synthesis ties the deployed Qiskit \texttt{StatePreparation} ($26$ native two-qubit gates each) and the two are the shallowest on six of the eight gate sets (bold), the M\"ott\"onen cascade leading on IonQ and Plesch--Brukner on Quantinuum.}
\label{tab:nisq_depth}
\resizebox{\textwidth}{!}{%
\begin{tabular}{lrrrrrrrr}
\hline
\textbf{Method} & Eagle & Heron & ECR & Rigetti & IonQ & Quant. & Syc. & Abstr. \\
\hline
QsiHT Fast Path Complex Synthesis & $\mathbf{161}$ & $\mathbf{308}$ & $\mathbf{279}$ & $\mathbf{308}$ & $191$ & $285$ & $\mathbf{522}$ & $\mathbf{53}$ \\
M\"ott\"onen (UCR) & $226$  & $576$  & $514$  & $522$  & $\mathbf{186}$  & $411$  & $1062$ & $118$  \\
Iten-style cascade     & $338$  & $674$  & $618$  & $674$  & $404$  & $620$  & $1146$ & $114$  \\
Qiskit \texttt{StatePreparation} & $\mathbf{161}$ & $\mathbf{308}$ & $\mathbf{279}$ & $\mathbf{308}$ & $191$ & $285$ & $\mathbf{522}$ & $\mathbf{53}$ \\
Multiplexor       & $338$  & $674$  & $618$  & $674$  & $404$  & $620$  & $1146$ & $114$  \\
General isometry   & $338$  & $674$  & $618$  & $674$  & $404$  & $620$  & $1146$ & $114$  \\
Plesch--Brukner   & $219$ & $395$ & $390$ & $353$ & $197$ & $\mathbf{282}$ & $783$ & $111$ \\
Quantum Shannon   & $3387$ & $6359$ & $6275$ & $5771$ & $3186$ & $4684$ & $12667$ & $1827$ \\
\hline
\end{tabular}}
\end{table}

\begin{table}[H]
\centering
\caption{Native-gate-set transpiled depth $d_{\text{T}}$ at $n=5$ on a \emph{real} Haar-random target across the gate sets of Table~\ref{tab:nisq_platforms}, under all-to-all connectivity (no routing SWAPs are inserted, so this isolates the gate-set effect). The QsiHT Real Synthesis carries $26$ native two-qubit gates, tied with the Complex Synthesis and Qiskit \texttt{StatePreparation} for the fewest, and its $R_y$-only single-qubit structure makes it the shallowest on every gate set (bold), uniquely leading on Rigetti, IonQ, and Quantinuum and tying the Complex Synthesis and \texttt{StatePreparation} on the other five. On a limited coupling map the ordering changes: routing at $n=3$ lifts the synthesis' $4$ abstract two-qubit gates to $7$, tied with \texttt{StatePreparation} (\S\ref{sec:noise}, \S\ref{sec:hw_limitations}), so this gate-set ranking does not carry over unchanged to a routed device.}
\label{tab:nisq_real}
\resizebox{\textwidth}{!}{%
\begin{tabular}{lrrrrrrrr}
\hline
\textbf{Method} & Eagle & Heron & ECR & Rigetti & IonQ & Quant. & Syc. & Abstr. \\
\hline
QsiHT Fast Path Real Synthesis    & $\mathbf{161}$ & $\mathbf{308}$ & $\mathbf{279}$ & $\mathbf{254}$ & $\mathbf{83}$ & $\mathbf{177}$ & $\mathbf{522}$ & $\mathbf{53}$ \\
QsiHT Fast Path Complex Synthesis & $\mathbf{161}$ & $\mathbf{308}$ & $\mathbf{279}$ & $308$ & $191$ & $285$ & $\mathbf{522}$ & $\mathbf{53}$ \\
M\"ott\"onen (UCR) & $216$ & $554$ & $494$ & $500$ & $172$ & $391$ & $1024$ & $108$ \\
M\"ott\"onen (sign-aware) & $165$ & $336$ & $304$ & $282$ & $93$ & $201$ & $582$ & $57$ \\
Iten-style cascade     & $303$ & $642$ & $582$ & $642$ & $366$ & $587$ & $1112$ & $107$ \\
Qiskit \texttt{StatePreparation} & $\mathbf{161}$ & $\mathbf{308}$ & $\mathbf{279}$ & $308$ & $191$ & $285$ & $\mathbf{522}$ & $\mathbf{53}$ \\
Multiplexor       & $303$ & $642$ & $582$ & $642$ & $366$ & $587$ & $1112$ & $107$ \\
General isometry   & $303$ & $642$ & $582$ & $642$ & $366$ & $587$ & $1112$ & $107$ \\
Plesch--Brukner   & $209$ & $386$ & $381$ & $346$ & $191$ & $277$ & $774$ & $105$ \\
Quantum Shannon   & $3387$ & $6359$ & $6275$ & $5771$ & $3186$ & $4684$ & $12667$ & $1827$ \\
\hline
\end{tabular}}
\end{table}

Three platform effects stand out. First, native gate sets that decompose the abstract CNOT into a longer sequence (the CZ-, ECR-, and especially iSWAP-based devices) inflate depth uniformly: Google Sycamore's iSWAP entangler also \emph{doubles} the realized two-qubit count for every method, e.g. Plesch--Brukner's $50$ abstract entanglers become $100$ iSWAPs, because a CNOT requires two iSWAP-family interactions. Second, the architectures with the broadest single-qubit sets and the most direct entanglers, IBM Eagle (direct CX) and IonQ (native XX), yield the shallowest realizations, the IonQ advantage coming from its broad single-qubit set and one-to-one XX-to-CNOT mapping rather than from connectivity, since this comparison inserts no routing SWAPs for any platform (\S\ref{sec:nisq}). Third, the \emph{coarse tier} is preserved across every platform: Quantum Shannon is roughly an order of magnitude deeper than the exact frontier on all eight gate sets, on both targets. The fine ordering, by contrast, is platform- and target-dependent rather than universal. On the real target (Table~\ref{tab:nisq_real} and Figure~\ref{fig:nisq_depth_real}), and on the gate-set axis alone with no routing applied, the QsiHT Real Synthesis, whose $26$ native two-qubit gates tie the deployed \texttt{StatePreparation} for the fewest, is the shallowest method on every gate set, uniquely leading on Rigetti, IonQ, and Quantinuum, where its $R_y$-only single-qubit structure compiles tighter than \texttt{StatePreparation}, and tying the Complex Synthesis and \texttt{StatePreparation} on the other five, including the abstract basis ($53$). On a limited coupling map this ordering does not survive unchanged, since routing lifts the synthesis' $4$ abstract two-qubit gates at $n=3$ to $7$, tied with \texttt{StatePreparation} (\S\ref{sec:noise}). On the complex target the lead is shared: the QsiHT Complex Synthesis and \texttt{StatePreparation} tie shallowest on six gate sets, the M\"ott\"onen cascade leads on IonQ, and Plesch--Brukner on Quantinuum. The hardware layer therefore rescales the costs without disturbing the frontier-versus-QSD split, and the abstract-basis comparison of~\S\ref{sec:scaling} predicts that coarse split reliably, but no single method holds the depth lead on every platform, and per-platform transpilation is what settles the fine order.

\section{Finite-Shot and Noisy Behavior}
\label{sec:noise}

\subsection{Sampling at Finite Shots}

\begin{figure}[H]
\centering
\includegraphics[width=0.82\textwidth]{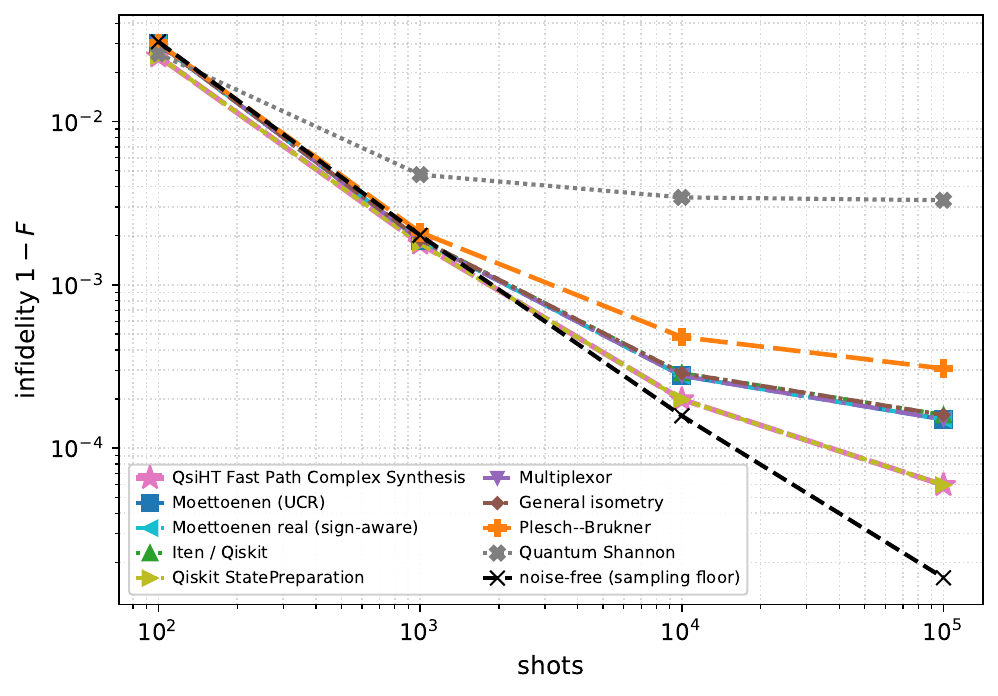}
\caption{Infidelity $1-F$ versus shot count (log--log) for the complex $n=3$ target. Each colored curve is one method under the seeded noise model, and the dashed black curve is the method-independent noise-free sampling floor. At low shot counts every method tracks the sampling floor. As the shots grow each plateaus at its own gate-error floor, QSD an order of magnitude above the frontier.}
\label{fig:infid_shots_complex}
\end{figure}

\begin{figure}[H]
\centering
\includegraphics[width=0.82\textwidth]{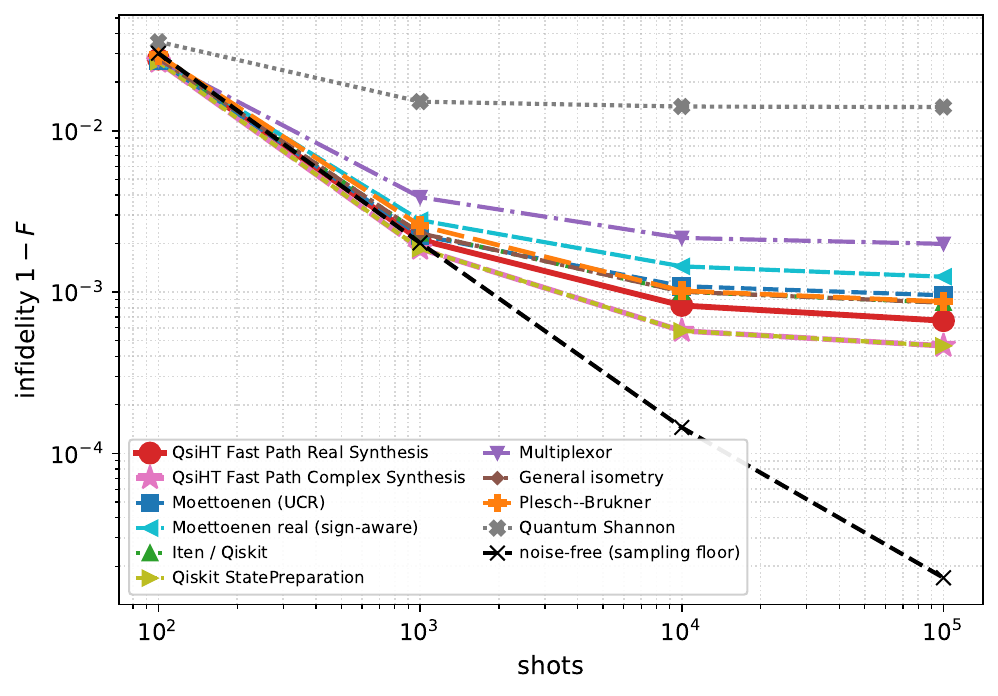}
\caption{Infidelity $1-F$ versus shot count (log--log) for the real $n=3$ target. Curves and floor as in Figure~\ref{fig:infid_shots_complex}. The QsiHT syntheses track the frontier cluster, and QSD again separates by an order of magnitude.}
\label{fig:infid_shots_real}
\end{figure}

Two contributions set the sampled error. Finite sampling alone produces a floor that every method shares, because each prepares the \emph{same} ideal distribution at unit fidelity (\S\ref{sec:exactness}), and gate noise adds a method-dependent term on top. Tables~\ref{tab:shot_complex} and~\ref{tab:shot_real} report the noise-free floor for the complex and real $n=3$ targets, the mean infidelity with its run-to-run variance and the MSE, as the shot count is swept. The floor falls steeply, by roughly an order of magnitude for every tenfold increase in shots (the inverse-shot-count law of these quadratic distribution distances), the infidelity dropping from $\approx 3\times10^{-2}$ at $100$ shots to $\approx 1.6\times10^{-5}$ at $10^5$. Figures~\ref{fig:infid_shots_complex} and~\ref{fig:infid_shots_real} overlay each method's infidelity under the seeded noise model on this floor. At low shot counts every method tracks the sampling floor, but as the shots grow each curve peels off and plateaus at its own gate-error floor. The frontier methods settle roughly an order of magnitude below QSD. This separates the finite-shot statistics from the gate-induced error that the next subsection compares method by method.

\begin{table}[H]
\centering
\caption{Noise-free per-shot error for the complex $n=3$ target (method-independent), averaged over $20$ runs: the mean infidelity $1-F$ with its run-to-run variance and the MSE~\eqref{eq:error_metrics}, versus shot count.}
\label{tab:shot_complex}
\begin{tabular}{lrrrr}
\hline
\textbf{Metric} & $10^2$ shots & $10^3$ shots & $10^4$ shots & $10^5$ shots \\
\hline
Infidelity & $3.082\times10^{-2}$ & $2.013\times10^{-3}$ & $1.580\times10^{-4}$ & $1.610\times10^{-5}$ \\
Variance & $8.675\times10^{-5}$ & $1.040\times10^{-6}$ & $4.758\times10^{-9}$ & $3.906\times10^{-11}$ \\
MSE & $1.380\times10^{-3}$ & $1.020\times10^{-4}$ & $8.491\times10^{-6}$ & $8.040\times10^{-7}$ \\
\hline
\end{tabular}
\end{table}

\begin{table}[H]
\centering
\caption{Noise-free per-shot error for the real $n=3$ target (method-independent), averaged over $20$ runs: the mean infidelity $1-F$ with its run-to-run variance and the MSE~\eqref{eq:error_metrics}, versus shot count.}
\label{tab:shot_real}
\begin{tabular}{lrrrr}
\hline
\textbf{Metric} & $10^2$ shots & $10^3$ shots & $10^4$ shots & $10^5$ shots \\
\hline
Infidelity & $3.017\times10^{-2}$ & $2.016\times10^{-3}$ & $1.448\times10^{-4}$ & $1.695\times10^{-5}$ \\
Variance & $1.118\times10^{-4}$ & $1.131\times10^{-6}$ & $4.576\times10^{-9}$ & $8.026\times10^{-11}$ \\
MSE & $9.561\times10^{-4}$ & $8.081\times10^{-5}$ & $7.083\times10^{-6}$ & $8.238\times10^{-7}$ \\
\hline
\end{tabular}
\end{table}

Figures~\ref{fig:infid_shots_complex} and~\ref{fig:infid_shots_real} show one counterintuitive feature: at the lowest shot counts nearly all of the noisy curves sit slightly \emph{below} the dashed noise-free floor, which a literal reading would call impossible, since gate noise can only add error. The crossing is an artifact of finite-shot estimation, not a real fidelity gain. The infidelity is a nonlinear functional of the sampled counts, the squared Bhattacharyya overlap of~\eqref{eq:error_metrics}, and at finite shots its estimate is biased upward by an amount that depends on the shape of the distribution being sampled. The ideal target distribution is peaked, its amplitudes spanning more than an order of magnitude, so its small-probability outcomes are poorly resolved at $100$ shots and inflate the floor. Gate noise flattens the output toward uniform, and the flatter distribution carries a smaller finite-shot bias, enough between $100$ and $1000$ shots to push a low-noise method below the floor measured on the peaked ideal. The sampling term dwarfs the tiny gate-error term, so the curves are statistically indistinguishable and their order is set by this estimator bias rather than by the noise. Once the shots reach about $10^4$ the sampling term drops below each method's gate-error floor, the bias becomes negligible, and the curves separate in the physically meaningful order, every method above the floor and the frontier roughly an order of magnitude below QSD.

\subsection{Error Under a Noise Model}

\begin{figure}[H]
\centering
\includegraphics[width=0.78\textwidth]{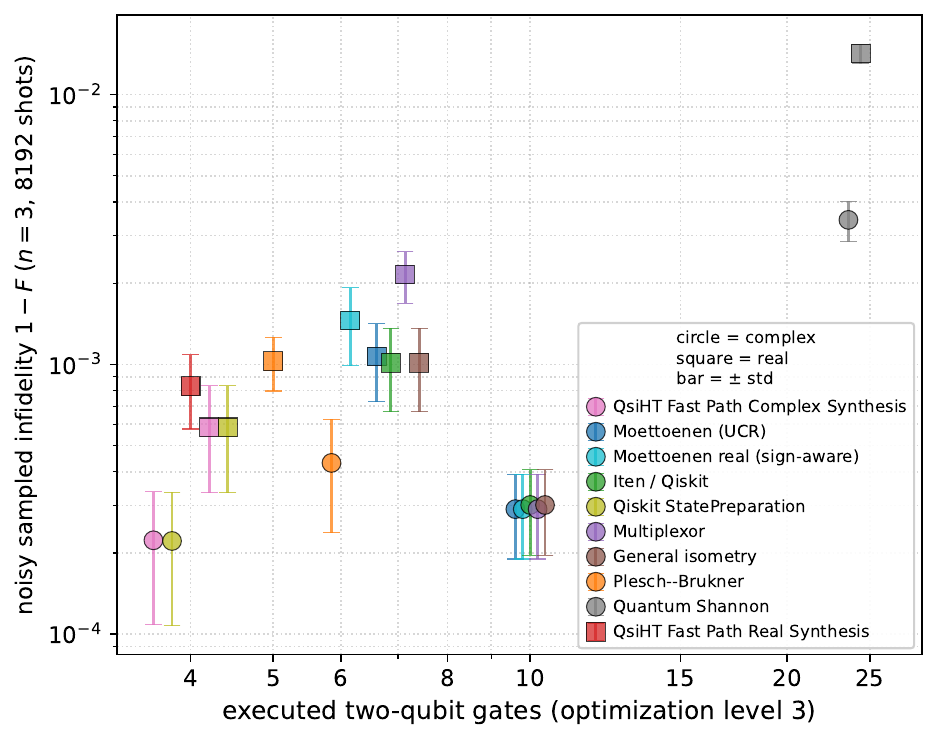}
\caption{Noisy sampled infidelity $1-F$ at $n=3$ ($8192$ shots, shared seeded noise model) against the number of two-qubit gates in the circuit that actually executed, after transpilation at optimization level~3 (log--log), for the complex (circles) and real (squares) targets, with error bars showing the run-to-run standard deviation over the $20$ runs. On this axis the two QsiHT syntheses and Qiskit \texttt{StatePreparation} share the low-gate edge at $4$ executed gates, one fewer than Plesch--Brukner ($5$ real, $6$ complex), so the syntheses now occupy the low-gate frontier alongside the deployed \texttt{StatePreparation}. The order-of-magnitude gap to Quantum Shannon ($24$ executed gates) survives the axis change, and the fidelity-decay model~\eqref{eq:fidelity_decay} appears as a coarse linear trend across the executed range (least-squares slope $7.0\times10^{-4}$ per gate at $R^2=0.98$).}
\label{fig:noise_vs_cnot}
\end{figure}

The comparison that matters most for method selection is under noise, where the gate count converts directly into output error through the first-order fidelity-decay model~\eqref{eq:fidelity_decay}, $F\approx 1-\sum_i\epsilon_i$. Tables~\ref{tab:noisy_complex} and~\ref{tab:noisy_real} report, for each builder at $n=3$ and $8192$ shots on a single shared, seeded \texttt{GenericBackendV2} noise model averaged over $20$ runs with the shot seeds paired across methods (\S\ref{sec:metrics_used}), the fidelity error (infidelity $1-F$) with its run-to-run variance and the MSE~\eqref{eq:error_metrics}, against both the abstract-basis CNOT count and the two-qubit count of the circuit that \emph{actually executed} after transpilation at optimization level~3 (Exec.). Pairing makes the tables internally consistent in a directly visible way: builders that execute the identical circuit print identical rows (the QsiHT Complex Synthesis and \texttt{StatePreparation} on the real target, the merged-cascade realizations on the complex one), so any difference between rows is the circuit, never a disjoint shot-noise block. Figure~\ref{fig:noise_vs_cnot} plots the infidelity against the executed count. That is the appropriate axis, because the noise acts on the executed circuit rather than the abstract one, and the two counts differ for the baselines: optimization level~3 compresses M\"ott\"onen's $12$ abstract CNOTs to $7$ executed and Plesch--Brukner's $9$ to $5$ on the real target, while the QsiHT syntheses and \texttt{StatePreparation} execute exactly their printed count ($4$ on both targets). On this axis the two syntheses share the low-gate edge with \texttt{StatePreparation} at $4$ executed gates, one fewer than Plesch--Brukner's $5$ (real) and $6$ (complex), a reversal of the original fast path, which sat at $6$ (real) and $12$ (complex). The coarse pattern, a frontier cluster separated from QSD by roughly an order of magnitude, holds on both targets: the non-QSD builders span $2.2$--$4.3\times10^{-4}$ on the complex target and $0.6$--$2.2\times10^{-3}$ on the real one, while QSD ($24$ executed gates) reaches $3.43\times10^{-3}$ and $1.42\times10^{-2}$ respectively. Within the frontier the low edge belongs to the fewest-executed-gate builders: on the complex target \texttt{StatePreparation} ($2.22\times10^{-4}$) and the QsiHT Complex Synthesis ($2.23\times10^{-4}$) are the two lowest, tied to within the run-to-run spread, and on the real target the QsiHT Complex Synthesis and \texttt{StatePreparation} print the identical lowest row ($5.84\times10^{-4}$), with the QsiHT Real Synthesis just above ($8.30\times10^{-4}$) and the $5$-to-$7$-gate cascade band (Iten $1.01\times10^{-3}$, Plesch--Brukner $1.03\times10^{-3}$, M\"ott\"onen $1.07\times10^{-3}$, multiplexor $2.15\times10^{-3}$) higher still. The infidelity and MSE columns do not rank the frontier identically: on the complex target Plesch--Brukner carries the highest frontier infidelity ($4.31\times10^{-4}$) yet an MSE ($1.85\times10^{-5}$) below the cascade builders' ($\approx 2.0\times10^{-5}$), while the joint-lowest MSE ($\approx 9.7\times10^{-6}$) belongs to the QsiHT Complex Synthesis and \texttt{StatePreparation}, because the two metrics are different norms on the same vector of probability errors (the decomposition is made precise in the hardware discussion below). The run-to-run spread is sizeable for the low-infidelity frontier methods, where finite-shot fluctuation is comparable to the small gate-error floor, but it stays well inside the gap to QSD, so the $20$-run mean ranks the coarse tiers unambiguously while leaving the fine within-frontier order unresolved at a single calibration. This is the empirical content of Chapter~\ref{ch:background}'s argument that minimizing the two-qubit budget is the key lever for high-fidelity preparation on near-term hardware, with the executed rather than the printed budget doing the predicting at the coarse level. On both targets the two syntheses now sit at that low executed-gate edge alongside \texttt{StatePreparation}, and the Complex Synthesis in particular ties it for the lowest sampled error.

\begin{table}[H]
\centering
\caption{Sampled error under a shared seeded \texttt{GenericBackendV2} noise model, complex $n=3$ target, $8192$ shots, averaged over $20$ runs, for all nine builders on this target. The per-run shot seeds are paired across methods, so every builder receives the same $20$ shot-noise draws and builders whose circuits transpile to the same executed circuit print identical rows. CNOTs is the abstract-basis count and Exec.\ the two-qubit count of the circuit that actually executed after transpilation at optimization level~3. The fidelity error (infidelity $1-F$) is reported with its run-to-run variance (Var.), alongside the mean-square error (MSE). The executed count drives the error. The lowest value in each column is shown in bold.}
\label{tab:noisy_complex}
\resizebox{\textwidth}{!}{%
\begin{tabular}{lrrrrrr}
\hline
\textbf{Method} & \textbf{CNOTs} & \textbf{Exec.} & \textbf{Depth} & \textbf{Infidelity} & \textbf{Var.} & \textbf{MSE} \\
\hline
QsiHT Fast Path Complex Synthesis & $\mathbf{4}$ & $\mathbf{4}$ & $\mathbf{9}$ & $2.228\times10^{-4}$ & $1.312\times10^{-8}$ & $\mathbf{9.723\times10^{-6}}$ \\
M\"ott\"onen (UCR) & $12$ & $10$ & $24$ & $2.908\times10^{-4}$ & $\mathbf{1.016\times10^{-8}}$ & $1.987\times10^{-5}$ \\
M\"ott\"onen (sign-aware) & $12$ & $10$ & $24$ & $2.908\times10^{-4}$ & $\mathbf{1.016\times10^{-8}}$ & $1.987\times10^{-5}$ \\
Iten-style cascade & $10$ & $10$ & $20$ & $3.012\times10^{-4}$ & $1.104\times10^{-8}$ & $2.108\times10^{-5}$ \\
Qiskit \texttt{StatePreparation} & $\mathbf{4}$ & $\mathbf{4}$ & $\mathbf{9}$ & $\mathbf{2.216\times10^{-4}}$ & $1.298\times10^{-8}$ & $9.727\times10^{-6}$ \\
Multiplexor & $10$ & $10$ & $20$ & $2.908\times10^{-4}$ & $\mathbf{1.016\times10^{-8}}$ & $1.987\times10^{-5}$ \\
General isometry & $10$ & $10$ & $20$ & $3.012\times10^{-4}$ & $1.104\times10^{-8}$ & $2.108\times10^{-5}$ \\
Plesch--Brukner & $9$ & $6$ & $26$ & $4.313\times10^{-4}$ & $3.766\times10^{-8}$ & $1.851\times10^{-5}$ \\
Quantum Shannon & $36$ & $24$ & $99$ & $3.432\times10^{-3}$ & $3.455\times10^{-7}$ & $1.149\times10^{-4}$ \\
\hline
\end{tabular}}
\end{table}

\begin{table}[H]
\centering
\caption{Sampled error under the shared seeded noise model, real $n=3$ target, $8192$ shots, $20$-run average, for all ten builders. Columns and paired shot seeding as in Table~\ref{tab:noisy_complex}: the QsiHT Complex Synthesis and \texttt{StatePreparation} emit the identical circuit on this target and therefore print identical rows. The two syntheses and \texttt{StatePreparation} share the fewest abstract and executed CNOTs ($4$), one below Plesch--Brukner's $5$ executed. The lowest value in each column is shown in bold.}
\label{tab:noisy_real}
\resizebox{\textwidth}{!}{%
\begin{tabular}{lrrrrrr}
\hline
\textbf{Method} & \textbf{CNOTs} & \textbf{Exec.} & \textbf{Depth} & \textbf{Infidelity} & \textbf{Var.} & \textbf{MSE} \\
\hline
QsiHT Fast Path Real Synthesis & $\mathbf{4}$ & $\mathbf{4}$ & $\mathbf{9}$ & $8.299\times10^{-4}$ & $6.440\times10^{-8}$ & $1.637\times10^{-5}$ \\
QsiHT Fast Path Complex Synthesis & $\mathbf{4}$ & $\mathbf{4}$ & $\mathbf{9}$ & $\mathbf{5.841\times10^{-4}}$ & $6.254\times10^{-8}$ & $\mathbf{1.328\times10^{-5}}$ \\
M\"ott\"onen (UCR) & $12$ & $7$ & $20$ & $1.070\times10^{-3}$ & $1.176\times10^{-7}$ & $2.623\times10^{-5}$ \\
M\"ott\"onen (sign-aware) & $6$ & $6$ & $11$ & $1.455\times10^{-3}$ & $2.172\times10^{-7}$ & $2.653\times10^{-5}$ \\
Iten-style cascade & $7$ & $7$ & $14$ & $1.014\times10^{-3}$ & $1.197\times10^{-7}$ & $2.458\times10^{-5}$ \\
Qiskit \texttt{StatePreparation} & $\mathbf{4}$ & $\mathbf{4}$ & $\mathbf{9}$ & $\mathbf{5.841\times10^{-4}}$ & $6.254\times10^{-8}$ & $\mathbf{1.328\times10^{-5}}$ \\
Multiplexor & $7$ & $7$ & $14$ & $2.153\times10^{-3}$ & $2.164\times10^{-7}$ & $4.513\times10^{-5}$ \\
General isometry & $7$ & $7$ & $14$ & $1.014\times10^{-3}$ & $1.197\times10^{-7}$ & $2.458\times10^{-5}$ \\
Plesch--Brukner & $9$ & $5$ & $24$ & $1.027\times10^{-3}$ & $\mathbf{5.319\times10^{-8}}$ & $1.593\times10^{-5}$ \\
Quantum Shannon & $36$ & $24$ & $99$ & $1.420\times10^{-2}$ & $1.392\times10^{-6}$ & $2.414\times10^{-4}$ \\
\hline
\end{tabular}}
\end{table}

\paragraph{Connectivity inside the noise model.} The shared model above is all-to-all at $n=3$, so no routing is ever needed and the executed count reflects only the optimizer. Real superconducting chips are not all-to-all, and the hardware runs of this section show routing deciding the on-device order, so the same paired-seed comparison is repeated on a \texttt{GenericBackendV2} built on the distance-3 heavy-hex coupling map ($19$ qubits), the connectivity pattern of the IBM Eagle and Heron processors. Table~\ref{tab:noisy_heavyhex} reports the result. Routing rearranges the ordering at $n=3$: the two syntheses and Qiskit \texttt{StatePreparation}, all at $4$ abstract CNOTs, route to $7$ executed gates on both targets (one inserted SWAP), while Plesch--Brukner's interaction structure routes from $9$ abstract CNOTs down to $5$ (real) or $6$ (complex), and the sign-aware M\"ott\"onen's $6$-CNOT triangle lifts to $9$. On the real target the infidelities follow the routed counts: Plesch--Brukner leads ($1.19\times10^{-3}$), the M\"ott\"onen cascade ($1.48\times10^{-3}$), the QsiHT Real Synthesis ($1.46\times10^{-3}$), and the QsiHT Complex Synthesis and \texttt{StatePreparation} pair ($1.58\times10^{-3}$) cluster just above, the higher-routed cascades trail, and QSD stays worst by an order of magnitude ($2.89\times10^{-2}$). On the complex target the QsiHT Complex Synthesis and \texttt{StatePreparation} lead ($\approx 2.5\times10^{-4}$). The routed ranking is, however, not stable across register sizes, and rather than read it off one pinned seed the routed two-qubit count was swept over $64$ transpiler seeds on the same heavy-hex map (the transpiler layout and routing passes are seeded, so a single pinned integer is one draw from a distribution over embeddings). The two syntheses route identically to \texttt{StatePreparation} at every size (median $7$, $13$, $36$, $80$ two-qubit gates at $n=3$--$6$), and their ordering against Plesch--Brukner reverses with $n$. At $n=3$ Plesch--Brukner routes to a median $5$ against the syntheses' median $7$ (they beat it on $0\%$ of the $64$ real-target seeds). At $n=4$ the syntheses take the lead, median $13$ (range $13$--$19$) against Plesch--Brukner's $16$ ($14$--$16$), winning $84\%$ of seeds, so the tie the pinned seed happened to show ($16$ each) is not the typical embedding. At $n=5$ they lead on every seed, median $36$ ($34$--$39$) against $51$ ($48$--$55$), win fraction $1.00$, and at $n=6$ they still lead on every seed, median $80$ ($77$--$89$) against $95$ ($88$--$100$). The $K_3$-triangle penalty that used to fall on the original fast path now falls on the sign-aware M\"ott\"onen cascade, whose $6$-CNOT circuit routes to a median $9$, $18$, $42$, $92$ at $n=3$--$6$ and beats Plesch--Brukner only from $n=5$ up. The per-seed spread is comparable to the median gaps at $n=6$, so a routed comparison at a single size does not extrapolate to a stable ordering, and the connectivity-fixed tables above remain the reference reading of the gate-count effect.

\begin{table}[H]
\centering
\caption{Sampled error under the paired-seed \texttt{GenericBackendV2} noise model on the distance-3 heavy-hex coupling map ($n=3$, $8192$ shots, $20$ runs). Routed is the two-qubit count of the circuit actually executed after optimization-level-3 transpilation onto the map with a pinned seed. Unlike the all-to-all baseline of Tables~\ref{tab:noisy_complex} and~\ref{tab:noisy_real}, routing overhead here acts inside the noise model: the two syntheses and \texttt{StatePreparation} route from $4$ abstract to $7$ on both targets, Plesch--Brukner compresses ($9\to5$ real, $9\to6$ complex) and takes the lowest real-target error, and the sign-aware M\"ott\"onen's $6$-CNOT triangle lifts to $9$. The Real Synthesis does not apply to complex targets (--). The lowest value in each column is shown in bold.}
\label{tab:noisy_heavyhex}
\resizebox{\textwidth}{!}{%
\begin{tabular}{lrrrrrr}
\hline
& \multicolumn{3}{c}{\textbf{Complex target}} & \multicolumn{3}{c}{\textbf{Real target}} \\
\textbf{Method} & \textbf{CNOTs} & \textbf{Routed} & \textbf{Infidelity} & \textbf{CNOTs} & \textbf{Routed} & \textbf{Infidelity} \\
\hline
QsiHT Fast Path Real Synthesis & -- & -- & -- & $\mathbf{4}$ & $7$ & $1.461\times10^{-3}$ \\
QsiHT Fast Path Complex Synthesis & $\mathbf{4}$ & $7$ & $2.527\times10^{-4}$ & $\mathbf{4}$ & $7$ & $1.584\times10^{-3}$ \\
M\"ott\"onen (UCR) & $12$ & $11$ & $3.625\times10^{-4}$ & $12$ & $10$ & $1.476\times10^{-3}$ \\
M\"ott\"onen (sign-aware) & $12$ & $11$ & $3.625\times10^{-4}$ & $6$ & $9$ & $3.584\times10^{-3}$ \\
Iten-style cascade & $10$ & $11$ & $3.625\times10^{-4}$ & $7$ & $10$ & $3.659\times10^{-3}$ \\
Qiskit \texttt{StatePreparation} & $\mathbf{4}$ & $7$ & $\mathbf{2.514\times10^{-4}}$ & $\mathbf{4}$ & $7$ & $1.584\times10^{-3}$ \\
Multiplexor & $10$ & $11$ & $3.625\times10^{-4}$ & $7$ & $8$ & $2.701\times10^{-3}$ \\
General isometry & $10$ & $11$ & $3.625\times10^{-4}$ & $7$ & $10$ & $3.659\times10^{-3}$ \\
Plesch--Brukner & $9$ & $\mathbf{6}$ & $8.381\times10^{-4}$ & $9$ & $\mathbf{5}$ & $\mathbf{1.192\times10^{-3}}$ \\
Quantum Shannon & $36$ & $34$ & $9.710\times10^{-3}$ & $36$ & $34$ & $2.887\times10^{-2}$ \\
\hline
\end{tabular}}
\end{table}

\begin{figure}[H]
\centering
\includegraphics[width=\textwidth]{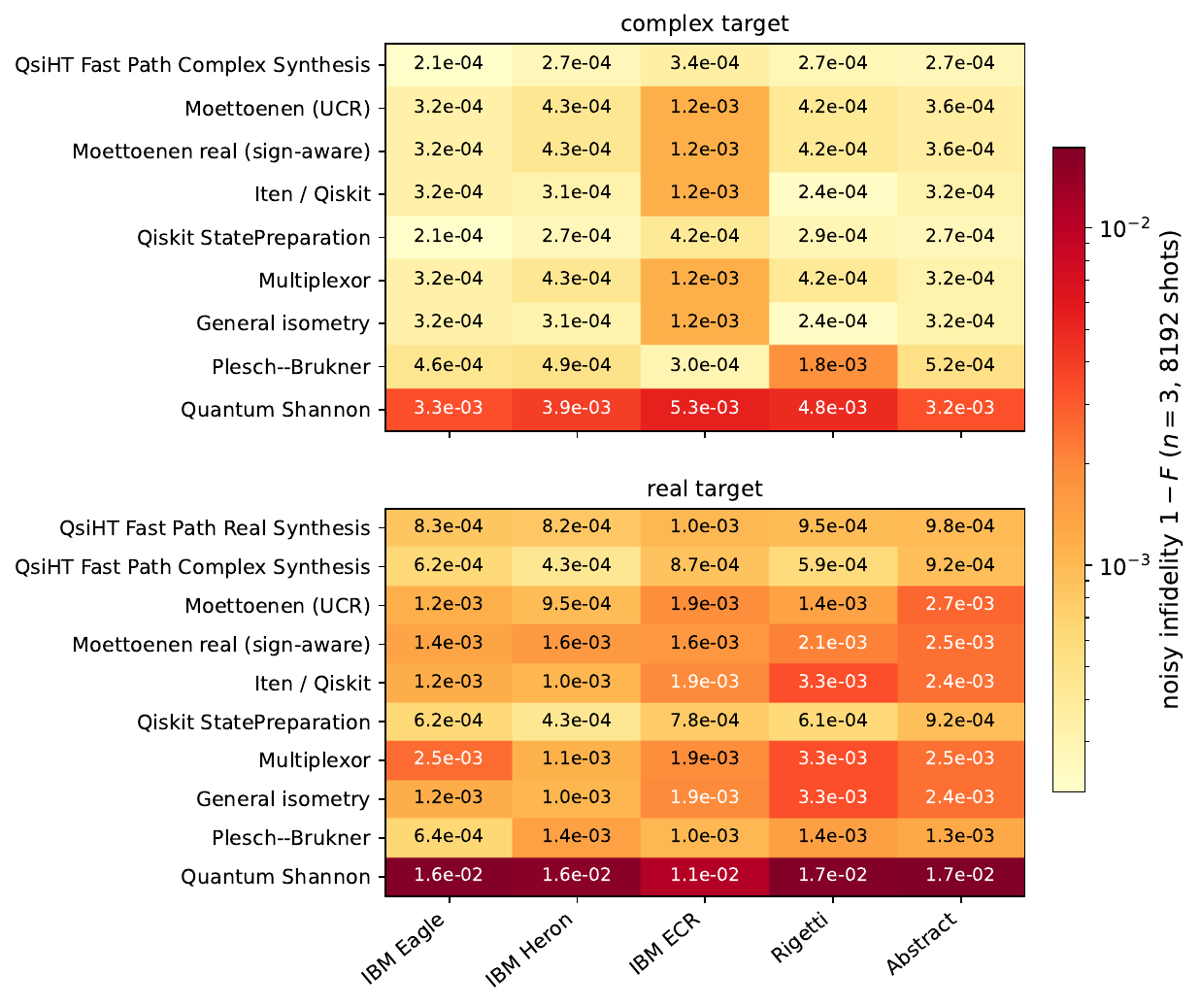}
\caption{Noisy infidelity $1-F$ (mean over $20$ runs, $8192$ shots, $n=3$) for every builder (rows) under the per-platform noise models (columns), complex target (left) and real target (right), on a shared logarithmic color scale. Quantum Shannon (bottom row) is the worst on every platform, while the frontier methods stay light. Only that coarse frontier-versus-QSD tier is stable across gate sets: the calibration ensemble of Figure~\ref{fig:noise_ensemble} shows that the fine per-method order reshuffles with the noise model.}
\label{fig:platform_infidelity}
\end{figure}

\paragraph{Across hardware noise models.} The comparison so far uses one shared noise model. To test whether the ranking survives a change of hardware, the per-method infidelity is recomputed under a separate \texttt{GenericBackendV2} noise model for each native gate set of Table~\ref{tab:nisq_platforms}, so that each platform's entangler and single-qubit set carry their own generated errors. Tables~\ref{tab:platforms_complex} and~\ref{tab:platforms_real} and Figure~\ref{fig:platform_infidelity} report the $20$-run mean infidelity at $8192$ shots for the complex and real targets. Three of the eight platforms are omitted: Google Sycamore's iSWAP entangler is not simulable in the Aer version used here, and the generic model assigns IonQ's and Quantinuum's trapped-ion $XX$ and $ZZ$ entanglers unrealistically large two-qubit errors that would misrepresent real trapped-ion hardware. The structural gate-set differences of all eight, the iSWAP doubling of the two-qubit count in particular, are already quantified in Table~\ref{tab:nisq_depth}.

The coarse tier is preserved on every platform. Quantum Shannon is the worst everywhere, separated from the best frontier member by an order of magnitude on most platforms and from the nearest (worst-frontier) member by a factor ranging from roughly $2.7$ to $9.8$ across platforms and targets (on the ECR set, complex, its $5.31\times10^{-3}$ against a frontier spanning $3.0\times10^{-4}$ to $1.2\times10^{-3}$). The platforms differ in absolute error level by up to a few-fold, but none reorders the frontier relative to QSD. Within each platform the shot seeds are paired across methods, so builders whose circuits coincide print identical cells (the merged-cascade family collapses to a single value on the complex target, and the QsiHT Complex Synthesis and \texttt{StatePreparation} coincide on several columns). Which frontier builder prints lowest, by contrast, moves from platform to platform on the complex target (the QsiHT Complex Synthesis or \texttt{StatePreparation} on the Eagle, Heron, and abstract columns, Plesch--Brukner on the ECR set, the Iten-style cascade on Rigetti), while on the real one the QsiHT Complex Synthesis and \texttt{StatePreparation} share the lead on three of the five columns (Eagle, Heron, and abstract), with the ECR set going to \texttt{StatePreparation} alone and Rigetti to the QsiHT Complex Synthesis alone. As with the structural cross-platform comparison of~\S\ref{sec:nisq}, the hardware noise rescales the costs without disturbing the coarse frontier-versus-QSD split, but the fine order is not a portable conclusion, and the calibration ensemble below quantifies how much of it is noise.

\begin{table}[H]
\centering
\caption{Noisy infidelity $1-F$ (mean over $20$ runs, $8192$ shots, $n=3$) for the complex target under a separate \texttt{GenericBackendV2} noise model per platform of Table~\ref{tab:nisq_platforms}, for all nine builders. Within each platform the per-run shot seeds are paired across methods, so builders whose circuits coincide print identical cells. Each cell lists the mean infidelity (top) and variance (bottom), with the lowest infidelity and lowest variance per platform (column) in bold. The coarse ranking, frontier versus Quantum Shannon, is preserved across platforms.}
\label{tab:platforms_complex}
\resizebox{\textwidth}{!}{%
\begin{tabular}{lccccc}
\hline
\textbf{Method} & \textbf{IBM Eagle} & \textbf{IBM Heron} & \textbf{IBM ECR} & \textbf{Rigetti} & \textbf{Abstract} \\
\hline
QsiHT Fast Path Complex Synthesis & \shortstack{$\mathbf{2.132\times10^{-4}}$\\$\mathbf{9.300\times10^{-9}}$} & \shortstack{$\mathbf{2.703\times10^{-4}}$\\$\mathbf{2.771\times10^{-8}}$} & \shortstack{$3.447\times10^{-4}$\\$\mathbf{1.469\times10^{-8}}$} & \shortstack{$2.720\times10^{-4}$\\$2.580\times10^{-8}$} & \shortstack{$\mathbf{2.682\times10^{-4}}$\\$\mathbf{2.159\times10^{-8}}$} \\
\hline
M\"ott\"onen (UCR) & \shortstack{$3.220\times10^{-4}$\\$1.581\times10^{-8}$} & \shortstack{$4.330\times10^{-4}$\\$5.486\times10^{-8}$} & \shortstack{$1.163\times10^{-3}$\\$9.408\times10^{-8}$} & \shortstack{$4.155\times10^{-4}$\\$3.046\times10^{-8}$} & \shortstack{$3.634\times10^{-4}$\\$2.466\times10^{-8}$} \\
\hline
M\"ott\"onen (sign-aware) & \shortstack{$3.220\times10^{-4}$\\$1.581\times10^{-8}$} & \shortstack{$4.330\times10^{-4}$\\$5.486\times10^{-8}$} & \shortstack{$1.163\times10^{-3}$\\$9.408\times10^{-8}$} & \shortstack{$4.155\times10^{-4}$\\$3.046\times10^{-8}$} & \shortstack{$3.634\times10^{-4}$\\$2.466\times10^{-8}$} \\
\hline
Iten-style cascade & \shortstack{$3.220\times10^{-4}$\\$1.581\times10^{-8}$} & \shortstack{$3.133\times10^{-4}$\\$3.450\times10^{-8}$} & \shortstack{$1.163\times10^{-3}$\\$9.408\times10^{-8}$} & \shortstack{$\mathbf{2.352\times10^{-4}}$\\$\mathbf{1.898\times10^{-8}}$} & \shortstack{$3.176\times10^{-4}$\\$2.529\times10^{-8}$} \\
\hline
Qiskit \texttt{StatePreparation} & \shortstack{$2.137\times10^{-4}$\\$9.314\times10^{-9}$} & \shortstack{$\mathbf{2.703\times10^{-4}}$\\$\mathbf{2.771\times10^{-8}}$} & \shortstack{$4.163\times10^{-4}$\\$1.624\times10^{-8}$} & \shortstack{$2.880\times10^{-4}$\\$2.805\times10^{-8}$} & \shortstack{$\mathbf{2.682\times10^{-4}}$\\$\mathbf{2.159\times10^{-8}}$} \\
\hline
Multiplexor & \shortstack{$3.220\times10^{-4}$\\$1.581\times10^{-8}$} & \shortstack{$4.330\times10^{-4}$\\$5.486\times10^{-8}$} & \shortstack{$1.163\times10^{-3}$\\$9.408\times10^{-8}$} & \shortstack{$4.155\times10^{-4}$\\$3.046\times10^{-8}$} & \shortstack{$3.176\times10^{-4}$\\$2.529\times10^{-8}$} \\
\hline
General isometry & \shortstack{$3.220\times10^{-4}$\\$1.581\times10^{-8}$} & \shortstack{$3.133\times10^{-4}$\\$3.450\times10^{-8}$} & \shortstack{$1.163\times10^{-3}$\\$9.408\times10^{-8}$} & \shortstack{$\mathbf{2.352\times10^{-4}}$\\$\mathbf{1.898\times10^{-8}}$} & \shortstack{$3.176\times10^{-4}$\\$2.529\times10^{-8}$} \\
\hline
Plesch--Brukner & \shortstack{$4.558\times10^{-4}$\\$2.725\times10^{-8}$} & \shortstack{$4.895\times10^{-4}$\\$5.116\times10^{-8}$} & \shortstack{$\mathbf{2.973\times10^{-4}}$\\$1.727\times10^{-8}$} & \shortstack{$1.800\times10^{-3}$\\$1.983\times10^{-7}$} & \shortstack{$5.181\times10^{-4}$\\$6.375\times10^{-8}$} \\
\hline
Quantum Shannon & \shortstack{$3.280\times10^{-3}$\\$5.844\times10^{-7}$} & \shortstack{$3.935\times10^{-3}$\\$6.289\times10^{-7}$} & \shortstack{$5.313\times10^{-3}$\\$6.653\times10^{-7}$} & \shortstack{$4.775\times10^{-3}$\\$5.141\times10^{-7}$} & \shortstack{$3.249\times10^{-3}$\\$3.477\times10^{-7}$} \\
\hline
\end{tabular}}
\end{table}

\begin{table}[H]
\centering
\caption{Noisy infidelity $1-F$ (mean over $20$ runs, $8192$ shots, $n=3$) for the real target under a separate \texttt{GenericBackendV2} noise model per platform, for all ten builders. Shot seeds are paired across methods within each platform, as in Table~\ref{tab:platforms_complex}: the QsiHT Complex Synthesis and Qiskit \texttt{StatePreparation} execute equivalent circuits on this target and print identical cells on the Eagle, Heron, and abstract gate sets (their ECR and Rigetti native compilations diverge). Each cell lists the mean infidelity (top) and variance (bottom), with the lowest infidelity and lowest variance per platform (column) in bold. As on the complex target, Quantum Shannon is the worst on every platform and the frontier methods cluster.}
\label{tab:platforms_real}
\resizebox{\textwidth}{!}{%
\begin{tabular}{lccccc}
\hline
\textbf{Method} & \textbf{IBM Eagle} & \textbf{IBM Heron} & \textbf{IBM ECR} & \textbf{Rigetti} & \textbf{Abstract} \\
\hline
QsiHT Fast Path Real Synthesis & \shortstack{$8.330\times10^{-4}$\\$1.129\times10^{-7}$} & \shortstack{$8.192\times10^{-4}$\\$6.763\times10^{-8}$} & \shortstack{$1.008\times10^{-3}$\\$1.583\times10^{-7}$} & \shortstack{$9.528\times10^{-4}$\\$6.458\times10^{-8}$} & \shortstack{$9.751\times10^{-4}$\\$1.248\times10^{-7}$} \\
\hline
QsiHT Fast Path Complex Synthesis & \shortstack{$\mathbf{6.215\times10^{-4}}$\\$\mathbf{5.978\times10^{-8}}$} & \shortstack{$\mathbf{4.314\times10^{-4}}$\\$\mathbf{2.608\times10^{-8}}$} & \shortstack{$8.660\times10^{-4}$\\$1.343\times10^{-7}$} & \shortstack{$\mathbf{5.900\times10^{-4}}$\\$\mathbf{4.660\times10^{-8}}$} & \shortstack{$\mathbf{9.186\times10^{-4}}$\\$1.162\times10^{-7}$} \\
\hline
M\"ott\"onen (UCR) & \shortstack{$1.164\times10^{-3}$\\$1.276\times10^{-7}$} & \shortstack{$9.512\times10^{-4}$\\$7.925\times10^{-8}$} & \shortstack{$1.871\times10^{-3}$\\$2.075\times10^{-7}$} & \shortstack{$1.389\times10^{-3}$\\$1.622\times10^{-7}$} & \shortstack{$2.673\times10^{-3}$\\$2.817\times10^{-7}$} \\
\hline
M\"ott\"onen (sign-aware) & \shortstack{$1.354\times10^{-3}$\\$1.729\times10^{-7}$} & \shortstack{$1.579\times10^{-3}$\\$2.770\times10^{-7}$} & \shortstack{$1.631\times10^{-3}$\\$1.485\times10^{-7}$} & \shortstack{$2.108\times10^{-3}$\\$2.073\times10^{-7}$} & \shortstack{$2.471\times10^{-3}$\\$1.926\times10^{-7}$} \\
\hline
Iten-style cascade & \shortstack{$1.170\times10^{-3}$\\$1.294\times10^{-7}$} & \shortstack{$1.049\times10^{-3}$\\$8.989\times10^{-8}$} & \shortstack{$1.943\times10^{-3}$\\$2.202\times10^{-7}$} & \shortstack{$3.309\times10^{-3}$\\$2.187\times10^{-7}$} & \shortstack{$2.443\times10^{-3}$\\$3.020\times10^{-7}$} \\
\hline
Qiskit \texttt{StatePreparation} & \shortstack{$\mathbf{6.215\times10^{-4}}$\\$\mathbf{5.978\times10^{-8}}$} & \shortstack{$\mathbf{4.314\times10^{-4}}$\\$\mathbf{2.608\times10^{-8}}$} & \shortstack{$\mathbf{7.804\times10^{-4}}$\\$1.203\times10^{-7}$} & \shortstack{$6.111\times10^{-4}$\\$6.292\times10^{-8}$} & \shortstack{$\mathbf{9.186\times10^{-4}}$\\$1.162\times10^{-7}$} \\
\hline
Multiplexor & \shortstack{$2.529\times10^{-3}$\\$1.713\times10^{-7}$} & \shortstack{$1.111\times10^{-3}$\\$9.644\times10^{-8}$} & \shortstack{$1.869\times10^{-3}$\\$2.493\times10^{-7}$} & \shortstack{$3.284\times10^{-3}$\\$2.042\times10^{-7}$} & \shortstack{$2.468\times10^{-3}$\\$2.274\times10^{-7}$} \\
\hline
General isometry & \shortstack{$1.170\times10^{-3}$\\$1.294\times10^{-7}$} & \shortstack{$1.049\times10^{-3}$\\$8.989\times10^{-8}$} & \shortstack{$1.943\times10^{-3}$\\$2.202\times10^{-7}$} & \shortstack{$3.309\times10^{-3}$\\$2.187\times10^{-7}$} & \shortstack{$2.443\times10^{-3}$\\$3.020\times10^{-7}$} \\
\hline
Plesch--Brukner & \shortstack{$6.403\times10^{-4}$\\$8.691\times10^{-8}$} & \shortstack{$1.421\times10^{-3}$\\$2.453\times10^{-7}$} & \shortstack{$1.003\times10^{-3}$\\$\mathbf{3.549\times10^{-8}}$} & \shortstack{$1.445\times10^{-3}$\\$1.711\times10^{-7}$} & \shortstack{$1.262\times10^{-3}$\\$\mathbf{9.058\times10^{-8}}$} \\
\hline
Quantum Shannon & \shortstack{$1.647\times10^{-2}$\\$1.971\times10^{-6}$} & \shortstack{$1.550\times10^{-2}$\\$2.876\times10^{-6}$} & \shortstack{$1.066\times10^{-2}$\\$1.687\times10^{-6}$} & \shortstack{$1.654\times10^{-2}$\\$1.976\times10^{-6}$} & \shortstack{$1.725\times10^{-2}$\\$1.554\times10^{-6}$} \\
\hline
\end{tabular}}
\end{table}

\begin{figure}[H]
\centering
\includegraphics[width=\textwidth]{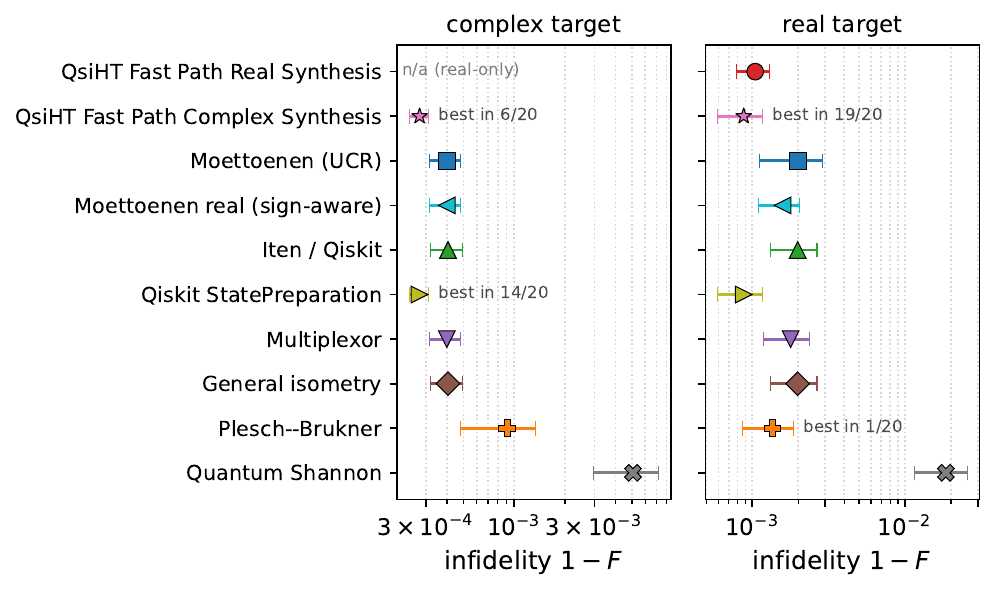}
\caption{Noisy infidelity per builder across $20$ calibration seeds of the $n=3$ \texttt{GenericBackendV2} noise model ($8192$ shots). Markers are the mean of the $20$ per-seed mean infidelities (each per-seed mean averages twenty sampling repeats with shot seeds paired across methods, so the seed-$12345$ calibration reproduces the single-seed Tables~\ref{tab:noisy_complex} and~\ref{tab:noisy_real} exactly). Bars are one standard deviation across the $20$ seeds, not the standard error of the mean and not the within-seed shot noise. Annotations count the seeds in which a builder ranked first. Only the coarse tiers are calibration-robust: Quantum Shannon is last in all $20$ seeds, and the QsiHT Complex Synthesis and Qiskit \texttt{StatePreparation}, an identical-circuit pair on the real target and a near-identical one on the complex, hold the rank-1 slot in all $20$ complex seeds and $19$ of $20$ real seeds (splitting $14$/$6$ complex and $19$/$0$ real by the deterministic tie-break, Plesch--Brukner taking the remaining real seed), while statistically identical builders (on the complex target the cascade family collapses to two byte-identical subgroups, a M\"ott\"onen/multiplexor triple and an Iten/isometry pair) still land several mean ranks apart, a deterministic arithmetic artifact of floating-point summation order below one unit in the last place (ULP) that shows the fine per-seed ranking at this scale carries no physical meaning.}
\label{fig:noise_ensemble}
\end{figure}

\paragraph{Calibration robustness.} The per-platform comparison above still rests on one calibration draw per platform, and the shared-model comparison of Tables~\ref{tab:noisy_complex} and~\ref{tab:noisy_real} on a single seed. To separate what is a property of the circuits from what is a property of a particular noise instance, the $n=3$ experiment is repeated across an ensemble of $20$ independently seeded \texttt{GenericBackendV2} calibrations, with twenty sampling repeats per calibration and the shot seeds \emph{paired} across methods: each repeat reuses the same simulator seed for every builder, so shot noise cannot reorder the field within a repeat. Figure~\ref{fig:noise_ensemble} reports each builder's mean infidelity with its spread across the $20$ calibrations and the induced rank distribution. The coarse tier is calibration-robust: Quantum Shannon is last in all $20$ calibrations on both targets. The low edge is now held by the QsiHT Complex Synthesis together with the deployed \texttt{StatePreparation}. On the complex target the two are near-identical but distinct circuits, \texttt{StatePreparation} ranking first in $14$ of $20$ and the QsiHT Complex Synthesis in the other $6$ (means $2.74\times10^{-4}$ and $2.75\times10^{-4}$). On the real target the QsiHT Complex Synthesis ranks first in $19$ of $20$ (mean $8.78\times10^{-4}$), its identical-circuit twin \texttt{StatePreparation} taking rank $2$, and Plesch--Brukner ($1.36\times10^{-3}$) the remaining one, while the QsiHT Real Synthesis, a distinct real-target circuit, sits mid-frontier ($1.04\times10^{-3}$, rank first in $0$ of $20$). The fine order among the higher-count builders is not robust: the cascade baselines, whose real-target means ($1.78$--$1.99\times10^{-3}$) exceed the leaders by less than the $5$--$9\times10^{-4}$ seed-to-seed spread of each method, interleave freely from calibration to calibration, so any within-frontier ordering at that scale is calibration noise. The complex target shows the reading can even invert: Plesch--Brukner carries the highest frontier mean infidelity there ($9.10\times10^{-4}$) despite executing fewer two-qubit gates ($6$) than the cascade family ($10$), the opposite of what the gate count predicts. The cross-platform statement these data support is the coarse one: the exact frontier sits well below QSD under every calibration drawn, with the QsiHT Complex Synthesis and the \texttt{StatePreparation} tied at its low edge, and nothing finer transfers.

\begin{figure}[H]
\centering
\includegraphics[width=0.78\textwidth]{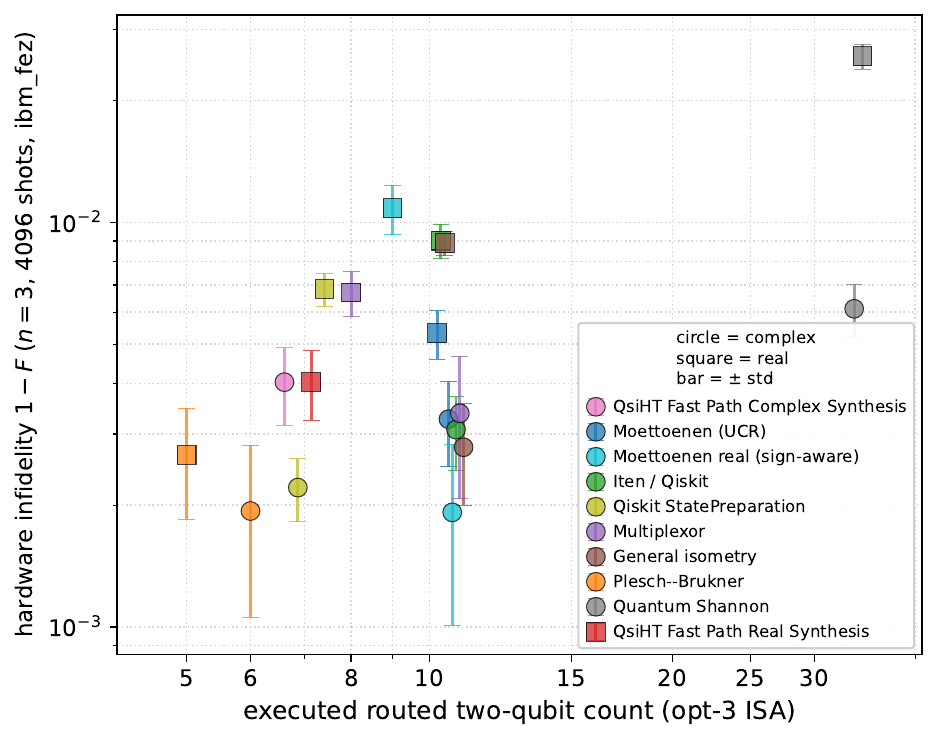}
\caption{Hardware sampled infidelity $1-F$ versus the executed routed two-qubit count (log--log) on \texttt{ibm\_fez}, $n=3$, $4096$ shots, for the complex (circles) and real (squares) targets, with error bars showing the run-to-run standard deviation over the $5$ runs. Each marker starts at the true integer routed count of the circuit that produced it, the Routed column of Tables~\ref{tab:hw_complex} and~\ref{tab:hw_real}, and only markers that would coincide there are spread slightly and symmetrically about the shared count so that no method is hidden behind another. Quantum Shannon ($34$ routed gates) sits well above the frontier on both targets, but within the frontier the realized infidelity does not order by the routed two-qubit count, for the reasons discussed above. The error bars are wider than in Figure~\ref{fig:noise_vs_cnot} because they reflect real device drift across the runs and the smaller $5$-run sample.}
\label{fig:hw_noise_vs_cnot}
\end{figure}

\paragraph{On a real quantum processor.} The noise models above are statistical reconstructions of hardware. As a final check the same $n=3$ circuits were executed on an actual IBM Quantum device, \texttt{ibm\_fez}, an IBM Heron r2 processor of $156$ fixed-frequency transmon qubits, through the Qiskit Runtime sampler. The original job ran six of the benchmarked methods (M\"ott\"onen, Iten, multiplexor, general isometry, Plesch--Brukner, and QSD) on both the real and complex targets, $5$ times at $4096$ shots, as a single submission. The sign-aware M\"ott\"onen and Qiskit \texttt{StatePreparation} were added in a later completion job on the same device at the same shots and runs, and the two QsiHT syntheses in a third, newest submission session, so those rows carry different calibration days, and any comparison spanning sessions compounds circuit differences with calibration drift. The infidelity is computed exactly as before, the statistical fidelity error against the ideal distribution. Tables~\ref{tab:hw_complex} and~\ref{tab:hw_real} report, per method, the mean infidelity, its run-to-run variance, and the mean-square error over the five runs, against the abstract-basis CNOT and depth of each circuit and the \emph{routed} two-qubit count (Routed): the number of native two-qubit gates in the instruction-set-architecture (ISA) circuit that actually executed, after the optimization-level-3 preset pass manager embeds each circuit onto the Heron heavy-hexagonal coupling map (reconstructed offline with the matching fake-backend snapshot and a fixed transpiler seed for the original job, and read directly from the submitted ISA circuits for the completion and synthesis jobs). Figure~\ref{fig:hw_noise_vs_cnot} plots these infidelities against the executed routed two-qubit count (the Routed column), with error-bar whiskers for the run-to-run spread.

The coarse prediction holds on real hardware. Quantum Shannon is the worst method by a wide margin on both targets, several times above the frontier on the real target ($2.574\times10^{-2}$ against a frontier spanning $2.7\times10^{-3}$ to $1.1\times10^{-2}$) and worst on MSE as well ($4.852\times10^{-4}$), exactly as its $36$ CNOTs and depth-$99$ circuit predict. The frontier methods cluster within a few times $10^{-3}$ ($1.9$--$3.4\times10^{-3}$ on the complex target, $2.7\times10^{-3}$--$1.1\times10^{-2}$ on the real one), with Plesch--Brukner attaining the lowest infidelity on the real target and the lowest within the single-job set on the complex one (the sign-aware M\"ott\"onen row from the completion job prints $1.920\times10^{-3}$ against Plesch--Brukner's $1.935\times10^{-3}$, a gap far inside the run-to-run spread and across calibration days). This differs from the simulated noise model, where Plesch--Brukner trailed the frontier on infidelity while sitting near the low end on MSE (Table~\ref{tab:noisy_complex}): which metric favors Plesch--Brukner flips between the simulated and hardware runs.

However, the ordering of the frontier methods is not reproduced through the CNOT ranking that the shared synthetic noise model gave. On the real target the two QsiHT syntheses and \texttt{StatePreparation} tie for the fewest abstract entanglers ($4$ CNOTs, against $6$ to $12$ for the cascades and $9$ for Plesch--Brukner) and the shallowest circuit (depth $9$), routing to $7$ two-qubit gates like the deployed \texttt{StatePreparation}, yet the QsiHT Real Synthesis realizes neither the lowest infidelity nor the lowest run-to-run variance. It sits in a separate submission session, so its $4.035\times10^{-3}$ against, say, \texttt{StatePreparation}'s $6.842\times10^{-3}$ cannot be read as a method difference, because the two rows were taken on different calibration days. Three hardware effects, none of which the gate count controls, account for the within-session decoupling. First, the transpiler must embed the logical circuit onto a specific triple of physical qubits and the couplers between them, and the two-qubit and readout error rates vary several-fold across the $156$-qubit chip, so two circuits with nearly equal CNOT counts can land on couplers of very different quality. The three $7$-CNOT methods that should be near-identical by gate count instead spread from $6.708\times10^{-3}$ (Multiplexor) to $9.015\times10^{-3}$ (Iten), a gap their identical gate counts cannot explain, while Plesch--Brukner with $9$ CNOTs and the deepest frontier circuit (depth $24$) takes the lowest infidelity of the original job ($2.660\times10^{-3}$). Second, measurement and state-preparation error is assigned per physical qubit at the ends of the circuit and does not scale with the gate count, so at these shallow depths it forms a large, roughly fixed floor set only by which qubits are read out, and the CNOT-minimizing method gains nothing against it. For example, the same Iten real circuit costs $1.014\times10^{-3}$ on the homogeneous simulated backend (Table~\ref{tab:noisy_real}) but $9.015\times10^{-3}$ on hardware, a gap the identical gate sequence cannot explain. Third, the reported depth is an abstract layer count, not physical duration. The realized time is the sum of native gate durations along the critical path plus scheduler idle padding, both of which vary across couplers, so equal abstract depth maps to unequal decoherence exposure. The Iten, Multiplexor, and General isometry trio illustrates this: they all carry depth $14$ yet produce $9.015\times10^{-3}$, $6.708\times10^{-3}$, and $8.882\times10^{-3}$, a spread consistent with their differing qubit placement, readout assignment, and physical duration rather than any one cause in isolation. The run-to-run variances are dominated by two further effects that are likewise independent of the gate count: the device calibration drifts over the minutes the job runs, and the variance is itself estimated from only $5$ runs. Both attach to the selected qubits and the timing of the job, not to the circuit body, which is why the variance does not order by CNOT count: the $10$-CNOT Multiplexor carries the worst frontier variance on the complex target ($1.688\times10^{-6}$) while the $9$-CNOT Plesch--Brukner sits well below it, with no gate-count basis. What the QsiHT syntheses' structural standing rests on is therefore the build-time count itself, which is device-independent: they emit $2^n-n-1$ CNOTs, tying the deployed \texttt{StatePreparation} floor (\S\ref{sec:scaling}), and unlike the original fast path they route to $7$ rather than $9$ at $n=3$, matching \texttt{StatePreparation} rather than paying a $K_3$-triangle SWAP. Whether that structural standing converts into a realized-error advantage is not established by these $n=3$ runs, where the frontier circuits differ by at most a few CNOTs and that term is small next to the fixed per-qubit floor, and where the syntheses were measured cross-session. The one method already in a different gate order, Quantum Shannon at $36$ CNOTs and depth $99$, is unambiguously the worst method on infidelity and MSE on both targets. What the hardware run establishes is the part of the ranking that drives method selection, that the exact frontier sits well below QSD on real hardware, and that the best frontier methods stay within a small factor of one another.

A separate point is that the method with the lowest infidelity does not need to be the method with the lowest MSE, because the two metrics are different norms on the same vector of per-state probability errors. Write the per-state error as $d_x=\hat p_x-p_x$ between the sampled distribution $\hat p_x$ and the ideal $p_x$, with $\sum_x d_x=0$ by normalization. Expanding the Bhattacharyya overlap to second order gives the leading-order infidelity $1-F\approx\tfrac14\sum_x d_x^2/p_x$, a probability-weighted divergence that divides each squared deviation by $p_x$, so it is governed by relative errors and is most sensitive to deviations on low-probability outcomes. The MSE is the uniform-weight average $\tfrac{1}{N}\sum_x d_x^2$ with $N=2^n$, which applies the same weight to every state regardless of $p_x$, so it tracks absolute deviations. Because the largest absolute sampling deviations typically fall on the high-probability outcomes, where the shot variance $p_x(1-p_x)$ peaks near $p_x=1/2$, the MSE is in practice driven by those outcomes. The two norms sum the same $d_x^2$ but weight them differently, the one by $1/p_x$ and the other uniformly, so their minimizers can differ. The hardware data show exactly this. On the real target (Table~\ref{tab:hw_real}) Plesch--Brukner has the lowest infidelity ($2.660\times10^{-3}$) while the QsiHT Real Synthesis has the lowest MSE ($6.711\times10^{-5}$), the two rankings parting. On the complex target (Table~\ref{tab:hw_complex}) Qiskit \texttt{StatePreparation} has the lowest MSE ($6.805\times10^{-5}$) while the lowest infidelities belong to the sign-aware M\"ott\"onen ($1.920\times10^{-3}$) and Plesch--Brukner ($1.935\times10^{-3}$), and Plesch--Brukner is at the same time the worst-MSE frontier method of the original job there ($1.415\times10^{-4}$, with the cross-session QsiHT Complex Synthesis row higher still at $1.445\times10^{-4}$). The gap flipping between the simulated and hardware runs on the same target confirms the divergence is a property of the two metrics being distinct norms, not a structural feature of any one circuit always concentrating its error on small or large amplitudes.

\begin{table}[H]
\centering
\caption[Sampled error, complex $n=3$, \texttt{ibm\_fez}]{Sampled error for the \emph{complex} $n=3$ target prepared on the IBM Quantum processor \texttt{ibm\_fez} (IBM Heron r2, $156$ qubits), over $5$ runs at $4096$ shots, all nine builders on this target. The QsiHT Complex Synthesis row comes from the newest submission session, the sign-aware M\"ott\"onen and \texttt{StatePreparation} rows from the earlier completion job, and the remaining six from the original job, so any comparison spanning sessions compounds circuit differences with calibration drift (\S\ref{sec:hw_limitations}). The fidelity error (infidelity $1-F$) is reported with its run-to-run variance (Var.) and the mean-square error (MSE), against the abstract-basis CNOT and depth counts and the routed two-qubit count of the executed opt-3 ISA circuit (Routed). The lowest routed count and lowest infidelity, the executed-axis columns, are shown in bold, and the other columns are left unbolded because at $n=3$ their minima are within-noise and carry no gate-count basis (\S\ref{sec:hw_limitations}).}
\label{tab:hw_complex}
\resizebox{\textwidth}{!}{%
\begin{tabular}{lrrrrrr}
\hline
\textbf{Method} & \textbf{CNOTs} & \textbf{Routed} & \textbf{Depth} & \textbf{Infidelity} & \textbf{Var.} & \textbf{MSE} \\
\hline
QsiHT Fast Path Complex Synthesis & $4$ & $7$ & $9$ & $4.028\times10^{-3}$ & $7.682\times10^{-7}$ & $1.445\times10^{-4}$ \\
M\"ott\"onen (UCR) & $12$ & $11$ & $24$ & $3.264\times10^{-3}$ & $5.973\times10^{-7}$ & $1.001\times10^{-4}$ \\
M\"ott\"onen (sign-aware) & $12$ & $11$ & $24$ & $\mathbf{1.920\times10^{-3}}$ & $8.316\times10^{-7}$ & $7.576\times10^{-5}$ \\
Iten-style cascade & $10$ & $11$ & $20$ & $3.076\times10^{-3}$ & $4.010\times10^{-7}$ & $1.125\times10^{-4}$ \\
Qiskit \texttt{StatePreparation} & $4$ & $7$ & $9$ & $2.211\times10^{-3}$ & $1.535\times10^{-7}$ & $6.805\times10^{-5}$ \\
Multiplexor & $10$ & $11$ & $20$ & $3.373\times10^{-3}$ & $1.688\times10^{-6}$ & $1.210\times10^{-4}$ \\
General isometry & $10$ & $11$ & $20$ & $2.782\times10^{-3}$ & $6.145\times10^{-7}$ & $8.092\times10^{-5}$ \\
Plesch--Brukner & $9$ & $\mathbf{6}$ & $26$ & $1.935\times10^{-3}$ & $7.684\times10^{-7}$ & $1.415\times10^{-4}$ \\
Quantum Shannon & $36$ & $34$ & $99$ & $6.117\times10^{-3}$ & $8.300\times10^{-7}$ & $2.555\times10^{-4}$ \\
\hline
\end{tabular}}
\end{table}

\begin{table}[H]
\centering
\caption[Sampled error, real $n=3$, \texttt{ibm\_fez}]{Sampled error for the \emph{real} $n=3$ target on \texttt{ibm\_fez}, over $5$ runs at $4096$ shots. The QsiHT Real Synthesis row is from the newest submission session, the sign-aware M\"ott\"onen and \texttt{StatePreparation} rows from the earlier completion job, and the remaining six from the original job. The infidelity $1-F$ is reported with its run-to-run variance (Var.) and the mean-square error (MSE), against the abstract-basis CNOT and depth counts and the routed two-qubit count of the executed opt-3 ISA circuit (Routed). The QsiHT Real Synthesis and \texttt{StatePreparation} route identically ($4$ printed, $7$ routed, depth $9$), while Plesch--Brukner's $9$ compile down to $5$ and the sign-aware M\"ott\"onen's $6$ lift to $9$ (\S\ref{sec:hw_limitations}). Quantum Shannon is an order of magnitude worse than the frontier. The lowest routed count and lowest infidelity, the executed-axis columns, are shown in bold, and the other columns are left unbolded because at $n=3$ their minima are within-noise and carry no gate-count basis (\S\ref{sec:hw_limitations}).}
\label{tab:hw_real}
\resizebox{\textwidth}{!}{%
\begin{tabular}{lrrrrrr}
\hline
\textbf{Method} & \textbf{CNOTs} & \textbf{Routed} & \textbf{Depth} & \textbf{Infidelity} & \textbf{Var.} & \textbf{MSE} \\
\hline
QsiHT Fast Path Real Synthesis & $4$ & $7$ & $9$ & $4.035\times10^{-3}$ & $6.200\times10^{-7}$ & $6.711\times10^{-5}$ \\
M\"ott\"onen (UCR) & $12$ & $10$ & $20$ & $5.321\times10^{-3}$ & $5.336\times10^{-7}$ & $1.693\times10^{-4}$ \\
M\"ott\"onen (sign-aware) & $6$ & $9$ & $11$ & $1.084\times10^{-2}$ & $2.304\times10^{-6}$ & $2.881\times10^{-4}$ \\
Iten-style cascade & $7$ & $10$ & $14$ & $9.015\times10^{-3}$ & $7.945\times10^{-7}$ & $4.251\times10^{-4}$ \\
Qiskit \texttt{StatePreparation} & $4$ & $7$ & $9$ & $6.842\times10^{-3}$ & $4.150\times10^{-7}$ & $1.291\times10^{-4}$ \\
Multiplexor & $7$ & $8$ & $14$ & $6.708\times10^{-3}$ & $7.440\times10^{-7}$ & $3.052\times10^{-4}$ \\
General isometry & $7$ & $10$ & $14$ & $8.882\times10^{-3}$ & $3.688\times10^{-7}$ & $3.971\times10^{-4}$ \\
Plesch--Brukner & $9$ & $\mathbf{5}$ & $24$ & $\mathbf{2.660\times10^{-3}}$ & $6.616\times10^{-7}$ & $1.077\times10^{-4}$ \\
Quantum Shannon & $36$ & $34$ & $99$ & $2.574\times10^{-2}$ & $3.428\times10^{-6}$ & $4.852\times10^{-4}$ \\
\hline
\end{tabular}}
\end{table}

\begin{figure}[H]
\centering
\includegraphics[width=0.78\textwidth]{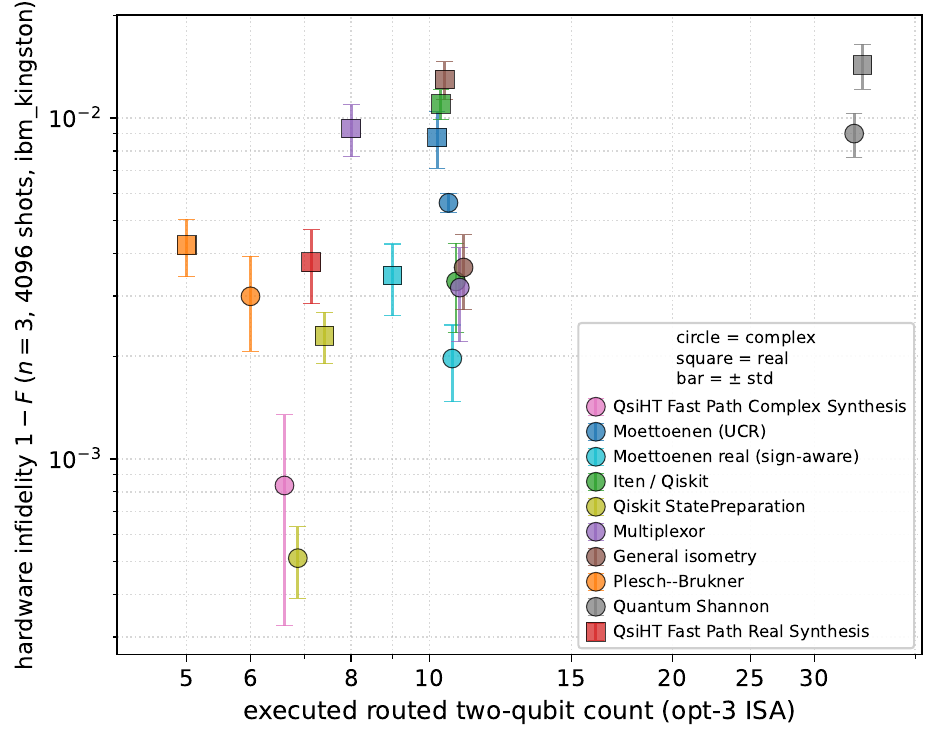}
\caption{Hardware sampled infidelity $1-F$ versus the executed routed two-qubit count (log--log) on \texttt{ibm\_kingston}, $n=3$, $4096$ shots, complex (circles) and real (squares), with error bars showing the run-to-run standard deviation over the $5$ runs. As in Figure~\ref{fig:hw_noise_vs_cnot}, each marker starts at its true integer routed count and coinciding markers are spread slightly about the shared count so that none is hidden. As on \texttt{ibm\_fez} (Figure~\ref{fig:hw_noise_vs_cnot}), Quantum Shannon is worst and Plesch--Brukner best, but the within-frontier order differs, since at $n=3$ the realized error is set by the qubits the job lands on rather than the gate count.}
\label{fig:hw_kingston_noise_vs_cnot}
\end{figure}

\paragraph{A second processor.} To check that these conclusions are not an artifact of one chip, the identical set of jobs was run on a second IBM Heron r2 device under the same $5$ runs at $4096$ shots. Tables~\ref{tab:hw_kingston_complex} and~\ref{tab:hw_kingston_real} and Figure~\ref{fig:hw_kingston_noise_vs_cnot} report the result. The coarse ranking is reproduced: Quantum Shannon is again the worst on infidelity ($8.992\times10^{-3}$ and $1.430\times10^{-2}$), Plesch--Brukner again attains the lowest infidelity within the original six-method job ($2.996\times10^{-3}$ complex, $4.232\times10^{-3}$ real), and the completion-job \texttt{StatePreparation} prints lower still on both targets ($5.111\times10^{-4}$ complex, $2.295\times10^{-3}$ real), consistent with its smallest routed footprint ($7$ gates, tied with the QsiHT syntheses). That the frontier-versus-QSD split reappears on a second, independently calibrated device is the strongest evidence the chapter offers that the separation is a property of the circuits and not of one chip's calibration.

The fine ordering, by contrast, reshuffles between the two processors, exactly as the per-qubit-calibration argument predicts. On \texttt{ibm\_kingston} the real target spreads the frontier so widely that the $7$-CNOT General isometry ($1.296\times10^{-2}$) almost reaches the $36$-CNOT Quantum Shannon ($1.430\times10^{-2}$), the opposite of what a gate-count reading predicts, while the QsiHT Real Synthesis ($3.772\times10^{-3}$) and \texttt{StatePreparation} ($2.295\times10^{-3}$), both at $4$ built and $7$ routed two-qubit gates, and Plesch--Brukner ($4.232\times10^{-3}$) hold the low edge, a different intra-frontier order than \texttt{ibm\_fez} gave. On the complex target the QsiHT Complex Synthesis prints an excellent $8.349\times10^{-4}$, second only to \texttt{StatePreparation} ($5.111\times10^{-4}$), where on \texttt{ibm\_fez} the same synthesis sat at the bottom of the frontier (the worst frontier method there, $4.028\times10^{-3}$), a swing of the same order between the two devices that is the size of the effect the fine ordering must be read against. The absolute error level differs too, with \texttt{ibm\_kingston} somewhat noisier than \texttt{ibm\_fez} on the real target. The metric divergence of the previous paragraphs repeats: on \texttt{ibm\_kingston} the real target gives Quantum Shannon the worst infidelity yet only a mid-range MSE ($2.570\times10^{-4}$, below four of the other methods), with General isometry instead the worst on MSE ($7.034\times10^{-4}$), one more case where the infidelity and MSE rankings part. Taken together, the two devices deliver the same verdict for method selection, the exact frontier well below QSD with the low-routed-count builders (\texttt{StatePreparation}, the QsiHT syntheses, and Plesch--Brukner) at its cheap edge, while the run-to-run order inside the frontier stays a property of the particular qubits and calibration day a job lands on rather than of the gate counts.

\begin{table}[H]
\centering
\caption[Sampled error, complex $n=3$, \texttt{ibm\_kingston}]{Sampled error for the \emph{complex} $n=3$ target on the second IBM Quantum processor \texttt{ibm\_kingston} (IBM Heron r2, $156$ qubits), over $5$ runs at $4096$ shots, all nine builders. Columns and two-job provenance as in Table~\ref{tab:hw_complex}. The lowest routed count and lowest infidelity, the executed-axis columns, are shown in bold, and the other columns are left unbolded because at $n=3$ their minima are within-noise and carry no gate-count basis (\S\ref{sec:hw_limitations}).}
\label{tab:hw_kingston_complex}
\resizebox{\textwidth}{!}{%
\begin{tabular}{lrrrrrr}
\hline
\textbf{Method} & \textbf{CNOTs} & \textbf{Routed} & \textbf{Depth} & \textbf{Infidelity} & \textbf{Var.} & \textbf{MSE} \\
\hline
QsiHT Fast Path Complex Synthesis & $4$ & $7$ & $9$ & $8.349\times10^{-4}$ & $2.609\times10^{-7}$ & $2.964\times10^{-5}$ \\
M\"ott\"onen (UCR) & $12$ & $11$ & $24$ & $5.634\times10^{-3}$ & $1.341\times10^{-7}$ & $4.356\times10^{-4}$ \\
M\"ott\"onen (sign-aware) & $12$ & $11$ & $24$ & $1.968\times10^{-3}$ & $2.502\times10^{-7}$ & $1.265\times10^{-4}$ \\
Iten-style cascade & $10$ & $11$ & $20$ & $3.312\times10^{-3}$ & $9.369\times10^{-7}$ & $1.900\times10^{-4}$ \\
Qiskit \texttt{StatePreparation} & $4$ & $7$ & $9$ & $\mathbf{5.111\times10^{-4}}$ & $1.465\times10^{-8}$ & $2.245\times10^{-5}$ \\
Multiplexor & $10$ & $11$ & $20$ & $3.177\times10^{-3}$ & $9.523\times10^{-7}$ & $1.623\times10^{-4}$ \\
General isometry & $10$ & $11$ & $20$ & $3.640\times10^{-3}$ & $8.105\times10^{-7}$ & $1.813\times10^{-4}$ \\
Plesch--Brukner & $9$ & $\mathbf{6}$ & $26$ & $2.996\times10^{-3}$ & $8.586\times10^{-7}$ & $1.429\times10^{-4}$ \\
Quantum Shannon & $36$ & $34$ & $99$ & $8.992\times10^{-3}$ & $1.789\times10^{-6}$ & $5.077\times10^{-4}$ \\
\hline
\end{tabular}}
\end{table}

\begin{table}[H]
\centering
\caption[Sampled error, real $n=3$, \texttt{ibm\_kingston}]{Sampled error for the \emph{real} $n=3$ target on \texttt{ibm\_kingston}, over $5$ runs at $4096$ shots, all nine builders. Columns and two-job provenance as in Table~\ref{tab:hw_real}. On this device the frontier spreads widely, the $7$-CNOT General isometry nearly matching the $36$-CNOT Quantum Shannon, while \texttt{StatePreparation}, the sign-aware M\"ott\"onen, Plesch--Brukner, and the QsiHT Real Synthesis hold the low edge. The lowest routed count and lowest infidelity, the executed-axis columns, are shown in bold, and the other columns are left unbolded because at $n=3$ their minima are within-noise and carry no gate-count basis (\S\ref{sec:hw_limitations}).}
\label{tab:hw_kingston_real}
\resizebox{\textwidth}{!}{%
\begin{tabular}{lrrrrrr}
\hline
\textbf{Method} & \textbf{CNOTs} & \textbf{Routed} & \textbf{Depth} & \textbf{Infidelity} & \textbf{Var.} & \textbf{MSE} \\
\hline
QsiHT Fast Path Real Synthesis & $4$ & $7$ & $9$ & $3.772\times10^{-3}$ & $8.462\times10^{-7}$ & $1.017\times10^{-4}$ \\
M\"ott\"onen (UCR) & $12$ & $10$ & $20$ & $8.760\times10^{-3}$ & $2.711\times10^{-6}$ & $4.661\times10^{-4}$ \\
M\"ott\"onen (sign-aware) & $6$ & $9$ & $11$ & $3.446\times10^{-3}$ & $6.720\times10^{-7}$ & $1.417\times10^{-4}$ \\
Iten-style cascade & $7$ & $10$ & $14$ & $1.098\times10^{-2}$ & $1.228\times10^{-6}$ & $5.922\times10^{-4}$ \\
Qiskit \texttt{StatePreparation} & $4$ & $7$ & $9$ & $\mathbf{2.295\times10^{-3}}$ & $1.547\times10^{-7}$ & $5.899\times10^{-5}$ \\
Multiplexor & $7$ & $8$ & $14$ & $9.308\times10^{-3}$ & $2.608\times10^{-6}$ & $3.184\times10^{-4}$ \\
General isometry & $7$ & $10$ & $14$ & $1.296\times10^{-2}$ & $2.723\times10^{-6}$ & $7.034\times10^{-4}$ \\
Plesch--Brukner & $9$ & $\mathbf{5}$ & $24$ & $4.232\times10^{-3}$ & $6.679\times10^{-7}$ & $9.457\times10^{-5}$ \\
Quantum Shannon & $36$ & $34$ & $99$ & $1.430\times10^{-2}$ & $4.682\times10^{-6}$ & $2.570\times10^{-4}$ \\
\hline
\end{tabular}}
\end{table}

\subsection{A Third Processor and Cross-Day Replication}
\label{sec:cross_session}

\begin{figure}[H]
\centering
\includegraphics[width=\textwidth]{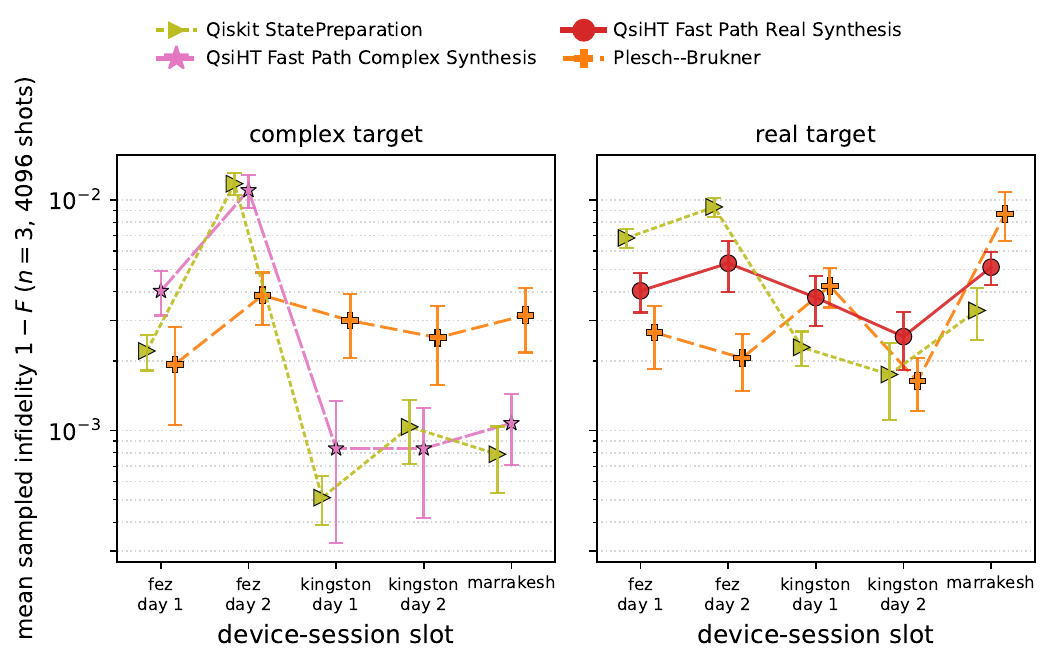}
\caption[Cross-session infidelity of four fixed circuits over five device-days]{Mean sampled infidelity $1-F$ of four fixed circuits (Qiskit \texttt{StatePreparation}, the two QsiHT syntheses, and Plesch--Brukner) across five single-session device-days: \texttt{ibm\_fez} and \texttt{ibm\_kingston} on two calibration days each and \texttt{ibm\_marrakesh} once, $n=3$, $4096$ shots, error bars one run-to-run standard deviation. The identical \texttt{StatePreparation} real circuit spans $1.7\times10^{-3}$ to $9.3\times10^{-3}$ (a factor of about $5$) across the slots while its within-run bars stay near $5\times10^{-4}$, and the Complex Synthesis on \texttt{ibm\_kingston} complex reproduces to $8.35\times10^{-4}$ on both days. The across-calibration scatter of one circuit exceeds the between-method spread within any one slot, which is why only the coarse frontier-versus-QSD tier transfers.}
\label{fig:hw_cross_session}
\end{figure}

The two devices above each mixed three submission sessions, so the QsiHT synthesis rows were cross-session and no within-frontier ordering could be resolved for them (\S\ref{sec:hw_limitations}). As a third check the same nine builders per target were run on a third $156$-qubit IBM Heron r2 processor, \texttt{ibm\_marrakesh}, for the first time, as a single submission, so that all nine share one calibration and within-session family-wide comparisons are legitimate for the first time. Table~\ref{tab:hw_marrakesh} reports the result and Figure~\ref{fig:hw_cross_session} places it beside the two-device runs. The coarse tier reproduces. Quantum Shannon is the worst method on both targets ($2.971\times10^{-2}$ real, $9.507\times10^{-3}$ complex), and after the Benjamini--Hochberg false-discovery-rate correction across the $\binom{9}{2}=36$ within-session pairwise tests per (target, metric) family it sits significantly below all eight other builders on both targets ($8/8$ pairs flagged in each family).

\begin{table}[H]
\centering
\caption[Sampled infidelity on the third processor \texttt{ibm\_marrakesh}, $n=3$]{Sampled infidelity $1-F$ for the $n=3$ targets on the third IBM Quantum processor \texttt{ibm\_marrakesh} (IBM Heron r2, $156$ qubits), over $5$ runs at $4096$ shots, all nine builders per target run as a single submission. Because the whole family shares one calibration, the within-session Benjamini--Hochberg tests over the $\binom{9}{2}=36$ pairwise comparisons are legitimate: Quantum Shannon is resolved below every other builder on both targets, the QsiHT Complex Synthesis is resolved below all five cascade builders and Plesch--Brukner on the complex target while tying \texttt{StatePreparation}, and on the real target Plesch--Brukner, the within-session leader on the other two devices, is instead the frontier's worst method, with the QsiHT Real Synthesis resolved below it. Routed is the executed two-qubit count of the opt-3 ISA circuit. The lowest routed count and lowest infidelity per target are bold.}
\label{tab:hw_marrakesh}
\resizebox{\textwidth}{!}{%
\begin{tabular}{lrrrr}
\hline
& \multicolumn{2}{c}{\textbf{Complex target}} & \multicolumn{2}{c}{\textbf{Real target}} \\
\textbf{Method} & \textbf{Routed} & \textbf{Infidelity} & \textbf{Routed} & \textbf{Infidelity} \\
\hline
QsiHT Fast Path Real Synthesis & -- & -- & $7$ & $5.107\times10^{-3}$ \\
QsiHT Fast Path Complex Synthesis & $7$ & $1.075\times10^{-3}$ & -- & -- \\
M\"ott\"onen (UCR) & $11$ & $3.465\times10^{-3}$ & $10$ & $7.772\times10^{-3}$ \\
M\"ott\"onen (sign-aware) & $11$ & $3.075\times10^{-3}$ & $9$ & $5.334\times10^{-3}$ \\
Iten-style cascade & $11$ & $5.466\times10^{-3}$ & $10$ & $6.513\times10^{-3}$ \\
Qiskit \texttt{StatePreparation} & $7$ & $\mathbf{7.882\times10^{-4}}$ & $7$ & $\mathbf{3.314\times10^{-3}}$ \\
Multiplexor & $11$ & $5.627\times10^{-3}$ & $8$ & $7.282\times10^{-3}$ \\
General isometry & $11$ & $6.173\times10^{-3}$ & $10$ & $6.627\times10^{-3}$ \\
Plesch--Brukner & $\mathbf{6}$ & $3.159\times10^{-3}$ & $\mathbf{5}$ & $8.699\times10^{-3}$ \\
Quantum Shannon & $34$ & $9.507\times10^{-3}$ & $34$ & $2.971\times10^{-2}$ \\
\hline
\end{tabular}}
\end{table}

On the complex target the low-routed builders lead. Qiskit \texttt{StatePreparation} ($7.882\times10^{-4}$) and the QsiHT Complex Synthesis ($1.075\times10^{-3}$) hold the low edge, and the Complex Synthesis is resolved (BH-FDR) below all five cascade builders, below Plesch--Brukner, and below Quantum Shannon, tying \texttt{StatePreparation}, the one pair that is not flagged. This is the first legitimate within-session frontier result the syntheses record on hardware. On the real target the low-routed \texttt{StatePreparation} leads ($3.314\times10^{-3}$) with the Real Synthesis next ($5.107\times10^{-3}$), but Plesch--Brukner, the within-session real-target leader on both \texttt{ibm\_fez} ($2.660\times10^{-3}$) and \texttt{ibm\_kingston} ($4.232\times10^{-3}$), is here the frontier's worst method ($8.699\times10^{-3}$), its five-routed circuit landing on poor couplers, and the Real Synthesis is resolved below it (adjusted $p=0.027$). The same five-routed Plesch--Brukner real circuit therefore swings from frontier best to frontier worst across the three devices ($2.66\times10^{-3}\to4.23\times10^{-3}\to8.70\times10^{-3}$), so the one within-frontier ordering the first two devices suggested does not transfer. This is direct on-device evidence for the calibration-dominance reading of \S\ref{sec:noise}.

\begin{sloppypar}
All nine builders were also rerun on \texttt{ibm\_fez} and \texttt{ibm\_kingston} on a second calibration day, again each as a single submission. The coarse tier reproduces on both days, Quantum Shannon worst and flagged below all eight builders in each within-session family. The cross-day drift of identical circuits is large. The same \texttt{StatePreparation} real circuit spans $1.7\times10^{-3}$ to $9.3\times10^{-3}$ across the three devices and two days (a factor of about $5$), while on \texttt{ibm\_kingston} the QsiHT Complex Synthesis reproduces to $8.35\times10^{-4}$ on the complex target across the two days, inside the run-to-run scatter (per-run standard deviation of $4$--$5\times10^{-4}$). Within a session the fine order reshuffles again, and in the opposite direction: on the \texttt{ibm\_fez} day-two complex job the low-routed Complex Synthesis ($1.102\times10^{-2}$) is resolved \emph{above} (worse than) the eleven-routed Iten, multiplexor, and general-isometry builders (near $1.0\times10^{-3}$) after BH-FDR, a full inversion of the gate-count reading and the opposite of the \texttt{ibm\_marrakesh} complex order. Within-session comparisons are now legitimate on three devices, and they agree only on the coarse frontier-versus-QSD tier, never on the within-frontier order, which is what the per-qubit-calibration argument of \S\ref{sec:noise} predicts.
\end{sloppypar}

\subsection{A Deep-Circuit Run at $n=8$}
\label{sec:hw_n8}

\begin{figure}[H]
\centering
\includegraphics[width=0.78\textwidth]{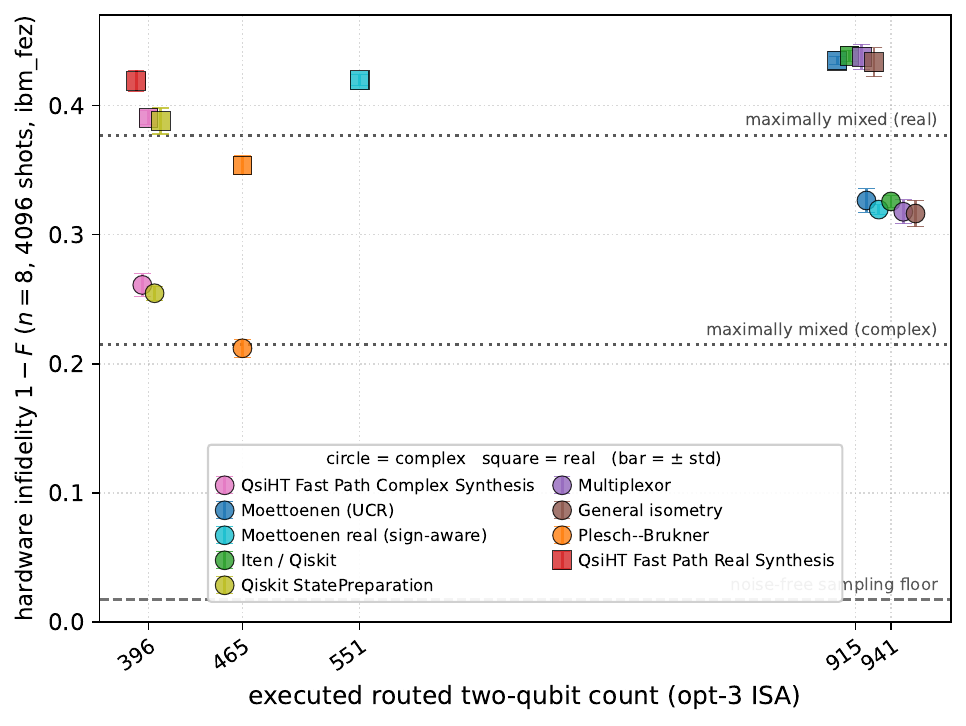}
\caption[Hardware infidelity versus routed two-qubit count at $n=8$]{Hardware sampled infidelity $1-F$ versus the executed routed two-qubit count on \texttt{ibm\_fez} at $n=8$ ($5$ runs, $4096$ shots), complex (circles) and real (squares), error bars one run-to-run standard deviation. Quantum Shannon is omitted: its recorded as-built abstract circuit contains $48{,}768$ two-qubit gates, and it was excluded before routing or hardware execution. Unlike the $n=3$ runs, the executed-count ordering re-emerges outside the error bars on the complex target, where the $396$-routed loaders (the Complex Synthesis and \texttt{StatePreparation}, $0.255$--$0.261$) and the shallower $465$-routed Plesch--Brukner ($0.212$) sit well below the $941$-routed cascades ($0.316$--$0.326$). The dashed line is the noise-free multinomial sampling floor at $4096$ shots ($1.8\times10^{-2}$ complex, $1.7\times10^{-2}$ real) and the dotted line the uniform-distribution reference ($0.215$ complex, $0.377$ real). The measured computational-basis distributions lie near or above that reference, so the figure reads the ordering but does not certify a maximally mixed state.}
\label{fig:hw_n8_noise_vs_cnot}
\end{figure}

Every hardware result above is at $n=3$, a readout-dominated regime where a few-gate two-qubit advantage provably cannot surface (\S\ref{sec:hw_limitations}). To reach the opposite regime, where the two-qubit budget rather than the fixed readout floor dominates, all builders except Quantum Shannon were run on \texttt{ibm\_fez} at $n=8$ ($5$ runs, $4096$ shots, single submission). Quantum Shannon was excluded deliberately: its recorded as-built abstract circuit at $n=8$ contains $48{,}768$ two-qubit gates. It was excluded before routing, so neither a routed count nor a hardware outcome was measured. The other builders route into five distinct executed clusters, $396$ (both syntheses and \texttt{StatePreparation}), $465$ (Plesch--Brukner), $551$ (the sign-aware M\"ott\"onen on the real target), $915$ (the real cascades), and $941$ (the complex cascades). Table~\ref{tab:hw_n8} and Figure~\ref{fig:hw_n8_noise_vs_cnot} report the result. There is no accompanying statistics file at this size, so no significance claim is made beyond the reported run-to-run standard deviations.

\begin{table}[H]
\centering
\caption[Sampled infidelity at $n=8$ on \texttt{ibm\_fez}]{Sampled infidelity $1-F$ for the $n=8$ targets on \texttt{ibm\_fez} ($5$ runs, $4096$ shots, single submission), all builders except Quantum Shannon, whose recorded as-built abstract circuit contains $48{,}768$ two-qubit gates at this size. QSD was excluded before routing, so no routed or hardware result is reported for it. Routed is the executed two-qubit count and Depth the ISA depth of the opt-3 circuit. At this size the executed-count ordering re-emerges outside the run-to-run spread on the complex target, the $396$- and $465$-routed loaders ($0.21$--$0.26$) sitting below the $941$-routed cascades ($0.32$). Plesch--Brukner's lowest infidelity is associated with its much shallower circuit (ISA depth $939$ against about $1460$ for the $396$-routed loaders), but this experiment does not isolate depth from mapping, gate composition, calibration, or session effects. The noise-free sampling floor at $4096$ shots is $1.8\times10^{-2}$ (complex) and $1.7\times10^{-2}$ (real), and the uniform-distribution reference is $0.215$ (complex) and $0.377$ (real). The measured computational-basis distributions lie near or above that reference and far above the sampling floor, so the comparison reads their ordering rather than certifying a maximally mixed state. The lowest infidelity per target is bold.}
\label{tab:hw_n8}
\resizebox{\textwidth}{!}{%
\begin{tabular}{lrrrrrr}
\hline
& \multicolumn{3}{c}{\textbf{Complex target}} & \multicolumn{3}{c}{\textbf{Real target}} \\
\textbf{Method} & \textbf{Routed} & \textbf{Depth} & \textbf{Infidelity} & \textbf{Routed} & \textbf{Depth} & \textbf{Infidelity} \\
\hline
QsiHT Fast Path Real Synthesis & -- & -- & -- & $396$ & $1415$ & $4.190\times10^{-1}$ \\
QsiHT Fast Path Complex Synthesis & $396$ & $1464$ & $2.611\times10^{-1}$ & $396$ & $1407$ & $3.903\times10^{-1}$ \\
M\"ott\"onen (UCR) & $941$ & $2868$ & $3.264\times10^{-1}$ & $915$ & $3034$ & $4.347\times10^{-1}$ \\
M\"ott\"onen (sign-aware) & $941$ & $2868$ & $3.194\times10^{-1}$ & $551$ & $1410$ & $4.195\times10^{-1}$ \\
Iten-style cascade & $941$ & $2868$ & $3.257\times10^{-1}$ & $915$ & $3032$ & $4.385\times10^{-1}$ \\
Qiskit \texttt{StatePreparation} & $396$ & $1462$ & $2.547\times10^{-1}$ & $396$ & $1411$ & $3.880\times10^{-1}$ \\
Multiplexor & $941$ & $2867$ & $3.177\times10^{-1}$ & $915$ & $3031$ & $4.377\times10^{-1}$ \\
General isometry & $941$ & $2868$ & $3.163\times10^{-1}$ & $915$ & $3032$ & $4.335\times10^{-1}$ \\
Plesch--Brukner & $465$ & $939$ & $\mathbf{2.120\times10^{-1}}$ & $465$ & $926$ & $\mathbf{3.536\times10^{-1}}$ \\
\hline
\end{tabular}}
\end{table}

On the complex target the executed-count ordering re-emerges cleanly, outside the error bars. The $396$-routed loaders sit at $0.255$ (\texttt{StatePreparation}) and $0.261$ (the Complex Synthesis), the $465$-routed Plesch--Brukner at $0.212$, and the $941$-routed cascades at $0.316$--$0.326$, a gap of about $0.06$ against run-to-run standard deviations of about $0.004$--$0.010$. Plesch--Brukner carries more routed two-qubit gates than the $396$-routed loaders yet the lowest infidelity, alongside a far shallower ISA circuit (depth $939$ against about $1460$ for the $396$-routed loaders and about $2870$ for the cascades). This is an association: the experiment does not isolate depth from mapping, gate composition, calibration, or session effects. On the real target the same low-versus-high split holds, Plesch--Brukner ($0.354$) and \texttt{StatePreparation} ($0.388$) at the low end and the cascades ($0.43$--$0.44$) at the high end, but compressed.

The comparison can be bounded from below and above. The noise-free multinomial sampling floor at $4096$ shots, computed on the exact seeded $n=8$ target (seed $12353$), is about $1.8\times10^{-2}$ (complex) and $1.7\times10^{-2}$ (real), so every measured point ($0.21$--$0.44$) is more than an order of magnitude above the finite-shot floor: the differences are gate error, not sampling. A uniform distribution over the $256$ outcomes gives the reference value $0.215$ (complex) and $0.377$ (real). The measured complex points straddle this value and the measured real points sit at or above it, which is consistent with near-uniform sampled computational-basis distributions. Because this comparison is sign- and phase-blind, it does not establish a maximally mixed density matrix or full decoherence. A decay-model check at the $396$-gate loaders sharpens the comparison. The first-order linear form $1-F\approx\epsilon N$ predicts $0.277$ at the simulated fitted slope $\epsilon=7.0\times10^{-4}$ and $0.317$ at the hardware two-point slope $\epsilon=8.0\times10^{-4}$, while the per-gate exponential $1-(1-\epsilon)^{N}$ predicts $0.242$ and $0.272$ respectively. Measured \texttt{StatePreparation} complex ($0.255$) sits inside this band and just above the $0.215$ uniform reference, whereas measured \texttt{StatePreparation} real ($0.388$) exceeds both extrapolations and the $0.377$ uniform reference, so the first-order model, valid only for $1-F\ll1$, has broken down on the real target. The $n=8$ run therefore confirms that the coarse executed-count ordering re-emerges once the two-qubit budget dominates the readout floor, but at this size the sampled distributions are already near the uniform reference, so the informative crossover is the intermediate $n=4$--$6$ sweep flagged in the follow-up plan below (\S\ref{sec:hw_limitations}).

\subsection{Scope of the Hardware Claims and a Follow-Up Experiment Plan}
\label{sec:hw_limitations}

The hardware evidence in this chapter is narrower than the simulated evidence, and its scope should be stated plainly. Everything above comes from one register size for the family-wide comparison ($n=3$, with a single deep-circuit probe at $n=8$ in \S\ref{sec:hw_n8}), one architecture (superconducting IBM Heron r2), three devices, and five runs of $4096$ shots per method and target, with the transpiler free to choose the physical qubits for each job. What this supports is the coarse tier, the exact frontier sitting well below QSD on all three devices and both targets, together with the build-time counts, which are device-independent. It supports no within-frontier ordering that transfers across the three devices. On the two original-job devices' real target, Plesch--Brukner is resolved as the leader of the original job: after the Benjamini--Hochberg false-discovery-rate correction across the $15$ within-session tests per (target, metric) family (the six methods run in the original job give $\binom{6}{2}=15$ pairwise tests), with no method data-selected before testing, its infidelity sits significantly below the four other frontier methods run in the same session and below QSD, consistent with its lowest executed routed count ($5$). That lead does not carry to the third device: on the single-session \texttt{ibm\_marrakesh} run (\S\ref{sec:cross_session}), where all nine builders share one calibration and the family-wide tests over $\binom{9}{2}=36$ pairs are legitimate, Plesch--Brukner is instead the frontier's worst real-target method ($8.699\times10^{-3}$) and the QsiHT Real Synthesis is resolved below it (adjusted $p=0.027$), so the single within-frontier ordering the first two devices suggested is not stable. The three later rows, the completion-job sign-aware M\"ott\"onen and \texttt{StatePreparation} and the newest-session QsiHT syntheses, sit on different calibration days, so they are compared descriptively but never folded into this family, and no within-frontier win is claimed for the QsiHT syntheses on these multi-session tables: the pipeline flags no cross-session pair as significant. The three single-session jobs of \S\ref{sec:cross_session} do resolve within-session QsiHT comparisons for the first time, the Complex Synthesis flagged below every other complex-target builder except \texttt{StatePreparation} on \texttt{ibm\_marrakesh} and flagged above the Iten, multiplexor, and general-isometry builders on the \texttt{ibm\_fez} day-two complex job, but because those verdicts disagree from device to device and day to day, no stable within-frontier ordering survives. Within the completion job's single test, \texttt{StatePreparation} is resolved below the sign-aware M\"ott\"onen on the real target of both devices (adjusted $p=0.002$ on \texttt{ibm\_fez}, $p=0.03$ on \texttt{ibm\_kingston}). At five runs the minimum detectable effect on infidelity at $80\%$ power is roughly $1.5\times10^{-3}$ per pairwise comparison on \texttt{ibm\_fez}, larger than the mid-frontier margins, so those gaps stay below the resolution of the experiment. The QsiHT syntheses' $4.035\times10^{-3}$ (real, \texttt{ibm\_fez}) and $8.349\times10^{-4}$ (complex, \texttt{ibm\_kingston}) sit at the cheap edge of the frontier, consistent with their $4$ built and $7$ routed two-qubit gates tying \texttt{StatePreparation}, but they were measured cross-session, so this thesis reads them as descriptive placements, not resolved wins. In addition, $n=3$ is a readout-dominated regime: the per-qubit measurement floor does not scale with the gate count, so a CNOT advantage of a few gates provably cannot surface here, and every fine hardware ordering at this size is confounded by layout and calibration.

\begin{sloppypar}
Three single-session multi-method jobs have since been run, the \texttt{ibm\_fez} and \texttt{ibm\_kingston} day-two reruns and the \texttt{ibm\_marrakesh} run (\S\ref{sec:cross_session}), each placing all nine builders per target on one calibration and removing the cross-session confound the original tables carried. They confirm the coarse frontier-versus-QSD tier within a single session and show that the within-frontier order does not reproduce across devices or days. What remains unrun is the fully rigorous version, specified here as a concrete experiment plan rather than as new data. (a) A layout-pinned rerun: fix one physical qubit triple, selected from the calibration data, via a fixed \texttt{initial\_layout} for every method, enable dynamical decoupling on idle wires, apply matrix-free measurement-error mitigation (M3), and interleave at least $20$ runs per method within one session so that calibration drift averages over methods rather than between them. This removes the placement, idle-time, and readout confounds identified above and lowers the detectable margin by roughly a factor of two. The single-session jobs already interleave the builders on one calibration, so what this rigorous rerun adds is the fixed layout, the idle-time and readout mitigation, and the larger run count, the combination the metered open-plan QPU budget could not fit during this analysis. Its comparison is pre-registered as a two-one-sided-tests (TOST) equivalence test against a margin fixed before data collection, so within-frontier methods that do not separate are certified as equivalent rather than left ambiguous. This is why every on-device statement here is held to the coarse frontier-versus-QSD tier, with no stable within-frontier ordering claimed. (b) A size sweep at $n=4$--$6$, where the frontier two-qubit budgets separate ($11$--$57$ for the QsiHT syntheses against $26$--$124$ for the cascades) and the accumulated gate error crosses over the roughly fixed readout floor. This is the smallest regime in which a build-time CNOT advantage could convert into a measurable fidelity advantage, and the crossover point is itself a quantity worth reporting. The $n=8$ run of \S\ref{sec:hw_n8} already shows the executed-count ordering returning once the two-qubit budget dominates, but at a size where the sampled computational-basis distributions lie near the uniform-distribution reference, so this intermediate window is where a usable advantage, if any, would surface. (c) A sign-sensitive validation: the sampled infidelity used here compares measured and ideal \emph{probability distributions}, so it is blind to sign and phase errors and cannot fully certify that a real (sign-bearing) target was prepared correctly. Direct fidelity estimation or state tomography on the pinned triple would close that gap for the real-target claim specifically. Until those runs exist, every $n=3$ hardware statement in this thesis should be read at that scope: the devices validate the coarse frontier-versus-QSD separation and the device-independent build-time counts, and nothing finer transfers, while the $n=8$ probe of \S\ref{sec:hw_n8} shows only that the coarse executed-count ordering returns once the two-qubit budget outweighs the readout floor, at a size where the sampled distributions are consistent with a near-uniform computational-basis distribution rather than certifying full decoherence.
\end{sloppypar}

\section{Discussion}
\label{sec:discussion}

The benchmarks confirm every qualitative prediction of Chapter~\ref{ch:theory}. All ten builders are exact and deterministic on their applicable target families. The exact-frontier methods (the two QsiHT syntheses, the UCR family, Plesch--Brukner, and the deployed \texttt{StatePreparation}) share $\Theta(2^n)$ CNOT and depth scaling. QSD pays the full $\Theta(4^n)$ price and is justified only when a complete unitary is independently required, and under the connectivity-fixed simulated noise model the realized error is governed by the executed two-qubit count, confirming its role as the primary cost metric, with the heavy-hex-routed model showing that on real connectivity it is the routed, not the abstract, count that does the governing.

Within the frontier, the methods are separated by their constant factors and the structure they exploit. The two QsiHT syntheses realize the multiplexor-family optimum $2^n-n-1$ ($57$ CNOTs and depth $115$ at $n=6$) on both real and complex data, tying the deployed \texttt{StatePreparation} and the theoretical $m{=}0$ isometry floor. The benchmarked General-isometry configuration instead aliases the Iten builder and measures $120$ CNOTs on real data and $122$ on complex data at $n=6$. On the as-built CNOT axis, the syntheses undercut every other from-scratch configuration, including the M\"ott\"onen cascade ($124$), the compiler-optimized cascades ($120$--$122$), and Plesch--Brukner ($81$). This is the substantive change from the original fast path, whose $2^n-2$ (real) and $2^{n+1}-4$ (complex) counts sat above the floor: the state-preparation reductions of \S\ref{sec:qsiht_real} and \S\ref{sec:qsiht_complex} defer one entangler per stage and land the syntheses on the deployed floor rather than above it. Plesch--Brukner keeps a marginally lower depth at $n=6$ ($111$ complex, $109$ real, against $115$) and, under optimization level~3, a marginally lower CNOT count ($56$ against the syntheses' opt-3-invariant $57$), because its half-register local unitaries cost $O(4^{n/2})$, so it remains the from-scratch method with the smallest constant while the syntheses match the deployed floor. The predicted SVD conditioning edge (\S\ref{sec:stability}) does not appear in the measurements, and if anything the measured direction reverses: the direct \texttt{arctan2} methods carry residuals at or below Plesch--Brukner's on the degenerate family (the QsiHT Real Synthesis peaking at $1.5\times10^{-15}$ across the grid against Plesch--Brukner's $5.0\times10^{-13}$), all at machine precision. Between the two syntheses, the Real Synthesis is the cheaper to synthesize, its FWHT-only pass running an order of magnitude or more below the merged-multiplexor and cascade routes at $n=11$--$12$ (\S\ref{sec:scaling}), and the shallowest across native gate sets on real targets (\S\ref{sec:nisq}), while the Complex Synthesis reaches the same gate counts on arbitrary targets at a build cost in the \texttt{StatePreparation} band and ties the deployed \texttt{StatePreparation} for the lowest simulated sampled error (\S\ref{sec:noise}). Both share a regular, data-independent circuit template convenient for compilation and reuse, and the \texttt{arctan2} angle computation they use is the numerically correct two-argument form that every direct method in the benchmark also uses, so it is a shared implementation choice rather than a QsiHT-specific conditioning edge (\S\ref{sec:stability}), producing no measurable accuracy separation on the benchmark targets, where every applicable method is exact to machine precision (\S\ref{sec:exactness}). Table~\ref{tab:method_verdict} consolidates the empirical verdict across all axes measured in this chapter.

\begin{table}[H]
\centering
\caption{Empirical verdict across the metrics of this chapter. ``Frontier'' denotes the $\Theta(2^n)$ exact minimum-CNOT band, and ``floor'' the multiplexor-family optimum $2^n-n-1$ that the two QsiHT syntheses tie with the deployed \texttt{StatePreparation}. ``Best const.'' marks the smallest measured constant among the from-scratch implementations. The synthesis-time column reflects the large-register build-time sweep (\S\ref{sec:scaling}). The residual column reports the measured statevector $2$-norm error at $n=6$ on the real family in place of a stability tier: the residuals reflect each implementation's numerical path, not the circuit length and not the conditioning analysis of \S\ref{sec:stability}, which produces no accuracy separation on the Haar families and separates the opt-3-peephole from as-built realizations only near a degenerate Schmidt spectrum or on the reproducibility suite's near-sparse (interior-zero) edge-case target (\S\ref{sec:exactness}).}
\label{tab:method_verdict}
\resizebox{\textwidth}{!}{%
\begin{tabular}{lcccc}
\hline
\textbf{Method} & \textbf{CNOT scaling} & \textbf{Noisy error} & \textbf{Synthesis time} & \textbf{Residual ($n{=}6$)} \\
\hline
QsiHT Fast Path Real Synthesis    & frontier (floor) & low & fast & $7.6\times10^{-16}$ \\
QsiHT Fast Path Complex Synthesis & frontier (floor) & low & moderate & $1.1\times10^{-14}$ \\
M\"ott\"onen (UCR) & frontier            & low       & moderate        & $7.5\times10^{-16}$ \\
Iten-style cascade     & frontier            & low       & moderate        & $1.2\times10^{-15}$ \\
Multiplexor       & frontier            & low       & moderate        & $1.7\times10^{-15}$ \\
General isometry   & frontier            & low       & moderate        & $1.2\times10^{-15}$ \\
Plesch--Brukner   & \textbf{best const.} & low & moderate & $2.7\times10^{-15}$ \\
Quantum Shannon   & $\Theta(4^n)$       & high      & slow            & $2.1\times10^{-14}$ \\
\hline
\end{tabular}}
\end{table}

These measurements complete the empirical program set out in Chapter~\ref{ch:background}: the resource metrics defined there, applied through the \texttt{measurement} framework to the benchmark configurations, reproduce the Chapter~\ref{ch:theory} theory and locate the two QsiHT syntheses precisely within it, on the $\Theta(2^n)$ frontier at the multiplexor-family floor $2^n-n-1$ on both real and complex dense data, tying the deployed \texttt{StatePreparation} and undercutting every other from-scratch builder on the as-built CNOT axis (\S\ref{sec:scaling}), with the Real Synthesis the cheapest of them to synthesize. The implications for method selection, the broader signal- and image-processing context, and directions for further work are discussed in Chapter~\ref{ch:conclusion}.

%% file: chapters/chapter6.tex
\chapter{Conclusions and Future Work}
\label{ch:conclusion}

This thesis set out to characterize and compare the family of exact, deterministic, ancilla-free methods for preparing arbitrary dense $n$-qubit quantum states. The work proceeded from formal foundations (Chapter~\ref{ch:background}), through a survey of the principal constructions (Chapter~\ref{ch:litreview}) and a unified theoretical comparison (Chapter~\ref{ch:theory}), to a reproducible benchmark and a measured comparison of the configurations on common targets (Chapter~\ref{ch:empirical}). Throughout this chapter, ``the field'' means the benchmarked method set whose youngest member is Iten's 2016 isometry method: the ancilla-assisted, sparse, and approximate lines of work surveyed in Section~\ref{sec:related_out_of_scope} were out of scope and are not covered by any claim below. This final chapter consolidates the contributions (\S\ref{sec:contributions}) and findings (\S\ref{sec:findings}), emphasizes the two QsiHT fast-path syntheses' results and their extensions to state-to-state and two-state maps (\S\ref{sec:qsiht_context}), discusses applications to quantum signal and image processing (\S\ref{sec:applications}), lays out concrete directions for future work (\S\ref{sec:future}), and closes with a summary assessment (\S\ref{sec:closing}).

\section{Summary of Contributions}
\label{sec:contributions}

This thesis makes four contributions.

\paragraph{A unified benchmark framework.} Chapter~\ref{ch:background} fixed the framework's measurement set, the seven per-circuit resource metrics (gate count, CNOT count, depth, ancilla count, compilation time, transpiled depth, and fidelity), together with the parameter-counting lower bound $\#\mathrm{CNOT}\ge\lceil\tfrac12(2^n-n-1)\rceil$, which~\eqref{eq:cnot_lb} establishes for complex targets and~\eqref{eq:cnot_lb_real} extends, at the same numerical value, to real targets under real-rotation circuits. Chapters~\ref{ch:litreview} and~\ref{ch:theory} wrapped that measurement set in the comparison criteria that delimit the method set, derived each method's construction with an explicit two-qubit worked example, and collected the asymptotic costs into a single comparison (Tables~\ref{tab:method_cost} and~\ref{tab:complexity}), and Chapter~\ref{ch:empirical} condensed the measurements into its verdict axes (Table~\ref{tab:method_verdict}). These are three layers of one framework, not three competing metric sets. Its discriminating power rests on the CNOT count: exactness, determinism, and ancilla use are common screens, while generality separates the real-only QsiHT Real Synthesis, gate count and depth track the CNOT count throughout the benchmark, and the framework functions as a CNOT-led screen followed by a Pareto view in which QSD is worse on every measured axis (counts, depth, total gates, and compilation time) at once on both target families, the compilation-time axis keeping the compiler-optimized cascades on the frontier on complex targets, while on real targets the Real Synthesis, also the fastest builder measured at the benchmarked sizes, dominates them on every measured axis (\S\ref{sec:framework}). This framework lets the benchmarked methods be compared on equal terms.

\paragraph{A reproducible implementation of the benchmark configurations.} The benchmark combines direct NumPy/Qiskit implementations with practical shared and Qiskit-realized configurations. The M\"ott\"onen, QSD, and Plesch--Brukner paths are implemented directly. The Iten row applies Qiskit optimization level~3 to the M\"ott\"onen cascade, the General-isometry row calls that same builder, the multiplexor row uses Qiskit's uniformly controlled rotation primitives, and the deployed \texttt{StatePreparation} is measured as shipped. The two QsiHT rows use the project's reference implementation and remain the syntheses evaluated as contributions. The benchmark code, built on the project's \texttt{measurement} package, computes the structural, cross-platform, and sampled-error metrics uniformly for every applicable configuration.

\paragraph{A heap-transform derivation route and the evaluation of the QsiHT fast path.} The DsiHT transform is Grigoryan's prior work~\cite{grigoryan2014heap}, and its realization as a state-preparation gate circuit is likewise prior published work~\cite{gomez2025qsiht, grigoryan2025stateprep}. What this thesis contributes is the evaluation, and, as the compilation step that enables it, the derivation route. Re-deriving the fast path as a cascade of uniformly controlled rotations compiled by the Gray-code decomposition places the heap-transform construction on the same primitive as the baselines, and carrying that route two steps further yields the two syntheses evaluated here: the \emph{QsiHT Fast Path Real Synthesis} for real targets, which truncates each stage's trailing CNOT and reorders the heap by the resulting control-basis permutation, and the \emph{QsiHT Fast Path Complex Synthesis} for arbitrary targets, which absorbs each disentangling stage's residual diagonal into the following stage. On that common footing the two syntheses are benchmarked against the other methods on identical targets, across qubit counts $n=2,\dots,6$, eight NISQ native gate sets, noise-free and noisy sampling, and execution on three real IBM Quantum processors at $n=3$. To our knowledge, this is the first head-to-head placement of the QsiHT fast-path syntheses against the standard direct, isometry, multiplexor, Plesch--Brukner, and Shannon methods under a single benchmark. The benchmark places both at the deployed multiplexor floor: each realizes $2^n-n-1$ CNOTs, matching Qiskit's \texttt{StatePreparation} exactly at every size and saturating the CNOT lower bound of~\eqref{eq:cnot_lb} at $n=2$, a bound that among the earlier baselines only Plesch--Brukner reached. The Real Synthesis reaches that count on real targets using $R_y$ and CNOT gates alone, with no separate phase stage (\S\ref{sec:scaling}).

\paragraph{An empirical confirmation and refinement of the theory.} Chapter~\ref{ch:empirical} verified that every applicable benchmark configuration is exact to machine precision, reproduced the predicted $\Theta(2^n)$-versus-$\Theta(4^n)$ split, measured the leading constants the theory left open (notably Plesch--Brukner's favorable bipartite constant and the two syntheses' arrival at the deployed multiplexor floor on both target families), and demonstrated that under a connectivity-fixed (all-to-all) simulated noise model the coarse error tier tracks the executed two-qubit count, while a heavy-hex-routed noise model and runs on three real IBM Quantum processors at $n=3$ show routing and per-qubit calibration deciding the fine order and only the coarse separation surviving (\S\ref{sec:hw_limitations}). The benchmark also quantified a feature invisible to the asymptotic analysis: at larger registers the Real Synthesis has the lightest recorded classical build, while the Complex Synthesis remains in the other frontier builders' time band. Both reference syntheses nevertheless reach the same $2^n-n-1$ entangling floor as the deployed \texttt{StatePreparation} ($57$ at $n=6$).

\section{Key Findings}
\label{sec:findings}

The empirical study in this thesis supports seven conclusions.

\begin{enumerate}
    \item \textbf{All methods are exact on their applicable targets. They differ only in cost.} Every arbitrary-complex configuration reproduced both target families at fidelity $1.000000$, and the QsiHT Real Synthesis did so on its real target family (\S\ref{sec:exactness}). They are distinguished not by whether they work within their stated domains but by what they cost, so the resource metrics of Chapter~\ref{ch:background}, and the CNOT count in particular, are what the comparison turns on.
    \item \textbf{The asymptotic separation is large.} The two QsiHT fast-path syntheses, the UCR family, and Plesch--Brukner stay on the $\Theta(2^n)$ frontier ($57$ to $124$ CNOTs at $n=6$), while QSD grows as $\Theta(4^n)$ ($2976$ CNOTs at $n=6$, a $24$-to-$52\times$ penalty over the frontier depending on the method compared against). Treating state preparation as full-unitary synthesis is, as predicted, wasteful unless the unitary is independently needed.
    \item \textbf{The two syntheses reach the deployed multiplexor floor on the as-built axis.} Among the from-scratch frontier implementations only the two QsiHT fast-path syntheses reach the as-built count $2^n-n-1$ ($26$ at $n=5$), matching Qiskit's deployed \texttt{StatePreparation} exactly on both target families and saturating the lower bound of~\eqref{eq:cnot_lb} at $n=2$. The other from-scratch configurations stay above that floor as built: the M\"ott\"onen amplitude-plus-phase realization and the Iten/multiplexor/isometry variants sit in the two-cascade band ($58$ to $60$ CNOTs at $n=5$ on complex data), and Plesch--Brukner's bipartite Schmidt split ($50$ at $n=5$ on complex, half-register unitaries at $O(4^{n/2})$) beats that band but still exceeds the as-built floor at every benchmarked size, dipping below $2^n-n-1$ in the asymptotic limit. Under optimization level~3, Plesch--Brukner reaches $56$ CNOTs at $n=6$, below the syntheses' invariant $57$. What separates the frontier methods as built at these sizes is whether the construction merges its residual diagonals down to the multiplexor floor, not its asymptotic order (\S\ref{sec:scaling}).
    \item \textbf{Hardware rescales the costs but preserves the coarse ranking.} Across eight native gate sets, depth inflated by platform-specific factors (most sharply on iSWAP-based Sycamore, which doubles the two-qubit count), and the coarse ranking, the frontier cluster against QSD, held on every device (\S\ref{sec:nisq}). The abstract-basis comparison is therefore a faithful predictor of relative on-device cost at that coarse level, while the fine order within the frontier shifts with the gate set and the target family.
    \item \textbf{Under connectivity-fixed noise, the coarse error tier tracks the executed two-qubit count.} On a shared seeded all-to-all noise model with shot seeds paired across methods, the sampled fidelity error (infidelity $1-F$) rose with the executed two-qubit budget at the coarse level: the frontier methods clustered below $2.2\times10^{-3}$ while QSD, with the most CNOTs, separated by roughly an order of magnitude, reaching $1.42\times10^{-2}$ on real data (\S\ref{sec:noise}). This confirms experimentally the first-order fidelity-decay model~\eqref{eq:fidelity_decay} in its CNOT-dominated form $F\approx 1-\epsilon_{\text{CNOT}}G_{\text{CNOT}}$, within a fixed connectivity. Within the frontier, by contrast, the sub-$2\times$ executed-gate differences are swamped by per-gate error heterogeneity and finite-shot noise and do not order cleanly: Plesch--Brukner carries the highest frontier infidelity on the complex target despite executing fewer two-qubit gates than the cascades ($6$ against $10$) (\S\ref{sec:noise}), so the confirmation is of the coarse tier separation, not of a fine gate-count ranking. The finding does not transfer across coupling maps unchanged: routing on a heavy-hex map reshuffles the frontier's executed two-qubit counts, so the low-error lead within the frontier changes hands and the routed order shifts again with register size, and the gate-count reading holds per connectivity map rather than across maps (\S\ref{sec:noise}).
    \item \textbf{The Real Synthesis has the lightest recorded large-register build.} Both syntheses compute their $2^n-1$ rotation angles through a single fast Walsh--Hadamard transform, an $O(N\log N)$ pass with $N=2^n$, against the dense $O(N^2)$ angle maps of the cascade baselines and the $O(N^{1.5})$ reshape SVD of Plesch--Brukner. At the benchmarked sizes, with a common optimization pass applied to every method, this leaves all frontier builds in one synthesis-time band of a few tens of milliseconds (\S\ref{sec:scaling}), and no finer classical-cost ranking among them is read off at $n\le 6$, but the asymptotic gap is real, and at larger registers the Real Synthesis pulls one to two orders of magnitude ahead of the Bergholm-style pipelines, including Qiskit's \texttt{StatePreparation}, while the Complex Synthesis, whose build is dominated by assembling the $U(2)$ block cascade rather than by the angle map, stays within their band. Since all of these routes land at the same $2^n-n-1$ entangling floor, the classical synthesis cost is the practical separator when one loader is compiled for many target vectors.
    \item \textbf{Real hardware preserves the coarse ranking but reshuffles the frontier.} On three 156-qubit IBM Heron processors (\texttt{ibm\_fez}, \texttt{ibm\_kingston}, and \texttt{ibm\_marrakesh}), the $n=3$ comparison kept the frontier methods well separated from QSD, with the low-count builders, now including the two QsiHT fast-path syntheses at the multiplexor floor and Qiskit \texttt{StatePreparation}, at the low-infidelity edge, Plesch--Brukner among them on two of the three devices. The fine order within the frontier, however, no longer followed the CNOT count, and the identical \texttt{StatePreparation} circuit measured across the three devices, the same circuit under a different device and calibration, differed by about a factor of $3.0$ in mean infidelity on the real target: at this scale, device calibration and readout dominate (\S\ref{sec:noise}).
\end{enumerate}

\section{The QsiHT Fast Path: Context and Extensions}
\label{sec:qsiht_context}

The benchmark locates the two QsiHT fast-path syntheses squarely on the exact $\Theta(2^n)$ frontier and distinguishes them from the other frontier methods on three axes.

\paragraph{Gate cost: both syntheses reach the deployed multiplexor floor.} On both target families the two QsiHT fast-path syntheses realize $2^n-n-1$ CNOTs ($57$ at $n=6$, $26$ at $n=5$), matching Qiskit's deployed \texttt{StatePreparation} exactly at every size and undercutting the M\"ott\"onen and Iten/multiplexor/isometry cascades, which stay in the two-cascade band ($124$ and $122$ at $n=6$ on complex data), by about $2.2\times$. The Real Synthesis reaches that count on real targets with $R_y$ and CNOT gates alone, folding the amplitude signs into the rotation cascade so that no separate phase stage appears, and the Complex Synthesis reaches it on arbitrary targets by absorbing each disentangling stage's residual diagonal into the following stage's uniformly controlled block (\S\ref{sec:scaling}). The floor they share is a family optimum, not the information-theoretic minimum: $2^n-n-1$ sits a factor of about $2.0$ above the parameter lower bound of~\eqref{eq:cnot_lb} ($29$ at $n=6$) and about $4.4$ above the weaker generic-gate floor ($13$ at $n=6$). Among the benchmarked methods only Plesch--Brukner reaches below that floor, and only asymptotically and marginally: its Schmidt bipartite split dips just under $2^n-n-1$ in the limit and, under optimization level~3, compresses to $56$ against the syntheses' opt-invariant $57$ at $n=6$, while at every other benchmarked size it stays above the floor. No benchmarked construction reaches the $29$ parameter floor, and no known bound excludes the gap.

\paragraph{Classical synthesis.} The DsiHT reduces the amplitude vector through $2^n-1$ Givens rotations, a $\Theta(N)$ heap, and maps the resulting angles to the uniformly controlled cascade through a fast Walsh--Hadamard transform, $O(N\log N)$. For the Real Synthesis, this produces the lightest recorded large-register build on the frontier. The cascade baselines compute that same Walsh--Hadamard angle map densely at $O(N^2)$, and Plesch--Brukner's reshape SVD costs $O(N^{1.5})$. At the benchmark sizes, with a common optimization pass timed on every method, this leaves all frontier builds in one synthesis-time band and no finer classical-cost ranking among them is claimed, but the asymptotic gap is real, and at larger registers the Real Synthesis pulls one to two orders of magnitude ahead of the Bergholm-style pipelines, including Qiskit's \texttt{StatePreparation}, while the Complex Synthesis, whose build is dominated by assembling the $U(2)$ block cascade rather than by the angle map, stays within their band. The substantive point is that the route is cheap both to specify and to compile, its synthesis cost sitting alongside that of the direct methods at these sizes and far below that of QSD's $\Theta(4^n)$ recursion.

\paragraph{Numerical robustness and regularity.} The route computes its angles with the two-argument \texttt{arctan2}, the numerically correct form that returns $\pm\pi/2$ rather than dividing by a vanishing amplitude. This is not a QsiHT-specific edge: every direct method in the benchmark uses the same two-argument arctangent for its amplitude angles, so it is a shared implementation choice rather than a discriminator (\S\ref{sec:stability}), and on the benchmark targets no conditioning behavior surfaced as an accuracy difference, since every method is exact to machine precision (\S\ref{sec:exactness}). The circuit template is regular and data-independent, convenient for hardware compilation and for reuse across many target states.

Together, these place the two QsiHT fast-path syntheses as compact, well-conditioned loaders at the deployed multiplexor floor, the Real Synthesis for real dense targets with $R_y$ and CNOT gates alone and the Complex Synthesis for arbitrary targets, with Plesch--Brukner relevant where the target is low-rank or the register large enough for its bipartite constant to overtake the floor (the low-rank saving belongs to the rank-adaptive Plesch--Brukner variant of future work, not the non-truncating implementation benchmarked here).

\paragraph{Beyond single-state preparation.} The published QsiHT construction reaches past one-shot preparation, a property established in the prior co-authored work this thesis evaluates~\cite{gomez2025qsiht} rather than a benchmarked finding of this study. Because the DsiHT is generated by the data it processes, recomputing its angles from a pair of targets realizes the general state-to-state map $\ket{x}\to\ket{y}$ in a single disentangling pass, at the same $2^n-1$ single-angle rotations as plain preparation rather than the roughly doubled cost of a naive two-stage route that clears both endpoints through the reference state (\S\ref{sec:qsiht_s2s})~\cite{gomez2025qsiht}. Preparation from the ground state is recovered as the special case $\ket{x}=\ket{0}^{\otimes n}$. A two-generator extension, the Quantum Two Signal-induced Heap Transform (Q2siHT), initializes two distinct superpositions within a single circuit at roughly twice the single-state cost (\S\ref{sec:q2siht})~\cite{grigoryan2025q2siht}, and its circuit is reusable in a strong sense: once the circuit is fixed, the second prepared state can be switched by retuning a single rotation angle, so a family of related states is reachable by one-parameter edits rather than full recompilation. This thesis benchmarks only single-state preparation, so the state-to-state and two-state maps are reported here as capabilities of the construction, not as measured advantages over the other methods, which are not evaluated on those tasks. The versatility nonetheless motivates the heap-transform route as a reusable primitive for the fused load-and-transform pipelines noted in~\S\ref{sec:applications} and~\S\ref{sec:future}.

\section{Applications to Quantum Signal and Image Processing}
\label{sec:applications}

The motivation for exact dense-state preparation is that it is the data-loading bottleneck of quantum signal and image processing, where a classical signal or image must be embedded into a quantum register before any transform or learning routine can act on it~\cite{grigoryan2025qip}. Amplitude encoding, mapping an $N$-sample signal onto the $2^n$ amplitudes of an $n=\lceil\log_2 N\rceil$-qubit state, is maximally qubit-efficient but carries the $\Omega(2^n)$ CNOT cost quantified throughout this thesis (\S\ref{sec:nisq_tradeoffs}). Aaronson's fine-print caveat should frame all three patterns below: an efficient loader is a necessary condition for an end-to-end quantum advantage on classical data, never a sufficient one~\cite{aaronson2015read}. Choosing the cheapest loader decides how the data gets in, not whether the surrounding pipeline beats its classical counterpart. The findings here bear directly on three application patterns.

\paragraph{Signal loading for spectral pipelines.} When a loaded state is immediately consumed by a quantum transform, such as a Fourier, cosine, or slant transform for filtering or compression, the preparation cost adds directly to the algorithm's depth budget~\cite{grigoryan2024qft, grigoryan2025qct}. The results recommend the two QsiHT fast-path syntheses for generic dense signals, since they reach the deployed multiplexor floor of $2^n-n-1$ CNOTs on both target families and so minimize the entangling overhead that dominates fidelity, the Complex Synthesis for complex-valued signals and the Real Synthesis, which uses $R_y$ and CNOT gates alone, for real-valued ones, while Plesch--Brukner becomes preferable when the signal is low-rank or the register large enough for its bipartite constant to fall below the floor. They carry a further motivation because the QsiHT route is drawn from the same signal-induced heap-transform methodology as those spectral routines. This shared origin opens the possibility of a fused load-and-transform pipeline whose Givens structure carries through to the transform stage.

\paragraph{Image encoding for pattern recognition.} Image-processing tasks such as feature extraction and enhancement load larger registers, where the depth wall is most acute~\cite{grigoryan2025edge}. The cross-platform results (\S\ref{sec:nisq}) are directly relevant since the choice of device entangler can double the realized two-qubit count, so the method and the hardware must be co-selected. Low-Schmidt-rank images, those with strong spatial correlations, are especially favorable to the rank-adaptive Plesch--Brukner variant (future work, \S\ref{sec:future}), whose cost falls with the Schmidt rank, while the non-truncating implementation benchmarked here does not realize that saving.

\paragraph{Data loading for quantum machine learning.} In variational and kernel-based quantum machine learning, the same state is often prepared many times across optimization iterations and data points. That repetition amortizes less than it may appear to. Only the classical synthesis is reusable: the angles are computed once per target and replayed, which favors the QsiHT route's reusable, data-independent circuit template. The quantum cost does not amortize at all, since the entangling cascade re-executes on every shot of every circuit evaluation, so the $2^n-n-1$ two-qubit budget, on real or complex targets alike, is paid in full at each execution and remains the dominant recurring cost. For that recurring budget the two syntheses match the deployed Qiskit \texttt{StatePreparation} exactly, all three sitting at the flat merged-multiplexor floor of $2^n-n-1$ ($57$ at $n=6$) independent of the target's Schmidt rank (\S\ref{sec:scaling}), so the syntheses' draw in this setting is that their classical angle synthesis is computed once per target through a single fast Walsh--Hadamard transform and replayed, their circuit template is data-independent, and they belong to the same signal-induced heap-transform family as the downstream transforms (\S\ref{sec:applications}), a reusability, lighter-synthesis, and methodological fit rather than a per-shot entangling-count edge. The angle pipeline uses the same two-argument \texttt{arctan2} as the other direct methods, sharing their correct handling of vanishing amplitudes rather than improving on it, and the cascade stays structurally dense, emitting its worst-case CNOT budget on full-support targets rather than pruning by the feature vector's support (Section~\ref{sec:related_out_of_scope}).

\section{Future Work}
\label{sec:future}

The findings of this work lay the groundwork for future research topics.

\paragraph{Constant-optimal QSD and complex-phase merging.} The QSD implementation used here is the standard recursion. Incorporating the diagonal-merging and CZ optimizations that achieve the $\tfrac{23}{48}4^n$ constant of~\eqref{eq:qsd_cost} would sharpen the upper-baseline comparison. Likewise, merging the amplitude and phase reductions of the direct methods into uniformly controlled single-qubit gates would realize the full Iten constant for complex targets, tightening the frontier-method comparison.

\paragraph{Ancilla--depth trade-offs.} All methods compared here are ancilla-free and sit at the qubit-frugal, depth-heavy end of the design space, at depth $\Theta(2^n)$, a factor of $n$ above the ancilla-free optimum $\Theta(2^n/n)$ of Sun \emph{et al.}~\cite{sun2023asymptotically}. Benchmarking the depth-optimal ancilla-free constructions, and the $\Theta(n)$-depth constructions that spend $O(2^n)$ ancillae~\cite{sun2023asymptotically, araujo2021divide, rosenthal2021query}, with the same benchmark would map the other end of the trade-off and is increasingly relevant as device qubit counts grow.

\paragraph{Exploiting sparsity and low rank.} Dense preparation is the worst case. Extending the benchmark to sparse states would bring in the dedicated sparse algorithms~\cite{gleinig2021sparse, malvetti2021sparse}, whose cost scales with the support $s$ rather than $2^n$, and extending it to explicitly low-Schmidt-rank targets would quantify the savings available to rank-adaptive Schmidt constructions in the Plesch--Brukner family. Neither regime plays to the QsiHT cascade as it stands, since it emits its worst-case $2^n-n-1$ CNOT budget on full-support, full-rank targets and reads neither support nor rank to prune it (Section~\ref{sec:related_out_of_scope}). Whether a heap-transform path can be made support-aware is an open question.

\paragraph{Transferring the floor mechanisms to the general maps.} The two syntheses reach the multiplexor floor through two operator identities applied to the disentangling cascade: deferring each stage's trailing CNOT as a control-basis permutation, which the Real Synthesis absorbs into a reordered heap, and folding each stage's residual diagonal into the following stage, which the Complex Synthesis applies on arbitrary targets. Both are derived here for single-state preparation. Carrying the trailing-CNOT deferral and the diagonal absorption to the state-to-state map $\ket{x}\to\ket{y}$ and to the two-generator Q2siHT construction (\S\ref{sec:qsiht_s2s} and~\S\ref{sec:q2siht}) would test whether those general maps admit the same floor, and is a concrete next step.

\paragraph{Broader hardware execution and error mitigation.} The hardware study here was limited to $n=3$ on three superconducting IBM Heron processors, together with a single deep-circuit run at $n=8$, a readout-dominated regime at $n=3$ that can validate only the coarse frontier-versus-QSD separation (\S\ref{sec:hw_limitations}). The concrete follow-up plan is written out there: a layout-pinned rerun with dynamical decoupling and measurement-error mitigation on a fixed physical qubit triple, a size sweep at $n=4$--$6$ through the gate-versus-readout crossover, and sign-sensitive tomography to certify the real-target claim beyond distribution overlap. Beyond that plan, extending to trapped-ion and other architectures, with error-mitigation techniques~\cite{temme2017error} applied, would test how far the CNOT-driven ordering carries on physical hardware. A preliminary $n=8$ deep-circuit run already shows the executed-count ordering reemerging once the two-qubit budget dominates, but only where the sampled computational-basis distributions lie near the uniform-distribution reference. Because that metric is sign- and phase-blind, it does not establish full decoherence. The informative on-device regime remains the intermediate $n=4$--$6$ sweep of the follow-up plan above, and larger simulated qubit counts would confirm the asymptotic trends directly.

\paragraph{Approximate and application-specific loading.} Finally, relaxing exactness, trading a controlled infidelity $\varepsilon$ for shallower circuits, as in variational or matrix-product-state preparation, would extend the comparison beyond the exact-deterministic scope of this thesis and connect it to the approximate loaders increasingly used in near-term applications.

\section{Closing Remarks}
\label{sec:closing}

Dense quantum state preparation is bounded below by an exponential CNOT cost that no exact method can escape, and the principal exact methods all approach that bound to within a constant factor, except when, like the Quantum Shannon Decomposition, they solve a strictly harder problem. This thesis built a uniform, reproducible benchmark spanning the principal ancilla-free exact dense methods (the field as it stood by 2016), and used it both to confirm the established theory and to place a recently published method, the QsiHT with its fast-path syntheses, within it. The two syntheses emerge as frontier methods: exact, ancilla-free, and, in their Real and Complex forms, reaching the deployed multiplexor floor of $2^n-n-1$ CNOTs on real and on arbitrary targets respectively, the Real Synthesis with $R_y$ and CNOT gates alone, matching Qiskit's \texttt{StatePreparation} at every size and saturating the CNOT lower bound at $n=2$, while recording the lightest large-register build on the frontier, with the same \texttt{arctan2} conditioning as the other direct methods. Carrying the comparison to larger registers on hardware and into the sparse and approximate regimes is the work that this thesis makes ready.

%% file: chapters/chapter1.tex
\chapter{Quantum Computing Preliminaries}
\label{ch:intro}

This appendix contains the quantum computing background assumed throughout the thesis. It defines qubits, unitary gates, and circuit notation~\cite{nielsen_chuang}, works through the Bell state and the SWAP decomposition referenced in the main text, and establishes the visual conventions used in all circuit diagrams. Readers already familiar with the quantum circuit model may treat this material as a notation reference.

\section{Quantum Circuits and Operations}

At the core of modern quantum computation is the quantum circuit model. A quantum circuit is a computational routine consisting of coherent operations on quantum states, applied sequentially or concurrently, culminating in the extraction of classical information. The primary components of a quantum circuit are qubits, quantum gates, and measurements.

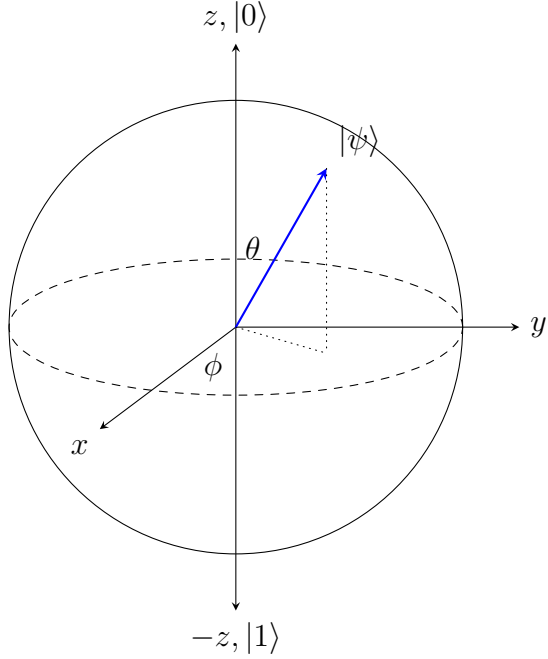
\begin{figure}[H]
\centering
\begin{tikzpicture}[line cap=round, line join=round, >=stealth, scale=1.5]
  \draw (0,0) circle (2cm);
  \draw [dash pattern=on 3pt off 3pt] (0,0) ellipse (2cm and 0.6cm);
  \draw [->] (0,0) -- (0,2.5) node[anchor=south] {$z, \ket{0}$};
  \draw [->] (0,0) -- (0,-2.5) node[anchor=north] {$-z, \ket{1}$};
  \draw [->] (0,0) -- (-1.2,-0.9) node[anchor=north east] {$x$};
  \draw [->] (0,0) -- (2.5,0) node[anchor=west] {$y$};
  \draw [->, thick, blue] (0,0) -- (0.8,1.4) node[anchor=south west, black] {$\ket{\psi}$};
  \draw [dotted] (0.8,1.4) -- (0.8,-0.23);
  \draw [dotted] (0,0) -- (0.8,-0.23);
  \draw (0.15,0.7) node {$\theta$};
  \draw (-0.2,-0.35) node {$\phi$};
\end{tikzpicture}
\caption{The Bloch sphere provides a geometric mapping of a pure single-qubit state space into three physical dimensions.}
\label{fig:bloch_sphere}
\end{figure}

A qubit is the fundamental unit of quantum information, mathematically described as a state vector $\ket{\psi}$ in a two-dimensional complex Hilbert space $\mathcal{H}$. In column vector representation, the computational basis states are defined as:
\begin{equation}
\ket{0} \equiv \begin{pmatrix} 1 \\ 0 \end{pmatrix}, \quad \ket{1} \equiv \begin{pmatrix} 0 \\ 1 \end{pmatrix}.
\end{equation}
Unlike a classical bit, a single qubit can exist in a linear combination, or superposition, of its computational basis states, $\ket{0}$ and $\ket{1}$, given by:
\begin{equation}
\ket{\psi} = \alpha\ket{0} + \beta\ket{1} = \cos\left(\frac{\theta}{2}\right)\ket{0} + e^{i\phi}\sin\left(\frac{\theta}{2}\right)\ket{1},
\label{eq:qubit}
\end{equation}
where $\alpha$ and $\beta$ are complex probability amplitudes such that $|\alpha|^2 + |\beta|^2 = 1$. The parameters $0 \le \theta \le \pi$ and $0 \le \phi < 2\pi$ define a point on the Bloch sphere, a geometric representation illustrated in Figure~\ref{fig:bloch_sphere}.

When scaled to an $n$-qubit register, the joint state is described by the tensor product space $\mathcal{H}^{\otimes n}$, and the state vector is given by:
\begin{equation}
\ket{\Psi} = \sum_{x=0}^{2^n-1} \alpha_x \ket{x}, \quad \text{with} \quad \sum_{x=0}^{2^n-1} |\alpha_x|^2 = 1,
\end{equation}
where each $\alpha_x$ is the complex amplitude of computational basis state $\ket{x}$ and the index $x$ runs over the integers $0$ to $2^n - 1$, spanning $2^n$ basis states $\ket{x_1 x_2 \dots x_n}$. The mathematical construction of these multi-qubit states relies on the Kronecker (tensor) product, denoted by $\otimes$. For instance, a two-qubit state is formed by taking the tensor product of two single-qubit column vectors:
\begin{equation}
\ket{0} \otimes \ket{1} \equiv \ket{01} = \begin{pmatrix} 1 \\ 0 \end{pmatrix} \otimes \begin{pmatrix} 0 \\ 1 \end{pmatrix} = \begin{pmatrix} 1 \cdot \begin{pmatrix} 0 \\ 1 \end{pmatrix} \\ 0 \cdot \begin{pmatrix} 0 \\ 1 \end{pmatrix} \end{pmatrix} = \begin{pmatrix} 0 \\ 1 \\ 0 \\ 0 \end{pmatrix}.
\end{equation}
Correspondingly, unitary operations applied to disjoint subsets of qubits combine through the same tensor product framework, producing $2^n \times 2^n$ matrices that act on the exponentially large joint state space.

Quantum operations, or quantum gates, take the form of reversible unitary matrices $U$ (where $UU^\dagger = I$, with $I$ the identity matrix) applied to qubits to manipulate their state. Single-qubit gates include the Pauli operators ($X, Y, Z$), which represent rotations of $\pi$ radians around the respective Bloch sphere axes:
\begin{equation}
X = \begin{pmatrix} 0 & 1 \\ 1 & 0 \end{pmatrix}, \quad Y = \begin{pmatrix} 0 & -i \\ i & 0 \end{pmatrix}, \quad Z = \begin{pmatrix} 1 & 0 \\ 0 & -1 \end{pmatrix}.
\end{equation}
Another ubiquitous single-qubit operation is the Hadamard ($H$) gate, which creates an equal superposition from a computational basis state by effectively rotating around the axis diagonal to $X$ and $Z$:
\begin{equation}
H = \frac{1}{\sqrt{2}} \begin{pmatrix} 1 & 1 \\ 1 & -1 \end{pmatrix}, \quad H\ket{0} = \frac{\ket{0} + \ket{1}}{\sqrt{2}} = \ket{+}.
\end{equation}
Multi-qubit operations, such as the Controlled-NOT (CNOT) gate, allow state manipulations conditioned on the state of a control qubit. Mathematically, the CNOT is represented by a $4 \times 4$ unitary matrix:
\begin{equation}
\text{CNOT} = \begin{pmatrix} 1 & 0 & 0 & 0 \\ 0 & 1 & 0 & 0 \\ 0 & 0 & 0 & 1 \\ 0 & 0 & 1 & 0 \end{pmatrix}.
\end{equation}
These multi-qubit gates are responsible for generating entanglement. Finally, non-unitary projective measurements project the final quantum state onto the computational basis, collapsing the superposition and yielding classical strings of 1s and 0s whose probabilities are governed by the Born rule, $p(x) = \lvert\langle x|\Psi\rangle\rvert^2$.

\subsection{Circuit Terminology and Visual Semantics}

To bridge the underlying mathematics with practical algorithm design, quantum operations are most commonly represented using quantum circuit diagrams. For a new reader, understanding the visual semantics of these diagrams is essential for analyzing quantum algorithms. Figures~\ref{fig:qubit-wire}, \ref{fig:unitary-gates}, \ref{fig:controlled-ops}, and~\ref{fig:measurement} illustrate each element:

\begin{itemize}
    \item \textbf{Qubit Wires:} Horizontal lines in a circuit represent individual qubits. Time flows from left to right across the diagram. The far left represents the initial state of the system (typically initialized to the ground state $\ket{0}$), and as one moves to the right across the wire, successive operations are applied over time.
    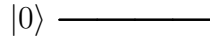
\begin{figure}[H]
    \centering
        \begin{quantikz}
        \lstick{$\ket{0}$} & \qw & \qw & \qw & \qw
        \end{quantikz}
    \caption{A single qubit wire demonstrating the passage of time from left to right.}
    \label{fig:qubit-wire}
    \end{figure}

    \item \textbf{Unitary Gates:} Single-qubit operations are depicted as solid boxes placed directly on a wire. These boxes contain a symbol or mathematical expression indicating the specific gate applied (e.g., an $H$ for a Hadamard gate, or $R_y(\theta)$ for a parameterized rotation about the Y-axis).
    \begin{figure}[H]
    \centering
        \begin{quantikz}
        \lstick{$\ket{\psi}$} & \gate{H} & \gate{R_y(\theta)} & \qw
        \end{quantikz}
    \caption{Unitary gates applied sequentially to a single qubit.}
    \label{fig:unitary-gates}
    \end{figure}
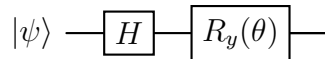

    \item \textbf{Controlled Operations:} Multi-qubit gates use vertical connecting lines to denote interactions between distinct qubits. A solid black dot ($\bullet$) indicates a ``control'' qubit, meaning the subsequent operation is only triggered if this control qubit is in the $\ket{1}$ state. The target of the controlled operation is denoted by a specific symbol, such as a crossed circle ($\oplus$), which generally represents a bit-flip (Pauli-X) operation applied to a target qubit.
    \begin{figure}[H]
    \centering
        \begin{quantikz}
        \lstick{Control ($q_0$)} & \ctrl{1} & \qw & \ctrl{1} & \qw \\
        \lstick{Target ($q_1$)}  & \targ{}  & \qw & \gate{U} & \qw
        \end{quantikz}
    \caption{Examples of controlled operations: a CNOT gate (left) and a generic Controlled-U gate (right).}
    \label{fig:controlled-ops}
    \end{figure}
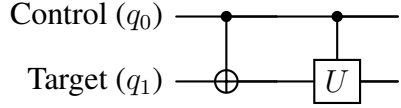

    \item \textbf{Measurements:} A gauge or meter symbol placed on a qubit wire signifies a measurement. This operation collapses the quantum superposition into a deterministic classical state. Mathematically, a measurement is defined by a set of measurement operators $\{M_m\}$, satisfying the completeness relation $\sum_m M_m^\dagger M_m = I$. For a system in state $\ket{\psi}$, the probability of outcome $m$ is given by $p(m) = \bra{\psi} M_m^\dagger M_m \ket{\psi}$. The standard computational basis measurement employs orthogonal projectors $M_0 = \ket{0}\bra{0}$ and $M_1 = \ket{1}\bra{1}$. In many circuit representations, the resulting classical information is subsequently carried forward on a ``double line,'' denoting a classical bit rather than a quantum wire.
    \begin{figure}[H]
    \centering
        \begin{quantikz}
        \lstick{$\alpha\ket{0} + \beta\ket{1}$} & \meter{} & \cw
        \end{quantikz}
    \caption{A measurement operation collapsing a qubit state into a classical bit (denoted by the double wire).}
    \label{fig:measurement}
    \end{figure}
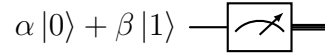
\end{itemize}

\section{Fundamental Quantum Circuits}

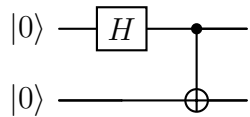
\begin{figure}[H]
\centering
    \begin{quantikz}
    \lstick{$\ket{0}$} & \gate{H} & \ctrl{1} & \qw \\
    \lstick{$\ket{0}$} & \qw      & \targ{}  & \qw
    \end{quantikz}

\caption{Quantum circuit generating the maximally entangled Bell state $\ket{\Phi^+}$. \label{fig:bell-state}}
\end{figure}

To illustrate the mechanics described above, consider the preparation of a maximally entangled bipartite state, known as the Bell state $\ket{\Phi^+} = \frac{1}{\sqrt{2}}(\ket{00} + \ket{11})$. The corresponding circuit applies a Hadamard gate to create superposition, followed by a CNOT gate to entangle the two qubits (Figure~\ref{fig:bell-state}).

The evolution of the system's state vector can be tracked step by step. The system is initialized in the joint state $\ket{\psi_0} = \ket{0} \otimes \ket{0} = \ket{00}$. Applying the Hadamard gate to the first qubit and the identity operation to the second yields the intermediate state $\ket{\psi_1}$:
\begin{equation}
\ket{\psi_1} = (H \otimes I) \ket{00} = \left[ \frac{1}{\sqrt{2}} (\ket{0} + \ket{1}) \right] \otimes \ket{0} = \frac{1}{\sqrt{2}} (\ket{00} + \ket{10}).
\label{eq:bell_step1}
\end{equation}
Next, the CNOT gate is applied, using the first qubit as the control and the second as the target. By definition, CNOT flips the target if and only if the control is $\ket{1}$. Distributing this linear operator over the superposition gives the final Bell state:
\begin{equation}
\ket{\psi_2} = \text{CNOT} \left[ \frac{1}{\sqrt{2}} (\ket{00} + \ket{10}) \right] = \frac{1}{\sqrt{2}} (\ket{00} + \ket{11}) \equiv \ket{\Phi^+}.
\label{eq:bell_step2}
\end{equation}

\begin{figure}[H]
\centering
    \begin{quantikz}
    \lstick{$\ket{\psi_a}$} & \swap{1} & \qw \\
    \lstick{$\ket{\psi_b}$} & \targX{} & \qw
    \end{quantikz}
    \quad = \quad
    \begin{quantikz}
    & \ctrl{1} & \targ{}   & \ctrl{1} & \qw \\
    & \targ{}  & \ctrl{-1} & \targ{}  & \qw
    \end{quantikz}
\caption{The standard SWAP gate and its equivalent logical decomposition using three CNOT gates.}
\label{fig:swap-circuit}
\end{figure}
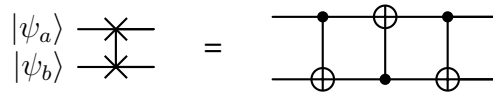

Another indispensable circuit abstraction is the SWAP gate, which exchanges the states of two qubits such that $\text{SWAP}\ket{\psi_a, \psi_b} = \ket{\psi_b, \psi_a}$. On hardware architectures lacking native SWAP operations, it is decomposed into a sequence of three alternating CNOT gates (Figure~\ref{fig:swap-circuit}).

Mathematically, the logical exchange can be verified by routing an arbitrary two-qubit computational basis state $\ket{x, y}$ (where $x, y \in \{0, 1\}$) through the three CNOTs. With modulo-2 addition ($\oplus$) denoting the target bit flips, and with $\text{CNOT}_{i \to j}$ denoting a CNOT with control qubit $i$ and target qubit $j$, the state evolves as:
\begin{equation}
\ket{x, y} \xrightarrow{\text{CNOT}_{0 \to 1}} \ket{x, x \oplus y} \xrightarrow{\text{CNOT}_{1 \to 0}} \ket{x \oplus (x \oplus y), x \oplus y} = \ket{y, x \oplus y},
\end{equation}
and applying the final CNOT recovers the exchanged variables:
\begin{equation}
\ket{y, x \oplus y} \xrightarrow{\text{CNOT}_{0 \to 1}} \ket{y, (x \oplus y) \oplus y} = \ket{y, x}.
\end{equation}
This decomposition matters for routing qubits across physical hardware topologies with limited nearest-neighbor connectivity.

%% file: chapters/appendix_gates.tex
\chapter{Quantum Gate Reference}
\label{ch:gates}

This appendix collects, in one place, the quantum gates used throughout the thesis together with their matrix definitions. The conceptual introduction to qubits, unitaries, and circuit notation is given in Appendix~\ref{ch:intro}. The conventions here match that appendix: each matrix acts on the computational basis ordered $\ket{00}, \ket{01}, \ket{10}, \ket{11}$ with the first (top) qubit most significant, angles follow the generator convention $R(\theta) = \exp(-i\theta G/2)$, and a two-qubit controlled gate takes the first listed qubit as its control unless stated otherwise~\cite{nielsen_chuang, qiskit2024}.

\section{Single-Qubit Gates}

The Pauli gates $X$, $Y$, $Z$ and the Hadamard gate $H$ are introduced in Appendix~\ref{ch:intro} and restated here for completeness:
\begin{equation}
X = \begin{pmatrix} 0 & 1 \\ 1 & 0 \end{pmatrix}, \quad
Y = \begin{pmatrix} 0 & -i \\ i & 0 \end{pmatrix}, \quad
Z = \begin{pmatrix} 1 & 0 \\ 0 & -1 \end{pmatrix}, \quad
H = \frac{1}{\sqrt{2}}\begin{pmatrix} 1 & 1 \\ 1 & -1 \end{pmatrix}.
\end{equation}
The parameterized rotation gates exponentiate the Pauli generators, with $R_a(\theta) = \exp(-i\theta\, a/2)$ for each $a \in \{X, Y, Z\}$. They appear in every state-preparation circuit here, since a uniformly controlled rotation reduces to a cascade of them:
\begin{equation}
R_x(\theta) = \begin{pmatrix} \cos\frac{\theta}{2} & -i\sin\frac{\theta}{2} \\[2pt] -i\sin\frac{\theta}{2} & \cos\frac{\theta}{2} \end{pmatrix}, \;
R_y(\theta) = \begin{pmatrix} \cos\frac{\theta}{2} & -\sin\frac{\theta}{2} \\[2pt] \sin\frac{\theta}{2} & \cos\frac{\theta}{2} \end{pmatrix}, \;
R_z(\theta) = \begin{pmatrix} e^{-i\theta/2} & 0 \\ 0 & e^{i\theta/2} \end{pmatrix}.
\label{eq:rotations}
\end{equation}
The real $R_y$ rotation carries the amplitude (magnitude) information of a target state, while $R_z$ imprints relative phases. This is the amplitude/phase split used throughout Chapter~\ref{ch:litreview}.

The phase gate $P(\lambda)$ applies a relative phase to $\ket{1}$ and equals $R_z$ up to a global phase, $P(\lambda) = e^{i\lambda/2} R_z(\lambda)$:
\begin{equation}
P(\lambda) = \begin{pmatrix} 1 & 0 \\ 0 & e^{i\lambda} \end{pmatrix}, \qquad S = P\!\left(\tfrac{\pi}{2}\right), \qquad T = P\!\left(\tfrac{\pi}{4}\right).
\end{equation}
The $\sqrt{X}$ gate ($\textsc{sx}$), a native single-qubit gate on several platforms in Chapter~\ref{ch:empirical}, is a half $X$ rotation, equal to $R_x(\pi/2)$ up to a global phase and satisfying $\textsc{sx}^2 = X$:
\begin{equation}
\textsc{sx} = \frac{1}{2}\begin{pmatrix} 1+i & 1-i \\ 1-i & 1+i \end{pmatrix}.
\end{equation}
The most general single-qubit gate is the three-parameter unitary $U(\theta, \phi, \lambda)$, which the abstract $\{u, \textsc{cx}\}$ basis of Chapter~\ref{ch:empirical} uses to absorb every single-qubit operation into one box:
\begin{equation}
U(\theta, \phi, \lambda) = \begin{pmatrix} \cos\frac{\theta}{2} & -e^{i\lambda}\sin\frac{\theta}{2} \\[2pt] e^{i\phi}\sin\frac{\theta}{2} & e^{i(\phi+\lambda)}\cos\frac{\theta}{2} \end{pmatrix}.
\end{equation}
Every gate above is a special case: $R_z(\theta) = U(0,0,\theta)$ up to a global phase, $R_y(\theta) = U(\theta, 0, 0)$, and $H = U(\pi/2, 0, \pi)$.

\section{Two-Qubit Gates}

The controlled-NOT (CNOT, also written CX) flips the target qubit when the control is $\ket{1}$. It is the canonical entangling gate and the primary cost metric of this thesis. The controlled-$Z$ (CZ) gate applies a phase of $-1$ to $\ket{11}$ and is symmetric in its two qubits. With the first qubit as control,
\begin{equation}
\text{CNOT} = \text{CX} = \begin{pmatrix} 1 & 0 & 0 & 0 \\ 0 & 1 & 0 & 0 \\ 0 & 0 & 0 & 1 \\ 0 & 0 & 1 & 0 \end{pmatrix}, \qquad
\text{CZ} = \begin{pmatrix} 1 & 0 & 0 & 0 \\ 0 & 1 & 0 & 0 \\ 0 & 0 & 1 & 0 \\ 0 & 0 & 0 & -1 \end{pmatrix}.
\end{equation}
The SWAP gate exchanges the two qubit states, and the imaginary SWAP (iSWAP) does the same while adding a phase of $i$ to the swapped amplitudes:
\begin{equation}
\text{SWAP} = \begin{pmatrix} 1 & 0 & 0 & 0 \\ 0 & 0 & 1 & 0 \\ 0 & 1 & 0 & 0 \\ 0 & 0 & 0 & 1 \end{pmatrix}, \qquad
\text{iSWAP} = \begin{pmatrix} 1 & 0 & 0 & 0 \\ 0 & 0 & i & 0 \\ 0 & i & 0 & 0 \\ 0 & 0 & 0 & 1 \end{pmatrix}.
\end{equation}
The remaining two-qubit gates are the hardware-native entanglers of the cross-platform comparison in Chapter~\ref{ch:empirical} (Table~\ref{tab:nisq_platforms}). The echoed cross-resonance (ECR) gate is the native two-qubit interaction on several IBM devices:
\begin{equation}
\text{ECR} = \frac{1}{\sqrt{2}} \begin{pmatrix} 0 & 0 & 1 & i \\ 0 & 0 & i & 1 \\ 1 & -i & 0 & 0 \\ -i & 1 & 0 & 0 \end{pmatrix}.
\end{equation}
Trapped-ion platforms expose parameterized Ising-type entanglers: the M{\o}lmer--S{\o}rensen $XX$ interaction $R_{xx}(\theta) = \exp(-i\theta\, X\!\otimes\! X/2)$ (IonQ) and the $ZZ$-phase gate $R_{zz}(\theta) = \exp(-i\theta\, Z\!\otimes\! Z/2)$ (Quantinuum),
\begin{equation}
R_{xx}(\theta) = \begin{pmatrix} c & 0 & 0 & -is \\ 0 & c & -is & 0 \\ 0 & -is & c & 0 \\ -is & 0 & 0 & c \end{pmatrix}, \qquad
R_{zz}(\theta) = \begin{pmatrix} e^{-i\theta/2} & 0 & 0 & 0 \\ 0 & e^{i\theta/2} & 0 & 0 \\ 0 & 0 & e^{i\theta/2} & 0 \\ 0 & 0 & 0 & e^{-i\theta/2} \end{pmatrix},
\end{equation}
with $c = \cos(\theta/2)$ and $s = \sin(\theta/2)$. A generic controlled-unitary, drawn with a control dot joined to a gate box in the circuit diagrams (Appendix~\ref{ch:intro}), applies an arbitrary single-qubit $U$ to the target conditioned on the control, and reduces to the CNOT when $U = X$.